\documentclass[11pt,a4paper]{article}
\usepackage[pagella]{preprint-style}
\usepackage{xcolor}
\definecolor{red}{RGB}{190,0,0}
\colorlet{manuscriptcaptioncolor}{black}
\AddToHook{cmd/caption/before}{\colorlet{manuscriptcaptioncolor}{.}}
\AddToHook{env/longtable/begin}{\colorlet{manuscriptcaptioncolor}{.}}
\DeclareCaptionFont{inheritcolor}{\color{manuscriptcaptioncolor}}
\usepackage{amsthm,changepage,needspace}
\usepackage{tcolorbox}
\tcbuselibrary{breakable,skins}
\newtcolorbox[auto counter,number within=section]{manuscriptalgorithm}[2]{enhanced,breakable,
  colback=white,colframe=black,frame hidden,boxrule=0pt,arc=0pt,left=5pt,right=5pt,
  borderline north={.3pt}{0pt}{black},borderline south={.3pt}{0pt}{black},
  fonttitle={\bfseries\raggedright\hyphenpenalty=10000\relax},colbacktitle=white,coltitle=black,
  titlerule=.3pt,titlerule style={black},
  title={Algorithm~\thetcbcounter\ #1},label={alg:#2},
  title after break={Algorithm~\thetcbcounter},
  before skip=\bigskipamount,after skip=\medskipamount}
\newtheoremstyle{authorstatement}{\medskipamount}{\medskipamount}{\itshape}{}{\bfseries}{.}{.5em}{\thmname{#1}\thmnumber{ #2}\thmnote{ \normalfont(#3)}}
\theoremstyle{authorstatement}
\newtheorem{assumption}{Assumption}[section]
\newtheorem{theorem}{Theorem}[section]
\newtheorem{lemma}{Lemma}[section]
\newtheorem{proposition}{Proposition}[section]
\newtheorem{corollary}{Corollary}[section]
\newtheorem{definition}{Definition}[section]
\newenvironment{wrappedtheorem}[1][]{%
  \refstepcounter{theorem}%
  \setlength{\topsep}{\medskipamount}%
  \setlength{\partopsep}{0pt}%
  \emergencystretch=2em
  \trivlist\item\relax
  {\bfseries Theorem~\thetheorem\ {\normalfont(#1)}.}\hspace{.5em}%
  \itshape\ignorespaces
}{\endtrivlist}
\theoremstyle{definition}
\newtheorem{remark}{Remark}[section]
\numberwithin{equation}{section}
\numberwithin{figure}{section}
\numberwithin{table}{section}
\usepackage{newunicodechar}
\newcommand{\manuscriptauthor}[1]{#1}
\newcommand{\manuscriptaffiliation}[1]{#1}
\usepackage[a4paper,margin=25mm]{geometry}
\newunicodechar{✓}{\ensuremath{\checkmark}}
\newunicodechar{β}{\ensuremath{\beta}}
\usepackage{graphicx,booktabs,longtable,array,calc,float,placeins}


\usepackage{titletoc}
\usepackage{xurl,hyperref}
\titlecontents{section}[0pt]{\addvspace{.8em}\bfseries}
  {Appendix \thecontentslabel\quad}{}{\titlerule*[.5pc]{.}\contentspage}
\titlecontents{subsection}[1.5em]{}
  {\thecontentslabel\quad}{}{\titlerule*[.5pc]{.}\contentspage}

\hypersetup{pdftitle={Adaptive experiments with observational data: Efficient experimental design and asymptotic time-uniform inference for long-term treatment effects},pdfauthor={Kentaro Kawato}}
\newcommand{\tightlist}{\setlength{\itemsep}{0pt}\setlength{\parskip}{0pt}}

\renewcommand{\thesubparagraph}{\theparagraph.\arabic{subparagraph}}
\titleformat{\paragraph}[block]{\normalfont\normalsize\itshape}{\theparagraph}{0.8em}{}
\titleformat{\subparagraph}[block]{\normalfont\normalsize\itshape}{\thesubparagraph}{0.8em}{}
\titlespacing*{\paragraph}{0pt}{1.6ex}{0.8ex}
\titlespacing*{\subparagraph}{0pt}{1.6ex}{0.8ex}
\begin{document}
\begin{frontmatter}
\title{Adaptive experiments with observational data: Efficient experimental design and asymptotic time-uniform inference for long-term treatment effects\tnoteref{authornote}}
\tnotetext[authornote]{Date: \today. I am grateful to Ryo Okui for his continued guidance and support. All errors are my own.}
\author{\manuscriptauthor{Kentaro Kawato}\fnref{contact}}
\address{\manuscriptaffiliation{Graduate School of Economics, The University of Tokyo}}
\fntext[contact]{\raggedright Graduate School of Economics, The University of Tokyo, 7-3-1 Hongo, Bunkyo-ku, Tokyo 113-0033, Japan.\\ \mbox{Email address: \textnormal{\href{mailto:kawato-kentaro380@g.ecc.u-tokyo.ac.jp}{kawato-kentaro380@g.ecc.u-tokyo.ac.jp}}}}
\begin{abstract}
Estimating long-term treatment effects in experiments requires waiting for outcomes to become observable, which can delay both the evaluation of an intervention and subsequent decisions. Recent econometric work has developed identification methods that combine experimental data with observational data containing long-term outcomes. This paper proposes using observational data available before the experiment begins not only for identification but also for designing adaptive experiments, in which experimental data are collected sequentially. To estimate the long-term average treatment effect as precisely as possible, we express its semiparametric efficiency bound as a function of the covariate sampling and treatment assignment rules and minimize the bound over these rules to characterize an optimal fixed experimental design. This experimental design depends on the treatment-specific conditional variances of the observational prediction of the long-term outcome, which can be learned only from experimental participants. We therefore construct a sequential procedure that learns the experimental design as experimental data accumulate and show that its estimator attains the same limiting variance as an oracle estimator implemented under the optimal fixed experimental design. We further construct an asymptotic confidence sequence whose coverage holds simultaneously over all rounds after a prespecified start, so that the experimenter can inspect the interval after each participant and stop once it excludes zero. Monte Carlo simulations show that the learned experimental design can improve precision and shorten the time to detection relative to baseline sampling and equal-probability assignment. An illustration using the Pakistan Graduation experiment shows how estimated conditional variance heterogeneity informs sampling and treatment assignment.
\end{abstract}
\end{frontmatter}
\section{Introduction}\label{section-1-introduction}

The time required to observe long-term outcomes in experiments can delay both the evaluation of an intervention and subsequent decisions. Consider a marketer offering coupons to new customers. The marketer may care not only about whether coupons increase purchases immediately, but also about whether they lead to a sustained increase in purchasing over time \citep{Huang2024}. Directly estimating these long-term effects requires following experimental participants over a much longer period than is needed to observe their initial purchases. By contrast, short-term outcomes such as purchase frequency, spending, and the composition of purchases shortly after the intervention can be observed relatively quickly. If observational data such as historical customer records link these early behaviors to subsequent purchasing, short-term information from a new experiment may help us learn about its long-term effects. This data combination idea extends beyond marketing to education, labor, and development, where intermediate outcomes can provide information about the eventual consequences of an intervention.

This paper asks how the experiment generating the short-term data should be designed when observational data are available before it begins. The question arises whenever a new experiment is planned to learn a long-term effect, because which participants are recruited and how treatment is assigned determine how precisely that effect can be estimated from the short-term data. The surrogate index approach of \citet{Athey2026} provides the basis for our analysis. Under suitable identification assumptions, it combines experimental data on how treatment changes short-term outcomes with observational data on the relationship between short- and long-term outcomes. Following \citet{Athey2026}, we call the conditional mean of the long-term outcome given short-term surrogates and pretreatment covariates in the observational population the surrogate index. A growing literature studies identification and estimation in such data combination settings \citep{Chen2023,Athey2025,Ghassami2025,Imbens2025,Kallus2025,Park2024,Park2026}. Their identification and efficiency results are the building blocks of our analysis, which uses them to design the experiment itself.

Learning and implementing this experimental design is a sequential problem. As we show below, the efficient experimental design depends on how much the surrogate index varies within each treatment arm conditional on covariates. Because the observational data do not record treatment, these conditional variances can be learned only from the participants of the experiment. The experimental design therefore draws on both data sources. The observational sample, available before the experiment, is used to estimate the target covariate distribution and the surrogate index, whereas the conditional variances are learned from earlier participants. At the same time, participants in such experiments are typically recruited over time and their short-term outcomes are observed quickly. The data from earlier participants can therefore be used to choose whom to sample and how to assign treatment next, and the estimate of the long-term effect can be updated and examined while the experiment is running. Our three contributions address this problem in turn. We characterize the experimental design to be learned, construct an estimator that remains valid when the experimental design responds to earlier data, and construct a confidence sequence that remains valid when it is examined after every round and used to stop the experiment.

Our first contribution is the optimal experimental design for estimating the long-term average treatment effect in the observational population. The experimental design consists of the covariate distribution from which experimental participants are sampled and the covariate-dependent treatment assignment probabilities, and we seek the experimental design under which this effect can be estimated with the smallest variance. As the experimental design criterion, we use the semiparametric efficiency bound, the smallest asymptotic variance that a regular estimator can attain. \citet{Chen2023} derive this bound in the data combination setting described above, with the two samples pooled. We write it for given experimental and observational sample sizes as a function of the experimental design, so that it separates into an experimental component that depends on the experimental design and an observational component that does not, and we minimize the experimental component over experimental designs. The experimental design that minimizes the bound, combined with an estimator that attains it, achieves the smallest asymptotic variance among all fixed experimental designs and regular estimators, and the sequential estimator we construct below attains the minimized bound while learning the experimental design. The resulting optimal experimental design depends on the covariate distribution in the target population and on the treatment- and control-specific conditional standard deviations of the surrogate index. The optimal experimental design recruits more participants from covariate regions that are common in the target population and where the surrogate index is more variable within the treatment arms. Conditional on covariates, the optimal assignment rule allocates a larger share of participants to the treatment arm with the larger conditional standard deviation. This assignment rule is a conditional Neyman allocation applied to the surrogate index \citep{Neyman1934,Hahn2011,Kato2024,Kato2025}.

Designing around the surrogate index rather than an individual short-term outcome has a direct practical implication. A large marketing literature distinguishes immediate responses to temporary price promotions, such as purchase acceleration and stockpiling, from longer-run changes in consumer purchasing behavior \citep[see][for a recent review]{Anderson2019}. In the coupon example, suppose that, among customers with high prior usage, a coupon makes immediate purchase quantities more variable and subsequent purchasing less variable than without a coupon, and that the opposite holds among customers with low prior usage, whose initial purchases under the coupon are more homogeneous and whose repeat purchasing at the regular price is more heterogeneous. Suppose also that the surrogate index for long-term purchasing depends mainly on subsequent purchasing. An experimental design that targets the conditional variance of immediate purchases then assigns a larger share of high-usage customers to the coupon, whereas an experimental design that targets the conditional variance of the surrogate index assigns a smaller share, and the reverse holds for low-usage customers. The two experimental designs generally also recruit the two customer types in different proportions. Directly optimizing a convenient short-term endpoint therefore need not yield an efficient experimental design for estimating the long-term effect.

Our second contribution is a sequential estimator that learns and implements this experimental design while the experiment is running. Before the experiment begins, we use the observational sample to estimate the target covariate distribution and the surrogate index, and these fits are used unchanged in every round. Before each experimental round, we use the accumulated experimental data to estimate the treatment-specific conditional means and variances of the fitted surrogate index, plug the variance estimates into the optimal sampling and treatment assignment rules, and compute the score of the new experimental observation using only quantities determined before it is collected. We combine these predictable experimental scores with an augmented correction evaluated on the observational sample, producing an estimate after every round. This predictability makes the leading experimental error a martingale average, so that a martingale central limit theorem applies even though the experimental design changes from round to round, while the leading observational error is an average over the observational sample. The same observational sample, however, also enters the experimental design and every experimental score, so the two errors are not computed from independent data. Because the observational sample is complete before the first round, we place its records at the start of the history and apply a single martingale central limit theorem to both errors. The two variance contributions add in the limit because the experimental increments are centered given all observational records and their predictable variance converges to a constant. When the observational and experimental sample sizes grow at comparable rates, the estimator is asymptotically normal and has the same limiting variance as an oracle procedure that knew the optimal fixed experimental design in advance.

Our third contribution is asymptotically time-uniform inference for the long-term treatment effect under this adaptive procedure. Sampling additional participants, implementing the intervention, and measuring outcomes all entail costs, so an experimenter may wish to stop once the available evidence is sufficiently informative. Repeatedly inspecting confidence intervals designed for a fixed time, however, does not generally preserve coverage over the monitoring period. We use the literature on anytime-valid inference and asymptotically time-uniform confidence sequences \citep{Johari2022,Ramdas2023,Ramdas2025,WaudbySmith2024} to construct an asymptotically time-uniform confidence sequence for the long-term average treatment effect that accommodates adaptive sampling and treatment assignment in this data combination setting.

A key complication in our setting is that the observational sample is collected before the experiment and reused at every monitoring time. Its leading estimation error is therefore shared across the sequence of estimates, whereas the experimental component of uncertainty evolves as new observations arrive. Asymptotically, however, the estimation errors over the monitoring period, each divided by its variance, behave like a single Brownian motion evaluated at the reciprocals of these variances. Our intervals therefore inflate the estimated standard error, which combines the experimental and observational variances, by a factor derived from Ville\textquotesingle s inequality for nonnegative supermartingales \citep{Ville1939,Howard2020,Howard2021}. A tuning variable chosen once at the monitoring start determines the round at which this factor is smallest, and the coverage guarantee holds for every tuning variable that satisfies our conditions. Our asymptotic analysis retains both sources of uncertainty as the observational sample size and the monitoring start time increase. Under the stated conditions, the probability that the intervals contain the true effect at every monitoring time from the prescribed start onward has an asymptotic lower limit of at least the nominal confidence level. This guarantee extends to data-dependent stopping times, including rules that stop when the interval first excludes a prespecified benchmark.

We complement the theoretical analysis with a Monte Carlo study calibrated to the coupon example. The calibration distinguishes immediate purchases, subsequent short-term repeat purchases, and long-term cumulative purchases and is constructed so that the immediate-purchase and surrogate-index criteria adjust treatment assignment in opposite directions within the same prior-usage strata. In this calibration, the experimental design that adapts both sampling and treatment assignment to the surrogate index reduces the Monte Carlo variance of the long-term effect estimator by 13.1 percent relative to a baseline experimental design that samples from the estimated target covariate distribution and assigns treatment with probability 1/2, whereas the corresponding experimental design based on immediate purchases increases it by 9.8 percent, although immediate purchases remain part of the surrogate vector used to estimate the effect. With the 90 percent time-uniform interval, the surrogate-index experimental design also raises the detection rate from 95.4 to 98.2 percent and reduces the mean stopping sample size from 2,062.4 to 1,846.1 participants.

We also provide a retrospective experimental design illustration using data from the Pakistan randomized evaluation of the Graduation program reported by \citet{Banerjee2015}, a multifaceted intervention for households in extreme poverty. Following the data construction of \citet{Chen2023}, we treat outcomes measured after two years as surrogates and monthly consumption per capita after three years as the long-term outcome. Because the experiment has already been completed, we repeatedly divide the realized experimental sample into two nonoverlapping, treatment-stratified samples and mask variables to reproduce the observational and experimental data structures of our framework. In each of 1,000 such splits, we estimate, for each of the 20 pretreatment variables, the experimental design whose assignment probability and sampling density ratio are constant within the quintile strata of that covariate. On average across the splits, the household asset index gives the largest reduction of the estimated experimental variance criterion, 8.23 percent relative to sampling from the target covariate distribution with equal-probability treatment assignment. The estimated experimental design for this covariate differs from this baseline mainly in sampling and recruits more households from the top stratum of the index, while the estimated experimental design for monthly nonfood consumption per capita differs mainly in treatment assignment.

\subsection{Related literature}\label{related-literature}

Our paper relates to three strands of literature. The first concerns causal inference through data combination involving experimental and observational data. The use of short-term endpoints has a long history in statistics for example \citet{Prentice1989} and \citet{Frangakis2002}. Our direct starting points are the surrogate index of \citet{Athey2026} and the semiparametric efficiency analysis of \citet{Chen2023}. We build on this identification and efficiency literature and extend the analysis to experimental design and sequential estimation.

The second strand concerns adaptive experimental design. The allocation principle of \citet{Neyman1934} has received attention in recent econometric research. \citet{Hahn2011} use Neyman allocation in the two-wave setting and improve treatment assignment for average treatment effect estimation. \citet{Kato2025} and \citet{Cook2024} study efficient estimation in multiple stage adaptive experiments.

The closest counterpart in experimental design is \citet{Kato2024}, who jointly choose the sampling distribution and treatment assignment probabilities when the outcome of interest is directly observed in the experiment. We share this joint optimization structure but combine observational and experimental data in a setting where the long-term outcome is not observed in the experiment. Consequently, the uncertainty in the observational data, which are used for both experimental design and estimation, enters the analysis.

Related work also incorporates surrogates into adaptive experimentation \citep{Anderer2022,Shi2024,Fan2025}. In these settings, surrogates supplement information about primary outcomes that are observed or arrive with delay. In our setting, the long-term outcome is not observed in the experiment, so the surrogates are needed to identify long-term treatment effects.

The third strand concerns time-uniform inference. \citet{Howard2020,Howard2021} develop time-uniform concentration bounds and confidence sequences based on nonnegative supermartingales. \citet{WaudbySmith2024} establish a time-uniform central limit theory and asymptotic confidence sequences \citep{WaudbySmith2024}, including applications with estimated nuisance functions. Applications to causal inference include \citet{Cook2024} and \citet{Dalal2024}. Our contribution is to accommodate an adaptive estimator combining observational and experimental data. The asymptotic confidence sequence controls accumulating experimental variation and shared observational variation according to their different structures.

\subsection{Paper organization}\label{paper-organization}

The remainder of the paper is organized as follows. Section 2 introduces the statistical surrogacy model and its identification result. Section 3 writes the efficiency bound as a function of the experimental design and derives the optimal experimental design. Section 4 establishes identification under adaptive experimental design, and Section 5 constructs the sequential estimator. Section 6 establishes asymptotic normality and attainment of the optimal fixed-design bound, and Section 7 develops asymptotic time-uniform inference. Sections 8 and 9 present the simulations and the experimental design illustration using the Graduation experiment, respectively. Section 10 concludes. The appendices contain the proofs and additional results, and Appendix D describes the supplementary material.

\section{Setting and identification}\label{section-2-setting-and-identification}

This section introduces the statistical surrogacy model and shows how it identifies the target parameter from the two data sources, which is the starting point for the efficiency bound in Section 3 and for the adaptive experimental design and the estimator in Sections 4 and 5. Section 2.1 describes the data, the target parameter, and the model, and Section 2.2 states the identifying assumptions and the resulting representations of the target.

\subsection{Statistical surrogacy model and target parameter}\label{statistical-surrogacy-model-and-target-parameter}

{
\emergencystretch=2em
This subsection describes the two samples, the target parameter, and the statistical surrogacy model that connects them. Let \(G_i\in\{\mathrm E,\mathrm O\}\) denote the data-source indicator, with \(\mathrm E\) and \(\mathrm O\) corresponding to the experimental and observational samples. \(X_i\) denotes the vector of baseline covariates, \(W_i\in\{0,1\}\) is treatment, \(S_i(w)\) is the potential short-term surrogate under treatment assignment \(w\), and \(Y_i(w)\) is the corresponding potential long-term outcome. The realized variables are denoted by \(S_i\) and \(Y_i\). The observational data contain \((X,S,Y)\) and the experimental data contain \((X,W,S)\), as shown in Table 2.1.
\par
}

This is the data structure of the statistical surrogacy model of \citet{Chen2023} and \citet{Athey2026}. Neither sample records treatment, the surrogate, and the long-term outcome jointly. The observational data reveal how the long-term outcome varies with the surrogate and baseline covariates, while the experiment reveals how treatment changes the conditional distribution of the surrogate. The model assumes that, conditional on the surrogate and baseline covariates, the long-term outcome is independent of treatment in the experimental population and has the same distribution in the two populations. These conditions allow the long-term treatment effect to be represented using the treatment-specific surrogate distributions and the observational outcome regression. Section 2.2 states these conditions formally and shows that they identify the target parameter defined below.

The target parameter is \[\tau := \mathbb{E}\bigl[Y_i(1)-Y_i(0)\mid G_i=\mathrm O\bigr].
\]

This is the long-term average treatment effect in the observational population. This population is a natural target in many applications. In the coupon example, the observational data are the marketer\textquotesingle s historical customer records, and the marketer wants to learn the effect for the broad customer population represented in those records. Participation in the experiment, by contrast, may be selective. The analysis below also applies when the observational population is not the target population, provided that the corresponding transportability and support conditions hold and the target covariate distribution is known or can be estimated at the rates required by the assumptions in the theoretical sections.

\begin{table}[tbp]
\centering
\small
\begin{tabular}{@{}
  >{\raggedright\arraybackslash}p{(\linewidth - 8\tabcolsep) * \real{0.2800}}
  >{\centering\arraybackslash}p{(\linewidth - 8\tabcolsep) * \real{0.1500}}
  >{\centering\arraybackslash}p{(\linewidth - 8\tabcolsep) * \real{0.1800}}
  >{\centering\arraybackslash}p{(\linewidth - 8\tabcolsep) * \real{0.1500}}
  >{\centering\arraybackslash}p{(\linewidth - 8\tabcolsep) * \real{0.2400}}@{}}
\toprule\noalign{}
\begin{minipage}[b]{\linewidth}\raggedright
\end{minipage} & \begin{minipage}[b]{\linewidth}\centering
Covariate\\
\(X\)\strut
\end{minipage} & \begin{minipage}[b]{\linewidth}\centering
Treatment\\
\(W\)\strut
\end{minipage} & \begin{minipage}[b]{\linewidth}\centering
Surrogate\\
\(S\)\strut
\end{minipage} & \begin{minipage}[b]{\linewidth}\centering
Long-term outcome\\
\(Y\)\strut
\end{minipage} \\
\midrule\noalign{}

Observational sample (\(G=\mathrm O\)) & ✓ & --- & ✓ & ✓ \\
Experimental sample (\(G=\mathrm E\)) & ✓ & ✓ & ✓ & --- \\
\bottomrule
\end{tabular}
\caption{Observed variables in the two data sources. A check mark indicates that the variable is observed in the sample, and a dash indicates that it is not observed.}
\end{table}

\subsection{Identification}\label{identification}

This subsection shows that the target parameter is identified from the two data structures in Table 2.1. We first state the identifying assumptions, which formalize the statistical surrogacy model of \citet{Chen2023} and \citet{Athey2026}, then introduce the nuisance functions that appear in the identification result, and finally state the result in Theorem 2.1.

\setcounter{assumption}{0}
\begin{assumption}[Consistency]

Almost surely, \[S_i=W_iS_i(1)+(1-W_i)S_i(0),
\qquad
Y_i=W_iY_i(1)+(1-W_i)Y_i(0).
\]

\end{assumption}

{
\emergencystretch=2em
Consistency links the observed variables to the potential variables under the assigned treatment. The potential variables are defined under no interference between units. Thus, unit \(i\)\textquotesingle s potential variables do not depend on the treatment assignments of other units.
\par
}

\setcounter{assumption}{1}
\begin{assumption}[Conditional treatment randomization]

\[W_i\mathrel{\perp\!\!\!\perp}\bigl(Y_i(0),Y_i(1),S_i(0),S_i(1)\bigr)
\mid X_i,G_i=\mathrm E.
\]

\end{assumption}

Within the experiment, treatment is randomized conditional on the baseline covariates.

\setcounter{assumption}{2}
\begin{assumption}[Statistical surrogacy]

\[W_i\mathrel{\perp\!\!\!\perp}Y_i\mid S_i,X_i,G_i=\mathrm E.
\]

\end{assumption}

Conditional on the surrogate and baseline covariates, the conditional distribution of the long-term outcome in the experimental population does not depend on treatment assignment. This restriction allows the long-term outcome mean under each treatment assignment, and hence the long-term treatment effect, to be represented using the corresponding distribution of the surrogate.

\setcounter{assumption}{3}
\begin{assumption}[Conditional external validity]

\[G_i\mathrel{\perp\!\!\!\perp}\bigl(Y_i(0),Y_i(1),S_i(0),S_i(1)\bigr)\mid X_i.
\]

\end{assumption}

Conditional on baseline covariates, the potential-outcome law is shared by the observational and experimental populations. The assumption excludes unmeasured sample selection that changes the potential variables.

\setcounter{assumption}{4}
\begin{assumption}[Conditional outcome comparability across samples]

\[G_i\mathrel{\perp\!\!\!\perp}Y_i\mid S_i,X_i.
\]

\end{assumption}

{
\emergencystretch=2em
The assumption rules out differences between the two populations in the conditional distribution of \(Y\) given \((S,X)\). It therefore allows the regression learned from observational outcomes to be applied to experimental surrogates.
\par
}

{
\emergencystretch=2em
To state the identification result, we next introduce the conditional distributions and nuisance functions that it involves. The conditional distributions used below have densities with respect to fixed reference measures \(\lambda_X\), \(\lambda_S\), and \(\lambda_Y\). The covariates, surrogates, and outcomes take values in Euclidean spaces. Write \(f_{X\mid G=\mathrm O}(x)\) and \(f_{X\mid G=\mathrm E}(x)\) for the covariate densities in the two populations, \(f_{S\mid X,G=\mathrm O}(\cdot\mid x)\) for the observational surrogate density, \(f_{S \mid W = w ,X,G=\mathrm E}(\cdot\mid x)\) for the conditional surrogate density under treatment assignment \(w\) in the experimental population, and \(f_{Y\mid S,X,G=\mathrm O}(\cdot\mid s,x)\) for the observational outcome density. Define \[\mu(s,x):=\mathbb E[Y\mid S=s,X=x,G=\mathrm O].
\]
\par
}

Following \citet{Athey2026}, we call this conditional mean the surrogate index. Integrating this function against the conditional surrogate distribution learned from the experimental data, we define \[\mu_w(x):= \int\mu(s,x)f_{S\mid W = w, X, G=\mathrm E}(s\mid x)d\lambda_S(s).
\]

We define the difference between the treatment and control groups by \[\tau(X) = \mu_1(X) - \mu_0(X).
\]

We also define the treatment assignment probability by \[p(x):=\mathbb P(W=1\mid X=x,G=\mathrm E).
\]

Finally, we define the surrogate representer by \[\alpha(S,X) := \frac{
f_{S\mid W=1, X,G=\mathrm E}(S\mid X)
-f_{S\mid W=0, X,G=\mathrm E}(S\mid X)
}{
f_{S\mid X,G=\mathrm O}(S\mid X)
}.
\]

Under the support and positivity conditions in Assumption 2.6 below, these functions are identified on the domains on which they are used.

\setcounter{assumption}{5}
\begin{assumption}[Covariate overlap, treatment positivity, and surrogate support]

{
\emergencystretch=2em
Every covariate value in the target support is also in the experimental support. Thus, \(f_{X\mid G=\mathrm O}(x)>0\) implies \(f_{X\mid G=\mathrm E}(x)>0\). There exists a constant \(\varepsilon\in(0,1/2)\) such that \(\varepsilon\leq p(x) \leq1-\varepsilon\) on the target support. For \(w\in\{0,1\}\) and every \(x\) in the target support, \(f_{S \mid W = w, X,G=\mathrm E}(s\mid x)>0\) implies \(f_{S\mid X,G=\mathrm O}(s\mid x)>0\) for \(\lambda_S\)-almost every \(s\), so the experimental surrogate law given \(W=w\) and \(X=x\) is absolutely continuous with respect to the observational surrogate law given \(X=x\).
\par
}

\end{assumption}

Target support ensures that every target covariate value can be sampled in the experiment, treatment positivity keeps both treatment assignments observable, and surrogate support ensures that the surrogate index in the observational population is evaluated only where it is identified.

Under these assumptions, the long-term treatment effect is identified as follows.

\setcounter{theorem}{0}
\begin{theorem}[Identification under the statistical surrogacy model]

\begin{align}
\tau &= \mathbb{E}\left[ \mu_1(X) - \mu_0(X) \mid G=\mathrm O \right]\label{eq:identification_regression} \\
\tau &=\mathbb E[\alpha(S,X) Y \mid G=\mathrm O]\label{eq:identification_observational_weighting} \\
\tau &=\mathbb E\left[
\frac{f_{X\mid G=\mathrm O}(X)}{f_{X\mid G=\mathrm E}(X)}
\left\{
\frac{W\mu(S,X)}{p(X)}
-\frac{(1-W)\mu(S,X)}{1-p(X)}
\right\}\middle|G=\mathrm E\right]\label{eq:identification_experimental_weighting}
\end{align}

\begin{equation}\label{eq:identification_augmented}
\begin{aligned}
\tau
&=\mathbb E\left[
\mu_1(X) - \mu_0(X) + \alpha(S,X)\{Y-\mu(S,X)\} \middle|G=\mathrm O\right] \\
&\quad+\mathbb E\Biggl[ \frac{f_{X\mid G=\mathrm O}(X)}{f_{X\mid G=\mathrm E}(X)} \Biggl\{
\frac{W\{\mu(S,X)-\mu_1(X)\}}{p(X)}-\frac{(1-W)\{\mu(S,X)-\mu_0(X)\}}{1-p(X)} \Biggr\}
\mathrel{\Big|}G=\mathrm E\Biggr].
\end{aligned}
\end{equation}

\end{theorem}

{
\emergencystretch=2em
In equation \eqref{eq:identification_regression}, the observational data identify \(\mu(s,x)\), while the randomized experiment identifies the two conditional surrogate distributions. Integrating the surrogate index against the conditional distribution of short-term outcomes in the treatment or control group identifies the corresponding conditional long-term mean. Averaging their difference over the target covariate distribution identifies the target effect. Equation \eqref{eq:identification_observational_weighting} rewrites the same target as a weighting representation for \(Y\). Equation \eqref{eq:identification_experimental_weighting} expresses the target in inverse-probability-weighted form in the experimental sample and transports it to the observational target population. Combining these representations yields equation \eqref{eq:identification_augmented}. The augmented identity combines the conditional effect and observational residual correction in one expectation under \(G=\mathrm O\), then adds the experimental weighted correction. We use equation \eqref{eq:identification_augmented} as the basis for the estimator developed below.
\par
}

Theorem 2.1 restates identification results that are known in the literature. Theorem 1 of \citet{Athey2026} gives the analogous representations for the average treatment effect in the experimental population, and Proposition 2.2 of \citet{Chen2023}, among others, gives the corresponding representation for the observational population. We restate them because the rest of the paper builds on them. The efficiency bound and the optimal experimental design in Section 3 are derived for the functional in equation \eqref{eq:identification_regression} in the fixed-design product model defined in Section 3.1, and the estimator in Section 5 is built on the augmented representation in equation \eqref{eq:identification_augmented}. The augmented score has an orthogonal, product-remainder structure. When its nuisance functions are replaced by estimates, the bias terms in the estimator decomposition of Lemma 6.2 are products of two estimation errors. For related robustness results with their own nuisance specifications, see Remark 6 of \citet{Athey2026} and Theorem 3.1(2) and Remark 3.2 of \citet{Chen2023}. The full proof is in Appendix A.1.

\section{Fixed-design efficiency and optimal experimental design}\label{section-3-fixed-design-efficiency-and-optimal-experimental-design}

This section derives the optimal experimental design in two steps. Section 3.1 expresses, as a function of a given experimental design, the semiparametric efficiency bound for the target parameter, which is the smallest asymptotic variance that a regular estimator can attain under that experimental design. Section 3.2 minimizes this bound over the experimental covariate density and the treatment assignment probability.

\subsection{Fixed-design efficiency bound}\label{fixed-design-efficiency-bound}

{
\emergencystretch=2em
This subsection writes the efficiency bound as a function of a given experimental design. The bound itself is known from \citet{Chen2023}, as explained after Theorem 3.1, and the experimental design problem requires its form for given experimental and observational sample sizes, in which the experimental design enters only through the experimental component. Recall that \(f_{X\mid G=\mathrm E} (x)\) is the experimental covariate density and \(p(x)\) is the treatment assignment probability conditional on \(X=x\). We define the experimental design as the pair \((f_{X\mid G=\mathrm E}(\cdot),p(\cdot))\). Thus, the experimental design determines the covariate distribution of experimental participants and the treatment assignment probability. In writing the bound, the experimental design \((f_{X\mid G=\mathrm E}(\cdot),p(\cdot))\) is treated as known.
\par
}

{
\emergencystretch=2em
Let the experimental and observational sample sizes be \(N_{\mathrm E}\) and \(N_{\mathrm O}\). From Section 4 onward, experimental participants are recruited one per round, so \(N_{\mathrm E}\) is also the number of rounds. For the experimental sample, write \(D_s^{\mathrm E}:=\allowbreak (X_s,W_s,W_sS_s(1),(1-W_s)S_s(0)), 1\leq s\leq N_{\mathrm E},\) where only the potential surrogate selected by \(W_s\) is observed. For the observational sample, write \(D_i^{\mathrm O}:=(X_i,S_i,Y_i), 1\leq i\leq N_{\mathrm O}.\)
\par
}

{
\emergencystretch=2em
Each collection is i.i.d. within sample in this fixed-design benchmark. Let \(P^{\rm exp}\) and \(P^{\rm obs}\) denote the laws of the experimental and observational data. The full dataset then has the product law \((P^{\rm exp})^{\otimes N_{\mathrm E}}\otimes (P^{\rm obs})^{\otimes N_{\mathrm O}}\), and we call the model of such product laws, with the experimental design held fixed, the fixed-design product model. The joint density of \((X,W,S)\) in the experimental population is \(f_{X\mid G=\mathrm E}(x)p(x)^w\{1-p(x)\}^{1-w}f_{S(1)\mid X,G=\mathrm E}(s\mid x)^wf_{S(0)\mid X,G=\mathrm E}(s\mid x)^{1-w},\) whereas the joint density of \((X,S,Y)\) in the observational population is \(f_{X\mid G=\mathrm O}(x)f_{S\mid X,G=\mathrm O}(s\mid x)f_{Y\mid S,X,G=\mathrm O}(y\mid s,x).\)
\par
}

{
\emergencystretch=2em
The efficiency bound is expressed through an experimental and an observational variance component, which we define first. For \(w\in\{0,1\}\), define \[\begin{aligned}
v_w(x)&:=\mathbb E[(\mu(S,x)-\mu_w(x))^2 \mid W = w, X=x,G=\mathrm E],\\
\sigma^2(s,x)&:=\mathbb E[(Y-\mu(S,X))^2\mid S=s,X=x,G=\mathrm O].
\end{aligned}
\]
\par
}

\begingroup
\setlength{\abovedisplayskip}{6pt}
\setlength{\belowdisplayskip}{6pt}
\setlength{\abovedisplayshortskip}{3pt}
\setlength{\belowdisplayshortskip}{3pt}
Define the experimental and observational variance components as follows. \[\begin{aligned}
V^{\rm exp}(f_{X\mid G=\mathrm E}(\cdot),p(\cdot))
&:=\mathbb E\left[
\frac{f_{X\mid G=\mathrm O}(X)}{f_{X\mid G=\mathrm E}(X)}
\left\{\frac{v_1(X)}{p(X)}+\frac{v_0(X)}{1-p(X)}\right\}
\middle|G=\mathrm O\right]\\
V^{\rm obs}
&:=\mathbb E[(\mu_1(X) - \mu_0(X) -\tau)^2\mid G=\mathrm O]
\\ &\quad+\mathbb E[(\alpha(S,X))^2\sigma^2(S,X)\mid G=\mathrm O].
\end{aligned}
\]
\endgroup

{
\emergencystretch=2em
The quantity \(v_w(x)\) is the variance of the surrogate index \(\mu(S,X)\) conditional on \(X=x\) in the treatment or control group. Residual variation in the long-term outcome after conditioning on \((S,X)\) instead enters \(V^{\rm obs}\) through \(\sigma^2(S,X)\). Using these quantities, the semiparametric efficiency bound can be stated as follows.
\par
}

\setcounter{theorem}{0}
\begin{wrappedtheorem}[Fixed-design efficiency bound]
{
\emergencystretch=2em
Suppose that Assumptions 2.1--2.6 and suitable moment conditions (Appendix A.2) hold and that \(N_{\mathrm E},N_{\mathrm O}\to\infty\) with \(N_{\mathrm E}/N_{\mathrm O}\to\kappa\in(0,\infty)\). For a fixed experimental design \((f_{X\mid G=\mathrm E}(\cdot),p(\cdot))\), the semiparametric efficiency bound for \(\tau\) on the \(\sqrt{N_{\mathrm E}}\) scale in the fixed-design product model is \[V^{\rm exp}(f_{X\mid G=\mathrm E}(\cdot),p(\cdot))+\kappa V^{\rm obs}.
\]
\par
}

{
\emergencystretch=2em
\begingroup
\setlength{\abovedisplayskip}{6pt}
\setlength{\belowdisplayskip}{6pt}
\setlength{\abovedisplayshortskip}{3pt}
\setlength{\belowdisplayshortskip}{3pt}
The quantities \(V^{\rm exp}(f_{X\mid G=\mathrm E}(\cdot),p(\cdot))\) and \(V^{\rm obs}\) are the second moments of the experimental and observational efficient influence functions, respectively, \[\phi^{\rm exp}(S,W,X)
=\frac{f_{X\mid G=\mathrm O}(X)}{f_{X\mid G=\mathrm E}(X)}
\Biggl[
\frac{W}{p(X)}\{\mu(S,X)-\mu_1(X)\} -\frac{1-W}{1-p(X)}\{\mu(S,X)-\mu_0(X)\}
\Biggr].
\]
\endgroup
\par
}

\nopagebreak[4]and\nopagebreak[4]

\[\phi^{\rm obs}(X,S,Y)
=\tau(X)-\tau+\alpha(S,X)\{Y-\mu(S,X)\}.
\]

Define the corresponding uncentered observational score by \[\psi^{\rm obs}(X,S,Y)
:=\tau(X)+\alpha(S,X)\{Y-\mu(S,X)\}.
\]

{
\emergencystretch=2em
Then \(\phi^{\rm obs}(X,S,Y)=\psi^{\rm obs}(X,S,Y)-\tau\). Their second moments under their respective population laws satisfy \(\mathbb E[\phi^{\rm exp}(S,W,X)^2\mid G=\allowbreak \mathrm E]=\allowbreak V^{\rm exp}(f_{X\mid G=\mathrm E}(\cdot),p(\cdot))\) and \(\mathbb E[\phi^{\rm obs}(X,S,Y)^2\mid G=\mathrm O]=V^{\rm obs}\).
\par
}

\end{wrappedtheorem}

{
\emergencystretch=2em
Only the experimental component of the bound depends on the experimental design, which the experimenter chooses. The term \(V^{\rm exp}(f_{X\mid G=\mathrm E}(\cdot),p(\cdot))\) changes with the covariate distribution from which participants are sampled and with the treatment assignment probability, whereas \(V^{\rm obs}\) is the variance contribution of the fixed observational law and does not depend on the experimental design. The experimenter can therefore lower the bound by choosing the experimental design, and Section 3.2 derives the experimental design that minimizes \(V^{\rm exp}(f_{X\mid G=\mathrm E}(\cdot),p(\cdot))\). The efficient influence functions serve to derive and write down the bound, and they also supply the scores from which the estimator in Section 5 is built.
\par
}

{
\emergencystretch=2em
The proof works in the fixed-design product model and does not pair observations across samples. The influence-function pair represents the unscaled pathwise derivative. In the limiting coordinates for the \(\sqrt{N_{\mathrm E}}\) scale, the pair is \((\phi^{\rm exp}(S,W,X),\sqrt\kappa\,\phi^{\rm obs}(X,S,Y))\), whose squared norm gives the displayed variance. After conditioning on \(G\), the two efficient influence functions agree, up to the constant factors that the pooled representation attaches to the two samples, with the experimental and observational pieces of the efficient influence function in \citet{Chen2023}, and the pooled-sample bound of \citet{Chen2023}, rescaled to the \(\sqrt{N_{\mathrm E}}\) scale, coincides with the bound in Theorem 3.1. Appendix A.2 verifies these efficient influence functions and states the formal convolution result, and Supplementary Material 2 derives the relation between the two efficiency bounds.
\par
}

\subsection{Optimal covariate sampling and treatment assignment}\label{optimal-covariate-sampling-and-treatment-assignment}

{
\emergencystretch=2em
This subsection derives the optimal experimental design. By Theorem 3.1, a fixed experimental design affects the bound only through \(V^{\rm exp}(f_{X\mid G=\mathrm E}(\cdot),p(\cdot))\), so the optimal experimental design minimizes this component over the experimental covariate density and the assignment probability. The next result gives the minimizer.
\par
}

\setcounter{theorem}{1}
\begin{theorem}[Optimal fixed experimental design]

{
\emergencystretch=2em
Suppose the regularity conditions in Appendix A.3 hold. In particular, the variances in the treatment and control groups are positive and finite, the normalizing expectation is positive and finite, and the assignment probability displayed below lies in \([\varepsilon,1-\varepsilon]\). Among experimental covariate densities covering the target support and assignment probabilities in this interval, the unique optimal assignment probability is \[p^\star(x)
=\frac{\sqrt{v_1(x)}}{\sqrt{v_1(x)}+\sqrt{v_0(x)}},
\]
\par
}

the unique optimal experimental covariate density is \[f_{X\mid G=\mathrm E}^\star(x)
=\frac{f_{X\mid G=\mathrm O}(x)\{\sqrt{v_1(x)}+\sqrt{v_0(x)}\}}
{\mathbb E[\sqrt{v_1(X)}+\sqrt{v_0(X)}\mid G=\mathrm O]},
\]

{
\emergencystretch=2em
so that the sampling density ratio is \(f_{X\mid G=\mathrm E}^\star(x)/\allowbreak f_{X\mid G=\mathrm O}(x)=\allowbreak \{\sqrt{v_1(x)}+\sqrt{v_0(x)}\}/\allowbreak \mathbb E[\sqrt{v_1(X)}+\sqrt{v_0(X)}\mid G=\allowbreak \mathrm O]\), and the minimum experimental variance is \(V^{{\rm exp},\star}=\allowbreak \allowbreak\left\{\mathbb E[\sqrt{v_1(X)}+\sqrt{v_0(X)}\mid G=\mathrm O]\right\}^2.\)
\par
}

\end{theorem}

{
\emergencystretch=2em
The assignment probability is the conditional standard deviation of the surrogate index in the treatment group divided by the sum of its conditional standard deviations in the treatment and control groups. Conditional on \(x\), the rule assigns more participants to the group with the larger conditional variance so that more observations are used to estimate its mean. This is the Neyman allocation, which originates in stratified sampling \citet{Neyman1934}.
\par
}

{
\emergencystretch=2em
Covariate sampling determines which covariate values are represented more heavily in the experimental sample. The optimal rule oversamples covariate regions where the surrogate index is more variable, as measured by the sum of the treatment- and control-specific conditional standard deviations, \(\sqrt{v_1(x)}+\sqrt{v_0(x)}\). Intuitively, these are the regions in which additional experimental observations are most valuable for reducing the variance of the long-term effect estimator. Because this changes the covariate distribution of the experiment relative to the target population, the estimator reweights experimental observations by the target-to-experimental density ratio. The direct analogue in experimental design is \citet{Kato2024}, who optimize the experimental covariate density jointly with treatment assignment when the outcome is observed in the experiment. Appendix A.3 gives the complete proof.
\par
}

\section{Identification under adaptive experimental design}\label{section-4-identification-under-adaptive-experimental-design}

{
\emergencystretch=2em
This section describes the adaptive experiment and shows that the identification result of Section 2 continues to hold under it. In the adaptive experiment, the experimenter recruits participants one at a time in rounds \(t=1,2,\ldots\), and the observational sample is available in full before the first round. Before each round, the experimenter chooses the covariate density from which the next participant is drawn and the treatment assignment probability, using the observational sample and the records of earlier participants. The participant\textquotesingle s covariates are then observed, treatment is randomized with the chosen probability, and the short-term outcomes are recorded, while the participant\textquotesingle s long-term outcome is never observed in the experiment. Because the experimental design responds to the accumulated data, it can change from round to round, and observations from different rounds are dependent.
\par
}

{
\emergencystretch=2em
To state identification under this experimental design, we use triangular-array notation indexed by the observational sample size \(N_{\mathrm O}\), and the experimental horizon \(N_{\mathrm E}\) may depend on \(N_{\mathrm O}\). Let \(\mathcal D_{N_{\mathrm O}}^{\rm obs}=\allowbreak \{(X_i,S_i,Y_i):1\leq i\leq N_{\mathrm O}\}\) be i.i.d. from the observational law. Define \(\mathcal F_{0,N_{\mathrm O}}:=\allowbreak \sigma(\mathcal D_{N_{\mathrm O}}^{\rm obs}),\mathcal F_{t,N_{\mathrm O}}:=\allowbreak \mathcal F_{0,N_{\mathrm O}}\vee\sigma\{(X_s,W_s,S_s):1\leq s\leq t\}.\) Thus \(\mathcal F_{t-1,N_{\mathrm O}}\) contains the observational sample and all experimental records available before round \(t\).
\par
}

{
\emergencystretch=2em
For each round \(t\), the conditional experimental covariate law admits a density \(f_{X\mid G=\mathrm E,t,N_{\mathrm O}}(x)\) satisfying, for every measurable set \(B\), \[\mathbb P(X_t\in B\mid\mathcal F_{t-1,N_{\mathrm O}})
=\int_B f_{X\mid G=\mathrm E,t,N_{\mathrm O}}(x)d\lambda_X(x).
\]
\par
}

The assignment probability is \[p_{t,N_{\mathrm O}}(x):=\mathbb P(W_t=1\mid X_t=x,\mathcal F_{t-1,N_{\mathrm O}}).
\]

{
\emergencystretch=2em
Both functions are chosen using the observational data and experimental records available before the participant at round \(t\) is observed. We impose the following conditions on the adaptive experiment.
\par
}

\setcounter{assumption}{0}
\begin{assumption}[Covariate overlap and treatment positivity]

{
\emergencystretch=2em
For every round \(t\), almost surely, \(f_{X\mid G=\mathrm O}(x)>0\) implies \(f_{X\mid G=\mathrm E,t,N_{\mathrm O}}(x)>0\) for \(\lambda_X\)-almost every \(x\). Moreover, for the constant \(\varepsilon\) in Assumption 2.6, almost surely, \(\varepsilon\leq p_{t,N_{\mathrm O}}(x)\leq1-\varepsilon\) for almost every \(x\) under the distribution with density \(f_{X\mid G=\mathrm E,t,N_{\mathrm O}}(x)\).
\par
}

\end{assumption}

This assumption requires the experimental covariate density to be positive throughout the support of the target population and keeps both the treatment and control groups available at every round.

For each round of the adaptive experiment, we impose the following counterpart of experimental treatment randomization.

\setcounter{assumption}{1}
\begin{assumption}[Sequential treatment randomization]

{
\emergencystretch=2em
For every round \(t\), \[W_t\mathrel{\perp\!\!\!\perp}(S_t(0),S_t(1),Y_t(0),Y_t(1))
\mid X_t,\mathcal F_{t-1,N_{\mathrm O}}.
\]
\par
}

\end{assumption}

{
\emergencystretch=2em
Before \(X_t\) is drawn, the recorded past determines the experimental covariate density and assignment probability. After \(X_t\) is observed, the assignment probability is evaluated and \(W_t\) is randomized conditional on the current covariate and recorded past. We further assume that the conditional surrogate distribution does not depend on the recorded history.
\par
}

\setcounter{assumption}{2}
\begin{assumption}[History-invariant conditional surrogate distributions]

{
\emergencystretch=2em
For every measurable set \(B\), every round \(t\), every \(w\in\{0,1\}\), and almost every \(x\) under the distribution with density \(f_{X\mid G=\mathrm O}(x)\), almost surely, \[\mathbb P\{S_t(w)\in B\mid X_t=x,\mathcal F_{t-1,N_{\mathrm O}}\}
=\mathbb P\{S\in B\mid W=w,X=x,G=\mathrm E\}.
\]
\par
}

\end{assumption}

{
\emergencystretch=2em
This condition rules out interference, carryover, or other history dependence insofar as it changes the displayed conditional distribution of \(S_t(w)\). Predictable randomization and conditional distributional equivalence of surrogates connect every adaptive round to the fixed experimental conditional distributions used in Section 2. Together with the identification assumptions, they yield an augmented representation under the realized adaptive experimental design. Here the fixed experiment of Section 2 is a reference experiment, and by Assumptions 2.1, 2.2, and 2.4 its treatment-specific surrogate laws are the conditional laws of \(S(w)\) given \(X\) that the populations share.
\par
}

\Needspace{7\baselineskip}

\setcounter{proposition}{0}
\begin{proposition}[Augmented representation under adaptive experimental design]

{
\emergencystretch=2em
Under Assumptions 2.1--2.6 and Assumptions 4.1--4.3, whenever the displayed terms are absolutely integrable, for every round \(t\), almost surely, \[\begin{aligned}
\tau
&=\mathbb E\left[
\mu_1(X)-\mu_0(X)
+\alpha(S,X)\{Y-\mu(S,X)\}
\middle|G=\mathrm O\right]\\
&\quad+\mathbb E\Biggl[
\frac{f_{X\mid G=\mathrm O}(X_t)}{f_{X\mid G=\mathrm E,t,N_{\mathrm O}}(X_t)} \Biggl\{
\frac{W_t\{\mu(S_t,X_t)-\mu_1(X_t)\}}
{p_{t,N_{\mathrm O}}(X_t)}
-\frac{(1-W_t)\{\mu(S_t,X_t)-\mu_0(X_t)\}}
{1-p_{t,N_{\mathrm O}}(X_t)}
\Biggr\}
\mathrel{\Big|}\mathcal F_{t-1,N_{\mathrm O}}\Biggr].
\end{aligned}
\]
\par
}

\end{proposition}

{
\emergencystretch=2em
This proposition is the counterpart of Theorem 2.1 under adaptive experimental design. Section 5 replaces the population nuisance functions by estimates and uses the experimental covariate density and treatment assignment probability used in round \(t\). Appendix A.4 proves the representation.
\par
}

\section{Adaptive experimental design and sequential estimation}\label{section-5-adaptive-experimental-design-and-sequential-estimation}

{
\emergencystretch=2em
This section constructs a sequential estimator of the target parameter \(\tau\) that learns the optimal experimental design of Section 3 while the experiment is running. Algorithm~\ref{alg:adaptive-sequential-estimation} defines the fitted nuisance functions, the plug-in experimental design, the estimated score components, and the time-\(t\) estimator in the order in which they are computed, and Remarks 5.1 and 5.2 discuss how the conditional moments and the surrogate representer can be estimated.
\par
}

{
\emergencystretch=2em
The observational sample \(\mathcal D_{N_{\mathrm O}}^{\rm obs}=\allowbreak \{(X_i,S_i,Y_i):1\leq i\leq N_{\mathrm O}\}\) is available before the experiment. It is used to estimate the target covariate density \(\widehat f_{X\mid G=\mathrm O,N_{\mathrm O}}(x)\) and the surrogate index \(\widehat\mu_{N_{\mathrm O}}(s,x)\), and these two fits are used unchanged throughout the experiment. The experimental rounds then estimate the treatment-specific conditional mean and variance of the fitted surrogate index. To state their population targets, we use the following notation. Conditional on the initial observational sample, define, for \(w\in\{0,1\}\), \[\widetilde\mu_{w,N_{\mathrm O}}(x)
:=
\int
\widehat\mu_{N_{\mathrm O}}(s,x)
f_{S\mid W=w,X,G=\mathrm E}(s\mid x)
\,d\lambda_S(s),
\]
\par
}

\nopagebreak[4]and\nopagebreak[4]

\[\widetilde v_{w,N_{\mathrm O}}(x)
:=
\int
\left\{
\widehat\mu_{N_{\mathrm O}}(s,x)-\widetilde\mu_{w,N_{\mathrm O}}(x)
\right\}^{2}
f_{S\mid W=w,X,G=\mathrm E}(s\mid x)
\,d\lambda_S(s).
\]

{
\emergencystretch=2em
Thus, \(\widetilde\mu_{w,N_{\mathrm O}}(x)\) and \(\widetilde v_{w,N_{\mathrm O}}(x)\) are, respectively, the conditional mean and variance of the fixed observational prediction \(\widehat\mu_{N_{\mathrm O}}(S,x)\) under the population surrogate law in treatment arm \(w\). The following algorithm formalizes the estimation procedure and the corresponding sequential estimator.
\par
}

\begin{manuscriptalgorithm}{Adaptive experimental design and sequential estimation}{adaptive-sequential-estimation}
\textbf{Initialization.} Before the first experimental round, use \(\mathcal D_{N_{\mathrm O}}^{\rm obs}\) to estimate \(\widehat f_{X\mid G=\mathrm O,N_{\mathrm O}}(x)\) and \(\widehat\mu_{N_{\mathrm O}}(s,x)\), and keep both fits fixed throughout the experiment.

For \(t=1,\ldots,N_{\mathrm E}\), proceed as follows.

\begin{enumerate}
\def\labelenumi{\arabic{enumi}.}
\item
  \textbf{Pre-round estimation.} Using the experimental history through round \(t-1\), construct \(\mathcal F_{t-1,N_{\mathrm O}}\)-measurable estimators \(\widehat\mu_{w,t,N_{\mathrm O}}(x)\) of \(\widetilde\mu_{w,N_{\mathrm O}}(x)\) and \(\widehat v_{w,t,N_{\mathrm O}}(x)\) of \(\widetilde v_{w,N_{\mathrm O}}(x)\) for \(w=0,1\). At \(t=1\), use the initial fits specified before the experiment. Define and record the pre-round conditional-effect function \(\widehat\tau_{t,N_{\mathrm O}}(x):=\allowbreak \widehat\mu_{1,t,N_{\mathrm O}}(x)-\widehat\mu_{0,t,N_{\mathrm O}}(x)\).
\item
  \textbf{Plug-in experimental design.} Using the pre-round conditional-variance estimates, define

  \[   \widehat p^\star_{t,N_{\mathrm O}}(x)
  :=
  \frac{
  \sqrt{\widehat v_{1,t,N_{\mathrm O}}(x)}
  }{
  \sqrt{\widehat v_{1,t,N_{\mathrm O}}(x)}
  +
  \sqrt{\widehat v_{0,t,N_{\mathrm O}}(x)}
  },
  \]

  and

  \[   \widehat f^\star_{X\mid G=\mathrm E,t,N_{\mathrm O}}(x)
  :=
  \frac{
  \widehat f_{X\mid G=\mathrm O,N_{\mathrm O}}(x)
  \left\{
  \sqrt{\widehat v_{1,t,N_{\mathrm O}}(x)}
  +
  \sqrt{\widehat v_{0,t,N_{\mathrm O}}(x)}
  \right\}
  }{
  \displaystyle
  \int
  \widehat f_{X\mid G=\mathrm O,N_{\mathrm O}}(u)
  \left\{
  \sqrt{\widehat v_{1,t,N_{\mathrm O}}(u)}
  +
  \sqrt{\widehat v_{0,t,N_{\mathrm O}}(u)}
  \right\}
  \,d\lambda_X(u)
  }.
  \]

  The denominators and the normalizing integral are required to be positive and finite. The resulting experimental design is assumed to satisfy Assumption 4.1. Set and record the actual round-\(t\) experimental design as

  \[   \begin{aligned}
  f_{X\mid G=\mathrm E,t,N_{\mathrm O}}(\cdot)
  &=
  \widehat f^\star_{X\mid G=\mathrm E,t,N_{\mathrm O}}(\cdot),\\
  p_{t,N_{\mathrm O}}(\cdot)
  &=
  \widehat p^\star_{t,N_{\mathrm O}}(\cdot).
  \end{aligned}
  \]
\item
  \textbf{Data collection.} Conditional on \(\mathcal F_{t-1,N_{\mathrm O}}\), draw \(X_t\) from the distribution with density \(f_{X\mid G=\mathrm E,t,N_{\mathrm O}}(x)\). Randomize \(W_t\) with assignment probability \(p_{t,N_{\mathrm O}}(X_t)\) and observe \(S_t=S_t(W_t)\).
\item
  \textbf{Experimental-score evaluation.} Using only the fixed observational-data fits, the pre-round treatment-specific mean and variance estimates from Step 1, and the recorded experimental design from Step 2, compute

  \[   \begin{aligned}
  \widehat\phi_{t,N_{\mathrm O}}^{\rm exp}(S_t,W_t,X_t)
  &:=
  \frac{
  \widehat f_{X\mid G=\mathrm O,N_{\mathrm O}}(X_t)
  }{
  \widehat f^\star_{X\mid G=\mathrm E,t,N_{\mathrm O}}(X_t)
  }
  \Biggl[
  \frac{
  W_t
  \left\{
  \widehat\mu_{N_{\mathrm O}}(S_t,X_t)
  -
  \widehat\mu_{1,t,N_{\mathrm O}}(X_t)
  \right\}
  }{
  \widehat p^\star_{t,N_{\mathrm O}}(X_t)
  }\\
  &\qquad\qquad-
  \frac{
  (1-W_t)
  \left\{
  \widehat\mu_{N_{\mathrm O}}(S_t,X_t)
  -
  \widehat\mu_{0,t,N_{\mathrm O}}(X_t)
  \right\}
  }{
  1-\widehat p^\star_{t,N_{\mathrm O}}(X_t)
  }
  \Biggr].
  \end{aligned}
  \]

  Record this score permanently. Experimental scores recorded in earlier rounds are not recomputed after later nuisance-function updates.
\item
  \textbf{Post-round representer update.} Incorporate \((X_t,W_t,S_t)\) into the experimental history. Using the initial observational sample and the experimental data through round \(t\), construct an \(\mathcal F_{t,N_{\mathrm O}}\)-measurable estimator \(\widehat\alpha_{t,N_{\mathrm O}}(s,x)\) of the surrogate representer \(\alpha(s,x)\) defined in Section 2.
\item
  \textbf{Estimator update.} Form the historical average of the conditional-effect functions recorded before their respective rounds, \(\overline\tau_{t,N_{\mathrm O}}(x):=\allowbreak t^{-1}\allowbreak\sum_{s=1}^t\widehat\tau_{s,N_{\mathrm O}}(x)\). Define the fitted uncentered observational score by

  \[   \widehat\psi_{t,N_{\mathrm O}}^{\rm obs}(X,S,Y)
  :=
  \overline\tau_{t,N_{\mathrm O}}(X)
  +
  \widehat\alpha_{t,N_{\mathrm O}}(S,X)
  \left\{
  Y-\widehat\mu_{N_{\mathrm O}}(S,X)
  \right\}.
  \]

  The time-\(t\) estimator is

  \[   \widehat\tau_{t,N_{\mathrm O}}
  :=
  \frac{1}{t}
  \sum_{s=1}^t
  \widehat\phi_{s,N_{\mathrm O}}^{\rm exp}(S_s,W_s,X_s)
  +
  \frac{1}{N_{\mathrm O}}
  \sum_{i=1}^{N_{\mathrm O}}
  \widehat\psi_{t,N_{\mathrm O}}^{\rm obs}(X_i,S_i,Y_i).
  \]

  At \(t=N_{\mathrm E}\), report \(\widehat\tau_{N_{\mathrm E},N_{\mathrm O}}\). An estimate is therefore available after every experimental round.
\end{enumerate}
\end{manuscriptalgorithm}

{
\emergencystretch=2em
Replacing the fitted quantities in the experimental score by their population counterparts recovers the experimental efficient influence function evaluated under the recorded experimental design. By Theorem 3.1, the observational efficient influence function satisfies \(\phi^{\rm obs}(X,S,Y)=\psi^{\rm obs}(X,S,Y)-\tau\), so \(\widehat\psi_{t,N_{\mathrm O}}^{\rm obs}\) estimates the corresponding uncentered observational score. The nuisance functions and experimental design quantities used in the round-\(t\) experimental score are determined before \(X_t\) is observed and are therefore \(\mathcal F_{t-1,N_{\mathrm O}}\)-measurable. This predictability is important for establishing the martingale structure in Lemma 6.1 of the next section. By contrast, \(\widehat\alpha_{t,N_{\mathrm O}}\) and the updated observational score may use the round-\(t\) observation and are \(\mathcal F_{t,N_{\mathrm O}}\)-measurable. In addition, the time average \(\overline\tau_{t,N_{\mathrm O}}(x):=\allowbreak t^{-1}\allowbreak\sum_{s=1}^t\widehat\tau_{s,N_{\mathrm O}}(x)\) is used in \(\widehat\psi_{t,N_{\mathrm O}}^{\rm obs}(X,S,Y)\) to obtain the decomposition established in Lemma 6.2 of the next section. The conditions required of the respective estimators are summarized in Sections 6 and 7.
\par
}

\setcounter{remark}{0}
\begin{remark}[Pre-round estimation of conditional moments]

{
\emergencystretch=2em
The conditional moments in Step 1 of Algorithm~\ref{alg:adaptive-sequential-estimation} may be estimated directly from the experimental history. In particular, within each treatment arm, one may regress \(\widehat\mu_{N_{\mathrm O}}(S_s,X_s)\) and \(\widehat\mu_{N_{\mathrm O}}(S_s,X_s)^2\) on \(X_s\) using the previous experimental observations \(s<t\). Alternatively, the same moments may be computed by integrating \(\widehat\mu_{N_{\mathrm O}}(s,x)\) under an estimated treatment-specific conditional surrogate density. This latter approach, however, requires estimation of the conditional density of \(S\) and may therefore become difficult when the surrogate vector \(S\) is high-dimensional.
\par
}

\end{remark}

\setcounter{remark}{1}
\begin{remark}[Estimation of the surrogate representer]

{
\emergencystretch=2em
The observational score involves the surrogate representer introduced in Section 2.2, \(\alpha(s,x)=\allowbreak \{f_{S\mid W=1,X,G=\mathrm E}(s\mid x)-f_{S\mid W=0,X,G=\mathrm E}(s\mid x)\}/\allowbreak f_{S\mid X,G=\mathrm O}(s\mid x)\). The representer may be estimated either through conditional-density estimation or through conditional-probability estimation.
\par
}

\begin{enumerate}
\def\labelenumi{\arabic{enumi}.}
\item
  \textbf{Density-based estimation.} Before the experiment begins, estimate \(\widehat f_{S\mid X,G=\mathrm O,N_{\mathrm O}}(s\mid x)\) from the observational sample and keep this estimate fixed. After observing round \(t\), use the experimental data through that round to estimate the treatment- and control-arm conditional surrogate densities, denoted by \(\widehat f_{S\mid W=w,X,G=\mathrm E,t}(s\mid x)\) for \(w=0,1\), and set \(\widehat\alpha_{t,N_{\mathrm O}}(s,x) =\allowbreak  \{\widehat f_{S\mid W=1,X,G=\mathrm E,t}(s\mid x) - \widehat f_{S\mid W=0,X,G=\mathrm E,t}(s\mid x)\}/\allowbreak \widehat f_{S\mid X,G=\mathrm O,N_{\mathrm O}}(s\mid x).\) Direct conditional-density estimation is, however, generally difficult when \(S\) is multidimensional.
\item
  \textbf{Binary-classification-based estimation.} Instead of estimating the conditional surrogate densities directly, the representer may be constructed from conditional treatment and data-source probabilities. To do so, embed the observational population law and a reference experimental law that preserves the treatment-specific conditional surrogate distributions \(f_{S\mid W=w,X,G=\mathrm E}\) into a common reference law \(Q\) that includes \(G\). Appendix A.5 gives the explicit construction of this reference law. Under the same reference law \(Q\), define
\end{enumerate}

\[   \begin{aligned}
   p_Q(x)
   &:=
   Q(W=1\mid X=x,G=\mathrm E),\\
   p_Q(s,x)
   &:=
   Q(W=1\mid S=s,X=x,G=\mathrm E),\\
   \gamma_Q(x)
   &:=
   Q(G=\mathrm O\mid X=x),\\
   \gamma_Q(s,x)
   &:=
   Q(G=\mathrm O\mid S=s,X=x).
   \end{aligned}
\]

By Bayes\textquotesingle{} rule, \[\alpha(s,x)
=
\frac{\gamma_Q(x)}{\gamma_Q(s,x)}
\frac{1-\gamma_Q(s,x)}{1-\gamma_Q(x)}
\frac{
p_Q(s,x)-p_Q(x)
}{
p_Q(x)\{1-p_Q(x)\}
}.
\]

{
\emergencystretch=2em
After observing round \(t\), estimate each component in this expression using \(\widehat p_Q(s,x)\), \(\widehat p_Q(x)\), \(\widehat\gamma_Q(s,x)\), and \(\widehat\gamma_Q(x)\), constructed to target the same reference law \(Q\), and replace the corresponding population quantities by these estimators to obtain \(\widehat\alpha_{t,N_{\mathrm O}}(s,x)\). This classification-based construction avoids direct estimation of conditional surrogate densities when constructing the representer.
\par
}

\end{remark}

\section{Asymptotic normality and attainment of the optimal fixed-design bound}\label{section-6-asymptotic-normality-and-attainment-of-the-optimal-fixed-design-bound}

{
\emergencystretch=2em
This section shows that the sequential estimator of Section 5 is asymptotically normal and attains the optimal fixed-design bound of Section 3 when \(N_{\mathrm O},N_{\mathrm E}\to\infty\). Section 6.1 centers the experimental scores to obtain martingale differences and decomposes the estimator into an experimental martingale average, an observational i.i.d. empirical average, and remainder terms, and Section 6.2 states the conditions under which the remainders are negligible and the estimator is asymptotically normal with the variance of the optimal fixed-design bound.
\par
}

\subsection{Martingale centering and estimator decomposition}\label{martingale-centering-and-estimator-decomposition}

{
\emergencystretch=2em
This subsection centers the experimental scores and decomposes the estimator. We use the following notation throughout this section. For a function \(g\) of \(X\), write \(P_{\widehat f_{X\mid G=\mathrm O,N_{\mathrm O}}}g:=\allowbreak \int g(x)\widehat f_{X\mid G=\mathrm O,N_{\mathrm O}}(x)d\lambda_X(x)\) and \(P_{f_{X\mid G=\mathrm O}}g:=\allowbreak \int g(x)f_{X\mid G=\mathrm O}(x)d\lambda_X(x)\). For a function \(g\) of \((X,S,Y)\), write \(P^{\rm obs}g:=\mathbb E[g(X,S,Y)\mid G=\mathrm O]\) and \(\mathbb P_{N_{\mathrm O}}^{\rm obs}g:=\allowbreak N_{\mathrm O}^{-1}\allowbreak\sum_{i=1}^{N_{\mathrm O}} g(X_i,S_i,Y_i)\). The corresponding norms are \(\|g\|_{2,{\rm obs}}:=\{P^{\rm obs}(g^2)\}^{1/2}\) and \(\|a\|_{2,X\mid G=\mathrm O}:=\allowbreak \{\mathbb E[a(X)^2\mid G=\allowbreak \mathrm O]\}^{1/2}\).
\par
}

{
\emergencystretch=2em
The estimation order in Algorithm~\ref{alg:adaptive-sequential-estimation} makes the experimental score predictable apart from the new round-\(t\) observation. In particular, \(\widehat\mu_{w,t,N_{\mathrm O}}(\cdot)\), \(\widehat\tau_{t,N_{\mathrm O}}(\cdot)\), and the recorded experimental design \((f_{X\mid G=\mathrm E,t,N_{\mathrm O}},p_{t,N_{\mathrm O}})\) are \(\mathcal F_{t-1,N_{\mathrm O}}\)-measurable when the round-\(t\) score is evaluated. Integrating first over \((W_t,S_t)\) conditional on \(X_t\) and then over the experimental covariate density gives the identity \(\mathbb E[\widehat\phi_{t,N_{\mathrm O}}^{\rm exp}(S_t,W_t,X_t)\mid\mathcal F_{t-1,N_{\mathrm O}}]=\allowbreak P_{\widehat f_{X\mid G=\mathrm O,N_{\mathrm O}}}\{\widetilde\tau_{N_{\mathrm O}}(\cdot)-\widehat\tau_{t,N_{\mathrm O}}(\cdot)\}\), where \(\widetilde\tau_{N_{\mathrm O}}(x) :=\allowbreak  \widetilde\mu_{1,N_{\mathrm O}}(x) - \widetilde\mu_{0,N_{\mathrm O}}(x)\). Accordingly, define \[\widetilde m_{t,N_{\mathrm O}}:=\widehat\phi_{t,N_{\mathrm O}}^{\rm exp}(S_t,W_t,X_t)+P_{\widehat f_{X\mid G=\mathrm O,N_{\mathrm O}}}\{\widehat\tau_{t,N_{\mathrm O}}(\cdot)-\widetilde\tau_{N_{\mathrm O}}(\cdot)\}.
\]
\par
}

The following lemma records the resulting martingale-difference property.

\setcounter{lemma}{0}
\begin{lemma}[Martingale centering]

{
\emergencystretch=2em
Under Assumption 2.1, Assumptions 4.1--4.3, and the common regularity conditions in Assumption A.6.1, for every applicable \(N_{\mathrm O}\) and \(t\) such that \(\mathbb E[\widetilde m_{t,N_{\mathrm O}}^2]<\infty\), \(\widetilde m_{t,N_{\mathrm O}}\) is \(\mathcal F_{t,N_{\mathrm O}}\)-measurable and \(\mathbb E[\widetilde m_{t,N_{\mathrm O}}\mid\mathcal F_{t-1,N_{\mathrm O}}]=\allowbreak 0.\)
\par
}

\end{lemma}

{
\emergencystretch=2em
Lemma 6.1 makes the adaptively collected experimental data mathematically tractable. Because the experimental design and the nuisance estimates change from round to round and depend on earlier observations, the experimental scores are neither independent nor identically distributed. After centering, however, they have mean zero conditional on the past, so a martingale central limit theorem applies to their average, as the ordinary central limit theorem does under a fixed experimental design with independent observations. This property requires every quantity in the round-\(t\) score to be fixed before the round-\(t\) participant is observed, which is why Algorithm~\ref{alg:adaptive-sequential-estimation} computes each score with the pre-round estimates and does not recompute it after later updates. The proof is given in Appendix A.6.2.
\par
}

We next give the decomposition needed to establish asymptotic normality, the main result of this section.

\setcounter{lemma}{1}
\begin{lemma}[Estimator decomposition]

{
\emergencystretch=2em
Under Assumptions 2.1--2.6, for every applicable \(N_{\mathrm O}\) and every \(t\ge1\), in particular \(1\leq t\leq N_{\mathrm E}\), whenever the displayed terms are integrable, \[\widehat\tau_{t,N_{\mathrm O}}
-
\tau
=
\frac1t
\sum_{s=1}^t
\widetilde m_{s,N_{\mathrm O}}
+
L_{N_{\mathrm O}}^{\rm obs}
+
R_{t,N_{\mathrm O}}^{\rm i}
+
R_{t,N_{\mathrm O}}^{\rm ii}
+
R_{t,N_{\mathrm O}}^{\rm iii},
\]
\par
}

where \[\begin{aligned}
L_{N_{\mathrm O}}^{\rm obs}
&:=
(\mathbb P_{N_{\mathrm O}}^{\rm obs}-P^{\rm obs})
\psi^{\rm obs}(\cdot),
\\
R_{t,N_{\mathrm O}}^{\rm i}
&:=
(\mathbb P_{N_{\mathrm O}}^{\rm obs}-P^{\rm obs})
\left\{
\widehat\psi_{t,N_{\mathrm O}}^{\rm obs}(\cdot)
-
\psi^{\rm obs}(\cdot)
\right\},
\\
R_{t,N_{\mathrm O}}^{\rm ii}
&:=
\left(
P_{\widehat f_{X\mid G=\mathrm O,N_{\mathrm O}}}
-
P_{f_{X\mid G=\mathrm O}}
\right)
\left\{
\widetilde\tau_{N_{\mathrm O}}(\cdot)
-
\overline\tau_{t,N_{\mathrm O}}(\cdot)
\right\},
\\
R_{t,N_{\mathrm O}}^{\rm iii}
&:=
P^{\rm obs}
\left[
\left\{
\widehat\alpha_{t,N_{\mathrm O}}(\cdot)
-
\alpha(\cdot)
\right\}
\left\{
\mu(\cdot)
-
\widehat\mu_{N_{\mathrm O}}(\cdot)
\right\}
\right].
\end{aligned}
\]

\end{lemma}

{
\emergencystretch=2em
The leading experimental term is a martingale average by Lemma 6.1, whereas \(L_{N_{\mathrm O}}^{\rm obs}\) is an i.i.d. empirical average based on the observational sample. The first remainder is the empirical-process error generated by replacing the population observational score with its fitted counterpart. The second is a product of the target-density error and the discrepancy between the fixed-fit population conditional effect and its historical estimate. The third is the product remainder generated by estimation of the surrogate index and the surrogate representer. Section 6.2 states conditions under which the three remainder terms are negligible at the \(N_{\mathrm E}^{-1/2}\) scale and the conditional variance of the experimental martingale average converges to the oracle fixed-design benchmark.
\par
}

\subsection{Limit distribution and attainment of the optimal fixed-design bound}\label{limit-distribution-and-attainment-of-the-optimal-fixed-design-bound}

{
\emergencystretch=2em
We establish asymptotic normality at time \(N_{\mathrm E}\) under the following assumptions and the regularity conditions in Appendix A.6. Those regularity conditions require the actual experimental design to coincide with the plug-in experimental design, the target and experimental covariate laws to have common support, and the true variances and, with probability tending to one, the fitted variance quantities in rounds \(1,\ldots,N_{\mathrm E}\) to be bounded above and away from zero.
\par
}

\setcounter{assumption}{0}
\begin{assumption}[Joint sample growth]

{
\emergencystretch=2em
As \(N_{\mathrm O}\to\infty\), \(N_{\mathrm E}\to\infty\) and \(N_{\mathrm E}/N_{\mathrm O}\to\kappa\in(0,\infty)\).
\par
}

\end{assumption}

{
\emergencystretch=2em
This regime keeps the experimental and observational components on the same \(N_{\mathrm E}^{-1/2}\) scale. It is therefore the normalization under which both variance contributions enter the limiting distribution.
\par
}

\setcounter{assumption}{1}
\begin{assumption}[Time-averaged consistency of the treatment-specific mean and variance estimates]

{
\emergencystretch=2em
For \(w=0,1\), with suprema over the target support, \[\frac1{N_{\mathrm E}}\sum_{t=1}^{N_{\mathrm E}}
\sup_x|\widehat\mu_{w,t,N_{\mathrm O}}(x)-\widetilde\mu_{w,N_{\mathrm O}}(x)|^2
\xrightarrow{p}0,
\qquad w=0,1,
\]
\par
}

\[\frac1{N_{\mathrm E}}\sum_{t=1}^{N_{\mathrm E}}
\sup_x\left|\widehat v_{w,t,N_{\mathrm O}}(x)-
\widetilde{v}_{w,N_{\mathrm O}}(x)\right|
\xrightarrow{p}0,
\qquad w=0,1.
\]

\end{assumption}

Assumption 6.2 allows the nuisance estimates to be inaccurate at the beginning of the experiment and requires their estimation errors to vanish on average over the experiment, so that errors by learning experimental designs do not persist in the asymptotic variance. Consistency suffices for this variance argument; Assumption 6.6 separately imposes the product rates needed to control bias.

\setcounter{assumption}{2}
\begin{assumption}[Martingale moment conditions]

{
\emergencystretch=2em
For every \(N_{\mathrm O}\) and \(1\le t\le N_{\mathrm E}\), \(\mathbb E[\widetilde m_{t,N_{\mathrm O}}^{2}]<\infty\). For some \(r>2\), \(N_{\mathrm E}^{-r/2}\allowbreak\sum_{t=1}^{N_{\mathrm E}}\mathbb E[|\widetilde m_{t,N_{\mathrm O}}|^r\mid\mathcal F_{t-1,N_{\mathrm O}}]\xrightarrow{p}0.\)
\par
}

\end{assumption}

These conditions are imposed to apply the martingale central limit theorem to the experimental component. The second condition is a conditional Lyapunov condition, which implies the conditional Lindeberg condition used in Corollary 3.1 of \citet{Hall1980}. It prevents a few extreme weighted observations from dominating the normalized estimator.

\setcounter{assumption}{3}
\begin{assumption}[Target-density and surrogate-index consistency]

{
\emergencystretch=2em
Assume \(\|\widehat\mu_{N_{\mathrm O}}(\cdot)\|_\infty=O_p(1)\). Also, for \(w=0,1\), \[\begin{aligned}
&\mathbb E_{X\mid G=\mathrm O}
\left[
\left|
\frac{\widehat f_{X\mid G=\mathrm O,N_{\mathrm O}}(X)}
{f_{X\mid G=\mathrm O}(X)}-1
\right|^2
\right]=o_p(1),\\
&\mathbb E_{X\mid G=\mathrm O}
\mathbb E_{S\mid W=w,X,G=\mathrm E}
[\{\widehat\mu_{N_{\mathrm O}}(S,X)-\mu(S,X)\}^2\mid X]
=o_p(1).
\end{aligned}
\]
\par
}

\end{assumption}

{
\emergencystretch=2em
These conditions require the displayed population \(L_2\)-risks of the density and surrogate index estimated from the observational data to converge to zero in probability.
\par
}

\setcounter{assumption}{4}
\begin{assumption}[Observational-score regularity at the terminal time]

{
\emergencystretch=2em
At \(t=N_{\mathrm E}\), \[\|\widehat\psi_{N_{\mathrm E},N_{\mathrm O}}^{\rm obs}(\cdot)-\psi^{\rm obs}(\cdot)\|_{2,{\rm obs}}=o_p(1),
\qquad
\sqrt{N_{\mathrm O}}(\mathbb P_{N_{\mathrm O}}^{\rm obs}-P^{\rm obs})
(\widehat\psi_{N_{\mathrm E},N_{\mathrm O}}^{\rm obs}(\cdot)-\psi^{\rm obs}(\cdot))=o_p(1).
\]
\par
}

{
\emergencystretch=2em
The observational score satisfies \(P^{\rm obs}\psi^{\rm obs}(\cdot)=\tau\) and \(\mathbb E^{\rm obs}[(\psi^{\rm obs}(X,S,Y)-\tau)^2]=\allowbreak V^{\rm obs}<\infty\).
\par
}

\end{assumption}

{
\emergencystretch=2em
The first condition requires consistency of the fitted observational score in the population \(L_2\) norm. The second makes the corresponding empirical-process error asymptotically negligible. A Donsker condition, see, for example, \citet{vanderVaart1998}, is sufficient.
\par
}

\setcounter{assumption}{5}
\begin{assumption}[Product-rate conditions at the terminal time]

\[\sqrt{N_{\mathrm O}}
\left\|
\frac{\widehat f_{X\mid G=\mathrm O,N_{\mathrm O}}(\cdot)}
{f_{X\mid G=\mathrm O}(\cdot)}-1
\right\|_{2,X\mid G=\mathrm O}
\|\widetilde\tau_{N_{\mathrm O}}(\cdot)-\overline\tau_{N_{\mathrm E},N_{\mathrm O}}(\cdot)\|_{2,X\mid G=\mathrm O}
=o_p(1).
\]

\[\sqrt{N_{\mathrm O}}\|\widehat\mu_{N_{\mathrm O}}(\cdot)-\mu(\cdot)\|_{2,{\rm obs}}
\|\widehat\alpha_{N_{\mathrm E},N_{\mathrm O}}(\cdot)-\alpha(\cdot)\|_{2,{\rm obs}}
=o_p(1).
\]

\end{assumption}

{
\emergencystretch=2em
The first product makes the difference between the fitted target law and the target law negligible at the \(N_{\mathrm E}^{-1/2}\) scale. The second prevents the fitted surrogate representer and surrogate index in the observational population from contributing a first-order bias at time \(N_{\mathrm E}\) . The two nuisance functions need not be estimated at the same rate. Faster estimation of one can allow a slower rate for the other. Such product-rate conditions arise in the canonical debiased machine learning framework of \citet{Chernozhukov2018}. Under the preceding assumptions, the estimator has the following limiting distribution.
\par
}

\setcounter{theorem}{0}
\begin{theorem}[Asymptotic normality and attainment of the optimal fixed-design bound]

Under Assumptions 2.1--2.6, Assumptions 4.1--4.3, Assumptions 6.1--6.6, and the Section 6 conditions in Appendix Assumption A.6.1(i)--(ix), \[\sqrt{N_{\mathrm E}}(\widehat\tau_{N_{\mathrm E},N_{\mathrm O}}-\tau)
\rightsquigarrow
\mathcal N(0,V^{{\rm exp},\star}+\kappa V^{\rm obs}).
\]

\end{theorem}

{
\emergencystretch=2em
Theorem 6.1 shows that learning the experimental design during the experiment does not increase the limiting variance. The limiting variance is the bound of Theorem 3.1 evaluated at the optimal experimental design of Theorem 3.2, that is, the variance that an oracle procedure knowing this experimental design in advance would attain. In the decomposition of Lemma 6.2, the three remainder terms are negligible at the \(N_{\mathrm E}^{-1/2}\) scale by Assumptions 6.5 and 6.6 and \(N_{\mathrm E}/N_{\mathrm O}\to\kappa\). Appendix A.6 proves the result by applying a single scalar martingale triangular-array central limit theorem to the sum of the observational and experimental terms, ordered with the observational records first.
\par
}

Supplementary Material 2 gives sufficient conditions for Assumptions 6.2, 6.4, and 6.6 under correctly specified finite-dimensional models.

\section{Asymptotic time-uniform inference}\label{section-7-asymptotic-time-uniform-inference}

This section develops an asymptotic confidence sequence for sequential monitoring of the target effect and establishes asymptotic simultaneous coverage from a prescribed monitoring time onward. Section 7.1 constructs the confidence sequence, Section 7.2 states the conditions for its coverage, and Section 7.3 establishes time-uniform coverage and its implication for data-dependent stopping.

\subsection{Monitoring framework and confidence-sequence construction}\label{monitoring-framework-and-confidence-sequence-construction}

{
\emergencystretch=2em
This subsection constructs the confidence sequence. Throughout this section, the rounds of Algorithm~\ref{alg:adaptive-sequential-estimation} continue without a terminal round, \(t=1,2,\ldots\), so that the estimator and the variance estimators below are defined at every round. We begin with the variance quantities that determine its width.
\par
}

{
\emergencystretch=2em
For \(t\ge1\), define the average conditional variance of the experimental martingale increments by \(V_{t,N_{\mathrm O}}^{\rm exp}:=\allowbreak t^{-1}\allowbreak\sum_{s=1}^{t}\mathbb E[\widetilde m_{s,N_{\mathrm O}}^{2}\mid\mathcal F_{s-1,N_{\mathrm O}}]\). Its feasible estimator is \[\begin{aligned}
\widehat V_{t,N_{\mathrm O}}^{\rm exp}
&:=
\frac1t
\sum_{s=1}^{t}
\Biggl[
\widehat\phi_{s,N_{\mathrm O}}^{\rm exp}(S_s,W_s,X_s)
+
P_{\widehat f_{X\mid G=\mathrm O,N_{\mathrm O}}}
\widehat\tau_{s,N_{\mathrm O}}(\cdot)\\
&\qquad-
\frac1t\sum_{\ell=1}^{t}
\left\{
\widehat\phi_{\ell,N_{\mathrm O}}^{\rm exp}(S_\ell,W_\ell,X_\ell)
+
P_{\widehat f_{X\mid G=\mathrm O,N_{\mathrm O}}}
\widehat\tau_{\ell,N_{\mathrm O}}(\cdot)
\right\}
\Biggr]^2,
\end{aligned}
\]
\par
}

and the observational variance estimator is \[\widehat V_{t,N_{\mathrm O}}^{\rm obs}
:=
\frac1{N_{\mathrm O}}\sum_{i=1}^{N_{\mathrm O}}
\left[
\widehat\psi_{t,N_{\mathrm O}}^{\rm obs}(X_i,S_i,Y_i)
-
\frac1{N_{\mathrm O}}\sum_{j=1}^{N_{\mathrm O}}
\widehat\psi_{t,N_{\mathrm O}}^{\rm obs}(X_j,S_j,Y_j)
\right]^2.
\]

{
\emergencystretch=2em
The two variance estimators account for uncertainty from the experimental and observational data, respectively. Since \(\widetilde m_{s,N_{\mathrm O}}\) has conditional mean zero given \(\mathcal F_{s-1,N_{\mathrm O}}\), \(\mathbb E[\widetilde m_{s,N_{\mathrm O}}^{2}\mid\mathcal F_{s-1,N_{\mathrm O}}]=\allowbreak \operatorname{Var}(\widetilde m_{s,N_{\mathrm O}}\mid\mathcal F_{s-1,N_{\mathrm O}})\). Thus, \(V_{t,N_{\mathrm O}}^{\rm exp}\) is the average of the conditional variances over the first \(t\) rounds. The experimental estimator \(\widehat V_{t,N_{\mathrm O}}^{\rm exp}\) is constructed to estimate this average conditional variance. The observational estimator \(\widehat V_{t,N_{\mathrm O}}^{\rm obs}\) is the empirical variance of the fitted observational scores evaluated at time \(t\) and estimates \(V^{\rm obs}\), the variance of the true observational score. Appendix A.7 establishes the uniform consistency of \(\widehat V_{t,N_{\mathrm O}}^{\rm exp}\) for \(V_{t,N_{\mathrm O}}^{\rm exp}\) and of \(\widehat V_{t,N_{\mathrm O}}^{\rm obs}\) for \(V^{\rm obs}\) uniformly over the rounds \(t\ge t_{0,N_{\mathrm O}}\), where the monitoring start \(t_{0,N_{\mathrm O}}\) is introduced below.
\par
}

{
\emergencystretch=2em
We next introduce two tuning quantities for constructing the confidence sequence. The monitoring start \(t_{0,N_{\mathrm O}}\) is a prespecified deterministic positive integer sequence that gives the round from which the intervals are reported. The tuning variable \(\omega_{N_{\mathrm O}}>0\) of the boundary of the intervals defined below is chosen once at round \(t_{0,N_{\mathrm O}}\) from the data available at that round and targets the round at which the boundary is narrowest relative to the pointwise standard error.
\par
}

\setcounter{assumption}{0}
\begin{assumption}[Monitoring start and tuning variable]

{
\emergencystretch=2em
For a fixed constant \(c_\omega\in(0,\infty)\), the following hold. (1) \(t_{0,N_{\mathrm O}}\rightarrow\infty\) and \(t_{0,N_{\mathrm O}}/N_{\mathrm O}\rightarrow0\). (2) \(N_{\mathrm O}\omega_{N_{\mathrm O}}\xrightarrow{p}c_\omega\).
\par
}

\end{assumption}

{
\emergencystretch=2em
Condition (1) concerns the monitoring start, and condition (2) concerns the tuning variable. Condition (1) lets the monitoring start diverge, so that the nuisance fits and the variance estimators rest on many rounds when the first interval is reported, while keeping it negligible relative to \(N_{\mathrm O}\). In particular, condition (1) adapts to the present setting the asymptotic framework of Definition 2.7 of \citet{WaudbySmith2024}.
\par
}

{
\emergencystretch=2em
Condition (2) requires \(N_{\mathrm O}\) times the tuning variable to converge in probability to a positive constant. Under it and the consistency of the variance estimators, \(\sqrt{N_{\mathrm O}}\) times the half-width defined below converges, as a function of the normalized time \(t/N_{\mathrm O}\), to a positive finite limit, and for given \(V^{\rm obs}\) and \(V^{{\rm exp},\star}\) the constant \(c_\omega\) determines this limit. Remark 7.1 gives a choice of the tuning variable that satisfies condition (2), and Remark 7.2 discusses the choice of the monitoring start. Condition (2) plays the role of the choice of the tuning parameter in Theorem 2.8 of \citet{WaudbySmith2024}.
\par
}

{
\emergencystretch=2em
With these quantities, we construct the confidence sequence as follows. Fix \(\alpha\in(0,1)\). The two variance estimators combine into the estimated variance of the estimator at round \(t\), \(\widehat\sigma_{t,N_{\mathrm O}}^2:=\allowbreak \widehat V_{t,N_{\mathrm O}}^{\rm exp}/\allowbreak t+\widehat V_{t,N_{\mathrm O}}^{\rm obs}/\allowbreak N_{\mathrm O}\), whose first term reflects the experimental average over \(t\) rounds and whose second term reflects the observational sample. At every round \(t\ge t_{0,N_{\mathrm O}}\), the proposed asymptotic time-uniform confidence sequence reports the interval \[C_{t,N_{\mathrm O}}(\alpha):=\left[\widehat\tau_{t,N_{\mathrm O}}\mathbin{\pm}\overline B_{t,N_{\mathrm O}}\right],
\qquad
\overline B_{t,N_{\mathrm O}}:=\widehat\sigma_{t,N_{\mathrm O}}\,h_\alpha\!\left(\frac{\omega_{N_{\mathrm O}}}{\widehat\sigma_{t,N_{\mathrm O}}^2}\right),
\]
\par
}

{
\emergencystretch=2em
where \(h_\alpha(x):=[\{(x+1)/x\}\log\{(x+1)/\alpha^2\}]^{1/2}\) for \(x>0\). If \(\widehat\sigma_{t,N_{\mathrm O}}^2=0\), we set \(\overline B_{t,N_{\mathrm O}}:=\infty\) and \(C_{t,N_{\mathrm O}}(\alpha):=\mathbb R\).
\par
}

{
\emergencystretch=2em
Each interval is centered at the current estimate, and its half-width is the pointwise standard error \(\widehat\sigma_{t,N_{\mathrm O}}\) multiplied by the factor \(h_\alpha(\omega_{N_{\mathrm O}}/\widehat\sigma_{t,N_{\mathrm O}}^2)\), which accounts for time-uniform monitoring. The factor comes from the two-sided normal mixture boundary of \citet{Howard2021}, stated in Appendix A.7.3. That boundary follows from Ville\textquotesingle s inequality, by which a nonnegative martingale with initial value one reaches \(1/\alpha\) at some time with probability at most \(\alpha\). Asymptotically, the estimation error \(\widehat\tau_{t,N_{\mathrm O}}-\tau\) divided by its variance behaves, over the monitoring rounds, like a standard Brownian motion evaluated at the reciprocal of this variance. For a Brownian motion, the Gaussian mixture of its exponential martingales with mixing variance \(\omega_{N_{\mathrm O}}\) is such a martingale. Ville\textquotesingle s inequality applied to it shows that, with probability at least \(1-\alpha\), the Brownian motion divided by the time stays at every time within its standard deviation multiplied by \(h_\alpha\) evaluated at the product of \(\omega_{N_{\mathrm O}}\) and the time. At time \(1/\widehat\sigma_{t,N_{\mathrm O}}^2\), this bound equals \(\widehat\sigma_{t,N_{\mathrm O}}h_\alpha(\omega_{N_{\mathrm O}}/\allowbreak \widehat\sigma_{t,N_{\mathrm O}}^2)=\allowbreak \overline B_{t,N_{\mathrm O}}\). Remark 7.1 below explains how the tuning variable shapes this factor over the monitoring period, and Appendix A.7.3.1 gives the details.
\par
}

{
\emergencystretch=2em
This confidence sequence permits data-dependent stopping. The experimenter can inspect the interval after every round from \(t_{0,N_{\mathrm O}}\) onward, and Theorem 7.1 shows that the intervals contain \(\tau\) at all of these rounds simultaneously with asymptotic probability at least \(1-\alpha\). Corollary 7.1 then shows that if the experimenter stops at a round that is chosen from the accumulated data, is no earlier than \(t_{0,N_{\mathrm O}}\), and is finite almost surely, such as the earlier of the first round at which the interval excludes zero and a prespecified maximum number of rounds, the interval reported at that round contains \(\tau\) with asymptotic probability at least \(1-\alpha\).
\par
}

\setcounter{remark}{0}
\begin{remark}[Choosing the tuning variable from a target round]

This remark describes how the tuning variable can be chosen. Theorem 7.1 holds for every tuning variable that is computed from the data up to the monitoring start and satisfies Assumption 7.1(2), so its coverage guarantee holds whichever such tuning variable is chosen. The choice determines only the round whose half-width the tuning variable minimizes.

{
\emergencystretch=2em
Fixing a target round determines the tuning variable that minimizes the half-width at that round. The half-width \(\overline B_{t,N_{\mathrm O}}\) is the product of the pointwise standard error \(\widehat\sigma_{t,N_{\mathrm O}}\) and the inflation factor \(h_\alpha(\omega_{N_{\mathrm O}}/\widehat\sigma_{t,N_{\mathrm O}}^2)\) associated with time-uniform monitoring, and only the inflation factor depends on the tuning variable. The function \(h_\alpha\) attains its minimum over \((0,\infty)\) only at \(x^\star\), the unique positive solution of \(x^\star=\log\{(1+x^\star)/\alpha^2\}\), which depends only on \(\alpha\).
\par
}

{
\emergencystretch=2em
Fix a prespecified deterministic target round \(t_{N_{\mathrm O}}^\star\) with \(t_{N_{\mathrm O}}^\star/N_{\mathrm O}\) converging to a positive constant. Because \(h_\alpha(\omega_{N_{\mathrm O}}/\widehat\sigma_{t,N_{\mathrm O}}^2)\), as a function of \(\omega_{N_{\mathrm O}}>0\), attains its minimum only at \(\omega_{N_{\mathrm O}}=x^\star\widehat\sigma_{t,N_{\mathrm O}}^2\), the half-width at round \(t=t_{N_{\mathrm O}}^\star\) is minimized over the tuning variable by the unique value \(x^\star\widehat\sigma_{t_{N_{\mathrm O}}^\star,N_{\mathrm O}}^2\). The estimated variance \(\widehat\sigma_{t_{N_{\mathrm O}}^\star,N_{\mathrm O}}^2\) is not available at the monitoring start, so we replace it by a prediction. Let \(V\in(0,\infty)\) be a planning value of \(V^{{\rm exp},\star}\) that is computed from the data up to the monitoring start, and set \[\omega_{N_{\mathrm O}}
:=
x^\star\left(\frac{V}{t_{N_{\mathrm O}}^\star}+\frac{\widehat V_{t_{0,N_{\mathrm O}},N_{\mathrm O}}^{\rm obs}}{N_{\mathrm O}}\right).
\]
\par
}

{
\emergencystretch=2em
A natural planning value is the experimental variance estimate at the monitoring start, \(V=\widehat V_{t_{0,N_{\mathrm O}},N_{\mathrm O}}^{\rm exp}\), which Section 8.4 uses. This tuning variable satisfies Assumption 7.1(2) whenever the planning value converges in probability to a positive constant \(\overline V\). For \(V=\widehat V_{t_{0,N_{\mathrm O}},N_{\mathrm O}}^{\rm exp}\), this requires that the experimental variance estimate at the monitoring start converges in probability to a positive constant. Under this condition, the planning value affects only how accurately the tuning variable minimizes the half-width at the target round. With the variance estimates replaced by their limits, \(\omega_{N_{\mathrm O}}\) equals \(x^\star\) times the variance \(V^{{\rm exp},\star}/t+V^{\rm obs}/N_{\mathrm O}\) of the estimator at round \(t=t_{N_{\mathrm O}}^\star V^{{\rm exp},\star}/\overline V\), so the tuning variable minimizes the half-width at that round. If \(\overline V\) exceeds \(V^{{\rm exp},\star}\), this round comes before the target round, and if \(\overline V\) is below \(V^{{\rm exp},\star}\), the round comes after it. Therefore, when the planning value is computed from the data up to the monitoring start and converges in probability to a positive constant, any deviation of its limit from \(V^{{\rm exp},\star}\) shifts this round away from the target round and leaves the coverage guarantee of Theorem 7.1 unchanged. Section 8.4 and Appendices B.2 and B.3 report results for several target rounds and planning values.
\par
}

\end{remark}

\subsection{Conditions for time-uniform coverage}\label{conditions-for-time-uniform-coverage}

This subsection states the remaining conditions under which the confidence sequence of Section 7.1 has asymptotic time-uniform coverage. The corresponding regularity conditions are given in Assumption A.6.1. They retain the plug-in sampling law and common supports, including the support of the fitted target law.

\setcounter{assumption}{1}
\begin{assumption}[Uniform martingale moment conditions]

{
\emergencystretch=2em
For some \(r>4\) and a finite deterministic constant \(C\), \(\mathbb E[|\widetilde m_{t,N_{\mathrm O}}|^r\mid\mathcal F_{t-1,N_{\mathrm O}}] \le C\) almost surely for every \(N_{\mathrm O}\) and \(t\ge1\).
\par
}

\end{assumption}

{
\emergencystretch=2em
Assumption 7.2 is sufficient for applying the martingale maximal inequality in Corollary 2.1 of \citet{Hall1980}, which the appendix uses for both partial sums of experimental scores and the variance estimator formed from their squares. Uniformity over \(N_{\mathrm O}\) and \(t\) controls square integrability, the jumps of the triangular martingale process, the feasible-variance error, and the infinite monitoring tail. It also implies the Lyapunov condition at time \(N_{\mathrm E}\) in Assumption 6.3 because \(N_{\mathrm E}^{-r/2} \allowbreak\sum_{t=1}^{N_{\mathrm E}} \mathbb E[|\widetilde m_{t,N_{\mathrm O}}|^r\mid\mathcal F_{t-1,N_{\mathrm O}}] \le CN_{\mathrm E}^{1-r/2} \longrightarrow0.\)
\par
}

\setcounter{assumption}{2}
\begin{assumption}[Uniform consistency of the treatment-specific mean and variance estimates]

{
\emergencystretch=2em
The bounded fitted-surrogate-index condition and both initial risk conditions in Assumption 6.4 are imposed. For \(w=0,1\), the treatment-specific mean and variance estimates satisfy \[\sup_{t\ge t_{0,N_{\mathrm O}}}
\sup_x
|\widehat\mu_{w,t,N_{\mathrm O}}(x)-\widetilde\mu_{w,N_{\mathrm O}}(x)|
=o_p(1)
\]
\par
}

\[\sup_{t\ge t_{0,N_{\mathrm O}}}
\sup_x
|\widehat v_{w,t,N_{\mathrm O}}(x)-\widetilde v_{w,N_{\mathrm O}}(x)|
=o_p(1).
\]

\end{assumption}

{
\emergencystretch=2em
The convergence requirements in Assumptions 6.2 and 7.3 control different time ranges and do not imply one another without additional restrictions. Assumption 6.2 controls average error through \(N_{\mathrm E}\), whereas Assumption 7.3 controls the largest error over all rounds after monitoring begins. Assumption 7.3 is therefore a uniform convergence requirement for nuisance estimates fitted using adaptively collected data.
\par
}

\setcounter{assumption}{3}
\begin{assumption}[Uniform observational-score regularity]

{
\emergencystretch=2em
As \(N_{\mathrm O}\to\infty\), \[\begin{aligned}
\sup_{t\ge t_{0,N_{\mathrm O}}}
\|\widehat\psi_{t,N_{\mathrm O}}^{\rm obs}(\cdot)-\psi^{\rm obs}(\cdot)\|_{2,{\rm obs}}
&=o_p(1),\\
\sqrt{N_{\mathrm O}}
\sup_{t\ge t_{0,N_{\mathrm O}}}
\left|
(\mathbb P_{N_{\mathrm O}}^{\rm obs}-P^{\rm obs})
\{\widehat\psi_{t,N_{\mathrm O}}^{\rm obs}(\cdot)-\psi^{\rm obs}(\cdot)\}
\right|
&=o_p(1),\\
\sup_{t\ge t_{0,N_{\mathrm O}}}
\left|
(\mathbb P_{N_{\mathrm O}}^{\rm obs}-P^{\rm obs})
\left[
\{\widehat\psi_{t,N_{\mathrm O}}^{\rm obs}(\cdot)-\psi^{\rm obs}(\cdot)\}^2
\right]
\right|
&=o_p(1).
\end{aligned}
\]
\par
}

\end{assumption}

{
\emergencystretch=2em
A sufficient set of conditions is that the first displayed convergence holds and, with probability tending to one, the true observational score and the entire fitted score path \(\{\widehat\psi_{t,N_{\mathrm O}}^{\rm obs}(\cdot):t\geq t_{0,N_{\mathrm O}}\}\) are contained in a fixed \(P^{\rm obs}\)-Donsker class \(\mathcal G\) with a square-integrable envelope. Suppose further that the class \(\{(g-\psi^{\rm obs})^2:g\in\mathcal G\}\) is \(P^{\rm obs}\)-Glivenko--Cantelli. Asymptotic equicontinuity of the empirical process then gives the second displayed convergence, while the uniform law of large numbers for the squared-difference class gives the third.
\par
}

\setcounter{assumption}{4}
\begin{assumption}[Uniform product-rate conditions]

The target-density remainder satisfies \[\sqrt{N_{\mathrm O}}
\left\|
\frac{\widehat f_{X\mid G=\mathrm O,N_{\mathrm O}}(\cdot)}
{f_{X\mid G=\mathrm O}(\cdot)}-1
\right\|_{2,X\mid G=\mathrm O}
\sup_{t\ge t_{0,N_{\mathrm O}}}
\|\widetilde\tau_{N_{\mathrm O}}(\cdot)-\overline\tau_{t,N_{\mathrm O}}(\cdot)\|_{2,X\mid G=\mathrm O}
=o_p(1).
\]

The representer and outcome-regression errors satisfy \[\sqrt{N_{\mathrm O}}
\|\widehat\mu_{N_{\mathrm O}}(\cdot)-\mu(\cdot)\|_{2,{\rm obs}}
\sup_{t\ge t_{0,N_{\mathrm O}}}
\|\widehat\alpha_{t,N_{\mathrm O}}(\cdot)-\alpha(\cdot)\|_{2,{\rm obs}}
=o_p(1).
\]

\end{assumption}

{
\emergencystretch=2em
The first product rate controls \(R_{t,N_{\mathrm O}}^{\rm ii}\), and the second controls \(R_{t,N_{\mathrm O}}^{\rm iii}\) by Cauchy--Schwarz. Assumption 7.5 requires the same \(\sqrt{N_{\mathrm O}}\) product-rate scale as Assumption 6.6, uniformly over the monitoring period.
\par
}

Supplementary Material 2 gives sufficient conditions for Assumptions 7.3 and 7.5 under correctly specified finite-dimensional models.

\setcounter{remark}{1}
\begin{remark}[Monitoring start and nuisance-rate trade-off]

{
\emergencystretch=2em
Maintain all conditions of Theorem 7.1 except for the second product-rate condition in Assumption 7.5. Suppose that, for some constants \(a,b>0\), \(\left\|\widehat\mu_{N_{\mathrm O}}(\cdot)-\mu(\cdot)\right\|_{2,\mathrm{obs}}=\allowbreak O_p(N_{\mathrm O}^{-a})\) and \(\sup_{t\geq t_{0,N_{\mathrm O}}}\allowbreak\left\|\widehat\alpha_{t,N_{\mathrm O}}(\cdot)-\alpha(\cdot)\right\|_{2,\mathrm{obs}}=\allowbreak O_p(t_{0,N_{\mathrm O}}^{-b}).\) Suppose further that \(t_{0,N_{\mathrm O}}\asymp N_{\mathrm O}^\gamma\) for some fixed \(\gamma\in(0,1)\). If \(a+\gamma b>1/2,\) then the conclusion of Theorem 7.1 holds. This makes explicit how the monitoring start interacts with nuisance-estimation accuracy. Under these separate polynomial rate bounds, monitoring must begin sufficiently late for the representer estimator to reach the accuracy required by the uniform remainder bound. A larger \(\gamma\) weakens the requirement on \(b\). One example of a practical choice is \(t_{0,N_{\mathrm O}}=\lceil N_{\mathrm O}^{3/4}\rceil\), which corresponds to \(\gamma=3/4\) and equals \(595\) when \(N_{\mathrm O}=5000\), close to the time \(500\) from which the simulation of a randomized experiment in Section 4 of \citet{WaudbySmith2024} computes widths and miscoverage rates.
\par
}

\end{remark}

\subsection{Time-uniform coverage and data-dependent stopping}\label{time-uniform-coverage-and-data-dependent-stopping}

These assumptions yield simultaneous coverage over the entire monitoring period.

\setcounter{theorem}{0}
\begin{theorem}[Asymptotic time-uniform coverage]

{
\emergencystretch=2em
Fix \(\alpha\in(0,1)\). Construct \(C_{t,N_{\mathrm O}}(\alpha)\) as in Section 7.1 using a deterministic sequence \(t_{0,N_{\mathrm O}}\) and an \(\mathcal F_{t_{0,N_{\mathrm O}},N_{\mathrm O}}\)-measurable tuning variable \(\omega_{N_{\mathrm O}}\) that satisfy Assumption 7.1. Suppose that the identification and experimental design assumptions of Sections 2--4, Assumptions 7.1--7.5, and the common and Section 7 regularity conditions in Assumption A.6.1 hold along this sequence. Then \[\liminf_{N_{\mathrm O}\to\infty}
\Pr\!\left\{
\tau\in C_{t,N_{\mathrm O}}(\alpha)
\text{ for all }t\ge t_{0,N_{\mathrm O}}
\right\}
\ge1-\alpha.
\]
\par
}

\end{theorem}

{
\emergencystretch=2em
Theorem 7.1 controls the probability that at least one interval fails to contain \(\tau\) after monitoring begins. Its asymptotic upper limit is at most \(\alpha\), even when the interval is inspected after every round and no finite monitoring horizon is specified. This is stronger than coverage at each fixed time separately and permits the experimenter to choose when to stop using the data accumulated during the experiment. Constructing intervals whose asymptotic simultaneous coverage equals the nominal level exactly is left for future research. Appendix A.7 gives the complete proof.
\par
}

\setcounter{corollary}{0}
\begin{corollary}[Coverage at data-dependent stopping times]

{
\emergencystretch=2em
Under the assumptions of Theorem 7.1, let \((T_{\mathrm{stop},N_{\mathrm O}})_{N_{\mathrm O}\ge1}\) be any sequence of integer-valued stopping times with respect to \((\mathcal F_{t,N_{\mathrm O}})_{t\ge0}\) satisfying \(t_{0,N_{\mathrm O}} \le T_{\mathrm{stop},N_{\mathrm O}} <\infty\) almost surely for every \(N_{\mathrm O}\). Then \(\liminf_{N_{\mathrm O}\to\infty} \Pr\!\allowbreak\left\{ \tau\in C_{T_{\mathrm{stop},N_{\mathrm O}},N_{\mathrm O}}(\alpha) \right\} \ge1-\alpha.\)
\par
}

\end{corollary}

{
\emergencystretch=2em
This result permits stopping once sufficient evidence has accumulated after monitoring begins while retaining the stated asymptotic coverage guarantee. For example, take the benchmark to be zero and fix a deterministic integer cap \(N_{\mathrm E}\ge t_{0,N_{\mathrm O}}\). Define \(T_{\mathrm{stop},N_{\mathrm O}}:=\allowbreak \min\allowbreak\left\{\inf\left\{t\ge t_{0,N_{\mathrm O}}:0\notin C_{t,N_{\mathrm O}}(\alpha)\right\},N_{\mathrm E}\right\}\), where the infimum is over integer times and \(\inf\varnothing=\infty\). This rule stops when the interval first excludes zero or when the maximum experimental sample size is reached. Because it is almost surely finite, Corollary 7.1 applies to the interval reported at that time. Appendix A.7.2 proves Corollary 7.1.
\par
}

\section{Monte Carlo simulations}\label{section-8-monte-carlo-simulations}

This section examines, in a simulated coupon experiment, whether designing the experiment around the long-term surrogate index, rather than around a short-term outcome, improves estimation of and inference on the long-term average treatment effect. Section 8.1 describes the data-generating process, Section 8.2 the compared experimental designs and their implementation, Section 8.3 the learned experimental designs and the precision of the long-term effect estimator, and Section 8.4 sequential monitoring and data-dependent stopping. Supplementary Material 1 gives the details of the computation, and Appendix B reports additional results.

\subsection{Data-generating process}\label{data-generating-process}

{
\emergencystretch=2em
This subsection specifies the data-generating process, a stylized benchmark motivated by a coupon experiment. The covariate \(X\) is a normalized score of prior usage, and the outcomes \(S^{(1)}\), \(S^{(2)}\), and \(Y\) represent immediate purchases, subsequent short-term repeat purchases, and long-term cumulative purchases. The observational sample records \((X,S^{(1)},S^{(2)},Y)\) for \(N_{\mathrm O}=5{,}000\) customers, whose outcomes are generated by a latent treatment label that is independent of the potential outcomes and \(X\) and equals each treatment state with probability \(1/2\). The experiment records \((X,W,S^{(1)}(W),S^{(2)}(W))\) for up to \(N_{\mathrm E}=5{,}000\) participants under the covariate density and assignment probability selected by each experimental design, and every experimental design uses the same estimator of the long-term effect. We use 2,000 Monte Carlo replications and report fixed-time performance at \(t=1{,}000,2{,}500,5{,}000\).
\par
}

{
\emergencystretch=2em
In the target population, \(X\mid G=\mathrm O\sim N(0,1)\). For each treatment state \(w\in\{0,1\}\), let \(S^{(1)}(w)\) and \(S^{(2)}(w)\) denote the immediate-purchase and short-term repeat-purchase potential outcomes, and let \(v_{j,w}^S(x)\) denote the conditional variance of \(S^{(j)}(w)\) given \(X=x\). The conditional variance functions are \[\begin{aligned}
v_{1,0}^S(x)
&=v_{2,0}^S(x)
=1+3x^2,\\
v_{1,1}^S(x)
&=1+3x^2+3x,\\
v_{2,1}^S(x)
&=1+3x^2-3x.
\end{aligned}
\]
\par
}

{
\emergencystretch=2em
The control-arm variance \(1+3x^2\) is U-shaped, so customers with unusually low or unusually high prior usage are more heterogeneous. Treatment redistributes short-term variation between the two outcomes, adding \(3x\) to the variance of \(S^{(1)}\) and subtracting \(3x\) from that of \(S^{(2)}\). Among high-usage customers, a coupon makes immediate purchases more variable and subsequent repeat purchases less variable, and the reverse holds among low-usage customers. Because \(\mathbb E[X\mid G=\mathrm O]=0\) and \(\mathbb E[X^2\mid G=\mathrm O]=1\), \(\mathbb E[v_{j,w}^S(X)\mid G=\mathrm O]=4\) for \(j=1,2\) and \(w=0,1\), so the construction changes the allocation of conditional variance across covariates without making one short-term outcome or treatment arm noisier on average.
\par
}

{
\emergencystretch=2em
The short-term potential outcomes are conditionally independent given \(X\), and they and the long-term outcome are generated as \[\begin{aligned}
&S^{(1)}(w)\mid X=x
\sim N\bigl(x+w,\ v_{1,w}^S(x)\bigr),\\
&S^{(2)}(w)\mid X=x
\sim N\bigl(x,\ v_{2,w}^S(x)\bigr),\\
&Y(w)
=S^{(1)}(w)+4S^{(2)}(w)+X+\nu_w,
\end{aligned}
\]
\par
}

{
\emergencystretch=2em
where \(\nu_w\sim N(0,64)\) is independent of \(X\) and of the short-term potential outcomes. Its variance implies that, for each \(w\), the surrogates \(S^{(1)}(w)\) and \(S^{(2)}(w)\) and the covariate \(X\) explain 61.9\% of the variance of \(Y(w)\) in the target population, and \(\nu_w\) accounts for the remaining 38.1\%. Treatment raises the mean of immediate purchases by one and leaves the mean of repeat purchases unchanged. Therefore, \(\mathbb E[Y(1)-Y(0)\mid X=x]=1\) for every \(x\), and the long-term average treatment effect is \(\tau=1\).
\par
}

The long-term effect is small relative to the variability of the outcome, as in many marketing field experiments. Both long-term potential outcomes have variance 168 in the target population, so the effect equals about 0.08 standard deviations. Individual purchases vary widely in such experiments, and \citet{Lewis2015} report that the standard deviation of individual sales in large retail advertising experiments is typically about ten times its mean. Expressed in the same units, field experiments on coupons and discounts imply effects of a similar order. Additional coupons raise purchases over ten weeks by about 0.10 standard deviations in \citet{Huang2024}, a 7-dollar coupon raises revenue per customer during a one-week campaign by about 0.09 standard deviations in \citet{Gopalakrishnan2021}, and emailed discount offers raise spending while the offers are valid by about 0.03 standard deviations in \citet{Sahni2017}. The calibrated effect lies within this range.

The long-term surrogate index learned from the observational data is \[\mu(s,x)
=s^{(1)}+4s^{(2)}+x.
\]

{
\emergencystretch=2em
The corresponding conditional variance of the long-term surrogate index is \(v_w^{\mathrm{LT}}(x) =\operatorname{Var}\{\mu(S(w),X)\mid X=x\}.\) Under the data-generating process, these conditional variances are \[\begin{aligned}
v_0^{\mathrm{LT}}(x)
&=v_{1,0}^S(x)+16v_{2,0}^S(x)
=17+51x^2,\\
v_1^{\mathrm{LT}}(x)
&=v_{1,1}^S(x)+16v_{2,1}^S(x)
=17+51x^2-45x.
\end{aligned}
\]
\par
}

{
\emergencystretch=2em
These expressions make the difference between the short-term and long-term design criteria explicit. For \(x>0\), \(v_{1,1}^S(x)>v_{1,0}^S(x)\), whereas \(v_{2,1}^S(x)<v_{2,0}^S(x)\) and \(v_1^{\mathrm{LT}}(x)<v_0^{\mathrm{LT}}(x)\). These inequalities reverse for \(x<0\). The \(S^{(1)}\) and surrogate-index criteria therefore move assignment probabilities in opposite directions within the same covariate strata. The \(S^{(2)}\) criterion is closer to the surrogate-index criterion because \(S^{(2)}\) receives the larger weight in the long-term surrogate index. Figure 8.1 plots the conditional variances used by the four criteria compared below.
\par
}

\begin{figure}[tbp]
\centering
\begin{minipage}[t]{.75\linewidth}
\centering
\includegraphics[width=\linewidth]{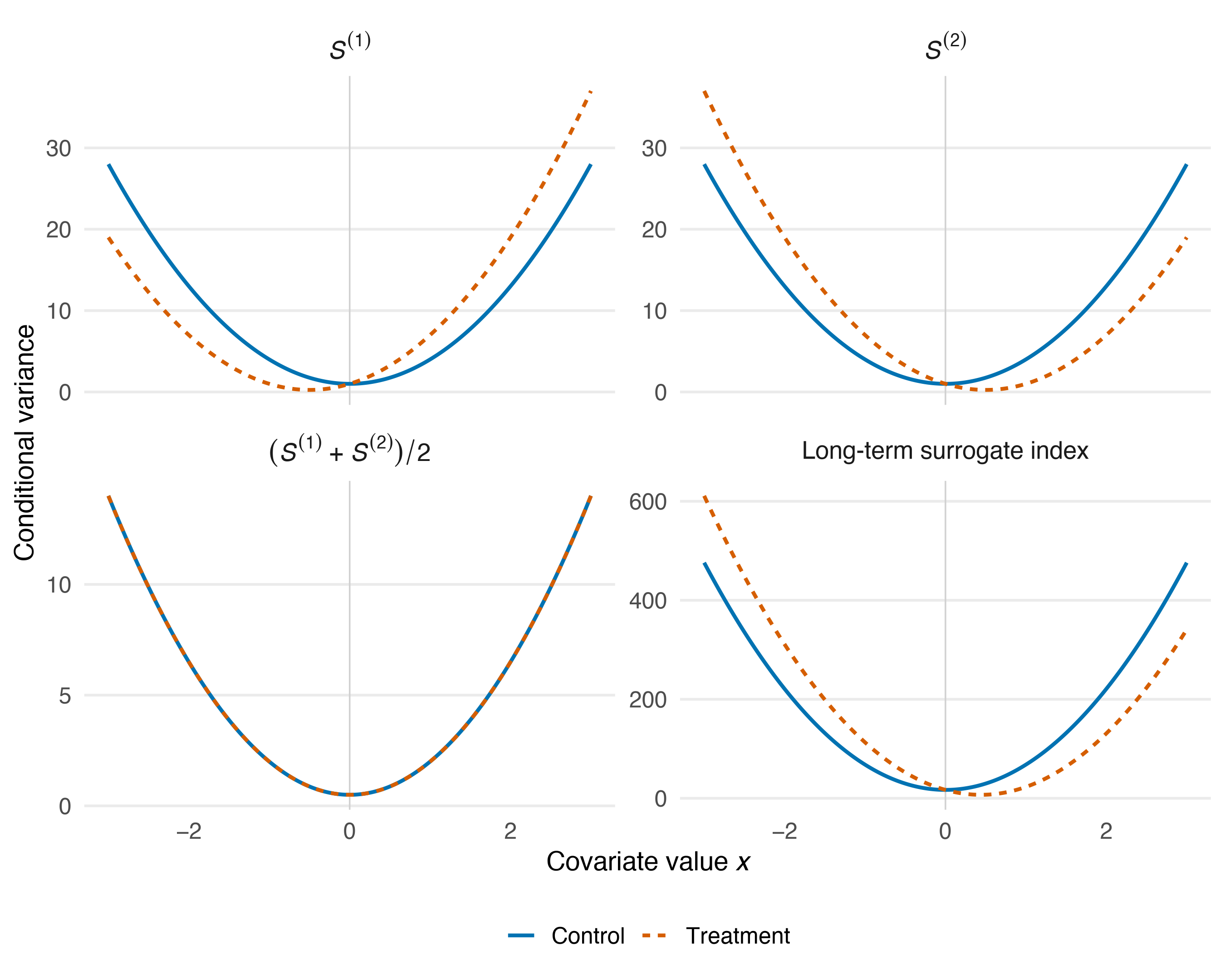}
\end{minipage}
\par\smallskip
\caption{Conditional variances of the two short-term outcomes, their average, and the long-term surrogate index in the control and treatment arms. The panels show the conditional variances of \(S^{(1)}\), \(S^{(2)}\), \((S^{(1)}+S^{(2)})/2\), and the long-term surrogate index given \(X=x\) in each treatment arm, each on its own vertical scale. These are population quantities under the data-generating process. For \(x>0\), the treatment-arm variance is larger for the first short-term outcome, while the ordering reverses for the second short-term outcome and the long-term surrogate index. The equally weighted average has the same variance in both treatment arms.}
\label{fig:8.1}
\end{figure}

\subsection{Experimental designs and implementation}\label{experimental-designs-and-implementation}

This subsection describes the experimental designs that we compare. All experimental designs use the same estimand, the same estimator of the long-term effect, and the same observational data, and they differ only in the criterion used to sample covariates and assign treatment. Performance differences therefore arise from where the experimental data are collected. The criteria differ in the outcome used to construct the design criterion and in whether adaptation is applied to treatment assignment, sampling, or both. We use SI as an abbreviation for surrogate index. Table 8.1 summarizes the eight experimental designs.

\begin{table}[tbp]
\centering
\small
\begin{tabular}{@{}
  >{\raggedright\arraybackslash}p{(\linewidth - 6\tabcolsep) * \real{0.2558}}
  >{\raggedright\arraybackslash}p{(\linewidth - 6\tabcolsep) * \real{0.3256}}
  >{\raggedright\arraybackslash}p{(\linewidth - 6\tabcolsep) * \real{0.1938}}
  >{\raggedright\arraybackslash}p{(\linewidth - 6\tabcolsep) * \real{0.2248}}@{}}
\toprule\noalign{}
\begin{minipage}[b]{\linewidth}\raggedright
Experimental design
\end{minipage} & \begin{minipage}[b]{\linewidth}\raggedright
Design criterion
\end{minipage} & \begin{minipage}[b]{\linewidth}\raggedright
Treatment assignment
\end{minipage} & \begin{minipage}[b]{\linewidth}\raggedright
Covariate sampling
\end{minipage} \\
\midrule\noalign{}

Baseline design & --- & Equal probability \((1/2)\) & Estimated target distribution \\
\(S^{(1)}\) joint design & \(S^{(1)}\) & Adaptive & Adaptive \\
\(S^{(2)}\) joint design & \(S^{(2)}\) & Adaptive & Adaptive \\
Surrogate-average joint design & \((S^{(1)}+S^{(2)})/2\) & Adaptive & Adaptive \\
SI assignment-only design & Long-term surrogate index & Adaptive & Estimated target distribution \\
SI sampling-only design & Long-term surrogate index & Equal probability \((1/2)\) & Adaptive \\
SI joint design & Long-term surrogate index & Adaptive & Adaptive \\
Stratified-oracle SI joint design & Long-term surrogate index (true variances) & Adaptive & Adaptive \\
\bottomrule
\end{tabular}
\caption{Experimental designs. The baseline design uses equal-probability treatment assignment and the reference covariate distribution. The \(S^{(1)}\), \(S^{(2)}\), and surrogate-average joint designs adapt both treatment assignment and covariate sampling using their respective short-term criteria. The three feasible surrogate-index designs use the long-term surrogate index and isolate the contribution of treatment assignment, sampling, and their combination. The stratified-oracle SI joint design replaces the learned conditional variances of the long-term surrogate index by the true conditional variances averaged within each of the 15 covariate strata used by the feasible experimental designs, and provides a benchmark for learning the design criterion.}
\end{table}

{
\emergencystretch=2em
The simulation follows Algorithm~\ref{alg:adaptive-sequential-estimation} and estimates each of its components by a simple method. The target covariate density \(f_{X\mid G=\mathrm O}\) is estimated by a Gaussian kernel density estimate of the observational covariates. Following the covariate discretization approach of \citet{Hahn2011}, the feasible experimental designs operate on 15 equiprobable strata of this estimate. The surrogate index \(\mu\) is estimated by least squares in the observational sample on a second-order polynomial in the standardized covariate and surrogates, the conditional means \(\mu_w\) by least squares in each treatment arm on a second-order polynomial in the standardized covariate, and the conditional variances used by the four criteria by averages of squared residuals within strata. The surrogate representer \(\alpha\) is estimated by the binary-classification-based construction of Remark 5.2, in which each of the four conditional probabilities is estimated by a single logistic regression on the same polynomials. All experimental designs begin with an initial period that assigns treatment with probability \(1/2\) and samples covariates from the estimated target distribution, and this period lasts for at least 200 participants and until each arm has at least 10 participants in every stratum. Adaptation therefore starts once each estimator has enough data to be reasonably stable, and the first adaptive round of the SI joint design has median 518 across the 2,000 replications.
\par
}

\subsection{Estimating the experimental designs and the long-term effect}\label{estimating-the-experimental-designs-and-the-long-term-effect}

{
\emergencystretch=2em
This subsection compares the learned experimental designs and the precision of the resulting long-term effect estimator. Figures 8.2 and 8.3 show the assignment probabilities and sampling density ratios that the \(S^{(1)}\) joint design and the SI joint design learn from the histories through \(t=1{,}000\) and \(t=5{,}000\), with their means and 10th to 90th percentiles across the 2,000 replications. The two experimental designs assign treatment in opposite directions across prior usage, and both learned experimental designs approach their population oracles as the history grows.
\par
}

\begin{figure}[tbp]
\centering
\includegraphics[width=.69\linewidth]{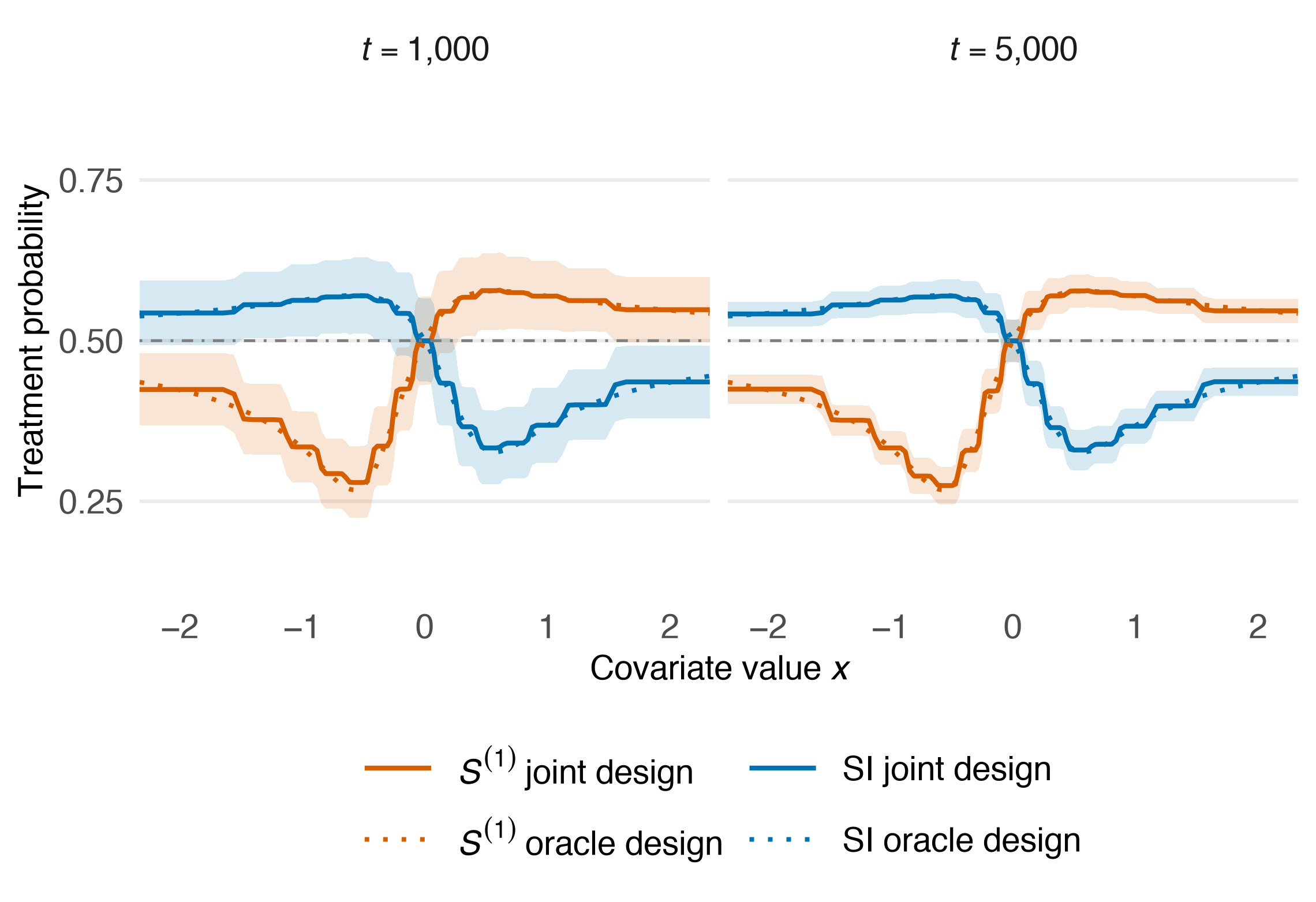}
\caption{Assignment probabilities under the \(S^{(1)}\) joint design and the SI joint design. The dot-dashed line marks the assignment probability \(1/2\). Dotted curves show the population oracle designs of the two criteria. The horizontal axis covers the 1st through 99th percentiles of the target covariate distribution.}
\label{fig:8.2}
\end{figure}

\begin{figure}[tbp]
\centering
\includegraphics[width=.69\linewidth]{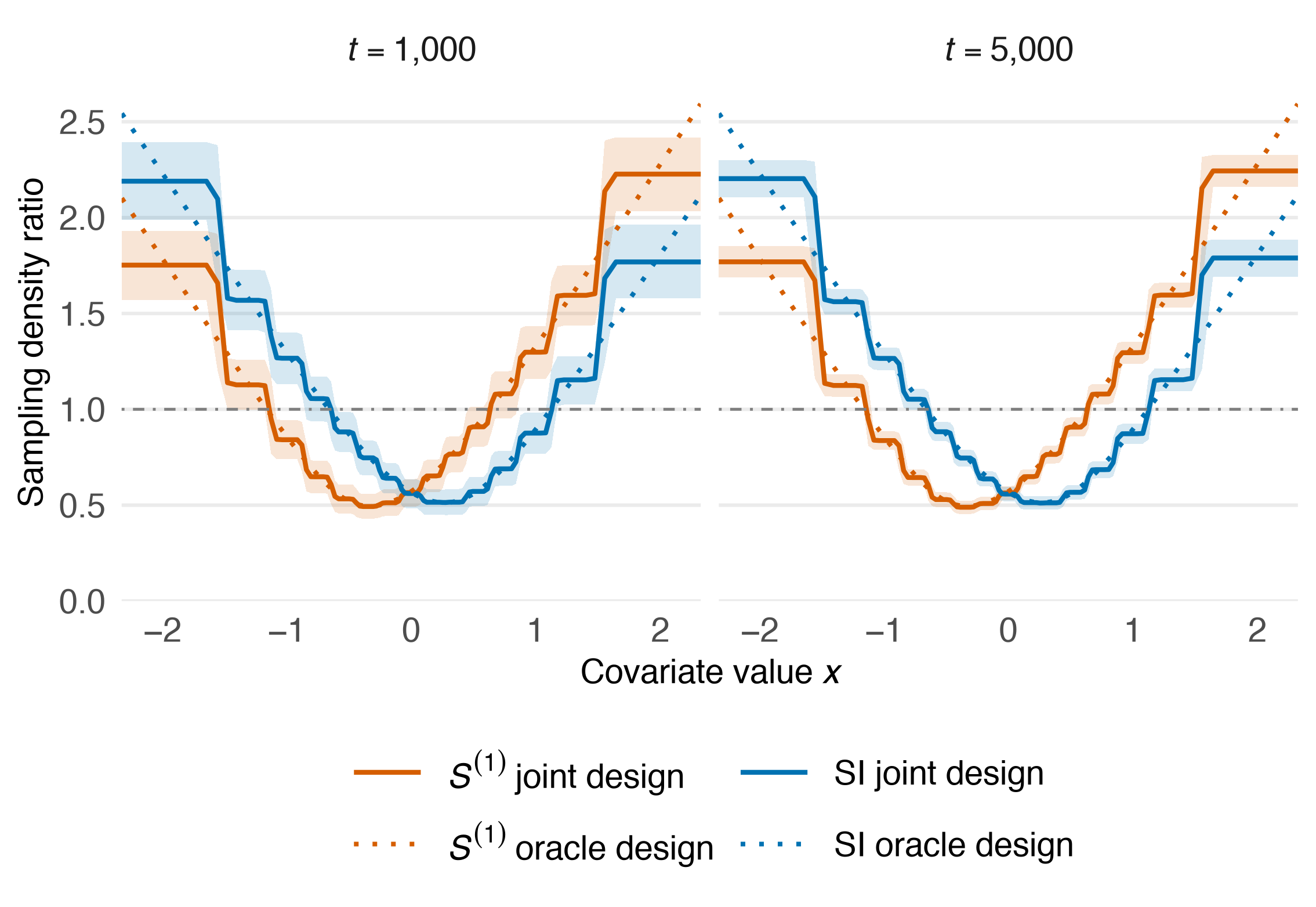}
\caption{Sampling density ratios under the \(S^{(1)}\) joint design and the SI joint design. The vertical axis shows the ratio of the experimental covariate density to a reference covariate density, not a density itself. The dot-dashed line at one is the equal-density reference that preserves the estimated reference covariate distribution. Dotted curves show the population oracle designs of the two criteria. Learned ratios are relative to the estimated reference covariate density, and the oracle ratio is relative to the target density. The horizontal axis covers the 1st through 99th percentiles of the target covariate distribution.}
\label{fig:8.3}
\end{figure}

Table 8.2 reports the bias, the coverage of the ordinary 95\% interval, and the Monte Carlo variance of the complete procedure, including estimation of the observational regression and of the target covariate density, the initial period, sequential learning of the experimental design and of the surrogate representer, and estimation of the long-term effect, for the eight experimental designs after 5,000 experimental participants.

{
\emergencystretch=2em
Table 8.2 also reports population benchmarks computed under the data-generating process. Each benchmark is \((V^{\rm exp}+V^{\rm obs})/5{,}000\), the variance that the efficiency bound of Theorem 3.1 gives for the long-term effect estimator with 5,000 experimental participants and \(N_{\mathrm O}=5{,}000\), where \(V^{\rm exp}\) is evaluated at the population version of the experimental design, which uses the true conditional variances of its design criterion. The observational component \(V^{\rm obs}\) does not depend on the experimental design and equals 61.868 in this data-generating process.
\par
}

\begin{table}[tbp]
\centering
\small
\begin{tabular}{@{}
  >{\raggedright\arraybackslash}p{(\linewidth - 10\tabcolsep) * \real{0.3100}}
  >{\raggedleft\arraybackslash}p{(\linewidth - 10\tabcolsep) * \real{0.1100}}
  >{\raggedleft\arraybackslash}p{(\linewidth - 10\tabcolsep) * \real{0.1300}}
  >{\raggedleft\arraybackslash}p{(\linewidth - 10\tabcolsep) * \real{0.1100}}
  >{\raggedleft\arraybackslash}p{(\linewidth - 10\tabcolsep) * \real{0.1600}}
  >{\raggedleft\arraybackslash}p{(\linewidth - 10\tabcolsep) * \real{0.1800}}@{}}
\toprule\noalign{}
\begin{minipage}[b]{\linewidth}\raggedright
Experimental design
\end{minipage} & \begin{minipage}[b]{\linewidth}\raggedleft
Bias
\end{minipage} & \begin{minipage}[b]{\linewidth}\raggedleft
Coverage
\end{minipage} & \begin{minipage}[b]{\linewidth}\raggedleft
MC var.
\end{minipage} & \begin{minipage}[b]{\linewidth}\raggedleft
MC var. reduction
\end{minipage} & \begin{minipage}[b]{\linewidth}\raggedleft
Population var. reduction
\end{minipage} \\
\midrule\noalign{}

Baseline design & -0.004 & 96.3\% & 0.054 & --- & --- \\
\(S^{(1)}\) joint design & -0.005 & 95.4\% & 0.059 & -9.8\% & -4.1\% \\
\(S^{(2)}\) joint design & -0.005 & 95.5\% & 0.047 & 12.7\% & 17.7\% \\
Surrogate-average joint design & -0.005 & 95.1\% & 0.048 & 11.2\% & 14.6\% \\
SI assignment-only design & -0.005 & 96.4\% & 0.054 & -1.1\% & 1.8\% \\
SI sampling-only design & -0.004 & 95.1\% & 0.046 & 13.2\% & 16.1\% \\
SI joint design & -0.006 & 95.8\% & 0.047 & 13.1\% & 17.9\% \\
Stratified-oracle SI joint design & -0.004 & 95.5\% & 0.046 & 13.1\% & 17.9\% \\
\bottomrule
\end{tabular}
\caption{Long-term average treatment effect estimates with 5,000 experimental participants and population benchmarks. Bias is measured relative to the true value 1, and Coverage is the coverage rate of the ordinary fixed-time 95\% interval. MC var. is the Monte Carlo (MC) variance, the sample variance of the 2,000 estimates. MC var. reduction is the percentage reduction in MC var. relative to the baseline design, and Population var. reduction is the percentage reduction in the population benchmark \((V^{\rm exp}+V^{\rm obs})/5{,}000\) relative to the baseline design, so a positive value indicates a reduction. The SI joint design and the stratified-oracle SI joint design have the same population benchmark because they evaluate the same population oracle rule.}
\end{table}

{
\emergencystretch=2em
Because the regression models for the surrogate index and for the conditional means of the short-term outcomes are correctly specified, the bias is close to zero for every experimental design and differs little across experimental designs, and the ordinary 95\% interval likewise has coverage close to 95\%, between 95.1\% and 96.4\%. The \(S^{(1)}\) joint design raises MC variance by 9.8\% relative to the baseline design, whereas the \(S^{(2)}\) and surrogate-average joint designs reduce it by 12.7\% and 11.2\%. These results preserve the main contrast of the population benchmarks. Targeting the first short-term outcome increases the MC variance of the long-term effect estimator, whereas criteria more closely aligned with the long-term surrogate index reduce it. The SI joint design has 20.8\% smaller MC variance than the \(S^{(1)}\) joint design.
\par
}

Under the surrogate-index criterion, treatment assignment alone leaves MC variance nearly unchanged, with an increase of 1.1\%, sampling alone reduces it by 13.2\%, and their combination reduces it by 13.1\%, so sampling is the main source of the reduction, as in the population benchmarks, in which assignment alone, sampling alone, and their combination reduce the variance by 1.8\%, 16.1\%, and 17.9\%. The stratified-oracle SI joint design yields a 13.1\% reduction.

\subsection{Sequential monitoring and data-dependent stopping}\label{sequential-monitoring-and-data-dependent-stopping}

{
\emergencystretch=2em
This subsection reports the simultaneous coverage and detection of the 90\% asymptotic time-uniform interval of Section 7 over the monitoring period \(t=595,\ldots,5{,}000\), which begins at \(t_{0,N_{\mathrm O}}=\lceil N_{\mathrm O}^{3/4}\rceil\). The tuning variable is chosen as in Remark 7.1, with target round \(t_{N_{\mathrm O}}^\star=2{,}500\) and with the planning value set to the experimental variance estimate \(\widehat V_{t_{0,N_{\mathrm O}},N_{\mathrm O}}^{\rm exp}\) of the monitored experiment at the monitoring start. Detection occurs when the lower endpoint first exceeds zero, and replications without a detection are assigned a stopping value of 5,000. Appendix B reports the 95\% interval and additional results for both confidence levels.
\par
}

\begin{table}[tbp]
\centering
\small
\begin{tabular}{@{}
  >{\raggedright\arraybackslash}p{(\linewidth - 10\tabcolsep) * \real{0.1900}}
  >{\raggedleft\arraybackslash}p{(\linewidth - 10\tabcolsep) * \real{0.1400}}
  >{\raggedleft\arraybackslash}p{(\linewidth - 10\tabcolsep) * \real{0.1400}}
  >{\raggedleft\arraybackslash}p{(\linewidth - 10\tabcolsep) * \real{0.1800}}
  >{\raggedleft\arraybackslash}p{(\linewidth - 10\tabcolsep) * \real{0.1400}}
  >{\raggedleft\arraybackslash}p{(\linewidth - 10\tabcolsep) * \real{0.2100}}@{}}
\toprule\noalign{}
\begin{minipage}[b]{\linewidth}\raggedright
Experimental design
\end{minipage} & \begin{minipage}[b]{\linewidth}\raggedleft
Simultaneous coverage
\end{minipage} & \begin{minipage}[b]{\linewidth}\raggedleft
Detection
\end{minipage} & \begin{minipage}[b]{\linewidth}\raggedleft
Mean stop
\end{minipage} & \begin{minipage}[b]{\linewidth}\raggedleft
Median stop
\end{minipage} & \begin{minipage}[b]{\linewidth}\raggedleft
Stop IQR
\end{minipage} \\
\midrule\noalign{}

Baseline design & 95.9\% & 95.4\% & 2,062.4 & 1,727.0 & 1,036.2--2,802.2 \\
\(S^{(1)}\) joint design & 95.7\% & 94.6\% & 2,090.1 & 1,792.5 & 1,032.0--2,875.5 \\
\(S^{(2)}\) joint design & 95.2\% & 97.7\% & 1,855.4 & 1,561.0 & 976.2--2,511.5 \\
Surrogate-average joint design & 95.6\% & 97.5\% & 1,905.5 & 1,620.0 & 998.0--2,572.0 \\
SI assignment-only design & 96.4\% & 95.8\% & 2,059.7 & 1,722.5 & 1,014.8--2,830.5 \\
SI sampling-only design & 95.8\% & 97.6\% & 1,873.2 & 1,607.0 & 989.0--2,506.0 \\
SI joint design & 95.6\% & 98.2\% & 1,846.1 & 1,564.5 & 984.0--2,461.5 \\
Stratified-oracle SI joint design & 95.2\% & 98.0\% & 1,821.4 & 1,530.5 & 971.5--2,415.2 \\
\bottomrule
\end{tabular}
\caption{Simultaneous coverage and detection for the 90\% time-uniform interval. Simultaneous coverage requires every 90\% time-uniform interval from participant 595 through participant 5,000 to contain the true value 1. Detection includes crossings at the horizon. Mean stop and Median stop are stopping sample sizes, with replications without detection recorded as 5,000. Stop IQR reports the 25th and 75th percentiles of those stopping sample sizes. Reported rates change in increments of 0.05 percentage points and are shown to one decimal place.}
\end{table}

{
\emergencystretch=2em
In Table 8.3, simultaneous coverage is between 95.2\% and 96.4\% for every experimental design, above the nominal 90\% level. The SI joint design raises the detection rate from 95.4\% under the baseline design to 98.2\% and reduces the mean stopping sample from 2,062.4 to 1,846.1 and the median from 1,727.0 to 1,564.5. The \(S^{(1)}\) joint design has a lower detection rate, 94.6\%, and a larger mean stopping sample, 2,090.1, than the baseline design, while the \(S^{(2)}\) and surrogate-average joint designs have stopping performance close to the SI joint design.
\par
}

{
\emergencystretch=2em
Recomputing the intervals on the saved paths with other tuning choices still leaves simultaneous coverage above the nominal level. With target rounds of 1,250, 3,750, and 5,000 participants, coverage of the 90\% interval lies between 95.0\% and 97.1\% across experimental designs (Table B.3). With the planning values of Appendix B.3, which are the experimental variance estimates at rounds 200 and 595 and the population value of \(V^{\rm exp}\) for each experimental design, it lies between 95.2\% and 96.4\% for the 90\% interval and between 97.8\% and 98.4\% for the 95\% interval. These results are consistent with Theorem 7.1, which gives a lower bound on the asymptotic simultaneous coverage, and intervals that attain the nominal level exactly are left for future research, as noted in Section 7.3.
\par
}

\section{Experimental design illustration using the Graduation experiment}\label{section-9-experimental-design-illustration-using-the-graduation-experiment}

This section uses the variance structure estimated from the completed Graduation experiment to illustrate how optimal treatment assignment and sampling differ from a baseline experimental design. The illustration takes the position of a researcher who plans a new experiment in the same population and uses the completed experiment to estimate the conditional variances on which the optimal rules of Theorem 3.2 depend. Section 9.1 describes the data and the construction of the two samples, Section 9.2 defines the estimated experimental designs and the variance criterion, and Section 9.3 reports the estimated experimental designs for a covariate measuring assets and a covariate measuring consumption. Supplementary Material 1 gives the details of the computation.

\subsection{The Graduation experiment}\label{the-graduation-experiment}

We use data from the randomized evaluation of the Graduation program in Pakistan by \citet{Banerjee2015}. \citet{Chen2023} cleaned and organized the data from this experiment and distribute the cleaned data with their R package \texttt{longterm}. The original data are available from \href{https://doi.org/10.7910/DVN/NHIXNT}{Harvard Dataverse}, and the cleaned data are included in the \href{https://github.com/DavidRitzwoller/longterm/blob/main/data/graduation.rda}{public repository of the package}.

In Pakistan, the Graduation program for households in extreme poverty combined a one-time grant of productive assets with technical training, cash consumption support during the first year, regular home visits and coaching for approximately two years, savings support, and basic health services, and the treatment assigned in this randomized evaluation is the entire package.

From the 1,299 households in the Pakistan sample before cleaning, \citet{Chen2023} excluded 399 households that had attrited before either post-treatment outcome measurement and 46 households for which all consumption variables were missing at either post-treatment time. This leaves an analysis sample of 854 households, 446 treated and 408 control, and full compliance with treatment assignment is reported. Supplemental Appendix D.1 of \citet{Chen2023} describes the imputation of the remaining missing values.

{
\emergencystretch=2em
Surveys were conducted before treatment and approximately two and three years afterward. The 20 pretreatment variables form \(X\) and the 21 two-year variables form \(S\). Following \citet{Chen2023}, the long-term outcome \(Y\) is the level of monthly total consumption per capita three years after treatment, measured in 2014 USD PPP.
\par
}

To reproduce the data structure of our framework, we stratify the 854 households by treatment group and divide them equally into two nonoverlapping samples, and we repeat this split 1,000 times to summarize sensitivity to the sample partition. In each split, we mask treatment assignment in the observational sample and the long-term outcome in the experimental sample. Table 9.1 reports the resulting sample sizes, which are the same in every split, and the information available in each sample.

\begin{table}[tbp]
\centering
\small
\begin{tabular}{@{}
  >{\raggedright\arraybackslash}p{(\linewidth - 8\tabcolsep) * \real{0.2200}}
  >{\raggedleft\arraybackslash}p{(\linewidth - 8\tabcolsep) * \real{0.1600}}
  >{\raggedleft\arraybackslash}p{(\linewidth - 8\tabcolsep) * \real{0.1800}}
  >{\raggedleft\arraybackslash}p{(\linewidth - 8\tabcolsep) * \real{0.1800}}
  >{\raggedright\arraybackslash}p{(\linewidth - 8\tabcolsep) * \real{0.2600}}@{}}
\toprule\noalign{}
\begin{minipage}[b]{\linewidth}\raggedright
Sample
\end{minipage} & \begin{minipage}[b]{\linewidth}\raggedleft
Total sample size
\end{minipage} & \begin{minipage}[b]{\linewidth}\raggedleft
Treatment-group sample size
\end{minipage} & \begin{minipage}[b]{\linewidth}\raggedleft
Control-group sample size
\end{minipage} & \begin{minipage}[b]{\linewidth}\raggedright
Variables
\end{minipage} \\
\midrule\noalign{}

Observational & 427 & 223 & 204 & \(X,S,Y\) \\
Experimental & 427 & 223 & 204 & \(X,W,S\) \\
\bottomrule
\end{tabular}
\caption{Sample construction. Each of the 1,000 splits divides the 854 households of the analysis sample, within each treatment group, into two nonoverlapping samples, so the sample sizes are the same in every split. Treatment- and control-group sample sizes count households by assigned treatment. Variables lists the variables used from each sample. Treatment assignment \(W\) is masked in the observational sample, and the long-term outcome \(Y\) is masked in the experimental sample.}
\end{table}

{
\emergencystretch=2em
The estimand is the long-term average treatment effect in the observational population, \(\tau=\mathbb E[Y(1)-Y(0)\mid G=\mathrm O]\), and in each split the empirical distribution \(\mathbb P_{N_{\mathrm O}}^{\rm obs}\) of the 427 observational households is the plug-in estimate of the target distribution \(P^{\rm obs}\).
\par
}

\subsection{Estimation method and the plug-in criterion}\label{estimation-method-and-the-plug-in-criterion}

{
\emergencystretch=2em
This subsection defines the experimental designs that we compare and the criterion used to compare them. We restrict the assignment probability and the sampling density ratio to be constant within strata of a single pretreatment covariate \(X_j\). As in the simulation, the strata divide \(X_j\) into five intervals of equal probability in the observational sample, and \(k(x_j)\) denotes the stratum that contains \(x_j\). Households with the same value of \(X_j\) belong to the same stratum, so a covariate with many tied values has fewer than five strata and a binary covariate has two. We use five strata because five is the largest number for which each arm of the experimental sample has at least 40 households in every stratum when the strata have equal mass, given that the control arm has 204 households.
\par
}

{
\emergencystretch=2em
The experimental designs use the conditional variance of the surrogate index in each arm, averaged within each stratum, and estimate it by the average squared residual. With \(\widehat\mu_{N_{\mathrm O}}(s,x)\) the least-squares surrogate index from the observational sample, \(\widehat\mu_{w,N_{\mathrm O}}(x)\) its least-squares fit on the covariates in arm \(w\) of the experimental sample, \(\mathcal I_w\) the experimental households with \(W_i=w\), and \(n_{w,k}\) the number of those in stratum \(k\), \[\widehat v_{w,j}(x_j)
:=\frac1{n_{w,k(x_j)}}\sum_{i\in\mathcal I_w:\,k(X_{ij})=k(x_j)}
\{\widehat\mu_{N_{\mathrm O}}(S_i,X_i)-\widehat\mu_{w,N_{\mathrm O}}(X_i)\}^{2}.
\]
\par
}

{
\emergencystretch=2em
The estimated experimental designs plug these variances into the optimal rules of Theorem 3.2, which give the estimated Neyman allocation and the optimal sampling density ratio, with the assignment probability kept in \([0.1,0.9]\) and the ratio in \([1/5,5]\). The criterion is the plug-in value of the experimental variance component \(V^{\rm exp}\) defined in Section 3.1.
\par
}

{
\emergencystretch=2em
We compare experimental designs using the variance estimate \(\widehat V_j(p(\cdot),r(\cdot))\) defined by \[\widehat V_j(p(\cdot),r(\cdot))
=\mathbb P_{N_{\mathrm O}}^{\rm obs}
\left[
\frac1{r(X_j)}
\left\{
\frac{\widehat v_{1,j}(X_j)}{p(X_j)}
+\frac{\widehat v_{0,j}(X_j)}{1-p(X_j)}
\right\}
\right].
\]
\par
}

{
\emergencystretch=2em
Here \(p(\cdot)\) and \(r(\cdot)\) are the assignment probability and the sampling density ratio of the experimental design being compared, and the criterion is a sum over the strata because \(p(\cdot)\), \(r(\cdot)\), and \(\widehat v_{w,j}\) are constant within strata. We compare four experimental designs. The baseline design assigns treatment with probability \(1/2\) and samples from the target covariate distribution, so that \(r(x_j)=1\). The assignment-only design uses the estimated Neyman allocation with \(r(x_j)=1\). The sampling-only design assigns treatment with probability \(1/2\) and uses the sampling density ratio that minimizes the criterion under this assignment probability. The joint design uses the estimated Neyman allocation together with the optimal sampling density ratio.
\par
}

\subsection{Experimental designs based on the household asset index and monthly nonfood consumption per capita}\label{experimental-designs-based-on-the-household-asset-index-and-monthly-nonfood-consumption-per-capita}

For each of the 20 pretreatment variables, we compute the reduction of the plug-in criterion under the joint design relative to the baseline design in each of the 1,000 splits and rank the variables by the mean reduction. Appendix C.1 reports the complete ranking. Among the highest-ranked variables, we choose the household asset index and monthly nonfood consumption per capita, which describe different characteristics of the participating households, their assets and their consumption.

{
\emergencystretch=2em
Figure 9.1 reports, for each stratum of the two covariates, the estimated conditional standard deviations of the surrogate index in the two groups, the estimated Neyman allocation, and the sampling density ratio of the joint design. The two experimental designs differ from the baseline design in different respects. For the household asset index, the experimental design differs from the baseline design mainly in the sampling density ratio. It samples more heavily the top stratum, which holds the households with the most household durables at baseline, with a median sampling density ratio of 1.24 that exceeds one in 98\% of the splits. For monthly nonfood consumption per capita, the experimental design differs from the baseline design mainly in treatment assignment. In the second stratum, it assigns treatment with a median probability of 0.61, which exceeds \(1/2\) in 97\% of the splits.
\par
}

\begin{figure}[tbp]
\centering
\includegraphics[width=\linewidth]{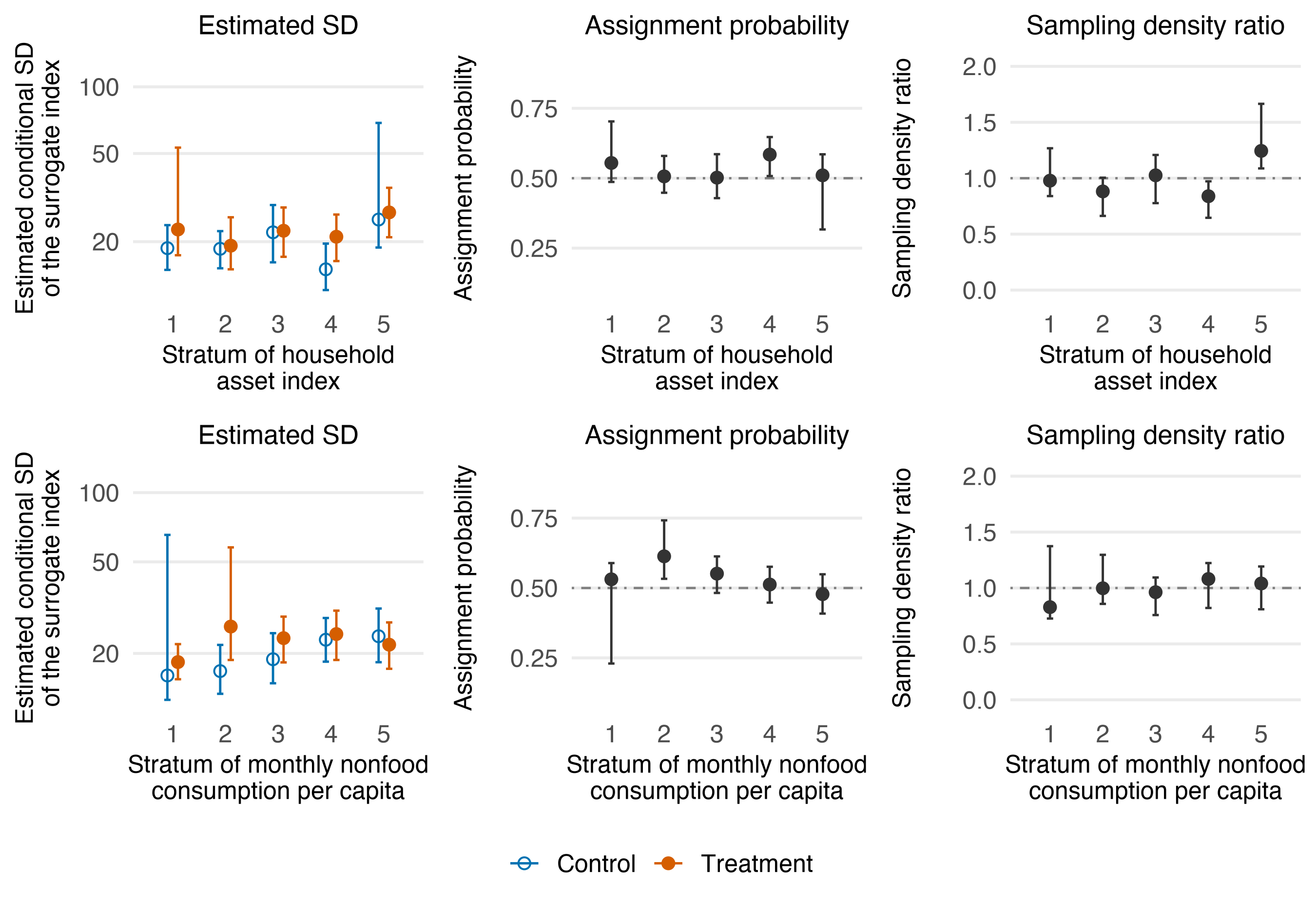}
\caption{Estimated conditional standard deviations, assignment probabilities, and sampling density ratios of the joint design by stratum of the household asset index and monthly nonfood consumption per capita. Points show medians and bars the 10th and 90th percentiles across 1,000 treatment-stratified splits of the same dataset. The strata are the five intervals of equal probability of each covariate in the observational sample and are numbered from the lowest values. The left panels show, on a logarithmic scale, the estimated conditional standard deviations of the surrogate index in the control and treatment groups, that is, the square roots of the stratum averages of the conditional variances, estimated by average squared residuals. The middle panels show the estimated Neyman allocation, and the right panels the sampling density ratio of the joint design. Dot-dashed lines mark probabilities of \(1/2\) and ratios of 1.}
\label{fig:9.1}
\end{figure}

Figure 9.2 shows the joint designs as allocations of 1,000 recruited households. For the household asset index, the largest change is in the top stratum, from which the joint design recruits 262 households instead of 200. For monthly nonfood consumption per capita, the largest change is in the second stratum, where it assigns 131 of 207 households to treatment, compared with half of 199 households under the baseline design. Appendix C.2 shows that both changes remain when the splits in which one household dominates the stratum are set aside and when seven strata are used.

\begin{figure}[tbp]
\centering
\includegraphics[width=\linewidth]{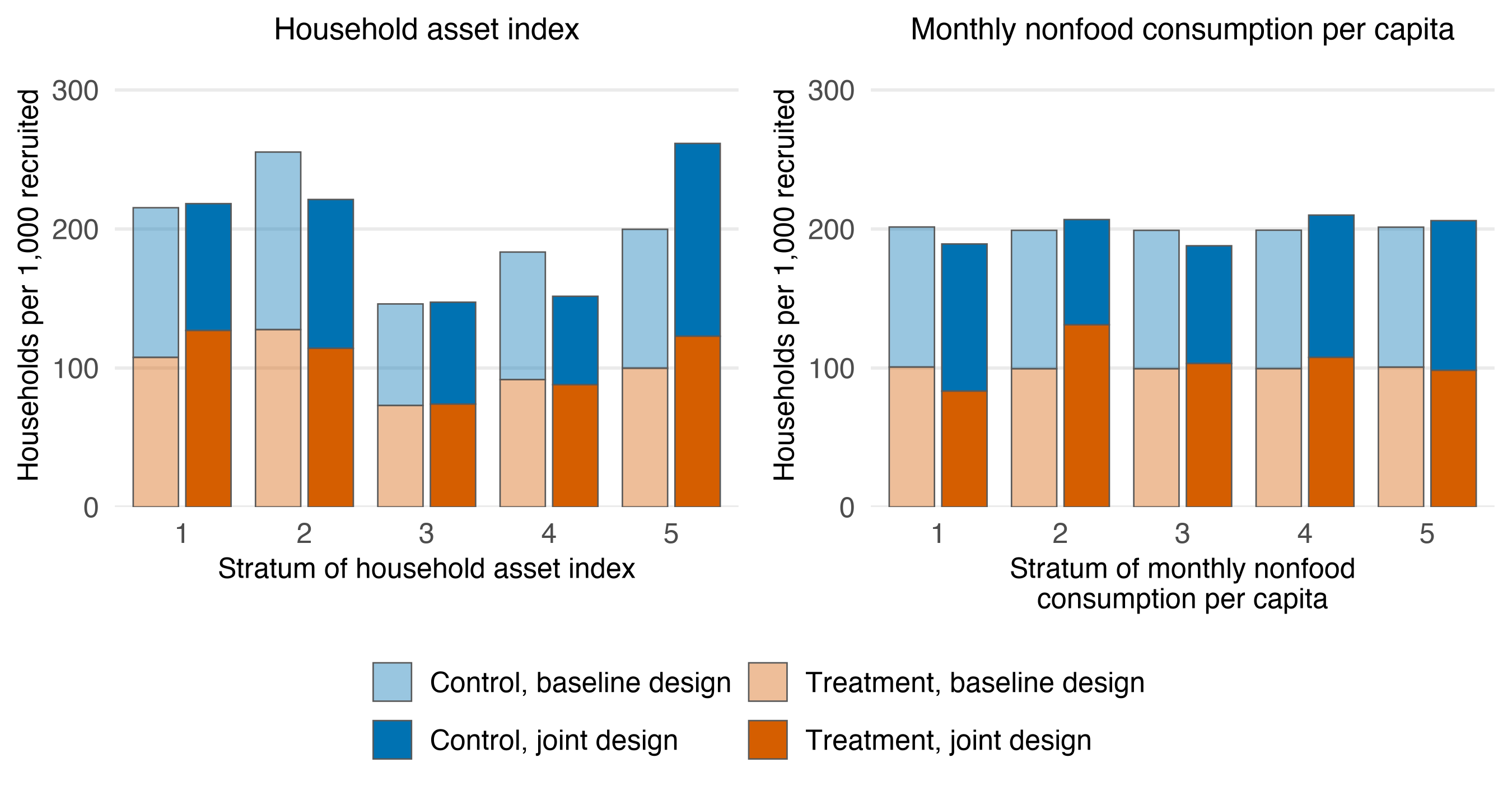}
\caption{Allocation of 1,000 recruited households by stratum under the baseline design and the joint design. Each pair of bars refers to one stratum, numbered from the lowest values of the covariate. The left bar, in light colors, shows the baseline design and the right bar, in dark colors, the joint design. The height of a bar is the number of households recruited from the stratum per 1,000 recruited households, and its lower and upper parts are the numbers assigned to treatment and to control. Under the baseline design, the number recruited from a stratum in each split is 1,000 times the share of the 427 observational households in that stratum, and half of them are assigned to treatment. Values are averages across the 1,000 splits.}
\label{fig:9.2}
\end{figure}

Table 9.2 reports the reductions of the plug-in criterion relative to the baseline design. Each experimental design is optimized and evaluated with the same estimated variances, so the reductions are nonnegative by construction. The reductions therefore show which part of the joint design, assignment or sampling, contributes more for each covariate. For the household asset index the sampling-only design gives the larger reduction, and for monthly nonfood consumption per capita the assignment-only design does, in line with Figure 9.1. With seven strata, the sampling-only design gives the larger mean reduction for both covariates (Appendix C.2).

\begin{table}[tbp]
\centering
\small
\begin{tabular}{@{}
  >{\raggedright\arraybackslash}p{(\linewidth - 10\tabcolsep) * \real{0.3000}}
  >{\raggedright\arraybackslash}p{(\linewidth - 10\tabcolsep) * \real{0.2200}}
  >{\raggedleft\arraybackslash}p{(\linewidth - 10\tabcolsep) * \real{0.1200}}
  >{\raggedleft\arraybackslash}p{(\linewidth - 10\tabcolsep) * \real{0.1200}}
  >{\raggedleft\arraybackslash}p{(\linewidth - 10\tabcolsep) * \real{0.1000}}
  >{\raggedleft\arraybackslash}p{(\linewidth - 10\tabcolsep) * \real{0.1400}}@{}}
\toprule\noalign{}
\begin{minipage}[b]{\linewidth}\raggedright
Covariate
\end{minipage} & \begin{minipage}[b]{\linewidth}\raggedright
Experimental design
\end{minipage} & \begin{minipage}[b]{\linewidth}\raggedleft
Mean reduction
\end{minipage} & \begin{minipage}[b]{\linewidth}\raggedleft
25th percentile
\end{minipage} & \begin{minipage}[b]{\linewidth}\raggedleft
Median
\end{minipage} & \begin{minipage}[b]{\linewidth}\raggedleft
75th percentile
\end{minipage} \\
\midrule\noalign{}

Household asset index & Assignment-only design & 3.66\% & 1.22\% & 2.12\% & 3.85\% \\
Household asset index & Sampling-only design & 5.29\% & 1.61\% & 2.92\% & 6.32\% \\
Household asset index & Joint design & 8.23\% & 3.21\% & 4.91\% & 9.16\% \\
Monthly nonfood consumption per capita & Assignment-only design & 4.89\% & 1.46\% & 2.68\% & 5.45\% \\
Monthly nonfood consumption per capita & Sampling-only design & 3.73\% & 1.46\% & 2.30\% & 3.94\% \\
Monthly nonfood consumption per capita & Joint design & 7.72\% & 3.32\% & 4.71\% & 8.33\% \\
\bottomrule
\end{tabular}
\caption{Reductions of the plug-in criterion relative to the baseline design for the household asset index and monthly nonfood consumption per capita. Entries are the mean and quartiles across the 1,000 splits of the percentage reduction of \(\widehat V_j\) relative to its value under the baseline design, with five strata.}
\end{table}

\section{Conclusion}\label{section-10-conclusion}

This paper uses observational data available before an experiment, together with data from earlier participants, to guide sampling and treatment assignment for estimating long-term treatment effects. The optimal experimental design depends on the conditional variation of the surrogate index in the treatment and control groups. The sequential estimator learns this experimental design and, under the stated conditions, is asymptotically normal and attains the optimal fixed-design bound as the sample sizes increase. The asymptotically time-uniform intervals account for both accumulating experimental variation and uncertainty shared across monitoring times through the observational data. Their coverage guarantee is asymptotic and applies from the prescribed monitoring start onward. The Graduation illustration shows how estimated variance heterogeneity translates into experimental designs for sampling and assignment.

\section*{Declaration of generative AI and AI-assisted technologies in the manuscript preparation process}\label{declaration-of-generative-ai-and-ai-assisted-technologies-in-the-manuscript-preparation-process}
\addcontentsline{toc}{section}{Declaration of generative AI and AI-assisted technologies in the manuscript preparation process}

During the preparation of this work the author used Claude (Anthropic), Codex (OpenAI), and Gemini (Google) to draft and revise text, translate parts of the manuscript into English, write and check code for the simulations and the empirical illustration, and check derivations and proofs. After using these tools, the author reviewed and edited the content as needed and takes full responsibility for the content of the published article.

\FloatBarrier
\clearpage
\begin{center}
{\Huge\bfseries Appendix\par}
\bigskip
{\Large\bfseries Adaptive experiments with observational data: Efficient experimental design and asymptotic time-uniform inference for long-term treatment effects\par}
\bigskip
Kentaro Kawato\par
Graduate School of Economics, The University of Tokyo\par
\end{center}
\startcontents[appendix]
\printcontents[appendix]{}{1}{\setcounter{tocdepth}{2}\section*{Contents}}
\clearpage
\appendix

\AppendixNumbering

\FloatBarrier

\section{Proofs}\label{appendix-a-proofs}

Appendix A proves the formal results of Sections 2--7. Appendix A.1 proves the identification result in Theorem 2.1, Appendix A.2 the efficiency bound in Theorem 3.1, Appendix A.3 the optimal experimental design in Theorem 3.2, and Appendix A.4 the representation under adaptive experimental design in Proposition 4.1. Appendix A.5 constructs the surrogate representer, Appendix A.6 proves Lemmas 6.1 and 6.2 and the asymptotic normality in Theorem 6.1, and Appendix A.7 proves the time-uniform coverage in Theorem 7.1 and Corollary 7.1.

\subsection{Proof of Theorem 2.1}\label{a.1-proof-of-theorem-2.1}

{
\emergencystretch=2em
Appendix A.1 proves Theorem 2.1. Throughout the proof, we impose suitable moment conditions so that every displayed conditional expectation and integral, including those involving \(Y(w)\), is well defined, and interpret conditional identities and density ratios on the denominator-positive supports specified in Assumption 2.6, with ratios set to zero off those supports.
\par
}

\textbf{Proof.} The proof derives \eqref{eq:identification_regression} from a conditional-mean identity for the potential outcomes and reduces the other three representations to (2.1). Step (i) proves the identity and (2.1), steps (ii) and (iii) compute the conditional means of the weighted terms in \eqref{eq:identification_experimental_weighting} and \eqref{eq:identification_observational_weighting} given the covariates, and step (iv) shows that the two terms that \eqref{eq:identification_augmented} adds to the right-hand side of (2.1) are zero.

\begin{enumerate}
\def\labelenumi{(\roman{enumi})}
\tightlist
\item
  We first show that \(\mathbb E[Y(w)\mid X=x,G=\mathrm O]=\mu_w(x)\) for every \(w\in\{0,1\}\) and every target-supported \(x\), and deduce (2.1). Fix such \(w\) and \(x\). Assumption 2.4 (conditional external validity), Assumption 2.2 (conditional treatment randomization), and Assumption 2.1 (consistency) give, in that order, \(\mathbb E[Y(w)\mid X=\allowbreak x,G=\allowbreak \mathrm O]=\allowbreak \mathbb E[Y(w)\mid X=\allowbreak x,G=\allowbreak \mathrm E]=\allowbreak \mathbb E[Y(w)\mid W=\allowbreak w,X=\allowbreak x,G=\allowbreak \mathrm E]=\allowbreak \mathbb E[Y\mid W=\allowbreak w,X=\allowbreak x,G=\allowbreak \mathrm E]\). Iterated expectations, Assumption 2.3 (statistical surrogacy), Assumption 2.5 (conditional outcome comparability across samples), and the nuisance-function definitions in Section 2.2 then give, in that order, \[\begin{aligned}
  &\mathbb E[Y\mid W=w,X=x,G=\mathrm E]\\
  &=\int\mathbb E[Y\mid W=w,S=s,X=x,G=\mathrm E]f_{S\mid W=w,X,G=\mathrm E}(s\mid x)d\lambda_S(s)\\
  &=\int\mathbb E[Y\mid S=s,X=x,G=\mathrm E]f_{S\mid W=w,X,G=\mathrm E}(s\mid x)d\lambda_S(s)\\
  &=\int\mu(s,x)f_{S\mid W=w,X,G=\mathrm E}(s\mid x)d\lambda_S(s)=\mu_w(x),
  \end{aligned}
  \]
\end{enumerate}

{
\emergencystretch=2em
which proves the identity. The identity, iterated expectations, and the definition of \(\tau\) then give (2.1), \[\begin{aligned}
\mathbb E[\mu_1(X)-\mu_0(X)\mid G=\mathrm O]
&=\mathbb E\!\left[\mathbb E[Y(1)-Y(0)\mid X,G=\mathrm O]\mid G=\mathrm O\right]\\
&=\mathbb E[Y(1)-Y(0)\mid G=\mathrm O]=\tau.
\end{aligned}
\]
\par
}

\begin{enumerate}
\def\labelenumi{(\roman{enumi})}
\setcounter{enumi}{1}
\item
  We next verify (2.3). For the treatment group, the definition of \(p\), iterated expectations, positivity in Assumption 2.6, and the definition of \(\mu_1\) give \(\mathbb E[W\mu(S,X)/\allowbreak p(X)\mid X=\allowbreak x,G=\allowbreak \mathrm E]=\allowbreak \{p(x)/\allowbreak p(x)\}\mathbb E[\mu(S,X)\mid W=\allowbreak 1,X=\allowbreak x,G=\allowbreak \mathrm E]=\allowbreak \mu_1(x)\), and the same calculation for the control group gives \(\mu_0(x)\). These two weighting identities, the importance-weight cancellation justified by Assumption 2.6, and (2.1) give \[\begin{aligned}
  &\mathbb E\Biggl[
  \frac{f_{X\mid G=\mathrm O}(X)}{f_{X\mid G=\mathrm E}(X)}
  \left\{\frac{W\mu(S,X)}{p(X)}
  -\frac{(1-W)\mu(S,X)}{1-p(X)}\right\}
  \mathrel{\Big|}G=\mathrm E\Biggr]\\
  &=\int\{\mu_1(x)-\mu_0(x)\}f_{X\mid G=\mathrm O}(x)d\lambda_X(x)
  =\tau.
  \end{aligned}
  \]
\item
  We next verify (2.2). Iterated expectations and the definition of \(\mu\) give \(\mathbb E[\alpha(S,X)Y\mid G=\allowbreak \mathrm O]=\allowbreak \mathbb E[\alpha(S,X)\mu(S,X)\mid G=\allowbreak \mathrm O]\). The definition of \(\alpha\), iterated expectations given \(X\), the surrogate-support condition in Assumption 2.6 together with the definition of \(\mu_w\), and (2.1) then give, in that order, \[\begin{aligned}
  &\mathbb E[\alpha(S,X)\mu(S,X)\mid G=\mathrm O]\\
  &=\mathbb E\left[
  \frac{f_{S\mid W=1,X,G=\mathrm E}(S\mid X)
  -f_{S\mid W=0,X,G=\mathrm E}(S\mid X)}
  {f_{S\mid X,G=\mathrm O}(S\mid X)}\mu(S,X)
  \middle|G=\mathrm O\right]\\
  &=\mathbb E\left[
  \int
  \frac{f_{S\mid W=1,X,G=\mathrm E}(s\mid X)
  -f_{S\mid W=0,X,G=\mathrm E}(s\mid X)}
  {f_{S\mid X,G=\mathrm O}(s\mid X)}
  \mu(s,X)f_{S\mid X,G=\mathrm O}(s\mid X)d\lambda_S(s)
  \middle|G=\mathrm O\right]\\
  &=\mathbb E[\mu_1(X)-\mu_0(X)\mid G=\mathrm O]=\tau.
  \end{aligned}
  \]
\item
  Finally, we verify (2.4). Iterated expectations, applied to the observational expectation given \((S,X)\) and to the experimental expectation given \(X\), and the importance-weight cancellation in Assumption 2.6 show that the right-hand side of (2.4) equals \[\begin{aligned}
  &\mathbb E[\mu_1(X)-\mu_0(X)\mid G=\mathrm O]
  +\mathbb E\!\left[\alpha(S,X)\mathbb E[Y-\mu(S,X)\mid S,X,G=\mathrm O]\middle|G=\mathrm O\right]\\
  &\quad+\int f_{X\mid G=\mathrm O}(x)\Bigl[\{\mu_1(x)-\mu_1(x)\}-\{\mu_0(x)-\mu_0(x)\}\Bigr]d\lambda_X(x)\\
  &=\mathbb E[\mu_1(X)-\mu_0(X)\mid G=\mathrm O]=\tau.
  \end{aligned}
  \]
\end{enumerate}

{
\emergencystretch=2em
The first equality in the display uses the definitions of \(\mu\) and \(\mu_w\), and the second is (2.1). This completes the proof of Theorem 2.1.
\par
}

\subsection{Proof of Theorem 3.1}\label{a.2-proof-of-theorem-3.1}

{
\emergencystretch=2em
The moment conditions in Theorem 3.1 are \(V^{\rm exp}(f_{X\mid G=\mathrm E}(\cdot),p(\cdot))<\infty\), \(V^{\rm obs}<\infty\), and \(\mathbb E[\int\mathbb E[|Y|\mid S=\allowbreak s,X,G=\allowbreak \mathrm O]\{f_{S\mid W=1,X,G=\mathrm E}(s\mid X)+f_{S\mid W=0,X,G=\mathrm E}(s\mid X)\}\,d\lambda_S(s)\mid G=\allowbreak \mathrm O]<\infty\).
\par
}

{
\emergencystretch=2em
The proof of Theorem 3.1 treats the complete data as the fixed-design product model \((P^{\rm exp})^{\otimes N_{\mathrm E}}\otimes (P^{\rm obs})^{\otimes N_{\mathrm O}}\) in which the design \((f_{X\mid G=\mathrm E},p)\) is known, and proceeds in two steps. Proposition A.2.1 shows that the pair \((\phi^{\rm exp},\phi^{\rm obs})\) displayed in Theorem 3.1 is the efficient influence-function pair of \(\tau\) in this model, and Appendix A.2.4 shows that the squared norm of the pair on the \(\sqrt{N_{\mathrm E}}\) scale converges to \(V^{\rm exp}(f_{X\mid G=\mathrm E}(\cdot),p(\cdot))+\kappa V^{\rm obs}\) and turns this limit into the efficiency bound by the convolution theorem. Two lemmas supply the calculations for these steps. Lemma A.2.1 computes the scores of the regular parametric submodels, which give the tangent space in Appendix A.2.1 and split the pathwise derivative of \(\tau\) in Appendix A.2.2 into the contributions of the observational covariate law, the observational outcome law, and the experimental surrogate laws. Lemma A.2.2 computes the moments of the pair. By its part (b), the observational component of the pair reproduces the first two contributions and the experimental component reproduces the third, which, together with a truncation argument that places both components in the tangent space, proves Proposition A.2.1 in Appendix A.2.3. Its part (c) gives the variance used in Appendix A.2.4. Appendix A.2.5 proves both lemmas.
\par
}

\subsubsection{Data laws, parametric submodels, and tangent spaces}\label{a.2.1-data-laws-parametric-submodels-and-tangent-spaces}

{
\emergencystretch=2em
This subsection describes the fixed-design product model, its regular parametric submodels and their scores, and the tangent space in which Appendix A.2.3 identifies the efficient influence-function pair. The stratified sampling scheme consists of two independent datasets, \(\{(X_t,W_t,S_t):t=1,\ldots,N_{\mathrm E}\}\) and \(\{(X_i,S_i,Y_i):i=1,\ldots,N_{\mathrm O}\}\), and the observations are i.i.d. within each sample. Under the fixed design, one experimental observation has density \(f_{X\mid G=\mathrm E}(x)p(x)^w\{1-p(x)\}^{1-w}f_{S\mid W=1,X,G=\mathrm E}(s\mid x)^wf_{S\mid W=0,X,G=\mathrm E}(s\mid x)^{1-w}\), and one observational observation has density \(f_{X\mid G=\mathrm O}(x)f_{S\mid X,G=\mathrm O}(s\mid x)f_{Y\mid S,X,G=\mathrm O}(y\mid s,x)\). These densities determine the one-observation laws \(P^{\rm exp}\) and \(P^{\rm obs}\) of Section 3.1, and \(\mathbb E^{\rm exp}\) and \(\mathbb E^{\rm obs}\) denote their expectation operators. The quantities \(v_w\), \(\sigma^2\), and \(V^{\rm obs}\) are as defined in Section 3.1, and from here on in Appendix A, \(V^{\rm exp}(f_{X\mid G=\mathrm E},p)\) abbreviates \(V^{\rm exp}(f_{X\mid G=\mathrm E}(\cdot),p(\cdot))\).
\par
}

{
\emergencystretch=2em
Consider a one-dimensional regular path indexed by \(r\) in a neighborhood of zero and passing through the true law at \(r=0\), along which each factor that moves is multiplied by \(1+r\) times a bounded function of its arguments whose conditional mean given the conditioning variables of that factor is zero. Along such a path, every integral below is a polynomial in \(r\) whose coefficients are finite under the moment conditions of Theorem 3.1, so the derivatives at \(r=0\) are taken without exchanging limits and integrals, and the bounded functions are the scores defined below. Bounded functions with these conditional-mean restrictions are dense in the corresponding spaces of square-integrable functions, so the tangent spaces below are the same as for the larger class of paths that are differentiable in quadratic mean. Theorem 3.1 gives the bound in this fixed-design product model. The assumptions of Section 2 also imply restrictions on the observed laws. For every measurable set \(B\) and every \(x\) in the target support, \(\mathbb P(S\in B\mid X=\allowbreak x,G=\allowbreak \mathrm O)\le\mathbb P(S\in B\mid W=\allowbreak 1,X=\allowbreak x,G=\allowbreak \mathrm E)+\mathbb P(S\in B\mid W=\allowbreak 0,X=\allowbreak x,G=\allowbreak \mathrm E)\), because \(S\in B\) requires \(S(1)\in B\) or \(S(0)\in B\) by Assumption 2.1, and Assumptions 2.2 and 2.4 give the right-hand side. The product model does not impose these inequality restrictions. Along the path, the design factors \(f_{X\mid G=\mathrm E}(x)\) and \(p(x)\) stay fixed, so the experimental density is \(f_{X\mid G=\mathrm E}(x)p(x)^w\{1-p(x)\}^{1-w}f_{S\mid W=1,X,G=\mathrm E,r}(s\mid x)^wf_{S\mid W=0,X,G=\mathrm E,r}(s\mid x)^{1-w}\), whereas the observational density is \(f_{X\mid G=\mathrm O,r}(x)f_{S\mid X,G=\mathrm O,r}(s\mid x)f_{Y\mid S,X,G=\mathrm O,r}(y\mid s,x)\). For \(w\in\{0,1\}\), define the scores of the factors that move along the path, all at \(r=0\), by \[\begin{aligned}
\ell^{\rm exp}_w(s,x)&:=\partial_r\log f_{S\mid W=w,X,G=\mathrm E,r}(s\mid x)|_{r=0}, & \ell^{\rm obs}_S(s,x)&:=\partial_r\log f_{S\mid X,G=\mathrm O,r}(s\mid x)|_{r=0},\\
\ell^{\rm obs}_X(x)&:=\partial_r\log f_{X\mid G=\mathrm O,r}(x)|_{r=0}, & \ell^{\rm obs}_Y(y,s,x)&:=\partial_r\log f_{Y\mid S,X,G=\mathrm O,r}(y\mid s,x)|_{r=0}.
\end{aligned}
\]
\par
}

{
\emergencystretch=2em
Along the same path, the conditional mean of the long-term outcome is \(\mu_r(s,x):=\allowbreak \int yf_{Y\mid S,X,G=\mathrm O,r}(y\mid s,x)\,d\lambda_Y(y)\). The next lemma collects the score calculations used below, and Appendix A.2.5 proves it.
\par
}

\expandafter\gdef\csname thelemma\endcsname{A.2.\arabic{lemma}}
\expandafter\gdef\csname theHlemma\endcsname{A.2.\arabic{lemma}}
\setcounter{lemma}{0}
\begin{lemma}[Scores of the regular paths]

Along every regular path above, the following hold.

\begin{enumerate}
\def\labelenumi{(\alph{enumi})}
\item
  The score of the complete observational and experimental data at \(r=0\) is \[\sum_{t=1}^{N_{\mathrm E}}\{W_t\ell^{\rm exp}_1(S_t,X_t)+(1-W_t)\ell^{\rm exp}_0(S_t,X_t)\}+\sum_{i=1}^{N_{\mathrm O}}\{\ell^{\rm obs}_X(X_i)+\ell^{\rm obs}_S(S_i,X_i)+\ell^{\rm obs}_Y(Y_i,S_i,X_i)\}.
  \]
\item
  \(\mathbb E[\ell^{\rm exp}_w(S,X)\mid X,W=w,G=\mathrm E]=0\) for \(w\in\{0,1\}\), \(\mathbb E[\ell^{\rm obs}_X(X)\mid G=\mathrm O]=0\), \(\mathbb E[\ell^{\rm obs}_S(S,X)\mid X,G=\mathrm O]=0\), and \(\mathbb E[\ell^{\rm obs}_Y(Y,S,X)\mid S,X,G=\mathrm O]=0\).
\item
  \(\partial_r\mu_r(s,x)|_{r=0}=\mathbb E[\{Y-\mu(S,X)\}\ell^{\rm obs}_Y(Y,S,X)\mid S=s,X=x,G=\mathrm O]\).
\end{enumerate}

\end{lemma}

{
\emergencystretch=2em
The tangent set for the complete observational and experimental data is the \(L_2((P^{\rm exp})^{\otimes N_{\mathrm E}}\otimes (P^{\rm obs})^{\otimes N_{\mathrm O}})\) closure of the scores in part (a), where the component scores satisfy the centering restrictions in part (b) and the complete score is square integrable. Equivalently, the one-observation experimental tangent space is the closure of \(W\ell^{\rm exp}_1(S,X)+(1-W)\ell^{\rm exp}_0(S,X)\), and the one-observation observational tangent space is the closure of \(\ell^{\rm obs}_X(X)+\ell^{\rm obs}_S(S,X)+\ell^{\rm obs}_Y(Y,S,X)\).
\par
}

\subsubsection{Pathwise derivative of the target parameter}\label{a.2.2-pathwise-derivative-of-the-target-parameter}

{
\emergencystretch=2em
This subsection computes the derivative of \(\tau\) along an arbitrary regular path and splits it into the three terms that Appendix A.2.3 matches with the influence-function pair. The target along the path is the regression representation \eqref{eq:identification_regression} of Theorem 2.1 evaluated at the path law, \[\tau(r)=\int f_{X\mid G=\mathrm O,r}(x)\int \mu_r(s,x)\Bigl\{f_{S\mid W=1,X,G=\mathrm E,r}(s\mid x)-f_{S\mid W=0,X,G=\mathrm E,r}(s\mid x)\Bigr\}\,d\lambda_S(s)d\lambda_X(x).
\]
\par
}

{
\emergencystretch=2em
Applying the product rule and evaluating at \(r=0\) gives one contribution from each of \(f_{X\mid G=\mathrm O,r}\), \(\mu_r\), and \(f_{S\mid W=w,X,G=\mathrm E,r}\), \[\begin{aligned}
&\left.\partial_r\tau(r)\right|_{r=0}\\
&=\int f_{X\mid G=\mathrm O}(x)\ell^{\rm obs}_X(x)\int \mu(s,x)\Bigl\{f_{S\mid W=1,X,G=\mathrm E}(s\mid x)-f_{S\mid W=0,X,G=\mathrm E}(s\mid x)\Bigr\}\,d\lambda_S(s)d\lambda_X(x)\\
&+\int f_{X\mid G=\mathrm O}(x)\int \left.\partial_r\mu_r(s,x)\right|_{r=0}\Bigl\{f_{S\mid W=1,X,G=\mathrm E}(s\mid x)-f_{S\mid W=0,X,G=\mathrm E}(s\mid x)\Bigr\}\,d\lambda_S(s)d\lambda_X(x)\\
&+\int f_{X\mid G=\mathrm O}(x)\int \mu(s,x)\Bigl\{f_{S\mid W=1,X,G=\mathrm E}(s\mid x)\ell^{\rm exp}_1(s,x)\\
&\qquad-f_{S\mid W=0,X,G=\mathrm E}(s\mid x)\ell^{\rm exp}_0(s,x)\Bigr\}\,d\lambda_S(s)d\lambda_X(x).
\end{aligned}
\]
\par
}

{
\emergencystretch=2em
The inner integral in the first contribution is \(\tau(x)\), so \(\mathbb E[\ell^{\rm obs}_X(X)\mid G=\mathrm O]=0\) from Lemma A.2.1(b) turns the first contribution into \(\int f_{X\mid G=\mathrm O}(x)\tau(x)\ell^{\rm obs}_X(x)\,d\lambda_X(x)=\allowbreak \mathbb E[\{\tau(X)-\tau\}\ell^{\rm obs}_X(X)\mid G=\allowbreak \mathrm O]\). The second contribution uses the density-ratio cancellation encoded by \(\alpha(S,X)\). By the definition of \(\alpha\) and the surrogate-support condition in Assumption 2.6, \(\alpha(s,x)f_{S\mid X,G=\mathrm O}(s\mid x)=\allowbreak f_{S\mid W=1,X,G=\mathrm E}(s\mid x)-f_{S\mid W=0,X,G=\mathrm E}(s\mid x)\), so writing \(\mathbb E[\alpha(S,X)\{\partial_r\mu_r(S,X)|_{r=0}\}\mid G=\mathrm O]\) as an integral against \(f_{X\mid G=\mathrm O}(x)f_{S\mid X,G=\mathrm O}(s\mid x)\) shows that this expectation equals the second contribution. Substituting Lemma A.2.1(c) and using iterated expectations, the second contribution becomes \(\mathbb E[\alpha(S,X)\{Y-\mu(S,X)\}\ell^{\rm obs}_Y(Y,S,X)\mid G=\allowbreak \mathrm O]\). Combining the three contributions yields \[\begin{aligned}
&\left.\partial_r\tau(r)\right|_{r=0}\\
&=\underbrace{\mathbb E[\{\tau(X)-\tau\}\ell^{\rm obs}_X(X)\mid G=\mathrm O]}_{\text{(I)}}
+\underbrace{\mathbb E[\alpha(S,X)\{Y-\mu(S,X)\}\ell^{\rm obs}_Y(Y,S,X)\mid G=\mathrm O]}_{\text{(II)}}\\
&+\underbrace{\begin{aligned}[t]&\int f_{X\mid G=\mathrm O}(x)\int \mu(s,x)\Bigl\{f_{S\mid W=1,X,G=\mathrm E}(s\mid x)\ell^{\rm exp}_1(s,x)\\
&\qquad-f_{S\mid W=0,X,G=\mathrm E}(s\mid x)\ell^{\rm exp}_0(s,x)\Bigr\}\,d\lambda_S(s)d\lambda_X(x)\end{aligned}}_{\text{(III)}}.
\end{aligned}
\]
\par
}

Term (I) is the contribution of the observational covariate law, Term (II) that of the observational outcome law, and Term (III) that of the experimental surrogate laws. Lemma A.2.2(b) below shows that the observational component of the pair in Theorem 3.1 reproduces Terms (I) and (II) and that the experimental component reproduces Term (III).

\subsubsection{Verification of the efficient influence-function pair}\label{a.2.3-verification-of-the-efficient-influence-function-pair}

This subsection proves Proposition A.2.1, which identifies the pair in Theorem 3.1 as the efficient influence-function pair on which Appendix A.2.4 bases the bound. Definition A.2.1 states the representation that the pair must satisfy, which is the pathwise derivative of Appendix A.2.2 written as an inner product with the scores.

\expandafter\gdef\csname thedefinition\endcsname{A.2.\arabic{definition}}
\expandafter\gdef\csname theHdefinition\endcsname{A.2.\arabic{definition}}
\setcounter{definition}{0}
\begin{definition}[Influence-function pairs and the efficient influence function]

{
\emergencystretch=2em
Let \(\ell^{\rm exp}(S,W,X)=\allowbreak W\ell^{\rm exp}_1(S,X)+(1-W)\ell^{\rm exp}_0(S,X)\) denote the score for one experimental observation, and let \(\ell^{\rm obs}(X,S,Y)=\allowbreak \ell^{\rm obs}_X(X)+\ell^{\rm obs}_S(S,X)+\ell^{\rm obs}_Y(Y,S,X)\) denote the score for one observational observation. An influence-function pair for \(\tau\) is a pair of square-integrable, mean-zero functions \((\phi^{\rm exp},\phi^{\rm obs})\) such that, for every regular parametric submodel through the true distribution, \[\left.\partial_r\tau(r)\right|_{r=0}
=\mathbb E^{\rm exp}\left[\phi^{\rm exp}(S,W,X)\ell^{\rm exp}(S,W,X)\right]
+\mathbb E^{\rm obs}\left[\phi^{\rm obs}(X,S,Y)\ell^{\rm obs}(X,S,Y)\right].
\]
\par
}

{
\emergencystretch=2em
The efficient influence-function pair is the influence-function pair whose components belong to the closures of the one-observation tangent spaces of Appendix A.2.1, that is, the Riesz representer of the pathwise derivative in the product tangent space. For samples of sizes \(N_{\mathrm E}\) and \(N_{\mathrm O}\), the full-sample influence function formed from an influence-function pair is \[\frac{1}{N_{\mathrm E}}\sum_{t=1}^{N_{\mathrm E}} \phi^{\rm exp}(S_t,W_t,X_t)
+\frac{1}{N_{\mathrm O}}\sum_{i=1}^{N_{\mathrm O}} \phi^{\rm obs}(X_i,S_i,Y_i).
\]
\par
}

\end{definition}

\expandafter\gdef\csname theproposition\endcsname{A.2.\arabic{proposition}}
\expandafter\gdef\csname theHproposition\endcsname{A.2.\arabic{proposition}}
\setcounter{proposition}{0}
\begin{proposition}[Efficient influence-function pair in the fixed-design product model]

{
\emergencystretch=2em
In the fixed-design product model, let \(\phi^{\rm exp}(S,W,X)\) and \(\phi^{\rm obs}(X,S,Y)\) be the functions displayed in Theorem 3.1. Then the following hold.
\par
}

\begin{enumerate}
\def\labelenumi{(\roman{enumi})}
\item
  The pair \((\phi^{\rm exp},\phi^{\rm obs})\) is an influence-function pair for \(\tau\) in the sense of Definition A.2.1, and \(\phi^{\rm obs}(X,S,Y)=\psi^{\rm obs}(X,S,Y)-\tau\) for the uncentered observational score \(\psi^{\rm obs}\) of Theorem 3.1.
\item
  Along every regular path, the product of the full-sample influence function formed from this pair and the score of the complete observational and experimental data in Lemma A.2.1(a) has expectation \(\partial_r\tau(r)|_{r=0}\).
\item
  Both components belong to the closures of the one-observation tangent spaces of Appendix A.2.1. Hence \(\phi^{\rm exp}\) and \(\phi^{\rm obs}\) are the experimental and observational efficient influence functions, and the full-sample influence function formed from them is the efficient influence function for the complete observational and experimental dataset.
\end{enumerate}

\end{proposition}

The proof uses the following properties of the pair, which Appendix A.2.5 proves. Part (a) gives the centering, part (b) matches the two score pairings with the terms of the decomposition in Appendix A.2.2, and part (c) gives the second moments used for square integrability here and for the variance in Appendix A.2.4.

\expandafter\gdef\csname thelemma\endcsname{A.2.\arabic{lemma}}
\expandafter\gdef\csname theHlemma\endcsname{A.2.\arabic{lemma}}
\setcounter{lemma}{1}
\begin{lemma}[Moments of the candidate influence-function pair]

{
\emergencystretch=2em
Let \(\phi^{\rm exp}\) and \(\phi^{\rm obs}\) be the functions displayed in Theorem 3.1.
\par
}

\begin{enumerate}
\def\labelenumi{(\alph{enumi})}
\item
  \(\mathbb E[\phi^{\rm exp}(S,W,X)\mid W,X,G=\mathrm E]=0\) and \(\mathbb E[\alpha(S,X)\{Y-\mu(S,X)\}\mid S,X,G=\mathrm O]=0\). In particular, \(\mathbb E^{\rm exp}[\phi^{\rm exp}(S,W,X)]=0\) and \(\mathbb E^{\rm obs}[\phi^{\rm obs}(X,S,Y)]=0\).
\item
  Along every regular path of Appendix A.2.1, with the terms of the decomposition in Appendix A.2.2, \(\mathbb E^{\rm exp}[\phi^{\rm exp}(S,W,X)\ell^{\rm exp}(S,W,X)]\) equals Term (III) and \(\mathbb E^{\rm obs}[\phi^{\rm obs}(X,S,Y)\ell^{\rm obs}(X,S,Y)]\) equals the sum of Terms (I) and (II).
\item
  \(\mathbb E^{\rm exp}[\phi^{\rm exp}(S,W,X)^2]=\allowbreak V^{\rm exp}(f_{X\mid G=\mathrm E},p)\) and \(\mathbb E^{\rm obs}[\phi^{\rm obs}(X,S,Y)^2]=V^{\rm obs}\), and both are finite under the moment conditions of Theorem 3.1.
\end{enumerate}

\end{lemma}

{
\emergencystretch=2em
\textbf{Proof of Proposition A.2.1.} Steps (i), (ii), and (iii) below prove parts (i), (ii), and (iii) of the proposition in turn. (i) We first verify the conditions of Definition A.2.1. Lemma A.2.2(a) shows that both functions have mean zero, and Lemma A.2.2(c) shows that they are square integrable under the moment conditions of Theorem 3.1. Along every regular path, Lemma A.2.2(b) shows that \(\mathbb E^{\rm exp}[\phi^{\rm exp}\ell^{\rm exp}]+\mathbb E^{\rm obs}[\phi^{\rm obs}\ell^{\rm obs}]\) is the sum of Terms (I)--(III), which is \(\partial_r\tau(r)|_{r=0}\) by the decomposition in Appendix A.2.2. Hence \((\phi^{\rm exp},\phi^{\rm obs})\) is an influence-function pair, and \(\phi^{\rm obs}=\psi^{\rm obs}-\tau\) follows from the definitions in Theorem 3.1.
\par
}

\begin{enumerate}
\def\labelenumi{(\roman{enumi})}
\setcounter{enumi}{1}
\item
  We next verify the derivative identity for the full-sample influence function and the complete-sample score of Lemma A.2.1(a), which by Definition A.2.1 is \(\sum_{s=1}^{N_{\mathrm E}}\ell^{\rm exp}(S_s,W_s,X_s)+\allowbreak\sum_{j=1}^{N_{\mathrm O}}\ell^{\rm obs}(X_j,S_j,Y_j)\). Expanding their product gives four double sums. In the experimental double sum, a term with \(t\neq s\) factors by independence into \(\mathbb E^{\rm exp}[\phi^{\rm exp}(S,W,X)]\mathbb E^{\rm exp}[\ell^{\rm exp}(S,W,X)]\), the mean of a centered influence function times the mean of a centered score, and is zero by Lemma A.2.2(a) and Lemma A.2.1(b). The same holds for \(i\neq j\) in the observational double sum, and the two mixed double sums vanish by independence of the two samples and centering. Only the diagonal terms remain, and their \(N_{\mathrm E}\) and \(N_{\mathrm O}\) multiplicities cancel the factors \(1/N_{\mathrm E}\) and \(1/N_{\mathrm O}\). By part (i), \[\begin{aligned}
  &\mathbb E\Biggl[\Biggl\{\frac1{N_{\mathrm E}}\sum_{t=1}^{N_{\mathrm E}}\phi^{\rm exp}(S_t,W_t,X_t)\\
  &\qquad+\frac1{N_{\mathrm O}}\sum_{i=1}^{N_{\mathrm O}}\phi^{\rm obs}(X_i,S_i,Y_i)\Biggr\}\Biggl\{\sum_{s=1}^{N_{\mathrm E}}\ell^{\rm exp}(S_s,W_s,X_s)+\sum_{j=1}^{N_{\mathrm O}}\ell^{\rm obs}(X_j,S_j,Y_j)\Biggr\}\Biggr]\\
  &=\mathbb E^{\rm exp}[\phi^{\rm exp}(S,W,X)\ell^{\rm exp}(S,W,X)]+\mathbb E^{\rm obs}[\phi^{\rm obs}(X,S,Y)\ell^{\rm obs}(X,S,Y)]=\left.\partial_r\tau(r)\right|_{r=0}.
  \end{aligned}
  \]
\item
  Finally, both components belong to the tangent-space closures of Appendix A.2.1. By Lemma A.2.2(a), the experimental component has conditional mean zero given treatment and covariates. The observational component is the sum of the mean-zero covariate function \(\tau(X)-\tau\) and the residual \(\alpha(S,X)\{Y-\mu(S,X)\}\), whose conditional mean given \((S,X)\) is zero. Truncating each square-integrable component and subtracting its corresponding conditional mean gives bounded centered approximations converging in \(L_2\). These bounded directions are scores of the factorwise submodels in the stated fixed-design product model. Their limits therefore lie in its tangent-space closures. By Definition A.2.1 and part (i), \((\phi^{\rm exp},\phi^{\rm obs})\) is the efficient influence-function pair. The full-sample influence function formed from it then lies in the tangent space of the complete data, and by part (ii) it is the efficient influence function for the complete data. This proves Proposition A.2.1.
\end{enumerate}

\subsubsection{Fixed-design semiparametric efficiency bound}\label{a.2.4-fixed-design-semiparametric-efficiency-bound}

{
\emergencystretch=2em
This subsection computes the variance of the efficient influence function and completes the proof of Theorem 3.1. By Lemma A.2.2(a) and (c), each \(\phi^{\rm exp}(S_t,W_t,X_t)\) has mean zero and variance \(V^{\rm exp}(f_{X\mid G=\mathrm E},p)\), and each \(\phi^{\rm obs}(X_i,S_i,Y_i)\) has mean zero and variance \(V^{\rm obs}\). The two sample averages in the full-sample efficient influence function of Proposition A.2.1(iii) are independent and centered, and unequal-index covariances within each sample vanish because the observations are i.i.d. within each sample. Hence the full-sample efficient influence function has variance \[\frac1{N_{\mathrm E}} V^{\rm exp}(f_{X\mid G=\mathrm E},p)+\frac1{N_{\mathrm O}} V^{\rm obs}.
\]
\par
}

{
\emergencystretch=2em
To apply the convolution theorem, consider the normalized coordinates on the \(\sqrt{N_{\mathrm E}}\) scale. The influence-function pair has the normalized coordinates \((\phi^{\rm exp}(S,W,X),\allowbreak \sqrt{N_{\mathrm E}/N_{\mathrm O}}\,\phi^{\rm obs}(X,S,Y))\), and the corresponding score coordinates are \((\ell^{\rm exp},\sqrt{N_{\mathrm O}/N_{\mathrm E}}\,\ell^{\rm obs})\). Their pairing preserves the unscaled pathwise derivative, because \(\mathbb E^{\rm exp}[\phi^{\rm exp}\ell^{\rm exp}]+\mathbb E^{\rm obs}[\{\sqrt{N_{\mathrm E}/N_{\mathrm O}}\,\phi^{\rm obs}\}\{\sqrt{N_{\mathrm O}/N_{\mathrm E}}\,\ell^{\rm obs}\}]=\allowbreak \partial_r\tau(r)|_{r=0}\) by Proposition A.2.1(i). By Lemma A.2.2(c), the squared norm of the normalized influence-function pair, which is \(N_{\mathrm E}\) times the preceding variance, is \[\begin{aligned}
\mathbb E^{\rm exp}[\phi^{\rm exp}(S,W,X)^2]+\frac{N_{\mathrm E}}{N_{\mathrm O}}\mathbb E^{\rm obs}[\phi^{\rm obs}(X,S,Y)^2]
&=V^{\rm exp}(f_{X\mid G=\mathrm E},p)+\frac{N_{\mathrm E}}{N_{\mathrm O}} V^{\rm obs}\\
&\longrightarrow V^{\rm exp}(f_{X\mid G=\mathrm E},p)+\kappa V^{\rm obs}.
\end{aligned}
\]
\par
}

{
\emergencystretch=2em
Proposition A.2.1 identifies the influence-function pair as the efficient influence-function pair in the stated model, and the normalized coordinates give its variance on the \(\sqrt{N_{\mathrm E}}\) scale. Theorem 25.20 of \citet{vanderVaart1998} is stated for one i.i.d. sample, so we apply its argument to the two independent samples. Along each path of Appendix A.2.1, and jointly along finitely many such paths, the log-likelihood ratio of the complete data is locally asymptotically normal. At the parameter value \(r/\sqrt{N_{\mathrm E}}\), it equals \(rN_{\mathrm E}^{-1/2}\{\allowbreak\sum_{t=1}^{N_{\mathrm E}}\ell^{\rm exp}(S_t,W_t,X_t)+\allowbreak\sum_{i=1}^{N_{\mathrm O}}\ell^{\rm obs}(X_i,S_i,Y_i)\}-(r^2/\allowbreak 2)\{\mathbb E^{\rm exp}[\ell^{\rm exp}(S,W,X)^2]+(N_{\mathrm O}/\allowbreak N_{\mathrm E})\mathbb E^{\rm obs}[\ell^{\rm obs}(X,S,Y)^2]\}\) plus a term that converges to zero in probability, by the quadratic-mean differentiability of each factor and the central limit theorem in each sample. The convolution argument in the proof of Theorem 25.20, applied to this locally asymptotically normal sequence of experiments with the normalized score coordinates above instead of to one i.i.d. sample, gives the convolution lower bound, and the displayed limit under \(N_{\mathrm E}/N_{\mathrm O}\to\kappa\) gives \(V^{\rm exp}(f_{X\mid G=\mathrm E},p)+\kappa V^{\rm obs}\) as the semiparametric efficiency bound for regular estimators of \(\tau\) on the \(\sqrt{N_{\mathrm E}}\) scale. If this limiting norm is zero, the lower bound is zero directly. Together with the efficient influence functions in Proposition A.2.1 and their second moments in Lemma A.2.2(c), this proves Theorem 3.1.
\par
}

\subsubsection{Proofs of Lemmas A.2.1 and A.2.2}\label{a.2.5-proofs-of-lemmas-a.2.1-and-a.2.2}

\textbf{Proof of Lemma A.2.1.} For part (a), the joint log-likelihood of the two independent samples along the path is \[\begin{aligned}
&\sum_{t=1}^{N_{\mathrm E}}\Bigl[\log f_{X\mid G=\mathrm E}(X_t)+W_t\log p(X_t)+(1-W_t)\log\{1-p(X_t)\}\\
&\qquad\quad+W_t\log f_{S\mid W=1,X,G=\mathrm E,r}(S_t\mid X_t)+(1-W_t)\log f_{S\mid W=0,X,G=\mathrm E,r}(S_t\mid X_t)\Bigr]\\
&\quad+\sum_{i=1}^{N_{\mathrm O}}\Bigl[\log f_{X\mid G=\mathrm O,r}(X_i)+\log f_{S\mid X,G=\mathrm O,r}(S_i\mid X_i)+\log f_{Y\mid S,X,G=\mathrm O,r}(Y_i\mid S_i,X_i)\Bigr].
\end{aligned}
\]

{
\emergencystretch=2em
Only the two surrogate terms of each experimental summand and the three terms of each observational summand depend on \(r\), so term-by-term differentiation at \(r=0\) gives the score in part (a). For part (b), differentiating the normalization of each density at \(r=0\) gives the four centering restrictions. For instance, \(0=\allowbreak \partial_r\{\int f_{S\mid W=w,X,G=\mathrm E,r}(s\mid x)\,d\lambda_S(s)\}|_{r=0}=\int f_{S\mid W=w,X,G=\mathrm E}(s\mid x)\ell^{\rm exp}_w(s,x)\,d\lambda_S(s)\). For part (c), differentiating \(\mu_r(s,x)\) under the integral gives \[\begin{aligned}
\left.\partial_r\mu_r(s,x)\right|_{r=0}
&=\int y f_{Y\mid S,X,G=\mathrm O}(y\mid s,x)\ell^{\rm obs}_Y(y,s,x)\,d\lambda_Y(y)\\
&=\int \{y-\mu(s,x)\}f_{Y\mid S,X,G=\mathrm O}(y\mid s,x)\ell^{\rm obs}_Y(y,s,x)\,d\lambda_Y(y)\\
&=\mathbb E[\{Y-\mu(S,X)\}\ell^{\rm obs}_Y(Y,S,X)\mid S=s,X=x,G=\mathrm O].
\end{aligned}
\]
\par
}

{
\emergencystretch=2em
The second equality subtracts \(\mu(s,x)\) multiplied by the conditional score mean, which is zero by part (b).
\par
}

{
\emergencystretch=2em
\textbf{Proof of Lemma A.2.2.} For part (a), let \(w\in\{0,1\}\). By the definition of \(\mu_w(x)\), \(\int\{\mu(s,x)-\mu_w(x)\}f_{S\mid W=w,X,G=\mathrm E}(s\mid x)\,d\lambda_S(s)=\allowbreak 0\). Given \(W=w\) and \(X=x\), \(\phi^{\rm exp}(S,W,X)\) is \(\mu(S,x)-\mu_w(x)\) times a factor that depends only on \(x\), so \(\mathbb E[\phi^{\rm exp}(S,W,X)\mid W,X,G=\mathrm E]=0\). Averaging over \(W\) given \(X=x\), the assignment probabilities cancel the inverse-assignment-probability factors, and \[\begin{aligned}
\mathbb E[\phi^{\rm exp}(S,W,X)\mid X=x,G=\mathrm E]
&=\frac{f_{X\mid G=\mathrm O}(x)}{f_{X\mid G=\mathrm E}(x)}\Biggl[\int\{\mu(s,x)-\mu_1(x)\}f_{S\mid W=1,X,G=\mathrm E}(s\mid x)\,d\lambda_S(s)\\
&\qquad-\int\{\mu(s,x)-\mu_0(x)\}f_{S\mid W=0,X,G=\mathrm E}(s\mid x)\,d\lambda_S(s)\Biggr]=0.
\end{aligned}
\]
\par
}

{
\emergencystretch=2em
Hence \(\mathbb E^{\rm exp}[\phi^{\rm exp}(S,W,X)]=0\). For the observational component, \(\mathbb E[\alpha(S,X)\{Y-\mu(S,X)\}\mid S,X,G=\allowbreak \mathrm O]=\allowbreak \alpha(S,X)\mathbb E[Y-\mu(S,X)\mid S,X,G=\allowbreak \mathrm O]=\allowbreak 0\) by the definition of \(\mu\). Since \(\mathbb E[\tau(X)\mid G=\mathrm O]=\tau\) by \eqref{eq:identification_regression}, it follows that \(\mathbb E^{\rm obs}[\phi^{\rm obs}(X,S,Y)]=\allowbreak \mathbb E[\tau(X)-\tau\mid G=\allowbreak \mathrm O]+\mathbb E[\alpha(S,X)\mathbb E[Y-\mu(S,X)\mid S,X,G=\allowbreak \mathrm O]\mid G=\allowbreak \mathrm O]=\allowbreak 0\).
\par
}

{
\emergencystretch=2em
For part (b), consider first the experimental block. Because \(W^2=W\), \((1-W)^2=1-W\), and \(W(1-W)=0\), \[\begin{aligned}
&\mathbb E^{\rm exp}\Bigl[\phi^{\rm exp}(S,W,X)\{W\ell^{\rm exp}_1(S,X)+(1-W)\ell^{\rm exp}_0(S,X)\}\Bigr]\\
&=\mathbb E^{\rm exp}\Biggl[\frac{f_{X\mid G=\mathrm O}(X)}{f_{X\mid G=\mathrm E}(X)}\frac{W}{p(X)}\{\mu(S,X)-\mu_1(X)\}\ell^{\rm exp}_1(S,X)\Biggr]\\
&-\mathbb E^{\rm exp}\Biggl[\frac{f_{X\mid G=\mathrm O}(X)}{f_{X\mid G=\mathrm E}(X)}\frac{1-W}{1-p(X)}\{\mu(S,X)-\mu_0(X)\}\ell^{\rm exp}_0(S,X)\Biggr].
\end{aligned}
\]
\par
}

Integration under the experimental law makes the density ratio and the appropriate assignment probability cancel in each treatment or control group, so the right side equals \[\begin{aligned}
&\int f_{X\mid G=\mathrm E}(x)p(x)\frac{f_{X\mid G=\mathrm O}(x)}{f_{X\mid G=\mathrm E}(x)}\frac1{p(x)}\int\{\mu(s,x)\\
&\qquad-\mu_1(x)\}\ell^{\rm exp}_1(s,x)f_{S\mid W=1,X,G=\mathrm E}(s\mid x)\,d\lambda_S(s)d\lambda_X(x)\\
&-\int f_{X\mid G=\mathrm E}(x)\{1-p(x)\}\frac{f_{X\mid G=\mathrm O}(x)}{f_{X\mid G=\mathrm E}(x)}\frac1{1-p(x)}\int\{\mu(s,x)\\
&\qquad-\mu_0(x)\}\ell^{\rm exp}_0(s,x)f_{S\mid W=0,X,G=\mathrm E}(s\mid x)\,d\lambda_S(s)d\lambda_X(x)\\
&=\int f_{X\mid G=\mathrm O}(x)\int \mu(s,x)\Bigl\{f_{S\mid W=1,X,G=\mathrm E}(s\mid x)\ell^{\rm exp}_1(s,x)\\
&\qquad-f_{S\mid W=0,X,G=\mathrm E}(s\mid x)\ell^{\rm exp}_0(s,x)\Bigr\}\,d\lambda_S(s)d\lambda_X(x).
\end{aligned}
\]

{
\emergencystretch=2em
The last equality uses \(\mathbb E[\ell^{\rm exp}_w(S,X)\mid X,W=w,G=\mathrm E]=0\) from Lemma A.2.1(b) to remove the terms containing \(\mu_w(x)\) for \(w\in\{0,1\}\), and the result is Term (III).
\par
}

{
\emergencystretch=2em
For the observational block, expanding \(\phi^{\rm obs}(X,S,Y)\{\ell^{\rm obs}_X(X)+\ell^{\rm obs}_S(S,X)+\ell^{\rm obs}_Y(Y,S,X)\}\) gives six products. The product \(\{\tau(X)-\tau\}\ell^{\rm obs}_S(S,X)\) has mean zero after conditioning on \(X\), because \(\mathbb E[\ell^{\rm obs}_S(S,X)\mid X,G=\mathrm O]=0\). The product \(\{\tau(X)-\tau\}\ell^{\rm obs}_Y(Y,S,X)\) has mean zero after conditioning on \((S,X)\), because \(\tau(X)-\tau\) is \(X\)-measurable and \(\mathbb E[\ell^{\rm obs}_Y(Y,S,X)\mid S,X,G=\mathrm O]=0\). The products \(\alpha(S,X)\{Y-\mu(S,X)\}\ell^{\rm obs}_X(X)\) and \(\alpha(S,X)\{Y-\mu(S,X)\}\ell^{\rm obs}_S(S,X)\) have mean zero because \(\mathbb E[Y-\mu(S,X)\mid S,X,G=\mathrm O]=0\). The remaining two products give \[\begin{aligned}
&\mathbb E^{\rm obs}\Bigl[\phi^{\rm obs}(X,S,Y)\{\ell^{\rm obs}_X(X)+\ell^{\rm obs}_S(S,X)+\ell^{\rm obs}_Y(Y,S,X)\}\Bigr]\\
&=\mathbb E[\{\tau(X)-\tau\}\ell^{\rm obs}_X(X)\mid G=\mathrm O]+\mathbb E[\alpha(S,X)\{Y-\mu(S,X)\}\ell^{\rm obs}_Y(Y,S,X)\mid G=\mathrm O],
\end{aligned}
\]
\par
}

which is the sum of Terms (I) and (II).

{
\emergencystretch=2em
For part (c), the cross-product of the treatment and control terms vanishes because \(W(1-W)=0\), and \(W^2=W\) and \((1-W)^2=1-W\) give \[\phi^{\rm exp}(S,W,X)^2
=\left\{\frac{f_{X\mid G=\mathrm O}(X)}{f_{X\mid G=\mathrm E}(X)}\right\}^2
\Biggl[\frac{W}{p(X)^2}\{\mu(S,X)-\mu_1(X)\}^2+\frac{1-W}{\{1-p(X)\}^2}\{\mu(S,X)-\mu_0(X)\}^2\Biggr].
\]
\par
}

{
\emergencystretch=2em
Conditional on \(X=x\), this gives \(\mathbb E[\phi^{\rm exp}(S,W,X)^2\mid X=\allowbreak x,G=\allowbreak \mathrm E]=\allowbreak \{f_{X\mid G=\mathrm O}(x)/\allowbreak f_{X\mid G=\mathrm E}(x)\}^2[v_1(x)/\allowbreak p(x)+v_0(x)/\allowbreak \{1-p(x)\}]\), and integrating under the experimental covariate law yields \[\begin{aligned}
\mathbb E^{\rm exp}[\phi^{\rm exp}(S,W,X)^2]
&=\int\frac{f_{X\mid G=\mathrm O}(x)^2}{f_{X\mid G=\mathrm E}(x)}\left\{\frac{v_1(x)}{p(x)}+\frac{v_0(x)}{1-p(x)}\right\}\,d\lambda_X(x)\\
&=\mathbb E\left[\frac{f_{X\mid G=\mathrm O}(X)}{f_{X\mid G=\mathrm E}(X)}\left\{\frac{v_1(X)}{p(X)}+\frac{v_0(X)}{1-p(X)}\right\}\middle|G=\mathrm O\right]
=V^{\rm exp}(f_{X\mid G=\mathrm E},p).
\end{aligned}
\]
\par
}

{
\emergencystretch=2em
For the observational component, \(\phi^{\rm obs}(X,S,Y)^2=\allowbreak \{\tau(X)-\tau\}^2+2\{\tau(X)-\tau\}\alpha(S,X)\{Y-\mu(S,X)\}+\alpha(S,X)^2\{Y-\mu(S,X)\}^2\). The cross term has mean zero because \(\mathbb E[\{\tau(X)-\tau\}\alpha(S,X)\{Y-\mu(S,X)\}\mid G=\allowbreak \mathrm O]=\allowbreak \mathbb E[\{\tau(X)-\tau\}\alpha(S,X)\mathbb E[Y-\mu(S,X)\mid S,X,G=\allowbreak \mathrm O]\mid G=\allowbreak \mathrm O]=\allowbreak 0\). Consequently, \(\mathbb E^{\rm obs}[\phi^{\rm obs}(X,S,Y)^2]=\allowbreak \mathbb E[\{\tau(X)-\tau\}^2\mid G=\allowbreak \mathrm O]+\mathbb E[\alpha(S,X)^2\sigma^2(S,X)\mid G=\allowbreak \mathrm O]=\allowbreak V^{\rm obs}\). Both second moments are finite under the moment conditions of Theorem 3.1.
\par
}

\subsection{Proof of Theorem 3.2}\label{a.3-proof-of-theorem-3.2}

{
\emergencystretch=2em
Appendix A.3 states the regularity conditions of Theorem 3.2 and then proves the theorem. Throughout this proof, for each \(w\in\{0,1\}\), \(0<v_w(x)<\infty\) for \(f_{X\mid G=\mathrm O}\)-almost every \(x\) and \(0<\mathbb E[\sqrt{v_1(X)}+\sqrt{v_0(X)}\mid G=\mathrm O]<\infty\). The comparison class consists of measurable assignment probabilities taking values in \([\varepsilon,1-\varepsilon]\) for \(f_{X\mid G=\mathrm O}\)-almost every \(x\), where \(\varepsilon\) is the overlap constant introduced in Assumption 2.6, and measurable nonnegative experimental covariate densities that integrate to one and satisfy \(f_{X\mid G=\mathrm O}\ll f_{X\mid G=\mathrm E}\). The assignment probability \(p^\star\) and the density \(f_{X\mid G=\mathrm E}^\star\) displayed in Theorem 3.2 belong to this comparison class. Assignment probabilities are identified up to \(f_{X\mid G=\mathrm O}\)-almost-everywhere equality and experimental covariate densities up to \(\lambda_X\)-almost-everywhere equality; uniqueness in Theorem 3.2 is understood in these senses.
\par
}

{
\emergencystretch=2em
\textbf{Proof.} The proof minimizes \(V^{\rm exp}(f_{X\mid G=\mathrm E},p)\) in two steps. Step (i) minimizes the integrand over the assignment probability at each \(x\), and step (ii) minimizes the resulting lower bound over the experimental covariate density by the Cauchy--Schwarz inequality and obtains uniqueness from the equality conditions.
\par
}

\begin{enumerate}
\def\labelenumi{(\roman{enumi})}
\item
  For each \(x\), the map \(r\mapsto v_1(x)/r+v_0(x)/(1-r)\) is strictly convex on \((0,1)\), and its first-order condition \(-v_1(x)/r^2+v_0(x)/(1-r)^2=0\) has the unique solution \(p^\star(x)\). Its maintained feasibility therefore makes it the unique pointwise minimizer, with \(v_1(x)/\allowbreak p^\star(x)+v_0(x)/\allowbreak \{1-p^\star(x)\}=\allowbreak \{\sqrt{v_1(x)}+\sqrt{v_0(x)}\}^2\).
\item
  After minimizing over the assignment probability as in step (i), the Cauchy--Schwarz inequality gives, for every feasible density, \[\begin{aligned}
  V^{\rm exp}(f_{X\mid G=\mathrm E},p)
  &\geq\int\frac{f_{X\mid G=\mathrm O}(x)^2
  \{\sqrt{v_1(x)}+\sqrt{v_0(x)}\}^2}
  {f_{X\mid G=\mathrm E}(x)}d\lambda_X(x)\\
  &\geq\left\{\int f_{X\mid G=\mathrm O}(x)
  \{\sqrt{v_1(x)}+\sqrt{v_0(x)}\}d\lambda_X(x)\right\}^2
  =V^{{\rm exp},\star}.
  \end{aligned}
  \]
\end{enumerate}

{
\emergencystretch=2em
The positive finite normalizing expectation makes \(f_{X\mid G=\mathrm E}^\star\) a density. Equality in the Cauchy--Schwarz inequality requires \(f_{X\mid G=\mathrm E}=\allowbreak f_{X\mid G=\mathrm O}\{\sqrt{v_1}+\sqrt{v_0}\}/\allowbreak \mathbb E[\sqrt{v_1(X)}+\sqrt{v_0(X)}\mid G=\allowbreak \mathrm O]\) almost everywhere. The positivity of \(v_1\) and \(v_0\) gives the target domination \(f_{X\mid G=\mathrm O}\ll f_{X\mid G=\mathrm E}^\star\), and this equality condition, together with the unique pointwise minimizer of step (i), proves uniqueness. This establishes all three formulas in Theorem 3.2.
\par
}

\subsection{Proof of Proposition 4.1}\label{a.4-proof-of-proposition-4.1}

{
\emergencystretch=2em
Appendix A.4 proves Proposition 4.1, the adaptive-design counterpart of Theorem 2.1. The proof uses Assumptions 2.1--2.6 and Assumptions 4.1--4.3, as stated in the proposition. The covariate density \(f_{X\mid G=\mathrm E,t,N_{\mathrm O}}\) and the assignment probability \(p_{t,N_{\mathrm O}}\) of Section 4 are jointly measurable in the covariate value and the history before round \(t\).
\par
}

{
\emergencystretch=2em
\textbf{Proof.} Fix a round \(t\). The representation to be proved is \[\begin{aligned}
\tau
&=\underbrace{\mathbb E\left[
\mu_1(X)-\mu_0(X)
+\alpha(S,X)\{Y-\mu(S,X)\}
\middle|G=\mathrm O\right]}_{\text{(I)}}\\
&\quad+\underbrace{\mathbb E\Biggl[
\frac{f_{X\mid G=\mathrm O}(X_t)}{f_{X\mid G=\mathrm E,t,N_{\mathrm O}}(X_t)}\Biggl\{
\frac{W_t\{\mu(S_t,X_t)-\mu_1(X_t)\}}
{p_{t,N_{\mathrm O}}(X_t)}
-\frac{(1-W_t)\{\mu(S_t,X_t)-\mu_0(X_t)\}}
{1-p_{t,N_{\mathrm O}}(X_t)}
\Biggr\}
\mathrel{\Big|}\mathcal F_{t-1,N_{\mathrm O}}\Biggr]}_{\text{(II)}}.
\end{aligned}
\]
\par
}

{
\emergencystretch=2em
Term (I) is the observational term of the augmented representation \eqref{eq:identification_augmented}, and step (i) shows that it equals \(\tau\) by the conditional-mean identity established in the proof of Theorem 2.1. Term (II) is the weighted experimental residual at round \(t\), and step (ii) shows that it is zero because Assumptions 4.2 and 4.3 connect round \(t\) to the fixed experimental conditional distributions of the surrogate used in Section 2.
\par
}

\begin{enumerate}
\def\labelenumi{(\roman{enumi})}
\item
  We first show that term (I) equals \(\tau\). Under Assumptions 2.1--2.6, step (i) of the proof of Theorem 2.1 in Appendix A.1 gives \(\mathbb E[Y(w)\mid X=x,G=\mathrm O]=\mu_w(x)\) for \(w\in\{0,1\}\) and target-supported \(x\). Because \(Y-\mu(S,X)\) has conditional mean zero given \((S,X)\) and \(G=\mathrm O\), this identity, iterated expectations, and the definition of \(\tau\) give \[\mathbb E\left[\mu_1(X)-\mu_0(X)+\alpha(S,X)\{Y-\mu(S,X)\}\middle|G=\mathrm O\right]
  =\mathbb E[\mu_1(X)-\mu_0(X)\mid G=\mathrm O]
  =\tau.
  \]
\item
  We next show that term (II) is zero. Because the surrogates and covariates take values in Euclidean spaces (Section 2.2), regular versions of the conditional distributions used below exist, and the joint measurability of the design functions makes the integrands below measurable. Fix \(w\in\{0,1\}\) and a target-supported covariate value \(x\). Assumption 2.1 (consistency), Assumption 4.2 (sequential treatment randomization), and Assumption 4.3 (history-invariant conditional surrogate distributions) give, for every measurable set \(B\), in that order, \[\begin{aligned}
  \mathbb P(S_t\in B\mid W_t=w,X_t=x,\mathcal F_{t-1,N_{\mathrm O}})
  &=\mathbb P(S_t(w)\in B\mid W_t=w,X_t=x,\mathcal F_{t-1,N_{\mathrm O}})\\
  &=\mathbb P(S_t(w)\in B\mid X_t=x,\mathcal F_{t-1,N_{\mathrm O}})\\
  &=\mathbb P(S\in B\mid W=w,X=x,G=\mathrm E).
  \end{aligned}
  \]
\end{enumerate}

{
\emergencystretch=2em
Integrating this conditional-law identity against \(\mu(\cdot,x)\) and using the definition of \(\mu_w\) give \(\mathbb E[\mu(S_t,X_t)\mid W_t=\allowbreak w,X_t=\allowbreak x,\mathcal F_{t-1,N_{\mathrm O}}]=\allowbreak \mu_w(x)\). For the treatment-group residual, sequential randomization, the treatment positivity in Assumption 4.1, and this conditional-mean identity yield \(\mathbb E[W_t\{\mu(S_t,X_t)-\mu_1(X_t)\}/\allowbreak p_{t,N_{\mathrm O}}(X_t)\mid X_t,\mathcal F_{t-1,N_{\mathrm O}}]=\allowbreak 0\), and likewise \(\mathbb E[(1-W_t)\{\mu(S_t,X_t)-\mu_0(X_t)\}/\allowbreak \{1-p_{t,N_{\mathrm O}}(X_t)\}\mid X_t,\mathcal F_{t-1,N_{\mathrm O}}]=\allowbreak 0\). Multiplying the difference of these two residuals by the density ratio \(f_{X\mid G=\mathrm O}(X_t)/f_{X\mid G=\mathrm E,t,N_{\mathrm O}}(X_t)\), whose denominator is positive on the target support by the covariate overlap in Assumption 4.1, and integrating \(X_t\) under its actual conditional density preserves the zero conditional mean, so term (II) is zero. Since term (I) equals \(\tau\) and term (II) is zero, the representation holds for every round \(t\), almost surely, which proves Proposition 4.1.
\par
}

\subsection{Construction of the surrogate representer}\label{a.5-construction-of-the-surrogate-representer}

This appendix derives the classification-based construction of the surrogate representer in Remark 5.2 for the rounds of the adaptive experiment.

{
\emergencystretch=2em
Fix round \(t\) and condition on \(\mathcal F_{t-1,N_{\mathrm O}}\). By consistency and Assumptions 4.2 and 4.3, for \(w\in\{0,1\}\), \(f_{S_t\mid W_t=w,X_t=x,\mathcal F_{t-1,N_{\mathrm O}}}(s)=\allowbreak f_{S\mid W=w,X,G=\mathrm E}(s\mid x)\) holds almost surely on the relevant support. Hence the surrogate density conditional on \(X_t=x\) is \(f_{S_t\mid X_t=x,\mathcal F_{t-1,N_{\mathrm O}}}(s)=\allowbreak p_{t,N_{\mathrm O}}(x)f_{S\mid W=1,X,G=\mathrm E}(s\mid x)+\{1-p_{t,N_{\mathrm O}}(x)\}f_{S\mid W=0,X,G=\mathrm E}(s\mid x)\). Let \(P^{\rm obs}\) be the observational population law of \((X,S,Y)\), fix an arbitrary \(\gamma\in(0,1)\), and define the auxiliary mixture law \(Q_{t,N_{\mathrm O}}\) as follows. \[\begin{aligned}
Q_{t,N_{\mathrm O}}(G=\mathrm O)
&=
\gamma,
\qquad
Q_{t,N_{\mathrm O}}(G=\mathrm E)
=
1-\gamma,\\
Q_{t,N_{\mathrm O}}\{(X,S,Y)\in A\mid G=\mathrm O\}
&=
P^{\rm obs}(A),\\
Q_{t,N_{\mathrm O}}\{(X,W,S)\in B\mid G=\mathrm E\}
&=
\mathbb P\{
(X_t,W_t,S_t)\in B
\mid
\mathcal F_{t-1,N_{\mathrm O}}
\}
\end{aligned}
\]
\par
}

{
\emergencystretch=2em
Here \(A\) and \(B\) are measurable sets of the corresponding observation spaces. The law \(Q_{t,N_{\mathrm O}}\) is an auxiliary law that serves to define conditional probabilities. Its experimental component is the conditional law of the next observation under the recorded design, and its observational component is the population law \(P^{\rm obs}\). Below, \(p_Q(x)\), \(p_Q(s,x)\), \(\gamma_Q(x)\), and \(\gamma_Q(s,x)\) of Remark 5.2 are evaluated under \(Q=Q_{t,N_{\mathrm O}}\). In particular, \(p_Q(x)=p_{t,N_{\mathrm O}}(x)\).
\par
}

{
\emergencystretch=2em
The calculations below are interpreted on the supports where the displayed denominators are positive. By Bayes\textquotesingle{} rule, the data-source odds satisfy \(\gamma_Q(x)/\allowbreak \{1-\gamma_Q(x)\}=\allowbreak \gamma f_{X\mid G=\mathrm O}(x)/\allowbreak \{(1-\gamma)f_{X\mid G=\mathrm E,t,N_{\mathrm O}}(x)\}\) and \(\gamma_Q(s,x)/\allowbreak \{1-\gamma_Q(s,x)\}=\allowbreak \gamma f_{X\mid G=\mathrm O}(x)f_{S\mid X,G=\mathrm O}(s\mid x)/\allowbreak \{(1-\gamma)f_{X\mid G=\mathrm E,t,N_{\mathrm O}}(x)f_{S_t\mid X_t=x,\mathcal F_{t-1,N_{\mathrm O}}}(s)\}\), and the treatment probabilities satisfy \(p_Q(s,x)/\allowbreak p_Q(x)=\allowbreak f_{S\mid W=1,X,G=\mathrm E}(s\mid x)/\allowbreak f_{S_t\mid X_t=x,\mathcal F_{t-1,N_{\mathrm O}}}(s)\) and \(\{1-p_Q(s,x)\}/\allowbreak \{1-p_Q(x)\}=\allowbreak f_{S\mid W=0,X,G=\mathrm E}(s\mid x)/\allowbreak f_{S_t\mid X_t=x,\mathcal F_{t-1,N_{\mathrm O}}}(s)\). Write the representer as a product of two factors. The first is the ratio of the experimental to the observational conditional surrogate density, and the second is the difference of the treatment-specific surrogate densities divided by \(f_{S_t\mid X_t=x,\mathcal F_{t-1,N_{\mathrm O}}}(s)\). Substituting the ratio of the two odds into the first factor and the difference of the two ratios of treatment probabilities into the second gives the following chain of equalities. \[\begin{aligned}
\alpha(s,x)
&=
\frac{f_{S_t\mid X_t=x,\mathcal F_{t-1,N_{\mathrm O}}}(s)}{f_{S\mid X,G=\mathrm O}(s\mid x)}
\frac{
f_{S\mid W=1,X,G=\mathrm E}(s\mid x)
-
f_{S\mid W=0,X,G=\mathrm E}(s\mid x)
}{
f_{S_t\mid X_t=x,\mathcal F_{t-1,N_{\mathrm O}}}(s)
}\\
&=
\frac{\gamma_Q(x)}{\gamma_Q(s,x)}
\frac{1-\gamma_Q(s,x)}{1-\gamma_Q(x)}
\left\{
\frac{p_Q(s,x)}{p_Q(x)}
-
\frac{1-p_Q(s,x)}{1-p_Q(x)}
\right\}\\
&=
\frac{\gamma_Q(x)}{\gamma_Q(s,x)}
\frac{1-\gamma_Q(s,x)}{1-\gamma_Q(x)}
\frac{
p_Q(s,x)-p_Q(x)
}{
p_Q(x)\{1-p_Q(x)\}
}
\end{aligned}
\]
\par
}

{
\emergencystretch=2em
This is the representation in Remark 5.2. At points where \(f_{S_t\mid X_t=x,\mathcal F_{t-1,N_{\mathrm O}}}(s)=0\) and \(f_{S\mid X,G=\mathrm O}(s\mid x)>0\), the treatment positivity in Assumption 4.1 implies that the treatment- and control-arm conditional surrogate densities are both zero, so that \(\alpha(s,x)=0\). The probability representation is also extended by zero at these points.
\par
}

{
\emergencystretch=2em
The mixture share \(\gamma\) and the recorded design \((f_{X\mid G=\mathrm E,t,N_{\mathrm O}},p_{t,N_{\mathrm O}})\) affect the four probabilities but cancel from the representation of \(\alpha(s,x)\). The derivation rests on the law under \(G=\mathrm O\) being \(P^{\rm obs}\), on the treatment-specific surrogate densities under \(G=\mathrm E\) being equal to \(f_{S\mid W=w,X,G=\mathrm E}\), and on covariate overlap and treatment positivity on the target support. The same representation therefore holds under any other reference law with these properties, such as a law that pools the observations of several rounds. All four probabilities on the right-hand side must, however, be defined under the same reference law.
\par
}

{
\emergencystretch=2em
To construct the estimator after round \(t\), let \(\widehat p_Q(s,x)\), \(\widehat p_Q(x)\), \(\widehat\gamma_Q(s,x)\), and \(\widehat\gamma_Q(x)\) be estimators of the four conditional probabilities defined under a common reference mixture law. Their dependence on \((t,N_{\mathrm O})\) is suppressed in the notation. On the region where the estimated denominators are positive, the representer estimator is defined as follows. \[\widehat\alpha_{t,N_{\mathrm O}}(s,x)
=
\frac{
\widehat\gamma_Q(x)
}{
\widehat\gamma_Q(s,x)
}
\frac{
1-\widehat\gamma_Q(s,x)
}{
1-\widehat\gamma_Q(x)
}
\frac{
\widehat p_Q(s,x)-\widehat p_Q(x)
}{
\widehat p_Q(x)\{1-\widehat p_Q(x)\}
}
\]
\par
}

{
\emergencystretch=2em
In the experimental data that represent the chosen reference law, the treatment probabilities can be estimated by binary probability models for \(W\) given \((S,X)\) and for \(W\) given \(X\). When \(Q_{t,N_{\mathrm O}}\) is taken as the reference law, the treatment propensity \(p_Q(x)=p_{t,N_{\mathrm O}}(x)\) is the recorded design and need not be estimated. In the pooled data, the data-source probabilities can be estimated by binary probability models for \(\mathbf 1\{G=\mathrm O\}\) given \((S,X)\) and for \(\mathbf 1\{G=\mathrm O\}\) given \(X\).
\par
}

{
\emergencystretch=2em
When these models are estimated from adaptive observations that have already been collected, the reference law must correspond to the pooling or reweighting method used in the estimation. In particular, the treatment propensity of a pooled experimental law need not coincide with \(p_{t,N_{\mathrm O}}(x)\), so probabilities estimated by pooling the observations of several rounds cannot in general be combined with the current-round treatment propensity \(p_{t,N_{\mathrm O}}(x)\). When \(Q_{t,N_{\mathrm O}}\) is used, the other probability estimates must target the same mixture law.
\par
}

{
\emergencystretch=2em
This construction removes the need to estimate conditional surrogate densities directly when estimating \(\widehat\alpha_{t,N_{\mathrm O}}\). Under the actual dependence structure generated by adaptive sampling, the estimated representer must satisfy the conditions imposed in Sections 6 and 7, namely the representer convergence conditions, the product-rate conditions, and the regularity conditions on the observational score.
\par
}

\subsection{Proofs for Section 6}\label{a.6-proofs-for-section-6}

This appendix proves the results of Section 6. Appendix A.6.1 explains the role of the historical average in Remark A.6.1, proves Lemma 6.2, and states the regularity conditions of Assumption A.6.1, Appendix A.6.2 proves Lemma 6.1, Appendix A.6.3 proves Theorem 6.1 by a martingale central limit theorem, and Appendix A.6.4 proves the lemma on the experimental predictable variance used in that proof.

\subsubsection{Proof of Lemma 6.2 and regularity conditions}\label{a.6.1-proof-of-lemma-6.2-and-regularity-conditions}

This subsection explains the role of the historical average in Remark A.6.1, proves Lemma 6.2, and states the regularity conditions used in Sections 6 and 7. Population operators integrate with the fitted functions held fixed.

\expandafter\gdef\csname theremark\endcsname{A.6.\arabic{remark}}
\expandafter\gdef\csname theHremark\endcsname{A.6.\arabic{remark}}
\setcounter{remark}{0}
\begin{remark}[Role of the historical average in the estimator decomposition]

{
\emergencystretch=2em
The historical average \(\overline\tau_{t,N_{\mathrm O}}\) gives the observational score the same average of fitted conditional effects that enters the experimental scores. The two terms enter with opposite signs, so their sum is the error from integrating the same function against the fitted and the true target densities.
\par
}

\end{remark}

{
\emergencystretch=2em
\textbf{Proof of Lemma 6.2.} The definition of the centered experimental score in Section 6.1 and the linearity of \(P_{\widehat f_{X\mid G=\mathrm O,N_{\mathrm O}}}\), which is fixed across rounds, give \[\begin{aligned}
&\widehat\phi_{s,N_{\mathrm O}}^{\rm exp}(S_s,W_s,X_s)
=\widetilde m_{s,N_{\mathrm O}}+P_{\widehat f_{X\mid G=\mathrm O,N_{\mathrm O}}}(\widetilde\tau_{N_{\mathrm O}}(\cdot)-\widehat\tau_{s,N_{\mathrm O}}(\cdot)),\\
&\frac1t\sum_{s=1}^t \widehat\phi_{s,N_{\mathrm O}}^{\rm exp}(S_s,W_s,X_s)
=\frac1t\sum_{s=1}^t \widetilde m_{s,N_{\mathrm O}}+P_{\widehat f_{X\mid G=\mathrm O,N_{\mathrm O}}}(\widetilde\tau_{N_{\mathrm O}}-\overline\tau_{t,N_{\mathrm O}}).
\end{aligned}
\]
\par
}

The second equality uses the historical average in Step 6 of Algorithm~\ref{alg:adaptive-sequential-estimation}, defined by \[\overline\tau_{t,N_{\mathrm O}}(x)=\frac1t\sum_{s=1}^t\widehat\tau_{s,N_{\mathrm O}}(x).
\]

{
\emergencystretch=2em
The representer \(\alpha\) connects the observational score to the conditional effect under the true target law. By its definition, the surrogate-support condition in Assumption 2.6, and the definitions of the two arm means, \[\begin{aligned}
&P^{\rm obs}[\alpha(\mu-\widehat\mu_{N_{\mathrm O}})]\\
&=\int f_{X\mid G=\mathrm O}(x)\int\{\mu(s,x)-\widehat\mu_{N_{\mathrm O}}(s,x)\}
 \{f_{S\mid W=1,X,G=\mathrm E}(s\mid x)-f_{S\mid W=0,X,G=\mathrm E}(s\mid x)\}\,d\lambda_S(s)\,d\lambda_X(x)\\
&=P_{f_{X\mid G=\mathrm O}}\{\tau(\cdot)-\widetilde\tau_{N_{\mathrm O}}\}.
\end{aligned}
\]
\par
}

{
\emergencystretch=2em
In Step 6, taking the population mean replaces \(Y\) by \(\mu(S,X)\). Adding and subtracting the representer term and using \(P_{f_{X\mid G=\mathrm O}}\tau(\cdot)=\tau\) from (2.1) yield \[\begin{aligned}
&P^{\rm obs}\widehat\psi_{t,N_{\mathrm O}}^{\rm obs}-\tau\\
&=P_{f_{X\mid G=\mathrm O}}\overline\tau_{t,N_{\mathrm O}}-\tau+P^{\rm obs}[\widehat\alpha_{t,N_{\mathrm O}}(\mu-\widehat\mu_{N_{\mathrm O}})]\\
&=P_{f_{X\mid G=\mathrm O}}(\overline\tau_{t,N_{\mathrm O}}-\widetilde\tau_{N_{\mathrm O}})+P^{\rm obs}[(\widehat\alpha_{t,N_{\mathrm O}}-\alpha)(\mu-\widehat\mu_{N_{\mathrm O}})].
\end{aligned}
\]
\par
}

Insert these identities into the estimator and split the empirical fluctuation into the true score and its estimation error. The conditional-effect terms combine as follows. \[P_{\widehat f_{X\mid G=\mathrm O,N_{\mathrm O}}}(\widetilde\tau_{N_{\mathrm O}}-\overline\tau_{t,N_{\mathrm O}})+P_{f_{X\mid G=\mathrm O}}(\overline\tau_{t,N_{\mathrm O}}-\widetilde\tau_{N_{\mathrm O}})
=(P_{\widehat f_{X\mid G=\mathrm O,N_{\mathrm O}}}-P_{f_{X\mid G=\mathrm O}})(\widetilde\tau_{N_{\mathrm O}}-\overline\tau_{t,N_{\mathrm O}}).
\]

Consequently, \[\begin{aligned}
&\widehat\tau_{t,N_{\mathrm O}}-\tau\\
&=\frac1t\sum_{s=1}^t \widetilde m_{s,N_{\mathrm O}}+P_{\widehat f_{X\mid G=\mathrm O,N_{\mathrm O}}}(\widetilde\tau_{N_{\mathrm O}}-\overline\tau_{t,N_{\mathrm O}})
+(\mathbb P_{N_{\mathrm O}}^{\rm obs}-P^{\rm obs})\widehat\psi_{t,N_{\mathrm O}}^{\rm obs}+P^{\rm obs}\widehat\psi_{t,N_{\mathrm O}}^{\rm obs}-\tau\\
&=\frac1t\sum_{s=1}^t \widetilde m_{s,N_{\mathrm O}}
+\underbrace{(\mathbb P_{N_{\mathrm O}}^{\rm obs}-P^{\rm obs})\psi^{\rm obs}}_{L_{N_{\mathrm O}}^{\rm obs}}
+\underbrace{(\mathbb P_{N_{\mathrm O}}^{\rm obs}-P^{\rm obs})(\widehat\psi_{t,N_{\mathrm O}}^{\rm obs}-\psi^{\rm obs})}_{R_{t,N_{\mathrm O}}^{\rm i}}\\
&\quad+\underbrace{(P_{\widehat f_{X\mid G=\mathrm O,N_{\mathrm O}}}-P_{f_{X\mid G=\mathrm O}})(\widetilde\tau_{N_{\mathrm O}}-\overline\tau_{t,N_{\mathrm O}})}_{R_{t,N_{\mathrm O}}^{\rm ii}}
+\underbrace{P^{\rm obs}[(\widehat\alpha_{t,N_{\mathrm O}}-\alpha)(\mu-\widehat\mu_{N_{\mathrm O}})]}_{R_{t,N_{\mathrm O}}^{\rm iii}}.
\end{aligned}
\]

{
\emergencystretch=2em
The identity holds for every applicable \(N_{\mathrm O}\) and every \(t\ge1\) for which the terms are integrable. This includes \(t=N_{\mathrm E}\) in Theorem 6.1 and every \(t\ge t_{0,N_{\mathrm O}}\) in Appendix A.7, and it proves Lemma 6.2.
\par
}

\expandafter\gdef\csname theassumption\endcsname{A.6.\arabic{assumption}}
\expandafter\gdef\csname theHassumption\endcsname{A.6.\arabic{assumption}}
\setcounter{assumption}{0}
\begin{assumption}[Regularity conditions]

{
\emergencystretch=2em
The scope of the conditions is summarized below. Common conditions hold for every \(N_{\mathrm O}\) and \(1\le t\le N_{\mathrm E}\) in Section 6, and for every \(N_{\mathrm O}\) and \(t\ge1\) in Section 7. The simultaneous variance bounds cover rounds \(1,\ldots,N_{\mathrm E}\) in Section 6 and the rounds from \(t_{0,N_{\mathrm O}}\) onward in Section 7, as specified in (viii) and (x).
\par
}

\begin{itemize}
\tightlist
\item
  Conditions (i)--(vii) are imposed in both Sections 6 and 7.
\item
  Conditions (viii) and (ix) are imposed only in Section 6.
\item
  Conditions (x) and (xi) are imposed only in Section 7.
\end{itemize}

{
\emergencystretch=2em
\(\mathrm{(i)}\) The observational sample is i.i.d. as specified in Section 4, and the population-score conditions \(P^{\rm obs}\psi^{\rm obs}=\tau\) and \(\mathbb E^{\rm obs}[(\psi^{\rm obs}-\tau)^2]=V^{\rm obs}<\infty\) from Assumption 6.5 hold. This does not impose the fitted-score conditions at time \(N_{\mathrm E}\) of Assumption 6.5 on Section 7.
\par
}

{
\emergencystretch=2em
\(\mathrm{(ii)}\) For each argument on the relevant support, \(\widehat\mu_{N_{\mathrm O}}(s,x)\), \(\widehat f_{X\mid G=\mathrm O,N_{\mathrm O}}(x)\), \(\widehat\mu_{w,t,N_{\mathrm O}}(x)\), \(\widehat v_{w,t,N_{\mathrm O}}(x)\), \(\widehat p_{t,N_{\mathrm O}}^{\star}(x)\), and \(\widehat f_{X\mid G=\mathrm E,t,N_{\mathrm O}}^{\star}(x)\) are \(\mathcal F_{t-1,N_{\mathrm O}}\)-measurable. The fixed observational-data fits \(\widehat\mu_{N_{\mathrm O}}\) and \(\widehat f_{X\mid G=\mathrm O,N_{\mathrm O}}\) and the initial experimental moment fits are \(\mathcal F_{0,N_{\mathrm O}}\)-measurable. Any additional observational fit required by the chosen representer construction has this same measurability. The actual design functions \(x\mapsto f_{X\mid G=\mathrm E,t,N_{\mathrm O}}(x)\) and \(x\mapsto p_{t,N_{\mathrm O}}(x)\) are \(\mathcal F_{t-1,N_{\mathrm O}}\)-measurable, as are the pre-observation fitted functions \(\widehat\tau_{t,N_{\mathrm O}}(\cdot)\) and \(\overline\tau_{t,N_{\mathrm O}}(\cdot)\). After round \(t\), the representer \(\widehat\alpha_{t,N_{\mathrm O}}\) is \(\mathcal F_{t,N_{\mathrm O}}\)-measurable; it need not be \(\mathcal F_{t-1,N_{\mathrm O}}\)-measurable. The recorded experimental score and the updated observational score are \(\mathcal F_{t,N_{\mathrm O}}\)-measurable. These requirements apply pointwise on the relevant supports, together with the measurability of functions and integrands in (iv).
\par
}

{
\emergencystretch=2em
\(\mathrm{(iii)}\) The actual conditional design laws satisfy \(f_{X_t\mid\mathcal F_{t-1,N_{\mathrm O}}}(x)=\allowbreak \widehat f_{X\mid G=\mathrm E,t,N_{\mathrm O}}^{\star}(x)\) and \(\Pr(W_t=\allowbreak 1\mid X_t=\allowbreak x,\mathcal F_{t-1,N_{\mathrm O}})=\allowbreak \widehat p_{t,N_{\mathrm O}}^{\star}(x)\), with \(\widehat f_{X\mid G=\mathrm E,t,N_{\mathrm O}}^{\star}(x)\ge0\), \(\int\widehat f_{X\mid G=\mathrm E,t,N_{\mathrm O}}^{\star}(x)\,d\lambda_X(x)=\allowbreak 1\), and \(0\le\widehat p_{t,N_{\mathrm O}}^{\star}(x)\le1\).
\par
}

{
\emergencystretch=2em
\(\mathrm{(iv)}\) Sequential randomization and conditional distributional equivalence of surrogates give \(f_{S_t\mid W_t=w,X_t=x,\mathcal F_{t-1,N_{\mathrm O}}}(s)=\allowbreak f_{S\mid W=w,X,G=\mathrm E}(s\mid x)\). The fitted functions, densities, weights, and score constituents are measurable, and all conditional expectations and integrals appearing in the proofs of Appendices A.6 and A.7 exist.
\par
}

{
\emergencystretch=2em
\(\mathrm{(v)}\) The target and experimental covariate laws have common support, \(\operatorname{supp}\widehat f_{X\mid G=\mathrm E,t,N_{\mathrm O}}^{\star}=\allowbreak \operatorname{supp}f_{X\mid G=\mathrm O}\). The fitted target law is a nonnegative density that integrates to one, satisfies \(\operatorname{supp}\widehat f_{X\mid G=\mathrm O,N_{\mathrm O}}=\allowbreak \operatorname{supp}f_{X\mid G=\mathrm O}\) and \(\widehat f_{X\mid G=\mathrm O,N_{\mathrm O}}\ll f_{X\mid G=\mathrm O}\), and has no mass outside this support.
\par
}

{
\emergencystretch=2em
\(\mathrm{(vi)}\) On the target support, for each applicable \(N_{\mathrm O},t\) and treatment assignment, \(0<\widehat p_{t,N_{\mathrm O}}^{\star}(x)<1\), \(0<\widehat f_{X\mid G=\mathrm E,t,N_{\mathrm O}}^{\star}(x)<\infty\), \(0<\widehat v_{w,t,N_{\mathrm O}}(x)<\infty\), and \(0<\int\widehat f_{X\mid G=\mathrm O,N_{\mathrm O}}(x)\allowbreak\sum_{w=0}^1\sqrt{\widehat v_{w,t,N_{\mathrm O}}(x)}\,d\lambda_X(x)<\infty\). All score denominators are positive and finite wherever used. These pointwise conditions do not impose uniform bounds over all rounds.
\par
}

{
\emergencystretch=2em
\(\mathrm{(vii)}\) The true surrogate laws are nonnegative densities on a common support. For every target-support \(x\), \(f_{S\mid W=w,X,G=\mathrm E}(s\mid x)\ge0\) and \(\int f_{S\mid W=w,X,G=\mathrm E}(s\mid x)\,d\lambda_S(s)=1\). A density-based construction additionally uses nonnegative fitted densities normalized to one; direct moment estimation does not require such fitted densities.
\par
}

{
\emergencystretch=2em
\(\mathrm{(viii)}\) For Section 6, the true variances satisfy \(\underline v\le v_w(x)\le\overline v\) on the target support. With probability tending to one, simultaneously for both treatment assignments, every target-support \(x\), and every \(1\le t\le N_{\mathrm E}\), \(\underline v\le\widehat v_{w,t,N_{\mathrm O}}(x)\le\overline v\) and \(\underline v\le\int\{\widehat\mu_{N_{\mathrm O}}(s,x)-\widetilde\mu_{w,N_{\mathrm O}}(x)\}^2f_{S\mid W=w,X,G=\mathrm E}(s\mid x)\,d\lambda_S(s)\le\overline v\).
\par
}

{
\emergencystretch=2em
\(\mathrm{(ix)}\) The constants in (viii) are fixed independently of \(N_{\mathrm O}\), with \(0<\underline v<\overline v<\infty\).
\par
}

{
\emergencystretch=2em
\(\mathrm{(x)}\) For Section 7, the variance bounds are imposed only from the monitoring start onward. There are fixed constants \(0<\underline v<\overline v<\infty\) such that the true variances \(v_w(x)\) in the treatment and control groups lie in \([\underline v,\overline v]\) on the target support. With probability tending to one, simultaneously for \(w=0,1\), every target-support value \(x\), and every \(t\ge t_{0,N_{\mathrm O}}\), \(\underline v\le\widehat v_{w,t,N_{\mathrm O}}(x)\le\overline v\) and \(\underline v\le\int\{\widehat\mu_{N_{\mathrm O}}(s,x)-\widetilde\mu_{w,N_{\mathrm O}}(x)\}^2f_{S\mid W=w,X,G=\mathrm E}(s\mid x)\,d\lambda_S(s)\le\overline v\). These uniform bounds make the square-root and inverse-square-root operations in the predictable-variance algebra stable.
\par
}

{
\emergencystretch=2em
\(\mathrm{(xi)}\) For Section 7, in addition to (i), \(V^{\rm obs}>0\) and \(\mathbb E_{X\mid G=\mathrm O}[\{\sum_{w=0}^1\sqrt{v_w(X)}\}^2]<\infty\). The risks for the target density and for the regressions in the treatment and control groups are imposed in Assumption 7.3. Assumption 7.2 supplies the score moments. No Cesàro or fitted-score condition at time \(N_{\mathrm E}\) is imposed on Section 7 here.
\par
}

\end{assumption}

\subsubsection{Proof of Lemma 6.1}\label{a.6.2-proof-of-lemma-6.1}

{
\emergencystretch=2em
\textbf{Proof.} The proof integrates the round-\(t\) experimental score over \(S_t\), \(W_t\), and \(X_t\) given \(\mathcal F_{t-1,N_{\mathrm O}}\) and shows that its conditional mean is the negative of the predictable centering term in the definition of \(\widetilde m_{t,N_{\mathrm O}}\). For \(w=0,1\), the pre-round measurability and the conditional distributional equivalence of surrogates in Assumption A.6.1(ii)--(iv) give \(\mathbb E[\widehat\mu_{N_{\mathrm O}}(S_t,X_t)\mid W_t=\allowbreak w,X_t=\allowbreak x,\mathcal F_{t-1,N_{\mathrm O}}]=\allowbreak \int\widehat\mu_{N_{\mathrm O}}(s,x)f_{S\mid W=w,X,G=\mathrm E}(s\mid x)\,d\lambda_S(s)=\allowbreak \widetilde\mu_{w,N_{\mathrm O}}(x)\). Since Assumption A.6.1(iii) makes the actual treatment probability and covariate density of round \(t\) equal to \(\widehat p_{t,N_{\mathrm O}}^{\star}\) and \(\widehat f_{X\mid G=\mathrm E,t,N_{\mathrm O}}^{\star}\), sequential randomization cancels the treatment weights and the actual experimental covariate density cancels its inverse weight. Hence \[\begin{aligned}
&\mathbb E[\widehat\phi_{t,N_{\mathrm O}}^{\rm exp}(S_t,W_t,X_t)\mid\mathcal F_{t-1,N_{\mathrm O}}]\\
&\quad=\int\widehat f_{X\mid G=\mathrm E,t,N_{\mathrm O}}^{\star}(x)\frac{\widehat f_{X\mid G=\mathrm O,N_{\mathrm O}}(x)}{\widehat f_{X\mid G=\mathrm E,t,N_{\mathrm O}}^{\star}(x)}\sum_{w=0}^1(2w-1)\{\widetilde\mu_{w,N_{\mathrm O}}(x)-\widehat\mu_{w,t,N_{\mathrm O}}(x)\}\,d\lambda_X(x)\\
&\quad=P_{\widehat f_{X\mid G=\mathrm O,N_{\mathrm O}}}(\widetilde\tau_{N_{\mathrm O}}(\cdot)-\widehat\tau_{t,N_{\mathrm O}}(\cdot)),
\end{aligned}
\]
\par
}

{
\emergencystretch=2em
and the definition of \(\widetilde m_{t,N_{\mathrm O}}\) gives \(\mathbb E[\widetilde m_{t,N_{\mathrm O}}\mid\mathcal F_{t-1,N_{\mathrm O}}]=\allowbreak P_{\widehat f_{X\mid G=\mathrm O,N_{\mathrm O}}}(\widetilde\tau_{N_{\mathrm O}}(\cdot)-\widehat\tau_{t,N_{\mathrm O}}(\cdot))+P_{\widehat f_{X\mid G=\mathrm O,N_{\mathrm O}}}(\widehat\tau_{t,N_{\mathrm O}}(\cdot)-\widetilde\tau_{N_{\mathrm O}}(\cdot))=\allowbreak 0\). The centered score \(\widetilde m_{t,N_{\mathrm O}}\) is \(\mathcal F_{t,N_{\mathrm O}}\)-measurable by its construction and Assumption A.6.1(ii). Together with the square integrability assumed in Lemma 6.1, this proves Lemma 6.1 for each applicable \(N_{\mathrm O}\) and \(t\). In Section 7, Assumption 7.2 supplies this square integrability, because conditional Jensen gives \(\mathbb E[\widetilde m_{t,N_{\mathrm O}}^2]\le C^{2/r}\).
\par
}

\subsubsection{Proof of Theorem 6.1}\label{a.6.3-proof-of-theorem-6.1}

{
\emergencystretch=2em
\textbf{Proof.} We evaluate the decomposition of Lemma 6.2 at \(t=N_{\mathrm E}\) on the \(\sqrt{N_{\mathrm E}}\) scale. Because the observational sample is used to choose the experimental design, the experimental martingale average and the observational empirical average are dependent, and we treat their sum as one ordered martingale-difference row in which the observational records come first. Appendix A.6.3.1 states the martingale central limit theorem for this row as Lemma A.6.1, Appendix A.6.3.2 states Lemma A.6.2 on the experimental predictable variance, and Appendix A.6.3.3 verifies the martingale-difference, conditional Lindeberg, and predictable-variance conditions of Lemma A.6.1. Appendix A.6.3.4 shows that the three remainders are negligible, and Appendix A.6.3.5 combines the two parts.
\par
}

{
\emergencystretch=2em
Multiplying the decomposition of Lemma 6.2 at \(t=N_{\mathrm E}\) by \(\sqrt{N_{\mathrm E}}\) and writing \(L_{N_{\mathrm O}}^{\rm obs}=\allowbreak N_{\mathrm O}^{-1}\allowbreak\sum_{i=1}^{N_{\mathrm O}}\{\psi^{\rm obs}(X_i,S_i,Y_i)-\tau\}\), which holds because \(P^{\rm obs}\psi^{\rm obs}=\tau\) by Assumption A.6.1(i), give \[\begin{aligned}
\sqrt{N_{\mathrm E}}(\widehat\tau_{N_{\mathrm E},N_{\mathrm O}}-\tau)
&=\frac1{\sqrt{N_{\mathrm E}}}\sum_{t=1}^{N_{\mathrm E}}\widetilde m_{t,N_{\mathrm O}}+\frac{\sqrt{N_{\mathrm E}}}{N_{\mathrm O}}\sum_{i=1}^{N_{\mathrm O}}\{\psi^{\rm obs}(X_i,S_i,Y_i)-\tau\}\\
&\quad+\sqrt{N_{\mathrm E}}R_{N_{\mathrm E},N_{\mathrm O}}^{\rm i}+\sqrt{N_{\mathrm E}}R_{N_{\mathrm E},N_{\mathrm O}}^{\rm ii}+\sqrt{N_{\mathrm E}}R_{N_{\mathrm E},N_{\mathrm O}}^{\rm iii}.
\end{aligned}
\]
\par
}

{
\emergencystretch=2em
The first two terms on the right are the sum of the ordered row and have a normal limit by the martingale central limit theorem. The last three are the remainders of Lemma 6.2, controlled by the empirical-process condition in Assumption 6.5 for \(R_{N_{\mathrm E},N_{\mathrm O}}^{\rm i}\) and by Cauchy--Schwarz and the product rates in Assumption 6.6 for \(R_{N_{\mathrm E},N_{\mathrm O}}^{\rm ii}\) and \(R_{N_{\mathrm E},N_{\mathrm O}}^{\rm iii}\).
\par
}

\paragraph{\texorpdfstring{Martingale central limit theorem of \citet{Hall1980}}{Martingale central limit theorem of }}\label{a.6.3.1-martingale-central-limit-theorem-of}

This step states the martingale central limit theorem used for the first two terms of the decomposition. Appendix A.6.3.3 verifies its three conditions.

\expandafter\gdef\csname thelemma\endcsname{A.6.\arabic{lemma}}
\expandafter\gdef\csname theHlemma\endcsname{A.6.\arabic{lemma}}
\setcounter{lemma}{0}
\begin{lemma}[Martingale central limit theorem for a scalar ordered row]

{
\emergencystretch=2em
We use Corollary 3.1 of \citet{Hall1980}, which follows their Theorem 3.2, specialized to one scalar ordered row. For a fixed \(N_{\mathrm O}\), the row consists of \(N_{\mathrm O}+N_{\mathrm E}\) increments. The \(N_{\mathrm O}\) observational records are revealed first, so that they generate \(\mathcal F_{0,N_{\mathrm O}}=\allowbreak \sigma(\mathcal D_{N_{\mathrm O}}^{\rm obs})\), and the \(N_{\mathrm E}\) experimental records follow according to \(\{\mathcal F_{t,N_{\mathrm O}}\}_{t=0}^{N_{\mathrm E}}\). The ordered increments are \[\begin{aligned}
&\frac{\sqrt{N_{\mathrm E}}}{N_{\mathrm O}}\{\psi^{\rm obs}(X_i,S_i,Y_i)-\tau\},&&i=1,\ldots,N_{\mathrm O},\\
&\frac{\widetilde m_{t,N_{\mathrm O}}}{\sqrt{N_{\mathrm E}}},&&t=1,\ldots,N_{\mathrm E}.
\end{aligned}
\]
\par
}

For this row, the corollary applies under the following conditions.

\begin{enumerate}
\def\labelenumi{\arabic{enumi}.}
\item
  The displayed increments form a square-integrable martingale-difference row.
\item
  For every \(\varepsilon>0\), the combined conditional Lindeberg sum satisfies

  \[   \begin{aligned}
  &\frac{N_{\mathrm E}}{N_{\mathrm O}^2}\sum_{i=1}^{N_{\mathrm O}}\mathbb E\!\Biggl[\{\psi^{\rm obs}(X_i,S_i,Y_i)\\
  &\qquad-\tau\}^2\mathbf 1\!\left\{\frac{\sqrt{N_{\mathrm E}}}{N_{\mathrm O}}|\psi^{\rm obs}(X_i,S_i,Y_i)-\tau|>\varepsilon\right\}\mathrel{\Big|}\sigma\{(X_j,S_j,Y_j):1\le j<i\}\Biggr]\\
  &+\frac1{N_{\mathrm E}}\sum_{t=1}^{N_{\mathrm E}}\mathbb E\!\left[\widetilde m_{t,N_{\mathrm O}}^2\mathbf 1\!\left\{|\widetilde m_{t,N_{\mathrm O}}|>\varepsilon\sqrt{N_{\mathrm E}}\right\}\mathrel{\Big|}\mathcal F_{t-1,N_{\mathrm O}}\right]\xrightarrow{p}0.
  \end{aligned}
  \]
\item
  The predictable conditional variance satisfies

  \[   \frac{N_{\mathrm E}}{N_{\mathrm O}}V^{\rm obs}+\frac1{N_{\mathrm E}}\sum_{t=1}^{N_{\mathrm E}}\mathbb E[\widetilde m_{t,N_{\mathrm O}}^{2}\mid\mathcal F_{t-1,N_{\mathrm O}}]\xrightarrow{p}\kappa V^{\rm obs}+V^{{\rm exp},\star}.
  \]
\end{enumerate}

{
\emergencystretch=2em
Because the limit in condition 3 is a constant, the remarks following Corollary 3.1 allow the nesting condition (3.21) on the \(\sigma\)-fields to be replaced by the measurability of this limit, which holds trivially, and the corollary then gives convergence in distribution without stability. Under conditions 1--3, the corollary therefore gives the scalar limit \[\frac{\sqrt{N_{\mathrm E}}}{N_{\mathrm O}}\sum_{i=1}^{N_{\mathrm O}}\{\psi^{\rm obs}(X_i,S_i,Y_i)-\tau\}+\frac1{\sqrt{N_{\mathrm E}}}\sum_{t=1}^{N_{\mathrm E}}\widetilde m_{t,N_{\mathrm O}}\rightsquigarrow\mathcal N(0,\kappa V^{\rm obs}+V^{{\rm exp},\star}).
\]
\par
}

\end{lemma}

\paragraph{Experimental predictable variance}\label{a.6.3.2-experimental-predictable-variance}

The experimental part of condition 3 needs a longer calculation, which the following lemma collects. Appendix A.6.3.3 uses it to verify condition 3, and Appendix A.6.4 proves it.

\expandafter\gdef\csname thelemma\endcsname{A.6.\arabic{lemma}}
\expandafter\gdef\csname theHlemma\endcsname{A.6.\arabic{lemma}}
\setcounter{lemma}{1}
\begin{lemma}[Experimental predictable variance]

Under the assumptions of Theorem 6.1, \[\frac1{N_{\mathrm E}}\sum_{t=1}^{N_{\mathrm E}}\mathbb E[\widetilde m_{t,N_{\mathrm O}}^{2}\mid\mathcal F_{t-1,N_{\mathrm O}}]\xrightarrow{p}V^{{\rm exp},\star}.
\]

\end{lemma}

\paragraph{Normal limit of the leading terms}\label{a.6.3.3-normal-limit-of-the-leading-terms}

{
\emergencystretch=2em
This step verifies the three conditions of Lemma A.6.1 for the first two terms of the decomposition. For condition 1, the observational records are i.i.d. and \(P^{\rm obs}\psi^{\rm obs}=\tau\), so revealing them sequentially gives \(\mathbb E[\sqrt{N_{\mathrm E}}N_{\mathrm O}^{-1}\{\psi^{\rm obs}(X_i,S_i,Y_i)-\tau\}\mid\sigma\{(X_j,S_j,Y_j):1\le j<i\}]=\allowbreak 0\), and Assumption A.6.1(i) makes these increments square integrable. Assumption 6.3 makes the experimental increments square integrable, so Lemma 6.1 shows that each \(\widetilde m_{t,N_{\mathrm O}}\) is \(\mathcal F_{t,N_{\mathrm O}}\)-measurable with \(\mathbb E[\widetilde m_{t,N_{\mathrm O}}\mid\mathcal F_{t-1,N_{\mathrm O}}]=\allowbreak 0\). Since the observational records generate \(\mathcal F_{0,N_{\mathrm O}}\), the ordered increments form one square-integrable martingale-difference row, which is condition 1.
\par
}

{
\emergencystretch=2em
For condition 2, the observational and experimental blocks are bounded separately. Because the observational records are i.i.d., the observational block equals \((N_{\mathrm E}/\allowbreak N_{\mathrm O})\mathbb E^{\rm obs}[(\psi^{\rm obs}-\tau)^2\mathbf 1\{|\psi^{\rm obs}-\tau|>\varepsilon N_{\mathrm O}/\allowbreak \sqrt{N_{\mathrm E}}\}]\), which converges to zero because Assumption 6.1 gives \(N_{\mathrm O}/\sqrt{N_{\mathrm E}}\asymp\sqrt{N_{\mathrm O}}\to\infty\) and \(N_{\mathrm E}/N_{\mathrm O}\to\kappa\) and Assumption A.6.1(i) gives integrability of the squared score. For the experimental block, the inequality \(\widetilde m_{t,N_{\mathrm O}}^2\mathbf 1\{|\widetilde m_{t,N_{\mathrm O}}|>\varepsilon\sqrt{N_{\mathrm E}}\}\le(\varepsilon\sqrt{N_{\mathrm E}})^{2-r}|\widetilde m_{t,N_{\mathrm O}}|^r\) and the conditional Lyapunov condition in Assumption 6.3 give \[\begin{aligned}
&\frac1{N_{\mathrm E}}\sum_{t=1}^{N_{\mathrm E}}\mathbb E\!\left[\widetilde m_{t,N_{\mathrm O}}^2\mathbf 1\!\left\{|\widetilde m_{t,N_{\mathrm O}}|>\varepsilon\sqrt{N_{\mathrm E}}\right\}\mathrel{\Big|}\mathcal F_{t-1,N_{\mathrm O}}\right]\\
&\qquad\le\frac1{\varepsilon^{r-2}N_{\mathrm E}^{r/2}}\sum_{t=1}^{N_{\mathrm E}}\mathbb E[|\widetilde m_{t,N_{\mathrm O}}|^r\mid\mathcal F_{t-1,N_{\mathrm O}}]\xrightarrow{p}0.
\end{aligned}
\]
\par
}

The sum of the two blockwise bounds verifies condition 2 for the entire ordered row.

{
\emergencystretch=2em
For condition 3, the i.i.d. observational records contribute exactly \(N_{\mathrm E}V^{\rm obs}/N_{\mathrm O}\) to the predictable conditional variance, which converges to \(\kappa V^{\rm obs}\) by Assumption 6.1. Lemma A.6.2 shows that the experimental contribution converges in probability to \(V^{{\rm exp},\star}\), so the sum converges in probability to \(\kappa V^{\rm obs}+V^{{\rm exp},\star}\), which is condition 3. Lemma A.6.1 therefore shows that the first two terms of the decomposition converge in distribution to \(\mathcal N(0,\kappa V^{\rm obs}+V^{{\rm exp},\star})\).
\par
}

\paragraph{Remainder terms}\label{a.6.3.4-remainder-terms}

{
\emergencystretch=2em
This step shows that the last three terms of the decomposition converge to zero in probability. By the definition of \(R_{N_{\mathrm E},N_{\mathrm O}}^{\rm i}\), the second condition in Assumption 6.5 states \(\sqrt{N_{\mathrm O}}R_{N_{\mathrm E},N_{\mathrm O}}^{\rm i}=o_p(1)\). For the two product remainders, Cauchy--Schwarz and Assumption 6.6 give \[\begin{aligned}
\sqrt{N_{\mathrm O}}|R_{N_{\mathrm E},N_{\mathrm O}}^{\rm ii}|
&\le\sqrt{N_{\mathrm O}}\left\|\frac{\widehat f_{X\mid G=\mathrm O,N_{\mathrm O}}}{f_{X\mid G=\mathrm O}}-1\right\|_{2,X\mid G=\mathrm O}\|\widetilde\tau_{N_{\mathrm O}}(\cdot)-\overline\tau_{N_{\mathrm E},N_{\mathrm O}}(\cdot)\|_{2,X\mid G=\mathrm O}=o_p(1),\\
\sqrt{N_{\mathrm O}}|R_{N_{\mathrm E},N_{\mathrm O}}^{\rm iii}|
&\le\sqrt{N_{\mathrm O}}\|\widehat\alpha_{N_{\mathrm E},N_{\mathrm O}}-\alpha\|_{2,{\rm obs}}\|\widehat\mu_{N_{\mathrm O}}-\mu\|_{2,{\rm obs}}=o_p(1).
\end{aligned}
\]
\par
}

{
\emergencystretch=2em
Hence each of the three remainders is \(o_p(N_{\mathrm O}^{-1/2})\). Since \(N_{\mathrm E}/N_{\mathrm O}\to\kappa\), \(\sqrt{N_{\mathrm E}}R_{N_{\mathrm E},N_{\mathrm O}}^{\rm i}=\allowbreak \sqrt{N_{\mathrm E}/N_{\mathrm O}}\,\sqrt{N_{\mathrm O}}R_{N_{\mathrm E},N_{\mathrm O}}^{\rm i}=\allowbreak o_p(1)\), and the same factorization gives \(\sqrt{N_{\mathrm E}}R_{N_{\mathrm E},N_{\mathrm O}}^{\rm ii}=o_p(1)\) and \(\sqrt{N_{\mathrm E}}R_{N_{\mathrm E},N_{\mathrm O}}^{\rm iii}=o_p(1)\).
\par
}

\paragraph{Completion of the proof}\label{a.6.3.5-completion-of-the-proof}

Appendix A.6.3.3 gives the normal limit of the first two terms of the decomposition, and Appendix A.6.3.4 shows that the last three terms converge to zero in probability. Substituting these limits into the decomposition and applying Slutsky\textquotesingle s theorem give \[\sqrt{N_{\mathrm E}}(\widehat\tau_{N_{\mathrm E},N_{\mathrm O}}-\tau)\rightsquigarrow\mathcal N(0,V^{{\rm exp},\star}+\kappa V^{\rm obs}).
\]

{
\emergencystretch=2em
This proves Theorem 6.1. \(\square\)
\par
}

\subsubsection{Proof of Lemma A.6.2}\label{a.6.4-proof-of-lemma-a.6.2}

\textbf{Proof of Lemma A.6.2.} We compare the conditional variance in each round with the square of an integral that depends only on the initial observational sample. Averaging this comparison over the rounds and applying Assumption 6.2 then proves the lemma.

\paragraph{Conditional variance in one round}\label{a.6.4.1-conditional-variance-in-one-round}

With the operators of Section 6.1, define \[\begin{aligned}
&A_{t,N_{\mathrm O}}:=P_{\widehat f_{X\mid G=\mathrm O,N_{\mathrm O}}}\!\left[\sum_{w=0}^1\sqrt{\widehat v_{w,t,N_{\mathrm O}}(X)}\right],\\
&A_{N_{\mathrm O}}^\circ:=P_{\widehat f_{X\mid G=\mathrm O,N_{\mathrm O}}}\!\left[\sum_{w=0}^1\sqrt{\widetilde v_{w,N_{\mathrm O}}(X)}\right].
\end{aligned}
\]

The normalization of the covariate density and the assignment probabilities in Step 2 of Algorithm~\ref{alg:adaptive-sequential-estimation} imply, on the target support, \[\begin{aligned}
&\widehat f^\star_{X\mid G=\mathrm E,t,N_{\mathrm O}}(x)\Pr(W_t=w\mid X_t=x,\mathcal F_{t-1,N_{\mathrm O}})\\
&=\frac{\widehat f_{X\mid G=\mathrm O,N_{\mathrm O}}(x)\sum_{j=0}^1\sqrt{\widehat v_{j,t,N_{\mathrm O}}(x)}}{A_{t,N_{\mathrm O}}}
  \frac{\sqrt{\widehat v_{w,t,N_{\mathrm O}}(x)}}{\sum_{j=0}^1\sqrt{\widehat v_{j,t,N_{\mathrm O}}(x)}}
=\frac{\widehat f_{X\mid G=\mathrm O,N_{\mathrm O}}(x)\sqrt{\widehat v_{w,t,N_{\mathrm O}}(x)}}{A_{t,N_{\mathrm O}}}.
\end{aligned}
\]

{
\emergencystretch=2em
Assumption A.6.1(ii)--(vi) identifies this as the conditional joint density of \((X_t,W_t)\) and makes the denominators positive on that support. Dividing the fitted target density by the joint density in Step 4 gives, on \(\{W_t=w\}\), \[\widehat\phi_{t,N_{\mathrm O}}^{\rm exp}(S_t,W_t,X_t)=(2w-1)\frac{A_{t,N_{\mathrm O}}}{\sqrt{\widehat v_{w,t,N_{\mathrm O}}(X_t)}}\{\widehat\mu_{N_{\mathrm O}}(S_t,X_t)-\widehat\mu_{w,t,N_{\mathrm O}}(X_t)\}.
\]
\par
}

{
\emergencystretch=2em
Conditional on \(X_t=x\), \(W_t=w\), and the past, the fitted index has mean \(\widetilde\mu_{w,N_{\mathrm O}}(x)\) and variance \(\widetilde v_{w,N_{\mathrm O}}(x)\). Expanding about that mean gives \[\int\{\widehat\mu_{N_{\mathrm O}}(s,x)-\widehat\mu_{w,t,N_{\mathrm O}}(x)\}^2 f_{S\mid W=w,X,G=\mathrm E}(s\mid x)\,d\lambda_S(s)
=\widetilde v_{w,N_{\mathrm O}}(x)+\{\widetilde\mu_{w,N_{\mathrm O}}(x)-\widehat\mu_{w,t,N_{\mathrm O}}(x)\}^2.
\]
\par
}

Squaring the score and integrating against the joint density therefore gives \[\begin{aligned}
&\mathbb E[(\widehat\phi_{t,N_{\mathrm O}}^{\rm exp}(S_t,W_t,X_t))^2\mid \mathcal F_{t-1,N_{\mathrm O}}]\\
&=\int\sum_{w=0}^1\frac{\widehat f_{X\mid G=\mathrm O,N_{\mathrm O}}(x)\sqrt{\widehat v_{w,t,N_{\mathrm O}}(x)}}{A_{t,N_{\mathrm O}}}
  \frac{A_{t,N_{\mathrm O}}^2}{\widehat v_{w,t,N_{\mathrm O}}(x)}
  \bigl[\widetilde v_{w,N_{\mathrm O}}(x)+\{\widetilde\mu_{w,N_{\mathrm O}}(x)-\widehat\mu_{w,t,N_{\mathrm O}}(x)\}^2\bigr]\,d\lambda_X(x)\\
&=A_{t,N_{\mathrm O}}\underbrace{P_{\widehat f_{X\mid G=\mathrm O,N_{\mathrm O}}}\!\left[\sum_{w=0}^1
\frac{\widetilde v_{w,N_{\mathrm O}}(X)+\{\widetilde\mu_{w,N_{\mathrm O}}(X)-\widehat\mu_{w,t,N_{\mathrm O}}(X)\}^2}{\sqrt{\widehat v_{w,t,N_{\mathrm O}}(X)}}\right]}_{B_{t,N_{\mathrm O}}}.
\end{aligned}
\]

Subtracting the squared conditional mean computed in the proof of Lemma 6.1 yields \[\mathbb E[\widetilde m_{t,N_{\mathrm O}}^2\mid \mathcal F_{t-1,N_{\mathrm O}}]=A_{t,N_{\mathrm O}}B_{t,N_{\mathrm O}}-\{P_{\widehat f_{X\mid G=\mathrm O,N_{\mathrm O}}}(\widetilde\tau_{N_{\mathrm O}}-\widehat\tau_{t,N_{\mathrm O}}(\cdot))\}^2.
\]

{
\emergencystretch=2em
Fix a round in which \(\widehat v_{w,t,N_{\mathrm O}}(x)\) and \(\widetilde v_{w,N_{\mathrm O}}(x)\) lie in \([\underline v,\overline v]\) for both arms and every target-support \(x\). For \(a,b\) in this interval and \(d\in\mathbb R\), \[\begin{aligned}
&|\sqrt a-\sqrt b|=\frac{|a-b|}{\sqrt a+\sqrt b}\le\frac{|a-b|}{2\sqrt{\underline v}},\\
&\left|\frac{b+d^2}{\sqrt a}-\sqrt b\right|
\le\frac{d^2}{\sqrt a}+\frac{\sqrt b}{\sqrt a}\,|\sqrt b-\sqrt a|
\le\frac{d^2}{\sqrt{\underline v}}+\frac{\sqrt{\overline v}}{2\underline v}|a-b|.
\end{aligned}
\]
\par
}

Integrating these bounds against the normalized fitted target density and taking suprema over the target support give \[\begin{aligned}
&|A_{t,N_{\mathrm O}}-A_{N_{\mathrm O}}^\circ|\le\frac1{2\sqrt{\underline v}}\sum_{w=0}^1\|\widehat v_{w,t,N_{\mathrm O}}-\widetilde v_{w,N_{\mathrm O}}\|_\infty,\\
&|B_{t,N_{\mathrm O}}-A_{N_{\mathrm O}}^\circ|\le\sum_{w=0}^1\left\{\frac{\|\widehat\mu_{w,t,N_{\mathrm O}}-\widetilde\mu_{w,N_{\mathrm O}}\|_\infty^2}{\sqrt{\underline v}}
+\frac{\sqrt{\overline v}}{2\underline v}\|\widehat v_{w,t,N_{\mathrm O}}-\widetilde v_{w,N_{\mathrm O}}\|_\infty\right\}.
\end{aligned}
\]

Jensen\textquotesingle s inequality and the treatment-minus-control definition of the conditional effect give \[\{P_{\widehat f_{X\mid G=\mathrm O,N_{\mathrm O}}}(\widetilde\tau_{N_{\mathrm O}}-\widehat\tau_{t,N_{\mathrm O}}(\cdot))\}^2
\le P_{\widehat f_{X\mid G=\mathrm O,N_{\mathrm O}}}[(\widetilde\tau_{N_{\mathrm O}}-\widehat\tau_{t,N_{\mathrm O}}(\cdot))^2]
\le2\sum_{w=0}^1\|\widehat\mu_{w,t,N_{\mathrm O}}-\widetilde\mu_{w,N_{\mathrm O}}\|_\infty^2.
\]

{
\emergencystretch=2em
Since \(0\le A_{t,N_{\mathrm O}},A_{N_{\mathrm O}}^\circ\le2\sqrt{\overline v}\) in this round, adding and subtracting \(A_{t,N_{\mathrm O}}A_{N_{\mathrm O}}^\circ\) gives the one-round bound \[\begin{aligned}
&\left|\mathbb E[\widetilde m_{t,N_{\mathrm O}}^2\mid \mathcal F_{t-1,N_{\mathrm O}}]-(A_{N_{\mathrm O}}^\circ)^2\right|\\
&\le A_{t,N_{\mathrm O}}|B_{t,N_{\mathrm O}}-A_{N_{\mathrm O}}^\circ|+A_{N_{\mathrm O}}^\circ|A_{t,N_{\mathrm O}}-A_{N_{\mathrm O}}^\circ|+\{P_{\widehat f_{X\mid G=\mathrm O,N_{\mathrm O}}}(\widetilde\tau_{N_{\mathrm O}}-\widehat\tau_{t,N_{\mathrm O}}(\cdot))\}^2\\
&\le \bigl(1+\sqrt{\overline v/\underline v}\bigr)^2\sum_{w=0}^1\left\{\|\widehat v_{w,t,N_{\mathrm O}}-\widetilde v_{w,N_{\mathrm O}}\|_\infty+\|\widehat\mu_{w,t,N_{\mathrm O}}-\widetilde\mu_{w,N_{\mathrm O}}\|_\infty^2\right\}.
\end{aligned}
\]
\par
}

\paragraph{Limit of the common fitted integral}\label{a.6.4.2-limit-of-the-common-fitted-integral}

{
\emergencystretch=2em
For each \(w,x\), the two conditional standard deviations are the \(L_2\) norms of the fitted and true indices centered under the same surrogate law. The reverse triangle inequality bounds their difference by the norm of the centered estimation error. Removing the centering gives \[\begin{aligned}
&\left|\sqrt{\widetilde v_{w,N_{\mathrm O}}(x)}-\sqrt{v_w(x)}\right|^2\\
&\le\int\bigl[\{\widehat\mu_{N_{\mathrm O}}(s,x)-\mu(s,x)\}-\{\widetilde\mu_{w,N_{\mathrm O}}(x)-\mu_w(x)\}\bigr]^2 f_{S\mid W=w,X,G=\mathrm E}(s\mid x)\,d\lambda_S(s)\\
&=\int\{\widehat\mu_{N_{\mathrm O}}(s,x)-\mu(s,x)\}^2 f_{S\mid W=w,X,G=\mathrm E}(s\mid x)\,d\lambda_S(s)-\{\widetilde\mu_{w,N_{\mathrm O}}(x)-\mu_w(x)\}^2\\
&\le\int\{\widehat\mu_{N_{\mathrm O}}(s,x)-\mu(s,x)\}^2 f_{S\mid W=w,X,G=\mathrm E}(s\mid x)\,d\lambda_S(s).
\end{aligned}
\]
\par
}

{
\emergencystretch=2em
By Theorem 3.2, \(\sqrt{V^{{\rm exp},\star}}=\allowbreak P_{f_{X\mid G=\mathrm O}}[\allowbreak\sum_{w=0}^1\sqrt{v_w(X)}]\). Add and subtract the integral of the fitted standard deviations under the true target law. When \(\widetilde v_{w,N_{\mathrm O}}(x)\le\overline v\) for both arms on the target support, absolute continuity in Assumption A.6.1(v) and Cauchy--Schwarz give \[\begin{aligned}
&|A_{N_{\mathrm O}}^\circ-\sqrt{V^{{\rm exp},\star}}|\\
&\le\left|P_{f_{X\mid G=\mathrm O}}\!\left[\left\{\frac{\widehat f_{X\mid G=\mathrm O,N_{\mathrm O}}(X)}{f_{X\mid G=\mathrm O}(X)}-1\right\}\sum_{w=0}^1\sqrt{\widetilde v_{w,N_{\mathrm O}}(X)}\right]\right|
+\sum_{w=0}^1 P_{f_{X\mid G=\mathrm O}}\!\left[\left|\sqrt{\widetilde v_{w,N_{\mathrm O}}(X)}-\sqrt{v_w(X)}\right|\right]\\
&\le2\sqrt{\overline v}\left\|\frac{\widehat f_{X\mid G=\mathrm O,N_{\mathrm O}}}{f_{X\mid G=\mathrm O}}-1\right\|_{2,X\mid G=\mathrm O}
+\sum_{w=0}^1\left\{P_{f_{X\mid G=\mathrm O}}\!\left[\int\{\widehat\mu_{N_{\mathrm O}}(s,X)-\mu(s,X)\}^2 f_{S\mid W=w,X,G=\mathrm E}(s\mid X)\,d\lambda_S(s)\right]\right\}^{1/2}.
\end{aligned}
\]
\par
}

\paragraph{Terminal average}\label{a.6.4.3-terminal-average}

{
\emergencystretch=2em
On the simultaneous event of Assumption A.6.1(viii), the variance bounds hold for both arms and all target-support \(x\) in rounds \(1,\ldots,N_{\mathrm E}\), and Assumption A.6.1(ix) fixes their constants independently of \(N_{\mathrm O}\). This event has probability tending to one. The initial risks in Assumption 6.4 and Appendix A.6.4.2 thus give \(A_{N_{\mathrm O}}^\circ\xrightarrow{p}\sqrt{V^{{\rm exp},\star}}\). On the same event, Appendix A.6.4.1 gives \[\begin{aligned}
&\left|\frac1{N_{\mathrm E}}\sum_{t=1}^{N_{\mathrm E}}\mathbb E[\widetilde m_{t,N_{\mathrm O}}^2\mid \mathcal F_{t-1,N_{\mathrm O}}]-V^{{\rm exp},\star}\right|\\
&\le\frac{\bigl(1+\sqrt{\overline v/\underline v}\bigr)^2}{N_{\mathrm E}}\sum_{t=1}^{N_{\mathrm E}}\sum_{w=0}^1
\left\{\|\widehat v_{w,t,N_{\mathrm O}}-\widetilde v_{w,N_{\mathrm O}}\|_\infty+\|\widehat\mu_{w,t,N_{\mathrm O}}-\widetilde\mu_{w,N_{\mathrm O}}\|_\infty^2\right\}
+\left|(A_{N_{\mathrm O}}^\circ)^2-V^{{\rm exp},\star}\right|.
\end{aligned}
\]
\par
}

{
\emergencystretch=2em
The average tends to zero in probability by Assumption 6.2, and the last term does so by the limit of \(A_{N_{\mathrm O}}^\circ\). Since the complement of the event has vanishing probability, this proves Lemma A.6.2.
\par
}

\subsection{Proofs for Section 7}\label{a.7-proofs-for-section-7}

Appendix A.7 proves Theorem 7.1 and Corollary 7.1. It first states Lemmas A.7.1--A.7.3 and proves Theorem 7.1 and Corollary 7.1 from them, then proves Lemmas A.7.1--A.7.3 one at a time, stating each further lemma at the start of the subsection whose proofs first use it. Figure A.1 shows how the results depend on one another. An arrow from one result to another means that the proof of the second uses the first, an arrow into the frame means that the proofs of all three lemmas in the frame use the result, the arrow from the frame means that the proof of Theorem 7.1 uses all three lemmas, and dashed boxes mark results from the literature. Theorem 7.1 and Corollary 7.1 are shown in darker blue, and Lemmas A.7.1--A.7.3, from which Theorem 7.1 is proved, are shown in lighter blue.

\begin{figure}[tbp]
\centering
\includegraphics[width=\linewidth]{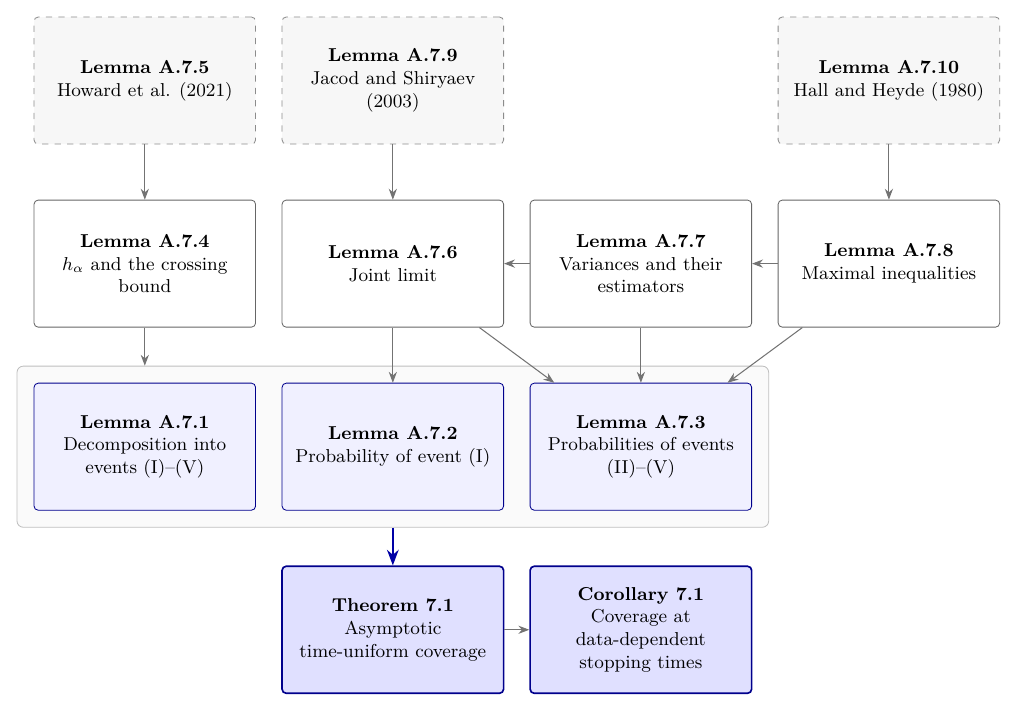}
\caption{Dependence of the results in the proof of Theorem 7.1}
\label{fig:A.1}
\end{figure}

\subsubsection{Statements of Lemmas A.7.1--A.7.3}\label{a.7.1-statements-of-lemmas-a.7.1a.7.3}

{
\emergencystretch=2em
This subsection records the conditions in force, introduces the notation of the proofs, and states Lemmas A.7.1--A.7.3. Section 7 uses the common conditions and the Section 7 conditions of Assumption A.6.1, Lemma 6.1 gives conditional centering, and Assumption 7.2 supplies square integrability for all \(N_{\mathrm O}\) and \(t\ge1\). Recall that \(\mathcal F_{0,N_{\mathrm O}}=\allowbreak \sigma(\mathcal D_{N_{\mathrm O}}^{\rm obs})\) is generated by the observational sample. Assumption A.6.1(xi) gives \(V^{\rm obs}>0\), and because Assumption A.6.1(x) places the true variances \(v_w(x)\) in \([\underline v,\overline v]\) on the target support, the form of the minimum experimental variance in Theorem 3.2 gives \(V^{{\rm exp},\star}=\allowbreak \{\mathbb E[\sqrt{v_1(X)}+\sqrt{v_0(X)}\mid G=\allowbreak \mathrm O]\}^2\ge4\underline v>0\). Assumption 7.2 and conditional Jensen\textquotesingle s inequality give \(\mathbb E[\widetilde m_{s,N_{\mathrm O}}^2\mid\mathcal F_{s-1,N_{\mathrm O}}]\le C^{2/r}\) for every \(N_{\mathrm O}\) and \(s\ge1\), and hence \(\mathbb E[\widetilde m_{s,N_{\mathrm O}}^2]\le C^{2/r}\).
\par
}

{
\emergencystretch=2em
The proofs start from the decomposition of the estimation error. Lemma 6.2, whose derivation in Appendix A.6.1 holds at every round \(t\ge1\), gives \[\sqrt{N_{\mathrm O}}(\widehat\tau_{t,N_{\mathrm O}}-\tau)
=
\sqrt{N_{\mathrm O}}\left(L_{N_{\mathrm O}}^{\rm obs}+\frac1t\sum_{s=1}^t\widetilde m_{s,N_{\mathrm O}}\right)
+
\sqrt{N_{\mathrm O}}\left(R_{t,N_{\mathrm O}}^{\rm i}+R_{t,N_{\mathrm O}}^{\rm ii}+R_{t,N_{\mathrm O}}^{\rm iii}\right).
\]
\par
}

{
\emergencystretch=2em
On the normalized time scale \(u=t/N_{\mathrm O}\), the first term on the right has asymptotic variance \[\sigma^2(u):=V^{\rm obs}+\frac{V^{{\rm exp},\star}}{u},\qquad u>0.
\]
\par
}

{
\emergencystretch=2em
The first lemma decomposes \(\{\exists t\ge t_{0,N_{\mathrm O}}:\tau\notin C_{t,N_{\mathrm O}}(\alpha)\}\) into five events, and Lemmas A.7.2 and A.7.3 bound their probabilities. Throughout, \(c_\omega\in(0,\infty)\) is the constant of Assumption 7.1(2). Lemmas A.7.4--A.7.10, which the explanations and proofs below cite, are stated at the start of the subsection whose proofs first use them.
\par
}

\expandafter\gdef\csname thelemma\endcsname{A.7.\arabic{lemma}}
\expandafter\gdef\csname theHlemma\endcsname{A.7.\arabic{lemma}}
\setcounter{lemma}{0}
\begin{lemma}[Decomposition into events (I)--(V)]

{
\emergencystretch=2em
Fix \(\eta\in(0,1)\), and put \[\varepsilon_\eta:=\frac\eta5\{2V^{\rm obs}\log(1/\alpha)\}^{1/2}.
\]
\par
}

{
\emergencystretch=2em
There is a constant \(K_0\ge2\), depending on \(\eta\) but not on \(K\) or \(N_{\mathrm O}\), such that, for every \(K\ge K_0\) and every \(N_{\mathrm O}\) with \(t_{0,N_{\mathrm O}}\le N_{\mathrm O}/K\), \[\{\exists t\ge t_{0,N_{\mathrm O}}:\tau\notin C_{t,N_{\mathrm O}}(\alpha)\}
\subseteq
\mathrm{(I)}\cup\mathrm{(II)}\cup\mathrm{(III)}\cup\mathrm{(IV)}\cup\mathrm{(V)},
\]
\par
}

{
\emergencystretch=2em
where \(t\) ranges over integers here and below and the five events are \[\begin{aligned}
\mathrm{(I)}&:=\left\{\max_{N_{\mathrm O}/K\le t\le KN_{\mathrm O}}\frac{\sqrt{N_{\mathrm O}}\,|L_{N_{\mathrm O}}^{\rm obs}+t^{-1}\sum_{s=1}^t\widetilde m_{s,N_{\mathrm O}}|}{\sigma(t/N_{\mathrm O})\,h_\alpha(c_\omega/\sigma^2(t/N_{\mathrm O}))}\ge1-\eta\right\},\\
\mathrm{(II)}&:=\{\exists t\in[t_{0,N_{\mathrm O}},N_{\mathrm O}/K):\tau\notin C_{t,N_{\mathrm O}}(\alpha)\},\\
\mathrm{(III)}&:=\left\{\sqrt{N_{\mathrm O}}\sup_{t\ge\lfloor KN_{\mathrm O}\rfloor}\left|\frac1t\sum_{s=1}^t\widetilde m_{s,N_{\mathrm O}}\right|>\varepsilon_\eta\right\},\\
\mathrm{(IV)}&:=\left\{\exists t\ge N_{\mathrm O}/K:\left|\sqrt{N_{\mathrm O}}\,\overline B_{t,N_{\mathrm O}}-\sigma(t/N_{\mathrm O})\,h_\alpha\!\left(\frac{c_\omega}{\sigma^2(t/N_{\mathrm O})}\right)\right|>\varepsilon_\eta\right\},\\
\mathrm{(V)}&:=\left\{\sqrt{N_{\mathrm O}}\sup_{t\ge t_{0,N_{\mathrm O}}}\left(|R_{t,N_{\mathrm O}}^{\rm i}|+|R_{t,N_{\mathrm O}}^{\rm ii}|+|R_{t,N_{\mathrm O}}^{\rm iii}|\right)>\varepsilon_\eta\right\}.
\end{aligned}
\]
\par
}

Consequently, \[\Pr\{\exists t\ge t_{0,N_{\mathrm O}}:\tau\notin C_{t,N_{\mathrm O}}(\alpha)\}
\le
\Pr\{\mathrm{(I)}\}+\Pr\{\mathrm{(II)}\}+\Pr\{\mathrm{(III)}\}+\Pr\{\mathrm{(IV)}\}+\Pr\{\mathrm{(V)}\}.
\]

\end{lemma}

\expandafter\gdef\csname thelemma\endcsname{A.7.\arabic{lemma}}
\expandafter\gdef\csname theHlemma\endcsname{A.7.\arabic{lemma}}
\setcounter{lemma}{1}
\begin{lemma}[Probability of event (I)]

{
\emergencystretch=2em
Under the conditions of Theorem 7.1, fix \(\eta\in(0,1)\), and let \(Z\) be a standard normal random variable and \(W\) a standard Brownian motion independent of \(Z\).
\par
}

\begin{enumerate}
\def\labelenumi{(\alph{enumi})}
\item
  For every \(K\ge K_0\), with \(K_0\) the constant of Lemma A.7.1, \[\limsup_{N_{\mathrm O}\to\infty}\Pr\{\mathrm{(I)}\}
  \le
  \Pr\left\{\sup_{u>0}\frac{|(V^{\rm obs})^{1/2}Z+(V^{{\rm exp},\star})^{1/2}W(u)/u|}{\sigma(u)\,h_\alpha(c_\omega/\sigma^2(u))}\ge1-\eta\right\}.
  \]
\item
  The right-hand side in part (a) does not depend on \(K\), and its limit as \(\eta\downarrow0\) is at most \(\alpha\).
\end{enumerate}

\end{lemma}

\expandafter\gdef\csname thelemma\endcsname{A.7.\arabic{lemma}}
\expandafter\gdef\csname theHlemma\endcsname{A.7.\arabic{lemma}}
\setcounter{lemma}{2}
\begin{lemma}[Probabilities of events (II)--(V)]

Under the conditions of Theorem 7.1, the following hold.

\begin{enumerate}
\def\labelenumi{(\alph{enumi})}
\item
  \(\sqrt{N_{\mathrm O}}\allowbreak\sup_{t\ge t_{0,N_{\mathrm O}}}(|R_{t,N_{\mathrm O}}^{\rm i}|+|R_{t,N_{\mathrm O}}^{\rm ii}|+|R_{t,N_{\mathrm O}}^{\rm iii}|)\xrightarrow{p}0\).
\item
  The probability of event (II) satisfies \[\lim_{K\to\infty}\limsup_{N_{\mathrm O}\to\infty}
  \Pr\{\exists t\in[t_{0,N_{\mathrm O}},N_{\mathrm O}/K):\tau\notin C_{t,N_{\mathrm O}}(\alpha)\}=0.
  \]
\item
  For every \(\varepsilon>0\), \[\lim_{K\to\infty}\limsup_{N_{\mathrm O}\to\infty}
  \Pr\left\{\sqrt{N_{\mathrm O}}\sup_{t\ge\lfloor KN_{\mathrm O}\rfloor}\left|\frac1t\sum_{s=1}^t\widetilde m_{s,N_{\mathrm O}}\right|>\varepsilon\right\}=0.
  \]
\item
  For every \(K\ge1\) and \(\varepsilon>0\), \[\lim_{N_{\mathrm O}\to\infty}
  \Pr\left\{\exists t\ge N_{\mathrm O}/K:\left|\sqrt{N_{\mathrm O}}\,\overline B_{t,N_{\mathrm O}}-\sigma(t/N_{\mathrm O})\,h_\alpha\!\left(\frac{c_\omega}{\sigma^2(t/N_{\mathrm O})}\right)\right|>\varepsilon\right\}=0.
  \]
\end{enumerate}

\end{lemma}

\subsubsection{Proofs of Theorem 7.1 and Corollary 7.1}\label{a.7.2-proofs-of-theorem-7.1-and-corollary-7.1}

This subsection proves Theorem 7.1 from Lemmas A.7.1--A.7.3 and then proves Corollary 7.1.

{
\emergencystretch=2em
\textbf{Proof of Theorem 7.1.} Fix \(\eta\in(0,1)\), and let \(K_0\) be the constant of Lemma A.7.1. For \(K\ge K_0\), the condition \(t_{0,N_{\mathrm O}}\le N_{\mathrm O}/K\) holds for all sufficiently large \(N_{\mathrm O}\), because \(t_{0,N_{\mathrm O}}/N_{\mathrm O}\to0\). Taking \(\limsup_{N_{\mathrm O}\to\infty}\) in the inequality of Lemma A.7.1 and applying Lemma A.7.2(a) to \(\Pr\{\mathrm{(I)}\}\), Lemma A.7.3(d) with \(\varepsilon=\varepsilon_\eta\) to \(\Pr\{\mathrm{(IV)}\}\), and Lemma A.7.3(a) to \(\Pr\{\mathrm{(V)}\}\) give, for every \(K\ge K_0\), \[\begin{aligned}
\limsup_{N_{\mathrm O}\to\infty}\Pr\{\exists t\ge t_{0,N_{\mathrm O}}:\tau\notin C_{t,N_{\mathrm O}}(\alpha)\}
&\le
\Pr\left\{\sup_{u>0}\frac{|(V^{\rm obs})^{1/2}Z+(V^{{\rm exp},\star})^{1/2}W(u)/u|}{\sigma(u)\,h_\alpha(c_\omega/\sigma^2(u))}\ge1-\eta\right\}\\
&\quad+\limsup_{N_{\mathrm O}\to\infty}\Pr\{\mathrm{(II)}\}+\limsup_{N_{\mathrm O}\to\infty}\Pr\{\mathrm{(III)}\}.
\end{aligned}
\]
\par
}

{
\emergencystretch=2em
The left-hand side depends on neither \(K\) nor \(\eta\), and the first term on the right does not depend on \(K\). Letting \(K\to\infty\), Lemma A.7.3(b) and Lemma A.7.3(c) with \(\varepsilon=\varepsilon_\eta\) remove the last two terms. Letting \(\eta\downarrow0\), Lemma A.7.2(b) bounds the first term by \(\alpha\). Hence \(\liminf_{N_{\mathrm O}\to\infty}\Pr\{\tau\in C_{t,N_{\mathrm O}}(\alpha)\text{ for all }t\ge t_{0,N_{\mathrm O}}\}\ge1-\alpha\), which proves Theorem 7.1.
\par
}

{
\emergencystretch=2em
\textbf{Proof of Corollary 7.1.} Because \(t_{0,N_{\mathrm O}}\le T_{\mathrm{stop},N_{\mathrm O}}<\infty\) almost surely, the event that \(\tau\in C_{t,N_{\mathrm O}}(\alpha)\) for all \(t\ge t_{0,N_{\mathrm O}}\) is contained, up to a null set, in the event \(\{\tau\in C_{T_{\mathrm{stop},N_{\mathrm O}},N_{\mathrm O}}(\alpha)\}\). Theorem 7.1 therefore gives \(\liminf_{N_{\mathrm O}\to\infty}\Pr\{\tau\in C_{T_{\mathrm{stop},N_{\mathrm O}},N_{\mathrm O}}(\alpha)\}\ge1-\alpha\), which proves Corollary 7.1. \(\square\)
\par
}

\subsubsection{Proof of Lemma A.7.1}\label{a.7.3-proof-of-lemma-a.7.1}

This subsection first states Lemmas A.7.4 and A.7.5, which the proofs in this subsection use, and then proves Lemma A.7.1 and Lemma A.7.4. The proof of Lemma A.7.1 uses Lemma A.7.4(c), and the proof of Lemma A.7.4(d) uses Lemma A.7.5. Lemmas A.7.5, A.7.9, and A.7.10 are results from the literature, stated in the forms used below.

\expandafter\gdef\csname thelemma\endcsname{A.7.\arabic{lemma}}
\expandafter\gdef\csname theHlemma\endcsname{A.7.\arabic{lemma}}
\setcounter{lemma}{3}
\begin{lemma}[Inflation factor and Gaussian-mixture crossing bound]

{
\emergencystretch=2em
Fix \(\alpha\in(0,1)\).
\par
}

\begin{enumerate}
\def\labelenumi{(\alph{enumi})}
\item
  For every \(x>0\), \(h_\alpha(x)^2\ge2\log(1/\alpha)\) and \(h_\alpha(x)^2\ge2\log(1/\alpha)/x\).
\item
  The map \((c,v)\mapsto v^{1/2}h_\alpha(c/v)\) is continuous on \((0,\infty)^2\).
\item
  For every \(c\in(0,\infty)\), the function \(u\mapsto\sigma(u)h_\alpha(c/\sigma^2(u))\) is continuous on \((0,\infty)\), is bounded below by \(\{2V^{\rm obs}\log(1/\alpha)\}^{1/2}\), and converges to a finite limit as \(u\to\infty\).
\item
  For every \(c\in(0,\infty)\), with \(\sigma^2(u)=V^{\rm obs}+V^{{\rm exp},\star}/u\) and \(Z\) and \(W\) as in Lemma A.7.2, \[\Pr\left\{\sup_{u>0}\frac{|(V^{\rm obs})^{1/2}Z+(V^{{\rm exp},\star})^{1/2}W(u)/u|}{\sigma(u)\,h_\alpha(c/\sigma^2(u))}\ge1\right\}\le\alpha.
  \]
\end{enumerate}

\end{lemma}

\expandafter\gdef\csname thelemma\endcsname{A.7.\arabic{lemma}}
\expandafter\gdef\csname theHlemma\endcsname{A.7.\arabic{lemma}}
\setcounter{lemma}{4}
\begin{lemma}[Two-sided normal mixture boundary]

{
\emergencystretch=2em
Equation (3.7) of \citet[Section 3.2]{Howard2021}, with details in their Appendix A.3, gives the two-sided normal mixture boundary for processes that satisfy their sub-Gaussian condition (Definition 1 with \(\psi(\lambda)=\lambda^2/2\) and \(l_0=1\)) in both directions. We use it in the following form. Let \((T_k)_{k\ge0}\) be a real-valued process adapted to a filtration \((\mathcal H_k)_{k\ge0}\), and let \((q_k)_{k\ge0}\) be a deterministic sequence in \([0,\infty)\) with \(T_0=q_0=0\). Suppose that, for every \(\lambda\in\mathbb R\), the process \((\exp\{\lambda T_k-\lambda^2q_k/2\})_{k\ge0}\) is a supermartingale with respect to \((\mathcal H_k)_{k\ge0}\). Then, for every \(\rho>0\) and \(\beta\in(0,1)\), \[\Pr\left[\exists k\ge1:|T_k|\ge\left\{(q_k+\rho)\log\frac{q_k+\rho}{\beta^2\rho}\right\}^{1/2}\right]\le\beta.
\]
\par
}

\end{lemma}

{
\emergencystretch=2em
\textbf{Proof of Lemma A.7.1.} Fix \(\eta\in(0,1)\). The boundary in Lemma A.7.4(c) has a finite limit at infinity. Choose \(K_0\ge2\), depending on \(\eta\) but independent of \(K\) and \(N_{\mathrm O}\), so that \[\left|\sigma(u)h_\alpha(c_\omega/\sigma^2(u))-\sigma(u')h_\alpha(c_\omega/\sigma^2(u'))\right|
\le\varepsilon_\eta\qquad (u,u'\ge K_0-1).
\]
\par
}

The lower bound in Lemma A.7.4(c) gives \[5\varepsilon_\eta=\eta\{2V^{\rm obs}\log(1/\alpha)\}^{1/2}
\le\eta\,\sigma(u)h_\alpha(c_\omega/\sigma^2(u))\qquad(u>0).
\]

{
\emergencystretch=2em
Fix any \(K\ge K_0\) and any \(N_{\mathrm O}\) with \(t_{0,N_{\mathrm O}}\le N_{\mathrm O}/K\). Suppose that \(\tau\notin C_{t,N_{\mathrm O}}(\alpha)\) at an integer \(t\ge t_{0,N_{\mathrm O}}\). If \(t<N_{\mathrm O}/K\), event (II) occurs. For \(t\ge N_{\mathrm O}/K\), suppose that (IV) and (V) do not occur. A miss implies \(|\widehat\tau_{t,N_{\mathrm O}}-\tau|>\overline B_{t,N_{\mathrm O}}\). Lemma 6.2 and the complements of (V) and (IV) then give \[\begin{aligned}
&\sqrt{N_{\mathrm O}}\left|L_{N_{\mathrm O}}^{\rm obs}+\frac1t\sum_{s=1}^t \widetilde m_{s,N_{\mathrm O}}\right|\\
&\ge\sqrt{N_{\mathrm O}}|\widehat\tau_{t,N_{\mathrm O}}-\tau|-\varepsilon_\eta
>\sqrt{N_{\mathrm O}}\,\overline B_{t,N_{\mathrm O}}-\varepsilon_\eta\\
&\ge\sigma(t/N_{\mathrm O})h_\alpha(c_\omega/\sigma^2(t/N_{\mathrm O}))-2\varepsilon_\eta.
\end{aligned}
\]
\par
}

{
\emergencystretch=2em
Compare the leading term at \(t\) with its value at the integer \[r:=\min\{t,\lfloor KN_{\mathrm O}\rfloor\}.
\]
\par
}

{
\emergencystretch=2em
Because \(N_{\mathrm O}/K\le N_{\mathrm O}\le\lfloor KN_{\mathrm O}\rfloor\), we have \(N_{\mathrm O}/K\le r\le KN_{\mathrm O}\). If \(r=t\), both the leading term and the boundary coincide at the two indices. If \(r<t\), then \(r=\lfloor KN_{\mathrm O}\rfloor\). In this case also suppose that (III) does not occur, so that \[\begin{aligned}
&\sqrt{N_{\mathrm O}}\left|\frac1t\sum_{s=1}^t \widetilde m_{s,N_{\mathrm O}}-\frac1r\sum_{s=1}^r \widetilde m_{s,N_{\mathrm O}}\right|\\
&\le2\sqrt{N_{\mathrm O}}\sup_{j\ge\lfloor KN_{\mathrm O}\rfloor}\left|\frac1j\sum_{s=1}^j \widetilde m_{s,N_{\mathrm O}}\right|
\le2\varepsilon_\eta.
\end{aligned}
\]
\par
}

{
\emergencystretch=2em
In this case \(t/N_{\mathrm O}\ge r/N_{\mathrm O}\ge K-1/N_{\mathrm O}\ge K_0-1\), so the choice of \(K_0\) bounds the difference of the boundary values by \(\varepsilon_\eta\). Since \(L_{N_{\mathrm O}}^{\rm obs}\) is common to both indices, in either case \[\begin{aligned}
&\sqrt{N_{\mathrm O}}\left|L_{N_{\mathrm O}}^{\rm obs}+\frac1r\sum_{s=1}^r \widetilde m_{s,N_{\mathrm O}}\right|\\
&\ge\sqrt{N_{\mathrm O}}\left|L_{N_{\mathrm O}}^{\rm obs}+\frac1t\sum_{s=1}^t \widetilde m_{s,N_{\mathrm O}}\right|-2\varepsilon_\eta\\
&>\sigma(r/N_{\mathrm O})h_\alpha(c_\omega/\sigma^2(r/N_{\mathrm O}))-5\varepsilon_\eta\\
&\ge(1-\eta)\,\sigma(r/N_{\mathrm O})h_\alpha(c_\omega/\sigma^2(r/N_{\mathrm O})).
\end{aligned}
\]
\par
}

{
\emergencystretch=2em
The boundary is positive, so event (I) occurs at \(r\). This proves the finite-sample inclusion \[\{\exists t\ge t_{0,N_{\mathrm O}}:\tau\notin C_{t,N_{\mathrm O}}(\alpha)\}
\subseteq\mathrm{(I)}\cup\mathrm{(II)}\cup\mathrm{(III)}\cup\mathrm{(IV)}\cup\mathrm{(V)}.
\]
\par
}

Taking probabilities and applying the union bound gives the stated inequality, which proves Lemma A.7.1.

\paragraph{Proof of Lemma A.7.4}\label{a.7.3.1-proof-of-lemma-a.7.4}

{
\emergencystretch=2em
Parts (a)--(c) follow from the definition of \(h_\alpha\). The proof of part (d) first represents the process in part (d), divided by its variance, as a Brownian motion evaluated at \(1/\sigma^2(u)\), then applies Lemma A.7.5 to this Brownian motion on dyadic grids, and passes from the grids to continuous time by path continuity.
\par
}

{
\emergencystretch=2em
\textbf{Proof of Lemma A.7.4(a)--(c).} (a) For \(x>0\), both \((x+1)/x\ge1\) and \((x+1)/x\ge1/x\) hold, and \(\log\{(x+1)/\alpha^2\}\ge\log(1/\alpha^2)=2\log(1/\alpha)\). Both bounds follow from \(h_\alpha(x)^2=\{(x+1)/x\}\log\{(x+1)/\alpha^2\}\).
\par
}

\begin{enumerate}
\def\labelenumi{(\alph{enumi})}
\setcounter{enumi}{1}
\tightlist
\item
  For \(c,v\in(0,\infty)\), the definition of \(h_\alpha\) gives \[v\,h_\alpha\!\left(\frac cv\right)^2=\frac{v(c+v)}{c}\log\left(\frac{c+v}{v\alpha^2}\right),
  \]
\end{enumerate}

{
\emergencystretch=2em
which is continuous and positive on \((0,\infty)^2\), and so is its square root.
\par
}

\begin{enumerate}
\def\labelenumi{(\alph{enumi})}
\setcounter{enumi}{2}
\tightlist
\item
  The function \(u\mapsto\sigma^2(u)\) is continuous and positive on \((0,\infty)\), satisfies \(\sigma^2(u)\ge V^{\rm obs}\), and converges to \(V^{\rm obs}\) as \(u\to\infty\). Part (b) therefore gives the continuity and the limit \((V^{\rm obs})^{1/2}h_\alpha(c/V^{\rm obs})\), and part (a) gives \(\sigma^2(u)h_\alpha(c/\sigma^2(u))^2\ge2V^{\rm obs}\log(1/\alpha)\). This proves parts (a)--(c).
\end{enumerate}

{
\emergencystretch=2em
\textbf{Proof of Lemma A.7.4(d).} Fix \(c\in(0,\infty)\), let \(\widetilde W\) be a standard Brownian motion, and write \[a(u):=\frac{1}{\sigma^2(u)}=\frac{u}{uV^{\rm obs}+V^{{\rm exp},\star}},\qquad u>0.
\]
\par
}

{
\emergencystretch=2em
Since \(V^{\rm obs}>0\) and \(V^{{\rm exp},\star}>0\) by Appendix A.7.1, the map \(u\mapsto a(u)\) is an increasing bijection from \((0,\infty)\) onto \((0,1/V^{\rm obs})\).
\par
}

{
\emergencystretch=2em
\textbf{Step 1.} We show that the process in part (d), divided by its variance, has the law of \(\widetilde W\) evaluated at \(a(u)\). Since \(Z\) is standard normal and independent of \(W\), the process \(\{(V^{\rm obs})^{1/2}Z+(V^{{\rm exp},\star})^{1/2}W(u)/u:u>0\}\) is a centered Gaussian process with continuous paths. Dividing it by \(\sigma^2(u)\) gives the normalized process, whose covariance at times \(u\) and \(u'\), with \(0<u\le u'\), is \[\begin{aligned}
&\operatorname{Cov}\left\{\frac{(V^{\rm obs})^{1/2}Z+(V^{{\rm exp},\star})^{1/2}W(u)/u}{\sigma^2(u)},\frac{(V^{\rm obs})^{1/2}Z+(V^{{\rm exp},\star})^{1/2}W(u')/u'}{\sigma^2(u')}\right\}\\
&=\frac{V^{\rm obs}+V^{{\rm exp},\star}u/(uu')}{\sigma^2(u)\sigma^2(u')}
=\frac{\sigma^2(u')}{\sigma^2(u)\sigma^2(u')}
=a(u)=\min\{a(u),a(u')\},
\end{aligned}
\]
\par
}

{
\emergencystretch=2em
where the last equality holds because \(a\) is increasing. This is the covariance of \(\widetilde W(a(u))\), which is also a centered Gaussian process with continuous paths on \((0,\infty)\). The two processes therefore have the same finite-dimensional laws. Because both have continuous paths, the supremum over \(u>0\) of each ratio below equals its supremum over positive rational \(u\), so the two suprema have the same law.
\par
}

{
\emergencystretch=2em
\textbf{Step 2.} We apply Lemma A.7.5 to \(\widetilde W\) on dyadic grids. Fix \(n\ge1\), and let \(T_k:=\widetilde W(k2^{-n})\), \(q_k:=k2^{-n}\), and \(\mathcal H_k:=\sigma\{\widetilde W(b):0\le b\le k2^{-n}\}\) for \(k\ge0\). Independent Gaussian increments give, for every \(\lambda\in\mathbb R\) and \(k\ge0\), \[\mathbb E\left[\exp\{\lambda T_{k+1}-\lambda^2q_{k+1}/2\}\mid\mathcal H_k\right]=\exp\{\lambda T_k-\lambda^2q_k/2\},
\]
\par
}

{
\emergencystretch=2em
so Lemma A.7.5 applies with \(\rho=c^{-1}\). For \(\beta\in(0,1)\), define \[\mathcal M_\beta(a):=\left\{(a+c^{-1})\log\frac{1+ca}{\beta^2}\right\}^{1/2},\qquad a\ge0.
\]
\par
}

{
\emergencystretch=2em
Since \((a+c^{-1})/(\beta^2c^{-1})=(1+ca)/\beta^2\), Lemma A.7.5 gives, for every \(n\ge1\) and \(\beta\in(0,1)\), \[\Pr\left\{\exists k\ge1:|\widetilde W(k2^{-n})|\ge\mathcal M_\beta(k2^{-n})\right\}\le\beta.
\]
\par
}

{
\emergencystretch=2em
\textbf{Step 3.} We match the boundary in part (d) with \(\mathcal M_\alpha\) and pass from the grids to continuous time. The definition of \(h_\alpha\) with \(x=c\,a(u)=c/\sigma^2(u)\) gives \[\frac{h_\alpha(c/\sigma^2(u))^2}{\sigma^2(u)}=a(u)\,\frac{c\,a(u)+1}{c\,a(u)}\log\frac{c\,a(u)+1}{\alpha^2}=\mathcal M_\alpha(a(u))^2.
\]
\par
}

{
\emergencystretch=2em
Dividing the numerator and the denominator of the ratio in part (d) by \(\sigma^2(u)\) and applying Step 1 therefore give \[\Pr\left\{\sup_{u>0}\frac{|(V^{\rm obs})^{1/2}Z+(V^{{\rm exp},\star})^{1/2}W(u)/u|}{\sigma(u)\,h_\alpha(c/\sigma^2(u))}\ge1\right\}
=\Pr\left\{\sup_{0<a<1/V^{\rm obs}}\frac{|\widetilde W(a)|}{\mathcal M_\alpha(a)}\ge1\right\}.
\]
\par
}

{
\emergencystretch=2em
The boundary \(\mathcal M_\alpha\) is continuous and strictly positive on \([0,\infty)\), and \(\widetilde W(0)=0\). The ratio \(|\widetilde W(a)|/\mathcal M_\alpha(a)\) is therefore continuous on \([0,1/V^{\rm obs}]\) and equals zero at \(a=0\), so its supremum over \((0,1/V^{\rm obs})\) equals its maximum over \([0,1/V^{\rm obs}]\). If this maximum is at least one, it is attained at a positive time.
\par
}

{
\emergencystretch=2em
Fix \(\beta\in(\alpha,1)\), so that \(\mathcal M_\beta<\mathcal M_\alpha\) on \([0,\infty)\). If \(|\widetilde W(a^*)|\ge\mathcal M_\alpha(a^*)\) at some \(a^*>0\), then \(|\widetilde W|>\mathcal M_\beta\) on a neighborhood of \(a^*\) by continuity, and this neighborhood contains a positive dyadic rational. The grid events in Step 2 increase with \(n\), so \[\Pr\left\{\max_{0\le a\le1/V^{\rm obs}}\frac{|\widetilde W(a)|}{\mathcal M_\alpha(a)}\ge1\right\}
\le\lim_{n\to\infty}\Pr\left\{\exists k\ge1:|\widetilde W(k2^{-n})|\ge\mathcal M_\beta(k2^{-n})\right\}\le\beta.
\]
\par
}

{
\emergencystretch=2em
Letting \(\beta\downarrow\alpha\) proves Lemma A.7.4(d) and completes the proof of Lemma A.7.4. \(\square\)
\par
}

\subsubsection{Proof of Lemma A.7.2}\label{a.7.4-proof-of-lemma-a.7.2}

This subsection first states Lemmas A.7.6--A.7.10, which the proofs in this subsection use, and then proves Lemma A.7.2 and Lemmas A.7.6--A.7.8. Lemma A.7.6 gives the limit used in part (a) of Lemma A.7.2, and Lemma A.7.4(d), proved in Appendix A.7.3.1, gives part (b). The proof of Lemma A.7.6 uses Lemma A.7.7(b) and Lemma A.7.9, the proof of Lemma A.7.7 uses Lemma A.7.8(b), and the proof of Lemma A.7.8 uses Lemma A.7.10.

\expandafter\gdef\csname thelemma\endcsname{A.7.\arabic{lemma}}
\expandafter\gdef\csname theHlemma\endcsname{A.7.\arabic{lemma}}
\setcounter{lemma}{5}
\begin{lemma}[Joint limit of the observational error and the experimental partial sums]

{
\emergencystretch=2em
Under the conditions of Theorem 7.1, for every fixed \(K<\infty\), \[\left(\sqrt{N_{\mathrm O}}L_{N_{\mathrm O}}^{\rm obs},\left\{\frac1{\sqrt{N_{\mathrm O}}}\sum_{s=1}^{\lfloor N_{\mathrm O}u\rfloor}\widetilde m_{s,N_{\mathrm O}}:0\le u\le K\right\}\right)
\rightsquigarrow
\left((V^{\rm obs})^{1/2}Z,\{(V^{{\rm exp},\star})^{1/2}W(u):0\le u\le K\}\right)
\]
\par
}

{
\emergencystretch=2em
in \(\mathbb R\times D([0,K])\), where \(Z\) and \(W\) are as in Lemma A.7.2 and \(D([0,K])\) carries its Skorokhod \(J_1\) topology.
\par
}

\end{lemma}

\expandafter\gdef\csname thelemma\endcsname{A.7.\arabic{lemma}}
\expandafter\gdef\csname theHlemma\endcsname{A.7.\arabic{lemma}}
\setcounter{lemma}{6}
\begin{lemma}[Variances of the experimental scores and the variance estimators]

Under the conditions of Theorem 7.1 other than Assumption 7.1(2), the following hold.

\begin{enumerate}
\def\labelenumi{(\alph{enumi})}
\item
  \(\sup_{t\ge t_{0,N_{\mathrm O}}}|\mathbb E[\widetilde m_{t,N_{\mathrm O}}^{2}\mid\mathcal F_{t-1,N_{\mathrm O}}]-V^{{\rm exp},\star}|\xrightarrow{p}0\).
\item
  For every fixed \(K\ge1\), \(\sup_{t\ge N_{\mathrm O}/K}|V_{t,N_{\mathrm O}}^{\rm exp}-V^{{\rm exp},\star}|\xrightarrow{p}0\).
\item
  \(\sup_{t\ge t_{0,N_{\mathrm O}}}|\widehat V_{t,N_{\mathrm O}}^{\rm exp}-V_{t,N_{\mathrm O}}^{\rm exp}|\xrightarrow{p}0\).
\item
  For every fixed \(\varepsilon\in(0,1)\), \(\Pr\{\inf_{t\ge t_{0,N_{\mathrm O}}}\widehat V_{t,N_{\mathrm O}}^{\rm exp}<(1-\varepsilon)V^{{\rm exp},\star}\}\to0\).
\item
  \(\sup_{t\ge t_{0,N_{\mathrm O}}}|\widehat V_{t,N_{\mathrm O}}^{\rm obs}-V^{\rm obs}|\xrightarrow{p}0\).
\end{enumerate}

\end{lemma}

\expandafter\gdef\csname thelemma\endcsname{A.7.\arabic{lemma}}
\expandafter\gdef\csname theHlemma\endcsname{A.7.\arabic{lemma}}
\setcounter{lemma}{7}
\begin{lemma}[Maximal inequalities for partial sums and running averages]

{
\emergencystretch=2em
For each fixed \(N_{\mathrm O}\), let \((U_{k,N_{\mathrm O}})_{k\ge1}\) be a martingale-difference sequence with respect to a filtration \((\mathcal H_{k,N_{\mathrm O}})_{k\ge0}\) such that \(\sup_{k\ge1}\mathbb E[U_{k,N_{\mathrm O}}^2]<\infty\).
\par
}

\begin{enumerate}
\def\labelenumi{(\alph{enumi})}
\item
  For every integer \(n\ge1\) and \(\lambda>0\), \(\Pr\{\max_{1\le k\le n}|\sum_{s=1}^kU_{s,N_{\mathrm O}}|>\lambda\}\le n\allowbreak\sup_{k\ge1}\mathbb E[U_{k,N_{\mathrm O}}^2]/\allowbreak \lambda^2\).
\item
  For every integer \(m\ge1\) and \(\lambda>0\), \[\Pr\left\{\sup_{t\ge m}\left|\frac1t\sum_{k=1}^tU_{k,N_{\mathrm O}}\right|>\lambda\right\}
  \le
  \frac{4\sup_{k\ge1}\mathbb E[U_{k,N_{\mathrm O}}^2]}{\lambda^2m}.
  \]
\end{enumerate}

\end{lemma}

\expandafter\gdef\csname thelemma\endcsname{A.7.\arabic{lemma}}
\expandafter\gdef\csname theHlemma\endcsname{A.7.\arabic{lemma}}
\setcounter{lemma}{8}
\begin{lemma}[Martingale functional central limit theorem]

{
\emergencystretch=2em
We use the sufficient direction of Theorem VIII.3.33 of \citet{Jacod2003} for scalar martingale-difference arrays whose partial sums are taken up to the deterministic indices \(\lfloor N_{\mathrm O}s\rfloor\), \(s\ge0\). For each \(N_{\mathrm O}\), let \((U_{k,N_{\mathrm O}})_{k\ge1}\) be a square-integrable martingale-difference sequence with respect to a filtration \((\mathcal H_{k,N_{\mathrm O}})_{k\ge0}\), and let \(\mathcal V:[0,\infty)\to[0,\infty)\) be a continuous nondecreasing deterministic function with \(\mathcal V(0)=0\). Suppose that
\par
}

\begin{enumerate}
\def\labelenumi{(\arabic{enumi})}
\item
  \(\sum_{k=1}^{\lfloor N_{\mathrm O}s\rfloor}\mathbb E[U_{k,N_{\mathrm O}}^2\mid\mathcal H_{k-1,N_{\mathrm O}}]\xrightarrow{p}\mathcal V(s)\) for every \(s\ge0\), and
\item
  \(\sum_{k=1}^{\lfloor N_{\mathrm O}s\rfloor}\mathbb E[U_{k,N_{\mathrm O}}^2\mathbf 1\{|U_{k,N_{\mathrm O}}|>\varepsilon\}\mid\mathcal H_{k-1,N_{\mathrm O}}]\xrightarrow{p}0\) for every \(s\ge0\) and \(\varepsilon>0\).
\end{enumerate}

Then \[\left\{\sum_{k=1}^{\lfloor N_{\mathrm O}s\rfloor}U_{k,N_{\mathrm O}}:s\ge0\right\}\rightsquigarrow\{\Gamma(s):s\ge0\}
\]

{
\emergencystretch=2em
in \(D([0,\infty))\) with its Skorokhod \(J_1\) topology, where \(\Gamma\) is a continuous centered Gaussian process with independent increments and \(\mathbb E[\Gamma(s)^2]=\mathcal V(s)\) for every \(s\ge0\).
\par
}

\end{lemma}

Condition (1) is the convergence of the predictable quadratic variation, and condition (2) is the conditional Lindeberg condition. Appendix A.7.4.1 verifies both conditions for the row of observational and experimental records.

\expandafter\gdef\csname thelemma\endcsname{A.7.\arabic{lemma}}
\expandafter\gdef\csname theHlemma\endcsname{A.7.\arabic{lemma}}
\setcounter{lemma}{9}
\begin{lemma}[Martingale maximal inequality]

{
\emergencystretch=2em
We use the \(p=2\) case of Corollary 2.1 of \citet{Hall1980}, which follows their Theorem 2.1. For each fixed \(N_{\mathrm O}\), let \((U_{k,N_{\mathrm O}})_{k\ge1}\) be a square-integrable martingale-difference sequence with respect to a filtration \((\mathcal H_{k,N_{\mathrm O}})_{k\ge0}\). For every finite \(n\) and \(\lambda>0\), \[\begin{aligned}
&\lambda^2\Pr\!\left\{\max_{1\le k\le n}\left|\sum_{s=1}^kU_{s,N_{\mathrm O}}\right|>\lambda\right\}\\
&\le\mathbb E\!\left[\left|\sum_{s=1}^nU_{s,N_{\mathrm O}}\right|^2\right]=\sum_{s=1}^n\mathbb E[U_{s,N_{\mathrm O}}^2],
\end{aligned}
\]
\par
}

{
\emergencystretch=2em
where the equality uses the \(L_2\)-orthogonality of martingale differences.
\par
}

\end{lemma}

{
\emergencystretch=2em
By Lemma 6.1 for the first sequence and by construction for the second, \((\widetilde m_{s,N_{\mathrm O}})_{s\ge1}\) and \((\widetilde m_{s,N_{\mathrm O}}^2-\mathbb E[\widetilde m_{s,N_{\mathrm O}}^2\mid\mathcal F_{s-1,N_{\mathrm O}}])_{s\ge1}\) are martingale-difference sequences with respect to \((\mathcal F_{s,N_{\mathrm O}})_{s\ge0}\), and Assumption 7.2, in which \(r>4\), makes them square integrable, so they are such sequences.
\par
}

{
\emergencystretch=2em
\textbf{Proof of Lemma A.7.2.} Recall from Lemma A.7.1 that, for each \(K\ge K_0\), \[\mathrm{(I)}=\left\{\max_{N_{\mathrm O}/K\le t\le KN_{\mathrm O}}\frac{\sqrt{N_{\mathrm O}}\,|L_{N_{\mathrm O}}^{\rm obs}+t^{-1}\sum_{s=1}^t\widetilde m_{s,N_{\mathrm O}}|}{\sigma(t/N_{\mathrm O})\,h_\alpha(c_\omega/\sigma^2(t/N_{\mathrm O}))}\ge1-\eta\right\}.
\]
\par
}

{
\emergencystretch=2em
We bound the maximum in this event by a continuous functional of the pair in Lemma A.7.6 and pass to the limit. (a) Fix \(K\ge K_0\), and let \(f\) be the path \(f(v)=\allowbreak N_{\mathrm O}^{-1/2}\allowbreak\sum_{s=1}^{\lfloor N_{\mathrm O}v\rfloor}\widetilde m_{s,N_{\mathrm O}}\), \(0\le v\le K\). For an integer \(t\in[N_{\mathrm O}/K,KN_{\mathrm O}]\), the point \(t/N_{\mathrm O}\) lies in \([1/K,K]\) and satisfies \(\lfloor N_{\mathrm O}(t/N_{\mathrm O})\rfloor=t\), so \(\sqrt{N_{\mathrm O}}(L_{N_{\mathrm O}}^{\rm obs}+t^{-1}\allowbreak\sum_{s=1}^t\widetilde m_{s,N_{\mathrm O}})\) equals \(z+f(t/N_{\mathrm O})/(t/N_{\mathrm O})\) with \(z=\sqrt{N_{\mathrm O}}L_{N_{\mathrm O}}^{\rm obs}\). The maximum in event (I) is therefore at most the value at this pair of the map \[(z,f)\longmapsto\sup_{1/K\le u\le K}\frac{|z+f(u)/u|}{\sigma(u)\,h_\alpha(c_\omega/\sigma^2(u))}
\]
\par
}

{
\emergencystretch=2em
from \(\mathbb R\times D([0,K])\) to \([0,\infty)\). By Lemma A.7.4(c), the denominator is at least \(\{2V^{\rm obs}\log(1/\alpha)\}^{1/2}\), and \(1/u\le K\) on \([1/K,K]\). Between two pairs \((z,f)\) and \((z',f')\), the map therefore changes by at most \((|z-z'|+K\allowbreak\sup_{0\le u\le K}|f(u)-f'(u)|)/\allowbreak \{2V^{\rm obs}\log(1/\allowbreak \alpha)\}^{1/2}\). It is continuous at every pair whose second coordinate is continuous, because a sequence in \(D([0,K])\) that converges in the Skorokhod \(J_1\) topology to a continuous limit converges uniformly on \([0,K]\), by Proposition VI.1.17(b) of \citet{Jacod2003} applied to the paths extended to \([0,\infty)\) by their values at \(K\), which converge in the Skorokhod topology of \(D([0,\infty))\) because the time changes of \([0,K]\) fix \(K\) and extend by the identity beyond \(K\). The limit in Lemma A.7.6 has continuous paths, so Lemma A.7.6 and the continuous mapping theorem show that the value of the map at \((\sqrt{N_{\mathrm O}}L_{N_{\mathrm O}}^{\rm obs},f)\) converges in law to the supremum over \(1/K\le u\le K\) of \(|(V^{\rm obs})^{1/2}Z+(V^{{\rm exp},\star})^{1/2}W(u)/u|/\allowbreak \{\sigma(u)h_\alpha(c_\omega/\allowbreak \sigma^2(u))\}\). The set \([1-\eta,\infty)\) is closed, so the Portmanteau theorem gives \[\limsup_{N_{\mathrm O}\to\infty}\Pr\{\mathrm{(I)}\}
\le\Pr\left\{\sup_{1/K\le u\le K}\frac{|(V^{\rm obs})^{1/2}Z+(V^{{\rm exp},\star})^{1/2}W(u)/u|}{\sigma(u)\,h_\alpha(c_\omega/\sigma^2(u))}\ge1-\eta\right\},
\]
\par
}

{
\emergencystretch=2em
and enlarging the range of \(u\) to \((0,\infty)\) gives part (a).
\par
}

\begin{enumerate}
\def\labelenumi{(\alph{enumi})}
\setcounter{enumi}{1}
\tightlist
\item
  The right-hand side in part (a) does not depend on \(K\). As \(\eta\downarrow0\), its events decrease to the event on which the supremum over \(u>0\) is at least one. Continuity from above and Lemma A.7.4(d) with \(c=c_\omega\) therefore bound the limit by \(\alpha\). This proves Lemma A.7.2.
\end{enumerate}

\paragraph{Proof of Lemma A.7.6}\label{a.7.4.1-proof-of-lemma-a.7.6}

\textbf{Proof.} Place the observational records first and the experimental rounds second in the martingale row \[\begin{aligned}
&U_{k,N_{\mathrm O}}:=\frac{\psi^{\rm obs}(X_k,S_k,Y_k)-\tau}{\sqrt{N_{\mathrm O}}},\qquad
\mathcal H_{k,N_{\mathrm O}}:=\sigma\{(X_i,S_i,Y_i):1\le i\le k\},
\qquad 1\le k\le N_{\mathrm O},\\
&U_{N_{\mathrm O}+t,N_{\mathrm O}}:=\frac{\widetilde m_{t,N_{\mathrm O}}}{\sqrt{N_{\mathrm O}}},\qquad
\mathcal H_{N_{\mathrm O}+t,N_{\mathrm O}}:=\mathcal F_{t,N_{\mathrm O}},\qquad t\ge1.
\end{aligned}
\]

{
\emergencystretch=2em
Take \(\mathcal H_{0,N_{\mathrm O}}\) trivial. At the join, \(\mathcal H_{N_{\mathrm O},N_{\mathrm O}}=\allowbreak \sigma(\mathcal D_{N_{\mathrm O}}^{\rm obs})=\allowbreak \mathcal F_{0,N_{\mathrm O}}\). Assumption A.6.1(i) makes the observational increments martingale differences with conditional variance \(V^{\rm obs}/N_{\mathrm O}\). Lemma 6.1 and Assumption 7.2 make the experimental increments square-integrable martingale differences for the same filtration.
\par
}

{
\emergencystretch=2em
For the experimental predictable variance, use Lemma A.7.7(b). Its proof in Appendix A.7.4.2.2 uses part (a), the moment bound in Assumption 7.2, and \(t_{0,N_{\mathrm O}}/N_{\mathrm O}\to0\) from Assumption 7.1(1). That argument establishes part (b) without invoking Lemma A.7.6. For fixed \(u>0\), eventually \(\lfloor N_{\mathrm O} u\rfloor\ge N_{\mathrm O} u/\allowbreak 2\ge N_{\mathrm O}/\allowbreak \max\{1,2/\allowbreak u\}\). Thus part (b) applies with the fixed constant \(\max\{1,2/u\}\), and \[\begin{aligned}
&\sum_{t=1}^{\lfloor N_{\mathrm O} u\rfloor}\mathbb E[U_{N_{\mathrm O}+t,N_{\mathrm O}}^2\mid \mathcal F_{t-1,N_{\mathrm O}}]\\
&=\frac{\lfloor N_{\mathrm O} u\rfloor}{N_{\mathrm O}}\,V_{\lfloor N_{\mathrm O} u\rfloor,N_{\mathrm O}}^{\rm exp}
\xrightarrow{p}uV^{{\rm exp},\star}.
\end{aligned}
\]
\par
}

{
\emergencystretch=2em
At \(u=0\) the sum is zero. Through row time \(s\ge0\), the observational contribution is \(V^{\rm obs}\min\{\lfloor N_{\mathrm O} s\rfloor,N_{\mathrm O}\}/\allowbreak N_{\mathrm O}\). Taking \(u=(s-1)^+\) gives the full predictable variance limit \[\sum_{k=1}^{\lfloor N_{\mathrm O} s\rfloor}\mathbb E[U_{k,N_{\mathrm O}}^2\mid\mathcal H_{k-1,N_{\mathrm O}}]
\xrightarrow{p}\mathcal V(s):=V^{\rm obs}\min(s,1)+V^{{\rm exp},\star}(s-1)^+.
\]
\par
}

{
\emergencystretch=2em
For \(\varepsilon>0\), the observational contribution to the conditional Lindeberg sum is bounded by the tail second moment of the true score. By independence and Assumption A.6.1(i), \[\begin{aligned}
&\sum_{k=1}^{\min\{\lfloor N_{\mathrm O} s\rfloor,N_{\mathrm O}\}}
\mathbb E[U_{k,N_{\mathrm O}}^2\mathbf 1\{|U_{k,N_{\mathrm O}}|>\varepsilon\}\mid\mathcal H_{k-1,N_{\mathrm O}}]\\
&\le P^{\rm obs}\!\left[(\psi^{\rm obs}-\tau)^2\mathbf 1\{|\psi^{\rm obs}-\tau|>\varepsilon\sqrt{N_{\mathrm O}}\}\right]\longrightarrow0
\end{aligned}
\]
\par
}

{
\emergencystretch=2em
by dominated convergence. For the experimental increments, \(z^2\mathbf 1\{|z|>\varepsilon\}\le\varepsilon^{2-r}|z|^r\). With \(u=(s-1)^+\), Assumption 7.2 gives \[\begin{aligned}
&\sum_{t=1}^{\lfloor N_{\mathrm O} u\rfloor}\mathbb E[U_{N_{\mathrm O}+t,N_{\mathrm O}}^2
\mathbf 1\{|U_{N_{\mathrm O}+t,N_{\mathrm O}}|>\varepsilon\}\mid \mathcal F_{t-1,N_{\mathrm O}}]\\
&\le\frac{\varepsilon^{2-r}}{N_{\mathrm O}^{r/2}}\sum_{t=1}^{\lfloor N_{\mathrm O} u\rfloor}
\mathbb E[|\widetilde m_{t,N_{\mathrm O}}|^r\mid \mathcal F_{t-1,N_{\mathrm O}}]
\le C u\,\varepsilon^{2-r}N_{\mathrm O}^{1-r/2}\longrightarrow0.
\end{aligned}
\]
\par
}

{
\emergencystretch=2em
Lemma A.7.9 now gives convergence of the row partial sums to the continuous centered Gaussian process \(\Gamma\) with independent increments and variance function \(\mathcal V\). For every fixed \(K<\infty\), the map \[\gamma\longmapsto\bigl(\gamma(1),\{\gamma(1+u)-\gamma(1):0\le u\le K\}\bigr)
\]
\par
}

{
\emergencystretch=2em
is continuous at continuous paths for the \(J_1\) topology, since convergence there is locally uniform by Proposition VI.1.17(b) of \citet{Jacod2003}. The two coordinates of the row under this map are \[\begin{aligned}
&\sum_{k=1}^{N_{\mathrm O}}U_{k,N_{\mathrm O}}=\sqrt{N_{\mathrm O}}\,L_{N_{\mathrm O}}^{\rm obs},\\
&\sum_{k=N_{\mathrm O}+1}^{\lfloor N_{\mathrm O}(1+u)\rfloor}U_{k,N_{\mathrm O}}
=\frac1{\sqrt{N_{\mathrm O}}}\sum_{t=1}^{\lfloor N_{\mathrm O} u\rfloor}\widetilde m_{t,N_{\mathrm O}}.
\end{aligned}
\]
\par
}

{
\emergencystretch=2em
By the continuous mapping theorem their joint limit is given by the same map applied to \(\Gamma\). Its first coordinate has law \(\mathcal N(0,V^{\rm obs})\) and is independent of the second by independent increments. For \(0\le u'\le u\le K\), \[\operatorname{Var}\{\Gamma(1+u)-\Gamma(1+u')\}
=\mathcal V(1+u)-\mathcal V(1+u')=V^{{\rm exp},\star}(u-u').
\]
\par
}

{
\emergencystretch=2em
The second coordinate starts at zero and has continuous paths and independent centered Gaussian increments, so it has the law of \((V^{{\rm exp},\star})^{1/2}W\) on \([0,K]\). This proves Lemma A.7.6.
\par
}

\paragraph{Proof of Lemma A.7.7}\label{a.7.4.2-proof-of-lemma-a.7.7}

{
\emergencystretch=2em
This subsection proves Lemma A.7.7. Appendix A.7.4.2.1 proves part (a) and the limit of \(A_{N_{\mathrm O}}^\circ\) defined in Appendix A.6.4, Appendix A.7.4.2.2 proves part (b), Appendix A.7.4.2.3 proves part (c), Appendix A.7.4.2.4 proves part (d), and Appendix A.7.4.2.5 proves part (e).
\par
}

\subparagraph{Proof of Lemma A.7.7(a)}\label{a.7.4.2.1-proof-of-lemma-a.7.7a}

{
\emergencystretch=2em
\textbf{Proof.} On the event in Assumption A.6.1(x), \(\widehat v_{w,t,N_{\mathrm O}}(x)\) and \(\widetilde v_{w,N_{\mathrm O}}(x)\) lie in \([\underline v,\overline v]\) simultaneously for both arms, all target-support \(x\), and every \(t\ge t_{0,N_{\mathrm O}}\). This event has probability tending to one. Among the conditions of Assumption A.6.1, the one-round bound in Appendix A.6.4.1 and the bound in Appendix A.6.4.2 use only the common conditions and variance bounds of this form, so they apply on this event. Since \(\widetilde v_{w,N_{\mathrm O}}\) is fixed across rounds, the bound in Appendix A.6.4.2 and the two initial risks in Assumption 7.3 give \[A_{N_{\mathrm O}}^\circ\xrightarrow{p}P_{f_{X\mid G=\mathrm O}}\!\left[\sum_{w=0}^1\sqrt{v_w(X)}\right]=\sqrt{V^{{\rm exp},\star}}.
\]
\par
}

On the same event, apply the one-round bound in Appendix A.6.4.1 at each monitored round. Then \[\begin{aligned}
&\sup_{t\ge t_{0,N_{\mathrm O}}}\left|\mathbb E[\widetilde m_{t,N_{\mathrm O}}^2\mid \mathcal F_{t-1,N_{\mathrm O}}]-V^{{\rm exp},\star}\right|\\
&\le \bigl(1+\sqrt{\overline v/\underline v}\bigr)^2\sum_{w=0}^1\left\{\sup_{t\ge t_{0,N_{\mathrm O}}}\|\widehat v_{w,t,N_{\mathrm O}}-\widetilde v_{w,N_{\mathrm O}}\|_\infty
+\left(\sup_{t\ge t_{0,N_{\mathrm O}}}\|\widehat\mu_{w,t,N_{\mathrm O}}-\widetilde\mu_{w,N_{\mathrm O}}\|_\infty\right)^2\right\}
+\left|(A_{N_{\mathrm O}}^\circ)^2-V^{{\rm exp},\star}\right|.
\end{aligned}
\]

The two suprema for each arm converge to zero in probability by Assumption 7.3. The last term converges to zero by the preceding limit. The probability of the exceptional event tends to zero, which proves part (a).

\subparagraph{Proof of Lemma A.7.7(b)}\label{a.7.4.2.2-proof-of-lemma-a.7.7b}

{
\emergencystretch=2em
This step proves part (b) of Lemma A.7.7. Fix \(K\ge1\). Because \(t_{0,N_{\mathrm O}}/N_{\mathrm O}\to0\), every \(t\ge N_{\mathrm O}/K\) satisfies \(t\ge t_{0,N_{\mathrm O}}\) for all sufficiently large \(N_{\mathrm O}\). Split the average \(V_{t,N_{\mathrm O}}^{\rm exp}=\allowbreak t^{-1}\allowbreak\sum_{s=1}^t\mathbb E[\widetilde m_{s,N_{\mathrm O}}^2\mid\mathcal F_{s-1,N_{\mathrm O}}]\) at \(t_{0,N_{\mathrm O}}\). The first \(t_{0,N_{\mathrm O}}\) summands deviate from \(V^{{\rm exp},\star}\) by at most \(C^{2/r}+V^{{\rm exp},\star}\) each, and the remaining summands deviate by at most the supremum in part (a). Hence, uniformly in \(t\ge N_{\mathrm O}/K\), \[\left|V_{t,N_{\mathrm O}}^{\rm exp}-V^{{\rm exp},\star}\right|\le\frac{Kt_{0,N_{\mathrm O}}}{N_{\mathrm O}}\left(C^{2/r}+V^{{\rm exp},\star}\right)+\sup_{s\ge t_{0,N_{\mathrm O}}}\left|\mathbb E[\widetilde m_{s,N_{\mathrm O}}^2\mid\mathcal F_{s-1,N_{\mathrm O}}]-V^{{\rm exp},\star}\right|=o_p(1),
\]
\par
}

which proves part (b).

\subparagraph{Proof of Lemma A.7.7(c)}\label{a.7.4.2.3-proof-of-lemma-a.7.7c}

\textbf{Proof.} Sample centering removes the common observational term in the feasible experimental scores. Indeed, Section 6.1 gives \[\widehat\phi_{s,N_{\mathrm O}}^{\rm exp}(S_s,W_s,X_s)+P_{\widehat f_{X\mid G=\mathrm O,N_{\mathrm O}}}\widehat\tau_{s,N_{\mathrm O}}(\cdot)
=\widetilde m_{s,N_{\mathrm O}}+P_{\widehat f_{X\mid G=\mathrm O,N_{\mathrm O}}}\widetilde\tau_{N_{\mathrm O}}.
\]

Expanding the sample variance and subtracting its predictable counterpart therefore give \[\begin{aligned}
&\widehat V_{t,N_{\mathrm O}}^{\rm exp}-V_{t,N_{\mathrm O}}^{\rm exp}\\
&=\frac1t\sum_{s=1}^t \widetilde m_{s,N_{\mathrm O}}^2-\left(\frac1t\sum_{s=1}^t \widetilde m_{s,N_{\mathrm O}}\right)^2-V_{t,N_{\mathrm O}}^{\rm exp}\\
&=\underbrace{\frac1t\sum_{s=1}^t\left\{\widetilde m_{s,N_{\mathrm O}}^2-\mathbb E[\widetilde m_{s,N_{\mathrm O}}^2\mid \mathcal F_{s-1,N_{\mathrm O}}]\right\}}_{\text{(i)}}
-\underbrace{\left(\frac1t\sum_{s=1}^t \widetilde m_{s,N_{\mathrm O}}\right)^2}_{\text{(ii)}}.
\end{aligned}
\]

Apply Lemma A.7.8(b) to the centered squares in (i) and the scores before squaring in (ii). The former are martingale differences by construction and the latter by Lemma 6.1. Their conditional second moments satisfy \[\begin{aligned}
&\mathbb E\!\left[\left\{\widetilde m_{s,N_{\mathrm O}}^2-\mathbb E[\widetilde m_{s,N_{\mathrm O}}^2\mid \mathcal F_{s-1,N_{\mathrm O}}]\right\}^2\mid \mathcal F_{s-1,N_{\mathrm O}}\right]
\le\mathbb E[\widetilde m_{s,N_{\mathrm O}}^4\mid \mathcal F_{s-1,N_{\mathrm O}}]\le C^{4/r},\\
&\mathbb E[\widetilde m_{s,N_{\mathrm O}}^2\mid \mathcal F_{s-1,N_{\mathrm O}}]\le C^{2/r}
\end{aligned}
\]

{
\emergencystretch=2em
by Assumption 7.2 and conditional Jensen. For fixed \(\varepsilon>0\), use thresholds \(\varepsilon/2\) for the centered-square average and \(\sqrt{\varepsilon/2}\) for the score average. The triangle inequality, the union bound, and Lemma A.7.8(b) with \(m=t_{0,N_{\mathrm O}}\) yield \[\Pr\!\left\{\sup_{t\ge t_{0,N_{\mathrm O}}}|\widehat V_{t,N_{\mathrm O}}^{\rm exp}-V_{t,N_{\mathrm O}}^{\rm exp}|>\varepsilon\right\}
\le\frac{4C^{4/r}}{t_{0,N_{\mathrm O}}(\varepsilon/2)^2}
+\frac{4C^{2/r}}{t_{0,N_{\mathrm O}}(\varepsilon/2)}
\longrightarrow0,
\]
\par
}

{
\emergencystretch=2em
since \(t_{0,N_{\mathrm O}}\to\infty\) by Assumption 7.1(1). This proves part (c).
\par
}

\subparagraph{Proof of Lemma A.7.7(d)}\label{a.7.4.2.4-proof-of-lemma-a.7.7d}

{
\emergencystretch=2em
\textbf{Proof.} The variation of the surrogate index within each arm gives a lower bound in every round \(t\ge1\). Given \(X_t=x\), \(W_t=w\), and the past, subtracting the centering term in Section 6.1 has no effect on the variance. The score formula in Appendix A.6.4.1 and the conditional surrogate law in Assumption A.6.1(iv) give \[\operatorname{Var}(\widetilde m_{t,N_{\mathrm O}}\mid X_t=x,W_t=w,\mathcal F_{t-1,N_{\mathrm O}})
=\frac{A_{t,N_{\mathrm O}}^2}{\widehat v_{w,t,N_{\mathrm O}}(x)}\,\widetilde v_{w,N_{\mathrm O}}(x).
\]
\par
}

{
\emergencystretch=2em
Lemma 6.1 centers the score given the past. The law of total variance and the joint density of \((X_t,W_t)\) in Appendix A.6.4.1 therefore give \[\begin{aligned}
&\mathbb E[\widetilde m_{t,N_{\mathrm O}}^2\mid \mathcal F_{t-1,N_{\mathrm O}}]
=\operatorname{Var}(\widetilde m_{t,N_{\mathrm O}}\mid \mathcal F_{t-1,N_{\mathrm O}})\\
&\ge\mathbb E[\operatorname{Var}(\widetilde m_{t,N_{\mathrm O}}\mid X_t,W_t,\mathcal F_{t-1,N_{\mathrm O}})\mid \mathcal F_{t-1,N_{\mathrm O}}]\\
&=\int\sum_{w=0}^1\frac{\widehat f_{X\mid G=\mathrm O,N_{\mathrm O}}(x)\sqrt{\widehat v_{w,t,N_{\mathrm O}}(x)}}{A_{t,N_{\mathrm O}}}\,
\frac{A_{t,N_{\mathrm O}}^2 \widetilde v_{w,N_{\mathrm O}}(x)}{\widehat v_{w,t,N_{\mathrm O}}(x)}\,d\lambda_X(x)
=A_{t,N_{\mathrm O}}\,P_{\widehat f_{X\mid G=\mathrm O,N_{\mathrm O}}}\!\left[\sum_{w=0}^1\frac{\widetilde v_{w,N_{\mathrm O}}(X)}{\sqrt{\widehat v_{w,t,N_{\mathrm O}}(X)}}\right].
\end{aligned}
\]
\par
}

{
\emergencystretch=2em
These identities use the common support and pointwise positivity in Assumption A.6.1(v) and (vi), which apply also before \(t_{0,N_{\mathrm O}}\).
\par
}

{
\emergencystretch=2em
Apply Cauchy--Schwarz over \(x\) and the two arms, with the fitted target density as the weight in \(x\). By the definitions in Appendix A.6.4.1, \[\begin{aligned}
&(A_{N_{\mathrm O}}^\circ)^2
=\left\{P_{\widehat f_{X\mid G=\mathrm O,N_{\mathrm O}}}\!\left[\sum_{w=0}^1
\widehat v_{w,t,N_{\mathrm O}}(X)^{1/4}\frac{\sqrt{\widetilde v_{w,N_{\mathrm O}}(X)}}{\widehat v_{w,t,N_{\mathrm O}}(X)^{1/4}}\right]\right\}^2\\
&\le P_{\widehat f_{X\mid G=\mathrm O,N_{\mathrm O}}}\!\left[\sum_{w=0}^1\sqrt{\widehat v_{w,t,N_{\mathrm O}}(X)}\right]\,
P_{\widehat f_{X\mid G=\mathrm O,N_{\mathrm O}}}\!\left[\sum_{w=0}^1\frac{\widetilde v_{w,N_{\mathrm O}}(X)}{\sqrt{\widehat v_{w,t,N_{\mathrm O}}(X)}}\right]
\le\mathbb E[\widetilde m_{t,N_{\mathrm O}}^2\mid \mathcal F_{t-1,N_{\mathrm O}}].
\end{aligned}
\]
\par
}

{
\emergencystretch=2em
The lower bound is common to all rounds. Averaging and using the definition of \(V_{t,N_{\mathrm O}}^{\rm exp}\) give \[\inf_{t\ge1}V_{t,N_{\mathrm O}}^{\rm exp}\ge(A_{N_{\mathrm O}}^\circ)^2.
\]
\par
}

{
\emergencystretch=2em
For fixed \(\varepsilon\in(0,1)\), split the tolerance between this bound and the error in the feasible variance. Then \[\begin{aligned}
&\Pr\!\left\{\inf_{t\ge t_{0,N_{\mathrm O}}}\widehat V_{t,N_{\mathrm O}}^{\rm exp}<(1-\varepsilon)V^{{\rm exp},\star}\right\}\\
&\le\Pr\!\left\{\sup_{t\ge t_{0,N_{\mathrm O}}}|\widehat V_{t,N_{\mathrm O}}^{\rm exp}-V_{t,N_{\mathrm O}}^{\rm exp}|>\varepsilon V^{{\rm exp},\star}/2\right\}
+\Pr\!\left\{(A_{N_{\mathrm O}}^\circ)^2<(1-\varepsilon/2)V^{{\rm exp},\star}\right\}
\longrightarrow0
\end{aligned}
\]
\par
}

{
\emergencystretch=2em
by part (c), the limit of \(A_{N_{\mathrm O}}^\circ\) in Appendix A.7.4.2.1, and \(V^{{\rm exp},\star}>0\). This proves part (d).
\par
}

\subparagraph{Proof of Lemma A.7.7(e)}\label{a.7.4.2.5-proof-of-lemma-a.7.7e}

{
\emergencystretch=2em
This step proves part (e) of Lemma A.7.7. We compare the empirical variance of the fitted observational scores with that of the true scores. Write the fitted observational score as the true score plus the difference \(\widehat\psi_{t,N_{\mathrm O}}^{\rm obs}-\psi^{\rm obs}\). When the centered square in the definition of \(\widehat V_{t,N_{\mathrm O}}^{\rm obs}\) is expanded, the estimator equals the empirical variance of \(\psi^{\rm obs}(X_i,S_i,Y_i)\), plus twice the empirical covariance of \(\psi^{\rm obs}(X_i,S_i,Y_i)\) and \((\widehat\psi_{t,N_{\mathrm O}}^{\rm obs}-\psi^{\rm obs})(X_i,S_i,Y_i)\), plus the empirical variance of the latter. The first term does not depend on \(t\) and converges in probability to \(V^{\rm obs}\) by the law of large numbers, because \(P^{\rm obs}\psi^{\rm obs}=\tau\) and \(\mathbb E^{\rm obs}[(\psi^{\rm obs}-\tau)^2]=V^{\rm obs}<\infty\) by Assumption A.6.1(i). The last term is at most \(\mathbb P_{N_{\mathrm O}}^{\rm obs}[\{\widehat\psi_{t,N_{\mathrm O}}^{\rm obs}-\psi^{\rm obs}\}^2]\), and by Cauchy--Schwarz the absolute value of the middle term is at most twice the square root of the first term times this bound. The third condition of Assumption 7.4 and the uniform \(L_2(P^{\rm obs})\) consistency in its first condition imply \[\begin{aligned}
&\sup_{t\ge t_{0,N_{\mathrm O}}}\mathbb P_{N_{\mathrm O}}^{\rm obs}[\{\widehat\psi_{t,N_{\mathrm O}}^{\rm obs}-\psi^{\rm obs}\}^2]\\
&\quad\le\sup_{t\ge t_{0,N_{\mathrm O}}}P^{\rm obs}[\{\widehat\psi_{t,N_{\mathrm O}}^{\rm obs}-\psi^{\rm obs}\}^2]+o_p(1)=o_p(1).
\end{aligned}
\]
\par
}

{
\emergencystretch=2em
Hence the middle and last terms are \(o_p(1)\) uniformly in \(t\ge t_{0,N_{\mathrm O}}\), and \(\sup_{t\ge t_{0,N_{\mathrm O}}}|\widehat V_{t,N_{\mathrm O}}^{\rm obs}-V^{\rm obs}|\xrightarrow{p}0\), which proves part (e) and completes the proof of Lemma A.7.7.
\par
}

\paragraph{Proof of Lemma A.7.8}\label{a.7.4.3-proof-of-lemma-a.7.8}

{
\emergencystretch=2em
\textbf{Proof.} (a) Lemma A.7.10 and \(\sum_{s=1}^n\mathbb E[U_{s,N_{\mathrm O}}^2]\le n\allowbreak\sup_{k\ge1}\mathbb E[U_{k,N_{\mathrm O}}^2]\) give part (a).
\par
}

\begin{enumerate}
\def\labelenumi{(\alph{enumi})}
\setcounter{enumi}{1}
\tightlist
\item
  Fix \(m\) and \(\lambda\). If \(2^jm\le t<2^{j+1}m\) for some \(j\ge0\) and \(|t^{-1}\sum_{k=1}^tU_{k,N_{\mathrm O}}|>\lambda\), then \(\max_{1\le k\le2^{j+1}m}|\sum_{s=1}^kU_{s,N_{\mathrm O}}|>\lambda2^jm\). Lemma A.7.10 with \(n=2^{j+1}m\) and the union bound over \(j\) give \[\begin{aligned}
  \Pr\!\left\{\sup_{t\ge m}\left|\frac1t\sum_{k=1}^tU_{k,N_{\mathrm O}}\right|>\lambda\right\}
  &\le\sum_{j=0}^{\infty}\frac{\sum_{s=1}^{2^{j+1}m}\mathbb E[U_{s,N_{\mathrm O}}^2]}{\lambda^22^{2j}m^2}\\
  &\le\sum_{j=0}^{\infty}\frac{2^{j+1}m\sup_{k\ge1}\mathbb E[U_{k,N_{\mathrm O}}^2]}{\lambda^22^{2j}m^2}=\frac{4\sup_{k\ge1}\mathbb E[U_{k,N_{\mathrm O}}^2]}{\lambda^2m}.
  \end{aligned}
  \]
\end{enumerate}

This proves Lemma A.7.8.

\subsubsection{Proof of Lemma A.7.3}\label{a.7.5-proof-of-lemma-a.7.3}

This subsection proves Lemma A.7.3, which bounds the probabilities of events (II)--(V).

{
\emergencystretch=2em
\textbf{Proof.} (a) Since \(R_{t,N_{\mathrm O}}^{\rm i}=\allowbreak (\mathbb P_{N_{\mathrm O}}^{\rm obs}-P^{\rm obs})\{\widehat\psi_{t,N_{\mathrm O}}^{\rm obs}(\cdot)-\psi^{\rm obs}(\cdot)\}\), the second condition of Assumption 7.4 is the first remainder bound \(\sqrt{N_{\mathrm O}}\allowbreak\sup_{t\ge t_{0,N_{\mathrm O}}}|R_{t,N_{\mathrm O}}^{\rm i}|=\allowbreak o_p(1)\). For the target-density remainder, Cauchy--Schwarz and the first product rate in Assumption 7.5 give \[\begin{aligned}
\sqrt{N_{\mathrm O}}\sup_{t\ge t_{0,N_{\mathrm O}}}|R_{t,N_{\mathrm O}}^{\rm ii}|
&\le\sqrt{N_{\mathrm O}}\left\|\frac{\widehat f_{X\mid G=\mathrm O,N_{\mathrm O}}}{f_{X\mid G=\mathrm O}}-1\right\|_{2,X\mid G=\mathrm O}\\
&\quad\times\sup_{t\ge t_{0,N_{\mathrm O}}}\|\widetilde\tau_{N_{\mathrm O}}(\cdot)-\overline\tau_{t,N_{\mathrm O}}(\cdot)\|_{2,X\mid G=\mathrm O}=o_p(1).
\end{aligned}
\]
\par
}

Similarly, Cauchy--Schwarz and the second product rate in Assumption 7.5 give \[\sqrt{N_{\mathrm O}}\sup_{t\ge t_{0,N_{\mathrm O}}}|R_{t,N_{\mathrm O}}^{\rm iii}|\le\sqrt{N_{\mathrm O}}\|\widehat\mu_{N_{\mathrm O}}-\mu\|_{2,{\rm obs}}\sup_{t\ge t_{0,N_{\mathrm O}}}\|\widehat\alpha_{t,N_{\mathrm O}}-\alpha\|_{2,{\rm obs}}=o_p(1).
\]

Summing the three bounds proves part (a).

\begin{enumerate}
\def\labelenumi{(\alph{enumi})}
\setcounter{enumi}{1}
\tightlist
\item
  We show that \(\tau\notin C_{t,N_{\mathrm O}}(\alpha)\) at a round \(t<N_{\mathrm O}/K\) forces a large partial sum of the experimental scores, and then bound its probability by a maximal inequality. Fix \(K\ge1\). Lemma A.7.4(a) with \(x=\omega_{N_{\mathrm O}}/\widehat\sigma_{t,N_{\mathrm O}}^2\) and the inequality \(\widehat\sigma_{t,N_{\mathrm O}}^2\ge\widehat V_{t,N_{\mathrm O}}^{\rm exp}/\allowbreak t\) give \[\sqrt{N_{\mathrm O}}\,\overline B_{t,N_{\mathrm O}}
  \ge\sqrt{N_{\mathrm O}}\,\widehat\sigma_{t,N_{\mathrm O}}^2\left\{\frac{2\log(1/\alpha)}{\omega_{N_{\mathrm O}}}\right\}^{1/2}
  \ge\frac{N_{\mathrm O}}t\,\widehat V_{t,N_{\mathrm O}}^{\rm exp}\left\{\frac{2\log(1/\alpha)}{N_{\mathrm O}\omega_{N_{\mathrm O}}}\right\}^{1/2}.
  \]
\end{enumerate}

{
\emergencystretch=2em
Put \(c_1:=(V^{{\rm exp},\star}/2)\{\log(1/\alpha)/c_\omega\}^{1/2}\), and consider the conditions \[\begin{aligned}
&\text{(C1)}\quad\inf_{t\ge t_{0,N_{\mathrm O}}}\widehat V_{t,N_{\mathrm O}}^{\rm exp}\ge\frac{V^{{\rm exp},\star}}2, \\
&\text{(C2)}\quad N_{\mathrm O}\omega_{N_{\mathrm O}}\le2c_\omega,\\
&\text{(C3)}\quad|\sqrt{N_{\mathrm O}}L_{N_{\mathrm O}}^{\rm obs}|+\sqrt{N_{\mathrm O}}\sup_{t\ge t_{0,N_{\mathrm O}}}\left(|R_{t,N_{\mathrm O}}^{\rm i}|+|R_{t,N_{\mathrm O}}^{\rm ii}|+|R_{t,N_{\mathrm O}}^{\rm iii}|\right)\le\frac{c_1K}2.
\end{aligned}
\]
\par
}

{
\emergencystretch=2em
If (C1) and (C2) hold, they and the lower bound on \(\sqrt{N_{\mathrm O}}\,\overline B_{t,N_{\mathrm O}}\) displayed above give \(\sqrt{N_{\mathrm O}}\,\overline B_{t,N_{\mathrm O}}\ge c_1N_{\mathrm O}/\allowbreak t\) for every \(t\ge t_{0,N_{\mathrm O}}\). The decomposition in Appendix A.7.1 gives \[\sqrt{N_{\mathrm O}}|\widehat\tau_{t,N_{\mathrm O}}-\tau|
\le\frac{N_{\mathrm O}}t\left|\frac1{\sqrt{N_{\mathrm O}}}\sum_{s=1}^t\widetilde m_{s,N_{\mathrm O}}\right|+|\sqrt{N_{\mathrm O}}L_{N_{\mathrm O}}^{\rm obs}|+\sqrt{N_{\mathrm O}}\left(|R_{t,N_{\mathrm O}}^{\rm i}|+|R_{t,N_{\mathrm O}}^{\rm ii}|+|R_{t,N_{\mathrm O}}^{\rm iii}|\right).
\]
\par
}

{
\emergencystretch=2em
Hence, if (C1)--(C3) hold and \(\tau\notin C_{t,N_{\mathrm O}}(\alpha)\) for some integer \(t\in[t_{0,N_{\mathrm O}},N_{\mathrm O}/K)\), then \(N_{\mathrm O}/t>K\) and (C3) give \[\frac{N_{\mathrm O}}t\left|\frac1{\sqrt{N_{\mathrm O}}}\sum_{s=1}^t\widetilde m_{s,N_{\mathrm O}}\right|
>\frac{c_1N_{\mathrm O}}t-\frac{c_1K}2
>\frac{c_1N_{\mathrm O}}{2t},
\]
\par
}

{
\emergencystretch=2em
and therefore \(|\sum_{s=1}^t\widetilde m_{s,N_{\mathrm O}}|>c_1N_{\mathrm O}^{1/2}/2\). Lemma A.7.8(a) with \(U_{s,N_{\mathrm O}}=\widetilde m_{s,N_{\mathrm O}}\), \(n=\lfloor N_{\mathrm O}/K\rfloor\), \(\lambda=c_1N_{\mathrm O}^{1/2}/2\), and \(\mathbb E[\widetilde m_{s,N_{\mathrm O}}^2]\le C^{2/r}\) gives \[\begin{aligned}
&\Pr\!\left[\{\exists t\in[t_{0,N_{\mathrm O}},N_{\mathrm O}/K):\tau\notin C_{t,N_{\mathrm O}}(\alpha)\}\cap\{\text{(C1)–(C3) hold}\}\right]\\
&\quad\le\Pr\!\left\{\max_{1\le k\le\lfloor N_{\mathrm O}/K\rfloor}\left|\sum_{s=1}^k\widetilde m_{s,N_{\mathrm O}}\right|>\frac{c_1N_{\mathrm O}^{1/2}}2\right\}\le\frac{4C^{2/r}}{c_1^2K}.
\end{aligned}
\]
\par
}

{
\emergencystretch=2em
For the complement, Lemma A.7.7(d) with \(\varepsilon=1/2\) and Assumption 7.1(2) show that (C1) and (C2) hold with probability tending to one. Lemma A.7.6 gives \(\sqrt{N_{\mathrm O}}L_{N_{\mathrm O}}^{\rm obs}\rightsquigarrow (V^{\rm obs})^{1/2}Z\), and part (a) shows that the uniform remainder in (C3) is \(o_p(1)\), so Slutsky\textquotesingle s lemma shows that the left-hand side of (C3) converges in law to \((V^{\rm obs})^{1/2}|Z|\). The Portmanteau theorem therefore bounds the \(\limsup\) of the probability that (C3) fails by \(\Pr\{(V^{\rm obs})^{1/2}|Z|\ge c_1K/2\}\). Combining the two bounds gives \[\limsup_{N_{\mathrm O}\to\infty}\Pr\!\left\{\exists t\in[t_{0,N_{\mathrm O}},N_{\mathrm O}/K):\tau\notin C_{t,N_{\mathrm O}}(\alpha)\right\}\le\frac{4C^{2/r}}{c_1^2K}+\Pr\!\left\{|Z|\ge\frac{c_1K}{2(V^{\rm obs})^{1/2}}\right\},
\]
\par
}

{
\emergencystretch=2em
which tends to zero as \(K\to\infty\). This proves part (b).
\par
}

{
\emergencystretch=2em
Part (c) applies Lemma A.7.8(b) after round \(\lfloor KN_{\mathrm O}\rfloor\), and part (d) combines the variance limits with the continuity of \(h_\alpha\). (c) Fix \(\varepsilon>0\) and \(K\ge1\). Lemma A.7.8(b) with \(U_{s,N_{\mathrm O}}=\widetilde m_{s,N_{\mathrm O}}\), \(m=\lfloor KN_{\mathrm O}\rfloor\), \(\lambda=\varepsilon/\sqrt{N_{\mathrm O}}\), and \(\mathbb E[\widetilde m_{s,N_{\mathrm O}}^2]\le C^{2/r}\) gives \[\Pr\left\{\sqrt{N_{\mathrm O}}\sup_{t\ge\lfloor KN_{\mathrm O}\rfloor}\left|\frac1t\sum_{s=1}^t\widetilde m_{s,N_{\mathrm O}}\right|>\varepsilon\right\}
\le\frac{4C^{2/r}N_{\mathrm O}}{\varepsilon^2\lfloor KN_{\mathrm O}\rfloor},
\]
\par
}

{
\emergencystretch=2em
which converges to \(4C^{2/r}/(\varepsilon^2K)\) as \(N_{\mathrm O}\to\infty\), because \(N_{\mathrm O}/\lfloor KN_{\mathrm O}\rfloor\to K^{-1}\). This limit tends to zero as \(K\to\infty\).
\par
}

\begin{enumerate}
\def\labelenumi{(\alph{enumi})}
\setcounter{enumi}{3}
\tightlist
\item
  Fix \(K\ge1\) and \(\varepsilon>0\). We first show that \(N_{\mathrm O}\widehat\sigma_{t,N_{\mathrm O}}^2\) is uniformly close to \(\sigma^2(t/N_{\mathrm O})\) after round \(N_{\mathrm O}/K\). By the definitions, \[N_{\mathrm O}\widehat\sigma_{t,N_{\mathrm O}}^2-\sigma^2(t/N_{\mathrm O})
  =\frac{N_{\mathrm O}}t\left(\widehat V_{t,N_{\mathrm O}}^{\rm exp}-V^{{\rm exp},\star}\right)+\left(\widehat V_{t,N_{\mathrm O}}^{\rm obs}-V^{\rm obs}\right),
  \]
\end{enumerate}

{
\emergencystretch=2em
and \(N_{\mathrm O}/t\le K\) for \(t\ge N_{\mathrm O}/K\). For all sufficiently large \(N_{\mathrm O}\), \(N_{\mathrm O}/K\ge t_{0,N_{\mathrm O}}\) because \(t_{0,N_{\mathrm O}}/N_{\mathrm O}\to0\), so Lemma A.7.7(c) and Lemma A.7.7(b) give \(\sup_{t\ge N_{\mathrm O}/K}|\widehat V_{t,N_{\mathrm O}}^{\rm exp}-V^{{\rm exp},\star}|\xrightarrow{p}0\), and Lemma A.7.7(e) gives \(\sup_{t\ge t_{0,N_{\mathrm O}}}|\widehat V_{t,N_{\mathrm O}}^{\rm obs}-V^{\rm obs}|\xrightarrow{p}0\). Hence \[\sup_{t\ge N_{\mathrm O}/K}\left|N_{\mathrm O}\widehat\sigma_{t,N_{\mathrm O}}^2-\sigma^2(t/N_{\mathrm O})\right|\xrightarrow{p}0.
\]
\par
}

{
\emergencystretch=2em
For \(t\ge N_{\mathrm O}/K\), \(\sigma^2(t/\allowbreak N_{\mathrm O})\in[V^{\rm obs},V^{\rm obs}+KV^{{\rm exp},\star}]\). By this convergence and Assumption 7.1(2), the event on which \(N_{\mathrm O}\omega_{N_{\mathrm O}}\in[c_\omega/2,2c_\omega]\) and \(N_{\mathrm O}\widehat\sigma_{t,N_{\mathrm O}}^2\in[V^{\rm obs}/\allowbreak 2,V^{\rm obs}+KV^{{\rm exp},\star}+1]\) for every \(t\ge N_{\mathrm O}/K\) has probability tending to one. On this event, \(\sqrt{N_{\mathrm O}}\,\overline B_{t,N_{\mathrm O}}=\allowbreak (N_{\mathrm O}\widehat\sigma_{t,N_{\mathrm O}}^2)^{1/2}h_\alpha(N_{\mathrm O}\omega_{N_{\mathrm O}}/\allowbreak (N_{\mathrm O}\widehat\sigma_{t,N_{\mathrm O}}^2))\) and \(\sigma(t/N_{\mathrm O})h_\alpha(c_\omega/\sigma^2(t/N_{\mathrm O}))\) are the values of the map \((c,v)\mapsto v^{1/2}h_\alpha(c/v)\) at two points of the compact set \([c_\omega/\allowbreak 2,2c_\omega]\times[V^{\rm obs}/\allowbreak 2,V^{\rm obs}+KV^{{\rm exp},\star}+1]\). By Lemma A.7.4(b), this map is uniformly continuous on the compact set. On this event, the difference between its two values is therefore at most its modulus of continuity evaluated at \(|N_{\mathrm O}\omega_{N_{\mathrm O}}-c_\omega|+\allowbreak\sup_{t\ge N_{\mathrm O}/K}|N_{\mathrm O}\widehat\sigma_{t,N_{\mathrm O}}^2-\sigma^2(t/N_{\mathrm O})|\), which converges to zero in probability by Assumption 7.1(2) and the preceding display. Because the event has probability tending to one, it follows that \[\sup_{t\ge N_{\mathrm O}/K}\left|\sqrt{N_{\mathrm O}}\,\overline B_{t,N_{\mathrm O}}-\sigma(t/N_{\mathrm O})\,h_\alpha\!\left(\frac{c_\omega}{\sigma^2(t/N_{\mathrm O})}\right)\right|\xrightarrow{p}0,
\]
\par
}

{
\emergencystretch=2em
and the probability in part (d), which is at most the probability of the complement of the event plus the probability that this supremum exceeds \(\varepsilon\), tends to zero. This proves part (d) and completes the proof of Lemma A.7.3.
\par
}

\FloatBarrier

\section{Additional simulation results}\label{appendix-b-additional-simulation-results}

Appendix B reports additional results of the simulations in Section 8. Appendix B.1 reports the complete fixed-time results, Appendix B.2 the results of the 95\% time-uniform interval and the sensitivity of the 90\% interval to the target round, and Appendix B.3 the sensitivity to the planning value.

\subsection{Fixed-time results}\label{b.1-fixed-time-results}

Table B.1 reports the fixed-time performance of the long-term average treatment effect estimator for the eight experimental designs at participants 1,000, 2,500, and 5,000, together with the reduction of the MC variance relative to the baseline design at the same round. Within each replication, all experimental designs share the observational sample and the random numbers that generate the experiment.

{
\small
\begin{longtable}[]{@{}
  >{\raggedright\arraybackslash}p{(\linewidth - 14\tabcolsep) * \real{0.2500}}
  >{\raggedleft\arraybackslash}p{(\linewidth - 14\tabcolsep) * \real{0.0900}}
  >{\raggedleft\arraybackslash}p{(\linewidth - 14\tabcolsep) * \real{0.1000}}
  >{\raggedleft\arraybackslash}p{(\linewidth - 14\tabcolsep) * \real{0.0900}}
  >{\raggedleft\arraybackslash}p{(\linewidth - 14\tabcolsep) * \real{0.0900}}
  >{\raggedleft\arraybackslash}p{(\linewidth - 14\tabcolsep) * \real{0.1000}}
  >{\raggedleft\arraybackslash}p{(\linewidth - 14\tabcolsep) * \real{0.1200}}
  >{\raggedleft\arraybackslash}p{(\linewidth - 14\tabcolsep) * \real{0.1600}}@{}}
\toprule\noalign{}
\begin{minipage}[b]{\linewidth}\raggedright
Experimental design
\end{minipage} & \begin{minipage}[b]{\linewidth}\raggedleft
Round
\end{minipage} & \begin{minipage}[b]{\linewidth}\raggedleft
Bias
\end{minipage} & \begin{minipage}[b]{\linewidth}\raggedleft
MC SD
\end{minipage} & \begin{minipage}[b]{\linewidth}\raggedleft
RMSE
\end{minipage} & \begin{minipage}[b]{\linewidth}\raggedleft
Mean SE
\end{minipage} & \begin{minipage}[b]{\linewidth}\raggedleft
Coverage
\end{minipage} & \begin{minipage}[b]{\linewidth}\raggedleft
MC var. reduction
\end{minipage} \\
\midrule\noalign{}
\endfirsthead
\toprule\noalign{}
\begin{minipage}[b]{\linewidth}\raggedright
Experimental design
\end{minipage} & \begin{minipage}[b]{\linewidth}\raggedleft
Round
\end{minipage} & \begin{minipage}[b]{\linewidth}\raggedleft
Bias
\end{minipage} & \begin{minipage}[b]{\linewidth}\raggedleft
MC SD
\end{minipage} & \begin{minipage}[b]{\linewidth}\raggedleft
RMSE
\end{minipage} & \begin{minipage}[b]{\linewidth}\raggedleft
Mean SE
\end{minipage} & \begin{minipage}[b]{\linewidth}\raggedleft
Coverage
\end{minipage} & \begin{minipage}[b]{\linewidth}\raggedleft
MC var. reduction
\end{minipage} \\
\midrule\noalign{}
\endhead
\bottomrule\noalign{}
\noalign{\vskip\abovecaptionskip}
\caption{Long-term average treatment effect estimates for the eight experimental designs at three rounds. Bias is measured relative to the true value 1. MC SD is the sample standard deviation of the 2,000 estimates, Mean SE is the average estimated standard error, and Coverage is the coverage rate of the ordinary fixed-time 95\% interval. MC var. reduction is the percentage reduction in the MC variance, the square of MC SD, relative to the baseline design at the same round, so a positive value indicates a reduction.}\tabularnewline
\endlastfoot
Baseline design & 1,000 & -0.0073 & 0.5238 & 0.5237 & 0.5354 & 95.3\% & --- \\
Baseline design & 2,500 & -0.0078 & 0.3307 & 0.3307 & 0.3411 & 95.5\% & --- \\
Baseline design & 5,000 & -0.0037 & 0.2313 & 0.2313 & 0.2441 & 96.3\% & --- \\
\(S^{(1)}\) joint design & 1,000 & -0.0064 & 0.5394 & 0.5393 & 0.5417 & 95.3\% & -6.0\% \\
\(S^{(1)}\) joint design & 2,500 & -0.0076 & 0.3353 & 0.3353 & 0.3464 & 95.0\% & -2.8\% \\
\(S^{(1)}\) joint design & 5,000 & -0.0047 & 0.2423 & 0.2423 & 0.2466 & 95.4\% & -9.8\% \\
\(S^{(2)}\) joint design & 1,000 & -0.0032 & 0.5072 & 0.5070 & 0.5090 & 95.0\% & 6.3\% \\
\(S^{(2)}\) joint design & 2,500 & -0.0051 & 0.3129 & 0.3128 & 0.3115 & 94.7\% & 10.5\% \\
\(S^{(2)}\) joint design & 5,000 & -0.0051 & 0.2161 & 0.2161 & 0.2189 & 95.5\% & 12.7\% \\
Surrogate-average joint design & 1,000 & -0.0073 & 0.5080 & 0.5079 & 0.5138 & 95.0\% & 5.9\% \\
Surrogate-average joint design & 2,500 & -0.0083 & 0.3113 & 0.3113 & 0.3166 & 95.0\% & 11.4\% \\
Surrogate-average joint design & 5,000 & -0.0049 & 0.2180 & 0.2180 & 0.2228 & 95.1\% & 11.2\% \\
SI assignment-only design & 1,000 & -0.0034 & 0.5199 & 0.5197 & 0.5343 & 95.2\% & 1.5\% \\
SI assignment-only design & 2,500 & -0.0056 & 0.3292 & 0.3292 & 0.3396 & 96.1\% & 0.9\% \\
SI assignment-only design & 5,000 & -0.0053 & 0.2326 & 0.2326 & 0.2429 & 96.4\% & -1.1\% \\
SI sampling-only design & 1,000 & -0.0077 & 0.5052 & 0.5051 & 0.5104 & 95.0\% & 7.0\% \\
SI sampling-only design & 2,500 & -0.0085 & 0.3102 & 0.3103 & 0.3134 & 94.7\% & 12.0\% \\
SI sampling-only design & 5,000 & -0.0036 & 0.2155 & 0.2155 & 0.2205 & 95.1\% & 13.2\% \\
SI joint design & 1,000 & -0.0039 & 0.5048 & 0.5047 & 0.5087 & 95.2\% & 7.1\% \\
SI joint design & 2,500 & -0.0061 & 0.3100 & 0.3100 & 0.3111 & 94.4\% & 12.1\% \\
SI joint design & 5,000 & -0.0055 & 0.2157 & 0.2157 & 0.2185 & 95.8\% & 13.1\% \\
Stratified-oracle SI joint design & 1,000 & -0.0012 & 0.5010 & 0.5009 & 0.5063 & 95.1\% & 8.5\% \\
Stratified-oracle SI joint design & 2,500 & -0.0034 & 0.3089 & 0.3089 & 0.3096 & 95.1\% & 12.7\% \\
Stratified-oracle SI joint design & 5,000 & -0.0041 & 0.2156 & 0.2156 & 0.2178 & 95.5\% & 13.1\% \\
\end{longtable}
}

\subsection{Sensitivity to the target round}\label{b.2-sensitivity-to-the-target-round}

Table B.2 reports the results for the 95\% time-uniform interval. As with the 90\% interval in Table 8.3, simultaneous coverage exceeds the nominal level for every experimental design.

\begin{table}[tbp]
\centering
\small
\begin{tabular}{@{}
  >{\raggedright\arraybackslash}p{(\linewidth - 10\tabcolsep) * \real{0.1900}}
  >{\raggedleft\arraybackslash}p{(\linewidth - 10\tabcolsep) * \real{0.1400}}
  >{\raggedleft\arraybackslash}p{(\linewidth - 10\tabcolsep) * \real{0.1400}}
  >{\raggedleft\arraybackslash}p{(\linewidth - 10\tabcolsep) * \real{0.1800}}
  >{\raggedleft\arraybackslash}p{(\linewidth - 10\tabcolsep) * \real{0.1400}}
  >{\raggedleft\arraybackslash}p{(\linewidth - 10\tabcolsep) * \real{0.2100}}@{}}
\toprule\noalign{}
\begin{minipage}[b]{\linewidth}\raggedright
Experimental design
\end{minipage} & \begin{minipage}[b]{\linewidth}\raggedleft
Simultaneous coverage
\end{minipage} & \begin{minipage}[b]{\linewidth}\raggedleft
Detection
\end{minipage} & \begin{minipage}[b]{\linewidth}\raggedleft
Mean stop
\end{minipage} & \begin{minipage}[b]{\linewidth}\raggedleft
Median stop
\end{minipage} & \begin{minipage}[b]{\linewidth}\raggedleft
Stop IQR
\end{minipage} \\
\midrule\noalign{}

Baseline design & 98.3\% & 92.0\% & 2,427.7 & 2,139.5 & 1,280.8--3,388.0 \\
\(S^{(1)}\) joint design & 97.8\% & 90.9\% & 2,478.1 & 2,216.0 & 1,282.8--3,508.0 \\
\(S^{(2)}\) joint design & 97.8\% & 96.0\% & 2,174.0 & 1,902.0 & 1,189.8--2,958.0 \\
Surrogate-average joint design & 98.1\% & 96.0\% & 2,202.7 & 1,909.0 & 1,197.8--2,952.2 \\
SI assignment-only design & 98.1\% & 91.7\% & 2,424.1 & 2,134.5 & 1,254.8--3,386.8 \\
SI sampling-only design & 98.0\% & 96.4\% & 2,170.1 & 1,879.5 & 1,186.0--2,935.2 \\
SI joint design & 97.9\% & 96.3\% & 2,167.4 & 1,906.0 & 1,181.5--2,933.2 \\
Stratified-oracle SI joint design & 98.0\% & 96.2\% & 2,135.1 & 1,869.0 & 1,178.5--2,895.8 \\
\bottomrule
\end{tabular}
\caption{Simultaneous coverage and detection for the 95\% time-uniform interval. Simultaneous coverage requires every 95\% time-uniform interval from participant 595 through participant 5,000 to contain the true value 1. Detection includes crossings at the horizon. Mean stop and Median stop are stopping sample sizes, with replications without detection recorded as 5,000. Stop IQR reports the 25th and 75th percentiles of those stopping sample sizes.}
\end{table}

Table B.3 reports the 90\% time-uniform interval with the target rounds 1,250, 3,750, and 5,000, in addition to the target round 2,500 of Table 8.3. For these target rounds, the effective target round of the SI joint design, which is the median over replications of the round at which its inflation factor is smallest, equals 1,150, 3,156, and 4,128, respectively.

{
\small
\begin{longtable}[]{@{}
  >{\raggedright\arraybackslash}p{(\linewidth - 12\tabcolsep) * \real{0.1000}}
  >{\raggedright\arraybackslash}p{(\linewidth - 12\tabcolsep) * \real{0.2500}}
  >{\raggedleft\arraybackslash}p{(\linewidth - 12\tabcolsep) * \real{0.1300}}
  >{\raggedleft\arraybackslash}p{(\linewidth - 12\tabcolsep) * \real{0.1100}}
  >{\raggedleft\arraybackslash}p{(\linewidth - 12\tabcolsep) * \real{0.1100}}
  >{\raggedleft\arraybackslash}p{(\linewidth - 12\tabcolsep) * \real{0.1100}}
  >{\raggedleft\arraybackslash}p{(\linewidth - 12\tabcolsep) * \real{0.1900}}@{}}
\toprule\noalign{}
\begin{minipage}[b]{\linewidth}\raggedright
Target round
\end{minipage} & \begin{minipage}[b]{\linewidth}\raggedright
Experimental design
\end{minipage} & \begin{minipage}[b]{\linewidth}\raggedleft
Simultaneous coverage
\end{minipage} & \begin{minipage}[b]{\linewidth}\raggedleft
Detection
\end{minipage} & \begin{minipage}[b]{\linewidth}\raggedleft
Mean stop
\end{minipage} & \begin{minipage}[b]{\linewidth}\raggedleft
Median stop
\end{minipage} & \begin{minipage}[b]{\linewidth}\raggedleft
Stop IQR
\end{minipage} \\
\midrule\noalign{}
\endfirsthead
\toprule\noalign{}
\begin{minipage}[b]{\linewidth}\raggedright
Target round
\end{minipage} & \begin{minipage}[b]{\linewidth}\raggedright
Experimental design
\end{minipage} & \begin{minipage}[b]{\linewidth}\raggedleft
Simultaneous coverage
\end{minipage} & \begin{minipage}[b]{\linewidth}\raggedleft
Detection
\end{minipage} & \begin{minipage}[b]{\linewidth}\raggedleft
Mean stop
\end{minipage} & \begin{minipage}[b]{\linewidth}\raggedleft
Median stop
\end{minipage} & \begin{minipage}[b]{\linewidth}\raggedleft
Stop IQR
\end{minipage} \\
\midrule\noalign{}
\endhead
\bottomrule\noalign{}
\noalign{\vskip\abovecaptionskip}
\caption{Sensitivity of the 90\% time-uniform interval to the target round. The intervals with each target round are recomputed on the same saved paths. In every setting, the planning value is the experimental variance estimate of the monitored experiment at the monitoring start. Every rate has denominator 2,000, and replications without a detection enter the stopping-size summaries at 5,000.}\tabularnewline
\endlastfoot
1,250 & Baseline design & 95.4\% & 94.9\% & 2,032.7 & 1,664.0 & 949.8--2,819.2 \\
1,250 & \(S^{(1)}\) joint design & 95.3\% & 94.1\% & 2,059.6 & 1,722.0 & 929.0--2,899.2 \\
1,250 & \(S^{(2)}\) joint design & 95.0\% & 97.4\% & 1,840.0 & 1,519.0 & 905.0--2,527.2 \\
1,250 & Surrogate-average joint design & 95.5\% & 97.2\% & 1,893.9 & 1,585.0 & 919.0--2,612.0 \\
1,250 & SI assignment-only design & 96.2\% & 95.4\% & 2,045.4 & 1,686.0 & 923.0--2,875.2 \\
1,250 & SI sampling-only design & 95.7\% & 97.3\% & 1,855.7 & 1,561.0 & 924.8--2,523.5 \\
1,250 & SI joint design & 95.4\% & 98.0\% & 1,828.6 & 1,516.0 & 904.0--2,479.0 \\
1,250 & Stratified-oracle SI joint design & 95.3\% & 97.5\% & 1,809.2 & 1,494.0 & 902.8--2,436.0 \\
3,750 & Baseline design & 96.5\% & 95.4\% & 2,137.3 & 1,826.0 & 1,137.5--2,863.2 \\
3,750 & \(S^{(1)}\) joint design & 96.0\% & 94.7\% & 2,154.1 & 1,859.0 & 1,131.0--2,907.2 \\
3,750 & \(S^{(2)}\) joint design & 95.6\% & 97.9\% & 1,910.9 & 1,638.5 & 1,063.0--2,536.5 \\
3,750 & Surrogate-average joint design & 96.1\% & 97.7\% & 1,952.2 & 1,693.0 & 1,081.8--2,609.2 \\
3,750 & SI assignment-only design & 96.7\% & 95.8\% & 2,133.4 & 1,823.0 & 1,126.2--2,921.8 \\
3,750 & SI sampling-only design & 96.1\% & 97.7\% & 1,921.9 & 1,659.5 & 1,065.8--2,551.2 \\
3,750 & SI joint design & 96.0\% & 98.1\% & 1,893.1 & 1,636.0 & 1,055.5--2,495.0 \\
3,750 & Stratified-oracle SI joint design & 95.7\% & 98.2\% & 1,868.9 & 1,606.0 & 1,052.0--2,443.2 \\
5,000 & Baseline design & 96.9\% & 95.4\% & 2,210.2 & 1,908.5 & 1,229.0--2,941.5 \\
5,000 & \(S^{(1)}\) joint design & 96.8\% & 94.6\% & 2,227.6 & 1,944.0 & 1,224.5--2,974.2 \\
5,000 & \(S^{(2)}\) joint design & 96.1\% & 98.0\% & 1,966.8 & 1,700.5 & 1,141.8--2,574.2 \\
5,000 & Surrogate-average joint design & 96.4\% & 97.7\% & 1,999.1 & 1,750.5 & 1,147.8--2,653.0 \\
5,000 & SI assignment-only design & 97.1\% & 95.6\% & 2,197.1 & 1,905.5 & 1,196.5--2,962.2 \\
5,000 & SI sampling-only design & 96.5\% & 97.7\% & 1,968.9 & 1,715.0 & 1,129.0--2,582.5 \\
5,000 & SI joint design & 96.5\% & 98.1\% & 1,951.4 & 1,694.5 & 1,121.0--2,563.0 \\
5,000 & Stratified-oracle SI joint design & 96.1\% & 98.2\% & 1,924.6 & 1,688.5 & 1,119.0--2,497.5 \\
\end{longtable}
}

\subsection{Sensitivity to the planning value}\label{b.3-sensitivity-to-the-planning-value}

{
\emergencystretch=2em
This subsection recomputes the intervals on the 2,000 saved paths of every experimental design and varies only the planning value \(V\) in the tuning variable, keeping the monitoring window \(t=595,\ldots,5{,}000\), the target round \(t_{N_{\mathrm O}}^\star=2{,}500\), and the estimates and variance estimates of the saved paths. The planning values are the experimental variance estimates of the monitored experiment at round 200 and at the monitoring start, round 595, and the population value of \(V^{\rm exp}\) for each experimental design. Table B.4 reports, for each planning value, the range of simultaneous coverage over the eight experimental designs and the results of the SI joint design. Simultaneous coverage changes little with the planning value.
\par
}

\begin{table}[tbp]
\centering
\small
\begin{tabular}{@{}
  >{\raggedright\arraybackslash}p{(\linewidth - 12\tabcolsep) * \real{0.1200}}
  >{\raggedright\arraybackslash}p{(\linewidth - 12\tabcolsep) * \real{0.2800}}
  >{\raggedleft\arraybackslash}p{(\linewidth - 12\tabcolsep) * \real{0.1300}}
  >{\raggedleft\arraybackslash}p{(\linewidth - 12\tabcolsep) * \real{0.1500}}
  >{\raggedleft\arraybackslash}p{(\linewidth - 12\tabcolsep) * \real{0.1000}}
  >{\raggedleft\arraybackslash}p{(\linewidth - 12\tabcolsep) * \real{0.1100}}
  >{\raggedleft\arraybackslash}p{(\linewidth - 12\tabcolsep) * \real{0.1100}}@{}}
\toprule\noalign{}
\begin{minipage}[b]{\linewidth}\raggedright
Nominal level
\end{minipage} & \begin{minipage}[b]{\linewidth}\raggedright
Planning value
\end{minipage} & \begin{minipage}[b]{\linewidth}\raggedleft
Effective target
\end{minipage} & \begin{minipage}[b]{\linewidth}\raggedleft
Coverage range
\end{minipage} & \begin{minipage}[b]{\linewidth}\raggedleft
Coverage
\end{minipage} & \begin{minipage}[b]{\linewidth}\raggedleft
Detection
\end{minipage} & \begin{minipage}[b]{\linewidth}\raggedleft
Mean stop
\end{minipage} \\
\midrule\noalign{}

90\% & Round 200 & 2,161 & 95.3\%--96.4\% & 95.6\% & 98.2\% & 1,848.3 \\
90\% & Monitoring start (round 595) & 2,163 & 95.2\%--96.4\% & 95.6\% & 98.2\% & 1,846.1 \\
90\% & Population variance & 2,731 & 95.4\%--96.4\% & 95.8\% & 98.2\% & 1,870.0 \\
95\% & Round 200 & 2,161 & 97.8\%--98.3\% & 97.9\% & 96.3\% & 2,169.1 \\
95\% & Monitoring start (round 595) & 2,163 & 97.8\%--98.3\% & 97.9\% & 96.3\% & 2,167.4 \\
95\% & Population variance & 2,731 & 97.8\%--98.4\% & 98.1\% & 96.4\% & 2,180.0 \\
\bottomrule
\end{tabular}
\caption{Sensitivity of the time-uniform interval to the planning value. All rates use 2,000 replications, and coverage is simultaneous over the entire monitoring window. Coverage range gives the smallest and largest simultaneous coverage among the eight experimental designs, and the other result columns refer to the SI joint design. Every planning value aims at the target round 2,500. Effective target is the median over replications of the round in the monitoring window at which the inflation factor of the SI joint design is smallest, so its difference from 2,500 shows how far the planning value moves this round away from the target round. Detection means that the lower endpoint first exceeds zero by 5,000 participants.}
\end{table}

\FloatBarrier

\section{Additional empirical results}\label{appendix-c-additional-empirical-results}

Appendix C reports the ranking of all 20 pretreatment variables and three sensitivity checks of the experimental designs in Section 9.

\subsection{Ranking of all 20 pretreatment variables}\label{c.1-ranking-of-all-20-pretreatment-variables}

Table C.1 reports, for each of the 20 pretreatment variables, the reduction of the plug-in criterion under the joint design relative to the baseline design. The variables are ranked by the mean over the splits in which the experimental design is computed (Step 3 of Section S1.2.2 of Supplementary Material 1), and the last column gives the rank with seven strata. Savings deposits and formal borrowing are nonzero for few households, so in some splits one arm of the experimental sample has no household in the stratum of nonzero values. In 29 of the 655 splits counted for formal borrowing, every observational household has zero formal borrowing, and the covariate forms a single stratum. The household asset index, the asset index, and monthly nonfood consumption per capita rank first, second, and third with both numbers of strata.

{
\small
\begin{longtable}[]{@{}
  >{\raggedleft\arraybackslash}p{(\linewidth - 14\tabcolsep) * \real{0.0700}}
  >{\raggedright\arraybackslash}p{(\linewidth - 14\tabcolsep) * \real{0.3400}}
  >{\raggedleft\arraybackslash}p{(\linewidth - 14\tabcolsep) * \real{0.0800}}
  >{\raggedleft\arraybackslash}p{(\linewidth - 14\tabcolsep) * \real{0.1000}}
  >{\raggedleft\arraybackslash}p{(\linewidth - 14\tabcolsep) * \real{0.1000}}
  >{\raggedleft\arraybackslash}p{(\linewidth - 14\tabcolsep) * \real{0.0800}}
  >{\raggedleft\arraybackslash}p{(\linewidth - 14\tabcolsep) * \real{0.1000}}
  >{\raggedleft\arraybackslash}p{(\linewidth - 14\tabcolsep) * \real{0.1300}}@{}}
\toprule\noalign{}
\begin{minipage}[b]{\linewidth}\raggedleft
Rank
\end{minipage} & \begin{minipage}[b]{\linewidth}\raggedright
Variable
\end{minipage} & \begin{minipage}[b]{\linewidth}\raggedleft
Splits
\end{minipage} & \begin{minipage}[b]{\linewidth}\raggedleft
Mean reduction
\end{minipage} & \begin{minipage}[b]{\linewidth}\raggedleft
25th percentile
\end{minipage} & \begin{minipage}[b]{\linewidth}\raggedleft
Median
\end{minipage} & \begin{minipage}[b]{\linewidth}\raggedleft
75th percentile
\end{minipage} & \begin{minipage}[b]{\linewidth}\raggedleft
Rank with seven strata
\end{minipage} \\
\midrule\noalign{}
\endfirsthead
\toprule\noalign{}
\begin{minipage}[b]{\linewidth}\raggedleft
Rank
\end{minipage} & \begin{minipage}[b]{\linewidth}\raggedright
Variable
\end{minipage} & \begin{minipage}[b]{\linewidth}\raggedleft
Splits
\end{minipage} & \begin{minipage}[b]{\linewidth}\raggedleft
Mean reduction
\end{minipage} & \begin{minipage}[b]{\linewidth}\raggedleft
25th percentile
\end{minipage} & \begin{minipage}[b]{\linewidth}\raggedleft
Median
\end{minipage} & \begin{minipage}[b]{\linewidth}\raggedleft
75th percentile
\end{minipage} & \begin{minipage}[b]{\linewidth}\raggedleft
Rank with seven strata
\end{minipage} \\
\midrule\noalign{}
\endhead
\bottomrule\noalign{}
\noalign{\vskip\abovecaptionskip}
\caption{Reductions of the plug-in criterion under the joint design for all 20 pretreatment variables. Splits is the number of the 1,000 splits in which the experimental design is computed, and the mean and quartiles are taken over these splits.}\tabularnewline
\endlastfoot
1 & \texttt{asset\_hh\_index\_bsl} & 1,000 & 8.23\% & 3.21\% & 4.91\% & 9.16\% & 1 \\
2 & \texttt{asset\_index\_bsl} & 1,000 & 7.97\% & 2.82\% & 4.26\% & 8.84\% & 2 \\
3 & \texttt{cnonfood\_pcmonth\_bsl} & 1,000 & 7.72\% & 3.32\% & 4.71\% & 8.33\% & 3 \\
4 & \texttt{ctotal\_pcmonth\_bsl} & 1,000 & 7.15\% & 3.21\% & 4.60\% & 7.43\% & 4 \\
5 & \texttt{index\_foodsecurity\_bsl} & 990 & 6.59\% & 2.46\% & 3.70\% & 6.79\% & 7 \\
6 & \texttt{percep\_econ\_bsl} & 1,000 & 6.49\% & 2.66\% & 3.97\% & 6.73\% & 6 \\
7 & \texttt{cfood\_pcmonth\_bsl} & 1,000 & 6.48\% & 2.95\% & 4.27\% & 6.46\% & 5 \\
8 & \texttt{cdurable\_pcmonth\_bsl} & 1,000 & 5.84\% & 2.99\% & 4.35\% & 6.54\% & 9 \\
9 & \texttt{asset\_prod\_index\_bsl} & 995 & 5.75\% & 1.82\% & 2.97\% & 6.29\% & 8 \\
10 & \texttt{loan\_informalamt\_bsl} & 991 & 4.92\% & 2.23\% & 3.38\% & 5.32\% & 11 \\
11 & \texttt{loan\_totalamt\_bsl} & 1,000 & 4.75\% & 1.94\% & 2.90\% & 5.09\% & 10 \\
12 & \texttt{fs\_enoughfood\_bsl} & 1,000 & 3.87\% & 0.81\% & 1.64\% & 4.39\% & 12 \\
13 & \texttt{fs\_adultskip\_bsl} & 1,000 & 3.15\% & 0.65\% & 1.44\% & 3.57\% & 13 \\
14 & \texttt{fs\_childskip\_bsl} & 1,000 & 3.09\% & 1.17\% & 2.04\% & 3.84\% & 14 \\
15 & \texttt{fs\_wholeday\_bsl} & 1,000 & 2.95\% & 0.73\% & 1.51\% & 3.34\% & 15 \\
16 & \texttt{fs\_twomeals\_bsl} & 1,000 & 1.90\% & 0.67\% & 1.30\% & 2.49\% & 16 \\
17 & \texttt{sav\_depositamt\_bsl} & 745 & 1.37\% & 0.61\% & 1.08\% & 1.83\% & 17 \\
18 & \texttt{iagri\_month\_bsl} & 997 & 1.30\% & 0.51\% & 0.96\% & 1.66\% & 18 \\
19 & \texttt{ranimals\_month\_bsl} & 1,000 & 1.22\% & 0.61\% & 1.03\% & 1.60\% & 19 \\
20 & \texttt{loan\_formalamt\_bsl} & 655 & 0.89\% & 0.36\% & 0.64\% & 1.21\% & 20 \\
\end{longtable}
}

\subsection{Sensitivity analysis}\label{c.2-sensitivity-analysis}

{
\emergencystretch=2em
This subsection reports three checks of the experimental designs in Section 9.3. First, each stratum contains few experimental households, so a single household can dominate the estimated variance of a stratum. For each stratum, we set aside the splits in which one household accounts for more than 60\% of the squared residuals of either group in that stratum. The two differences from the baseline design reported in Section 9.3 remain. The joint design samples the top stratum of the household asset index more heavily, with a median sampling density ratio of 1.20 that exceeds one in 98\% of the remaining splits, and it assigns treatment in the second stratum of monthly nonfood consumption per capita with a median probability of 0.60, which exceeds \(1/2\) in 97\% of the remaining splits.
\par
}

Second, Figure C.1 repeats Figure 9.1 with seven strata, and the joint design again differs from the baseline design in the same places. It samples the top stratum of the household asset index more heavily, with a median sampling density ratio of 1.28 that exceeds one in 97\% of the splits, and its largest median assignment probability for monthly nonfood consumption per capita, 0.58, is in the third stratum, which covers part of the second stratum with five strata. Third, with seven strata the sampling-only design gives a larger mean reduction of the plug-in criterion than the assignment-only design for both covariates, 6.43\% against 4.84\% for the household asset index and 5.68\% against 3.48\% for monthly nonfood consumption per capita, so the ordering of these two designs for monthly nonfood consumption per capita in Table 9.2 depends on the number of strata.

\begin{figure}[tbp]
\centering
\includegraphics[width=\linewidth]{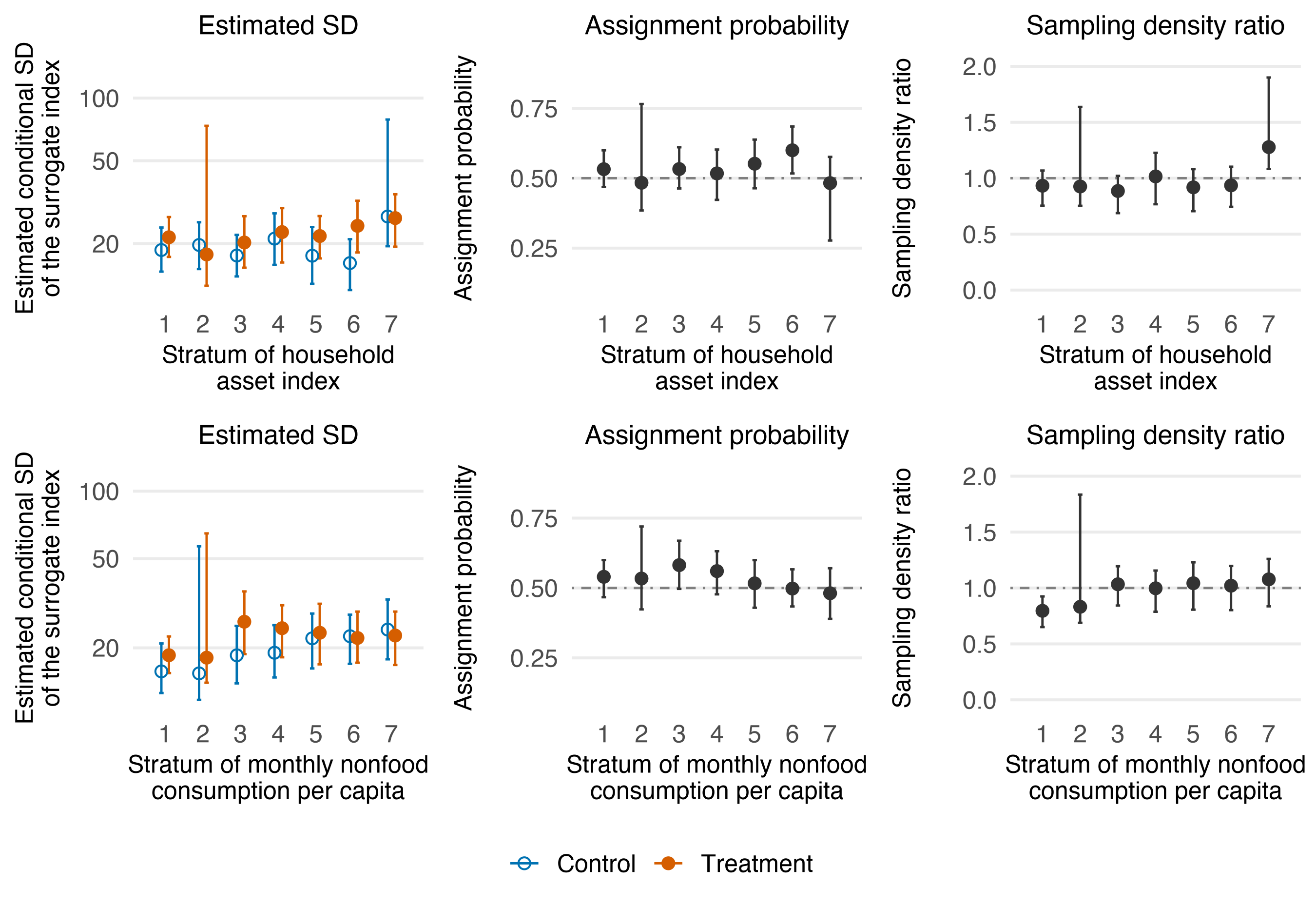}
\caption{Estimated conditional standard deviations, assignment probabilities, and sampling density ratios of the joint design with seven strata. The panels correspond to those of Figure 9.1, with the strata given by the seven intervals of equal probability of each covariate in the observational sample.}
\label{fig:C.1}
\end{figure}

\FloatBarrier

\section{Supplementary material}\label{appendix-d-supplementary-material}

This appendix describes the two supplementary materials that accompany this paper, which will be made available at \url{https://kkawato.github.io/}.

Supplementary Material 1 (Implementation Note) describes how the numerical results of Sections 8 and 9 are computed. For the simulations in Section 8, it specifies the estimators, experimental designs, and scores used in each step of Algorithm~\ref{alg:adaptive-sequential-estimation}. For the illustration in Section 9, it lists the variables of the Graduation data and describes how the experimental designs are estimated in each split of the data. It ends with R code for the simulations and the illustration.

Supplementary Material 2 (Additional Theoretical Results) relates the efficiency bound of Theorem 3.1 to the pooled-sample bound of \citet{Chen2023}. It also gives sufficient conditions, under correctly specified finite-dimensional models, for Assumptions 6.2, 6.4, 6.6, 7.3, and 7.5.

\stopcontents[appendix]

\clearpage
\begin{thebibliography}{99}
\bibitem[Anderer et al.(2022)]{Anderer2022}
Anderer, A., Bastani, H., and Silberholz, J. (2022). Adaptive Clinical Trial Designs with Surrogates: When Should We Bother? \emph{Management Science}, 68(3), 1982--2002.

\bibitem[Anderson and Fox(2019)]{Anderson2019}
Anderson, E. T., and Fox, E. J. (2019). How Price Promotions Work: A Review of Practice and Theory. In Jean-Pierre Dubé and Peter E. Rossi (eds.), \emph{Handbook of the Economics of Marketing}, 1, 497--552. North-Holland.

\bibitem[Athey et al.(2025)]{Athey2025}
Athey, S., Chetty, R., and Imbens, G. W. (2025). The Experimental Selection Correction Estimator: Using Experiments to Remove Biases in Observational Estimates. NBER Working Paper No. 33817.

\bibitem[Athey et al.(2026)]{Athey2026}
Athey, S., Chetty, R., Imbens, G. W., and Kang, H. (2026). The Surrogate Index: Combining Short-Term Proxies to Estimate Long-Term Treatment Effects More Rapidly and Precisely. \emph{The Review of Economic Studies}, 93(4), 2284--2312.

\bibitem[Banerjee et al.(2015)]{Banerjee2015}
Banerjee, A., Duflo, E., Goldberg, N., Karlan, D., Osei, R., Parienté, W., Shapiro, J., Thuysbaert, B., and Udry, C. (2015). A multifaceted program causes lasting progress for the very poor: Evidence from six countries. \emph{Science}, 348(6236), 1260799.

\bibitem[Chen and Ritzwoller(2023)]{Chen2023}
Chen, J., and Ritzwoller, D. M. (2023). Semiparametric estimation of long-term treatment effects. \emph{Journal of Econometrics}, 237(2, Part A), 105545.

\bibitem[Chernozhukov et al.(2018)]{Chernozhukov2018}
Chernozhukov, V., Chetverikov, D., Demirer, M., Duflo, E., Hansen, C., Newey, W., and Robins, J. (2018). Double/Debiased Machine Learning for Treatment and Structural Parameters. \emph{The Econometrics Journal}, 21(1), C1--C68.

\bibitem[Cook et al.(2024)]{Cook2024}
Cook, T., Mishler, A., and Ramdas, A. (2024). Semiparametric Efficient Inference in Adaptive Experiments. In \emph{Proceedings of the Third Conference on Causal Learning and Reasoning}, Proceedings of Machine Learning Research, 236, 1033--1064.

\bibitem[Dalal et al.(2024)]{Dalal2024}
Dalal, A., Blöbaum, P., Kasiviswanathan, S., and Ramdas, A. (2024). Anytime-Valid Inference for Double/Debiased Machine Learning of Causal Parameters. arXiv:2408.09598v2.

\bibitem[Fan et al.(2025)]{Fan2025}
Fan, Y., Wang, Z., and Wei, W. (2025). SLOACI: Surrogate-Leveraged Online Adaptive Causal Inference. arXiv:2512.06872v1.

\bibitem[Frangakis and Rubin(2002)]{Frangakis2002}
Frangakis, C. E., and Rubin, D. B. (2002). Principal Stratification in Causal Inference. \emph{Biometrics}, 58(1), 21--29.

\bibitem[Ghassami et al.(2025)]{Ghassami2025}
Ghassami, A., Liu, C., Yang, A., Richardson, D., Shpitser, I., and Tchetgen Tchetgen, E. (2025). Combining Experimental and Observational Data for Identification and Estimation of Long-Term Causal Effects. arXiv:2201.10743v4. Revised September 2025; first circulated in 2022.

\bibitem[Gopalakrishnan and Park(2021)]{Gopalakrishnan2021}
Gopalakrishnan, A., and Park, Y.-H. (2021). The Impact of Coupons on the Visit-to-Purchase Funnel. \emph{Marketing Science}, 40(1), 48--61.

\bibitem[Hahn et al.(2011)]{Hahn2011}
Hahn, J., Hirano, K., and Karlan, D. (2011). Adaptive Experimental Design Using the Propensity Score. \emph{Journal of Business \& Economic Statistics}, 29(1), 96--108.

\bibitem[Hall and Heyde(1980)]{Hall1980}
Hall, P., and Heyde, C. C. (1980). \emph{Martingale Limit Theory and Its Application}. Academic Press.

\bibitem[Howard et al.(2020)]{Howard2020}
Howard, S. R., Ramdas, A., McAuliffe, J., and Sekhon, J. (2020). Time-uniform Chernoff bounds via nonnegative supermartingales. \emph{Probability Surveys}, 17, 257--317.

\bibitem[Howard et al.(2021)]{Howard2021}
Howard, S. R., Ramdas, A., McAuliffe, J., and Sekhon, J. (2021). Time-uniform, nonparametric, nonasymptotic confidence sequences. \emph{The Annals of Statistics}, 49(2), 1055--1080.

\bibitem[Huang and Ascarza(2024)]{Huang2024}
Huang, T.-W., and Ascarza, E. (2024). Doing More with Less: Overcoming Ineffective Long-Term Targeting Using Short-Term Signals. \emph{Marketing Science}, 43(4), 863--884.

\bibitem[Imbens et al.(2025)]{Imbens2025}
Imbens, G. W., Kallus, N., Mao, X., and Wang, Y. (2025). Long-term causal inference under persistent confounding via data combination. \emph{Journal of the Royal Statistical Society Series B: Statistical Methodology}, 87(2), 362--388.

\bibitem[Jacod and Shiryaev(2003)]{Jacod2003}
Jacod, J., and Shiryaev, A. N. (2003). \emph{Limit Theorems for Stochastic Processes}, 2nd ed. Grundlehren der mathematischen Wissenschaften 288. Springer.

\bibitem[Johari et al.(2022)]{Johari2022}
Johari, R., Koomen, P., Pekelis, L., and Walsh, D. (2022). Always Valid Inference: Continuous Monitoring of A/B Tests. \emph{Operations Research}, 70(3), 1806--1821.

\bibitem[Kallus and Mao(2025)]{Kallus2025}
Kallus, N., and Mao, X. (2025). On the role of surrogates in the efficient estimation of treatment effects with limited outcome data. \emph{Journal of the Royal Statistical Society Series B: Statistical Methodology}, 87(2), 480--509.

\bibitem[Kato et al.(2024)]{Kato2024}
Kato, M., Oga, A., Komatsubara, W., and Inokuchi, R. (2024). Active Adaptive Experimental Design for Treatment Effect Estimation with Covariate Choice. In \emph{Proceedings of the 41st International Conference on Machine Learning}, Proceedings of Machine Learning Research, 235, 23291--23323.

\bibitem[Kato et al.(2025)]{Kato2025}
Kato, M., Ishihara, T., Honda, J., and Narita, Y. (2025). Efficient Adaptive Experimental Design for Average Treatment Effect Estimation. arXiv:2002.05308v7. Revised February 2025; first circulated in 2020.

\bibitem[Lewis and Rao(2015)]{Lewis2015}
Lewis, R. A., and Rao, J. M. (2015). The Unfavorable Economics of Measuring the Returns to Advertising. \emph{The Quarterly Journal of Economics}, 130(4), 1941--1973.

\bibitem[Neyman(1934)]{Neyman1934}
Neyman, J. (1934). On the Two Different Aspects of the Representative Method: The Method of Stratified Sampling and the Method of Purposive Selection. \emph{Journal of the Royal Statistical Society}, 97(4), 558--606.

\bibitem[Park and Sasaki(2024)]{Park2024}
Park, Y., and Sasaki, Y. (2024). A Bracketing Relationship for Long-Term Policy Evaluation with Combined Experimental and Observational Data. arXiv:2401.12050v1.

\bibitem[Park and Sasaki(2026)]{Park2026}
Park, Y., and Sasaki, Y. (2026). The informativeness of combined experimental and observational data under dynamic selection. \emph{Journal of Econometrics}, 254(Part B), 106219.

\bibitem[Prentice(1989)]{Prentice1989}
Prentice, R. L. (1989). Surrogate endpoints in clinical trials: Definition and operational criteria. \emph{Statistics in Medicine}, 8(4), 431--440.

\bibitem[Ramdas and Wang(2025)]{Ramdas2025}
Ramdas, A., and Wang, R. (2025). Hypothesis Testing with E-values. \emph{Foundations and Trends in Statistics}, 1(1--2), 1--390.

\bibitem[Ramdas et al.(2023)]{Ramdas2023}
Ramdas, A., Grünwald, P., Vovk, V., and Shafer, G. (2023). Game-Theoretic Statistics and Safe Anytime-Valid Inference. \emph{Statistical Science}, 38(4), 576--601.

\bibitem[Sahni et al.(2017)]{Sahni2017}
Sahni, N. S., Zou, D., and Chintagunta, P. K. (2017). Do Targeted Discount Offers Serve as Advertising? Evidence from 70 Field Experiments. \emph{Management Science}, 63(8), 2688--2705.

\bibitem[Shi et al.(2024)]{Shi2024}
Shi, L., Wei, W., and Wang, J. (2024). Using Surrogates in Covariate-adjusted Response-adaptive Randomization Experiments with Delayed Outcomes. \emph{Advances in Neural Information Processing Systems}, 37, 71085--71121.

\bibitem[van der Vaart(1998)]{vanderVaart1998}
van der Vaart, Aad W. (1998). \emph{Asymptotic Statistics}. Cambridge Series in Statistical and Probabilistic Mathematics 3. Cambridge University Press.

\bibitem[Ville(1939)]{Ville1939}
Ville, J. (1939). \emph{Étude critique de la notion de collectif}. Paris: Gauthier-Villars.

\bibitem[Waudby-Smith et al.(2024)]{WaudbySmith2024}
Waudby-Smith, I., Arbour, D., Sinha, R., Kennedy, E. H., and Ramdas, A. (2024). Time-uniform central limit theory and asymptotic confidence sequences. \emph{The Annals of Statistics}, 52(6), 2613--2640.
\end{thebibliography}
\end{document}